\documentclass[11pt,a4paper]{article}
\pdfoutput=1
\usepackage{jcapmod}

\usepackage[utf8]{inputenc} 
\usepackage{amsfonts, amsmath}
\usepackage{booktabs}
\usepackage[english]{babel}
\usepackage{colortbl}
\usepackage{graphicx}
\usepackage{hyperref}
\usepackage{framed}
\usepackage{simplewick}
\usepackage{siunitx}
\usepackage{bm}
\usepackage{enumitem}
\usepackage{upgreek}
\usepackage{wasysym}
\usepackage[bottom]{footmisc}
\usepackage{epstopdf}

\graphicspath{{Figures/}} 
\RequirePackage{color}
\usepackage{colortbl}

\definecolor{rp}{cmyk}{0.2, 1, 0.6, 0}
\definecolor{green2}{cmyk}{0, 1, 0.5, 0}
\definecolor{lightgreen}{cmyk}{0.2, 0, 0.2, 0.2}
\definecolor{lightgray}{cmyk}{0.1,0.2,0,0.1}
\definecolor{lightgray2}{cmyk}{0.4,0.4,0,0.8}
\definecolor{black}{cmyk}{1.0,1.0,1.0,1.0}
\definecolor{NewRed}{RGB}{200,37,6}
\definecolor{NewOrange}{RGB}{222,106,16}
\definecolor{NewGreen}{RGB}{0,136,43}

\allowdisplaybreaks[1]

\makeatletter
\newlength{\apb@width}
\newcommand{\autoparbox}[2][c]{\settowidth{\apb@width}{#2}\parbox[#1]{\apb@width}{#2}}

\makeatother

\numberwithin{equation}{section}

\definecolor{olive}{rgb}{0.5, 0.5, 0.0}

\def\p{\partial}
\def\na{\nabla}

\def\dif{\textrm{d}}

\def\eps{\epsilon}

\def\si{\sigma}

\renewcommand{\vec}[1]{\boldsymbol{#1}}

\newcommand{\ben}{\begin{enumerate}}
\newcommand{\een}{\end{enumerate}}
\newcommand{\Tr}{\mbox{tr}}

\newcommand{\leqsim}{\,\mbox{{\scriptsize $\stackrel{<}{\sim}$}}\,}
\newcommand{\geqsim}{\,\mbox{{\scriptsize $\stackrel{>}{\sim}$}}\,}

\newcommand{\del}{\partial}

\def\be{\begin{equation}}
\def\ee{\end{equation}}
\newcommand{\beq}{\begin{eqnarray}}
\newcommand{\eeq}{\end{eqnarray}} 
\newcommand{\ba}{\begin{align}}
\newcommand{\ea}{\end{align}}

\begin{document}

\begin{titlepage}

\baselineskip=15.5pt \thispagestyle{empty}

\begin{center}
{\fontsize{15}{30}\selectfont University of Belgrade 
\\[7pt] 
Faculty of Physics} 
\end{center}

\bigskip

\vspace{1cm}
\begin{center}
{\fontsize{22}{22} \bf Parametric-resonance based phenomenology \\ of gravitating axion configurations}
\end{center}

\bigskip

\vspace{0.2cm}
\begin{center}
{\fontsize{15}{30}\selectfont  Mateja Bošković} 
\\[10pt]
{\fontsize{15}{30}\selectfont  \textit{Master’s thesis}} 
\end{center}

\mbox{}
\vfill

\begin{center}
{\fontsize{15}{30}Belgrade, 2019.}
\end{center}

\vspace{0.6cm}
\end{titlepage}

\begin{titlepage}

\baselineskip=15.5pt \thispagestyle{empty}

\begin{center}
{\fontsize{15}{30}\selectfont Univerzitet u Beogradu
\\[7pt] 
Fizički fakultet} 
\end{center}

\bigskip

\vspace{1cm}
\begin{center}
{\fontsize{19}{19} \bf Fenomenologija gravitirajućih konfiguracija aksiona \\ zasnovana na parametarskim rezonancama}
\end{center}

\bigskip

\vspace{0.2cm}
\begin{center}
{\fontsize{15}{30}\selectfont  Mateja Bošković} 
\\[10pt]
{\fontsize{15}{30}\selectfont  \textit{Master rad}} 
\end{center}

\mbox{}
\vfill

\begin{center}
Beograd, 2019.
\end{center}

\vspace{0.6cm}
\end{titlepage}


\pagenumbering{gobble}

\part*{Abstract}

One of the most compelling candidates for Dark Matter (DM) are light pseudo-scalar particles (axions), motivated by the strong CP problem and axiverse scenario in string theory. Depending on their mass and type of self-interaction, these particles can build self-gravitating configurations such as compact objects, DM clumps or even galactic DM halos. On the other hand, superradiant instabilities can produce long-living extended configurations (scalar clouds) gravitating around Black Holes (BHs). As these scalars are real and harmonic, their interaction with the other matter components can induce a parametric resonance that might lead to their observable signatures. First, we consider the orbital dynamics of test particles in these axion configurations, show when resonances can occur and discuss the secular evolution of the orbital elements. This scenario can lead to observable consequences for binary pulsars or S stars around the supermassive BH in our Galaxy. Secondly, we discuss electromagnetic (EM) field instabilities in homogeneous axion configurations as well as scalar clouds around Kerr BHs. These axion-photon resonances can quench superradiant instabilities, while producing an observable signature in the EM sector. We give an analytical estimate of the rate of these processes that have good agreement with the fully relativistic numerical simulations and discuss the impact of plasma in the vicinity of BHs on these instabilities.

\vskip 4pt
\noindent
Subjects: Classical Field Theory $\ast$ General Relativity $\ast$ Astroparticle Physics $\ast$ Celestial Mechanics

\newpage


\part*{Apstrakt}

Među najsnažnijim kandidatima za česticu tamne materije su laki pseudo-skalari (aksioni), motivisani jakim CP problemom i scenariom aksiverzuma u teoriji struna. U zavisnosti od mase ovih hipotetičkih čestica i tipa njihove samo-interakcije, one mogu graditi samo-gravitirajuće konfiguracije kao što su kompaktni objekti, grudve tamne materije kao i delovih tamnih haloa galaksija. Sa druge strane, superradijantne nestabilnosti mogu da dovedu do formiranja dugo-živećih konfiguracija oko crnih rupa - skalarnih oblaka. S obzirom da je aksionsko polje realno i harmonijsko, njegova interakcija sa vidljivom materijom može da indukuje parametarske rezonance, koje zauzvrat mogu pružiti posmatrački potpis aksiona. Prvo ćemo posmatrati orbitalnu dinamiku probnih čestica u ovakvim konfiguracijama, pokazati kada može doći do rezonanci i diskutovati sekularnu evoluciju njihovih orbitalnih elemenata. Ovakav scenario može dovesti do posmatračkih posledica kod dvojnih pulsara ili S zvezda oko supermacivne crne rupe u centru Galaksije. Zatim ćemo diskutovati nestabilnosti elektromagnetnog polja kod homogenih konfiguracija aksiona kao i skalarnih oblaka oko Kerovih crnih rupa. Ove rezonance između aksiona i fotona mogu da prekinu superradijantne nestabilnosti, proizvodeći signal u elektromagnetnom sektoru. Daćemo analitičke procene brzine ovih procesa, koje su u dobrom saglasju sa relativističkim numeričkim simulacijama. Takođe ćemo diskutovati uticaj plazme u blizini crnih rupa na ove nestabilnosti.

\vskip 4pt
\noindent
Oblasti: Klasična teorija polja $\ast$ Opšta teorija relativnosti $\ast$ Astročestična fizika $\ast$ Nebeska mehanika

\newpage


\vspace*{\fill}
\begin{flushright} 

\textit{We need a dream-world in order to discover \\ the features of the real world we think we inhabit.}  \\

Paul Feyerabend, \textit{Against Method}
\end{flushright} 
\vspace*{\fill}

\newpage


\part*{Declaration and Acknowledgements}

Most of the original parts of this thesis were done in the period July 2017 - November 2018, under the supervision and the coordination of Prof. Vitor Cardoso. Parts of this work have been done during the visit to Gravitation in Tehnico (GRIT), Instituto Superior Técnico, University of Lisbon in October 2017 and October 2018, with the support of GWverse COST Action CA16104 and GRIT. Most of the original results presented in this thesis were obtained in collaboration with Richard Brito, Vitor Cardoso, Francisco Duque, Miguel C. Ferreira, Taishi Ikeda, Filipe S. Miguel and Helvi Witek and were published in Ref. \cite{Boskovic:2018rub} and Ref. \cite{Boskovic:2018lkj}. In this sense ``we'' in this thesis is not a consequence of the passive voice in the academic writing. However, I have mostly focused exposition on the parts of these works that I have contributed significantly, when applicable, while referring to the original papers for further details on other aspects. These papers have been cited (as of 25.6.) 9 (5) and 11 (7) times, respectively (without self-citations).

\vskip 4pt
\noindent
Other material in this thesis provides deeper context and some further elaboration of the previously mentioned results. I would like to thank all of my collaborators for the collective effort on this interesting research and would also like to thank Ana-Marija Ćeranić, Luka Jevtović, Milica Stepanović, Nikola Savić and Vladan Đukić for the discussion on several topics related to this work, Aleksandra Arsovski for the help in preparing this thesis and Prof. Marija Dimitrijević-Ćirić for several comments and suggestions that improved the original draft of the thesis. Ana-Marija and Milica have reproduced several results from Part \ref{ch:orbital} as a part of their student project at the Department of Astronomy, Petnica Science Center as well as expanded some of the results.

\vskip 4pt
\noindent
I have presented some of the results described in this thesis previously:
\begin{itemize}
    \item Talk \textit{Parametric-resonance based phenomenology of gravitating axion configurations} at\footnote{With the support of GWverse COST Action CA16104 and Faculty of Physics, University of Belgrade.} the \textit{Athens 2019: Gravitational Waves, Black Holes and Fundamental Physics} conference (January 2019)
    \item Lecture \textit{Gravitation and fingerprints of new particles} as a part\footnote{Inivited by Prof. Marija Dimitrijević-Ćirić.} of \textit{Contemporary Physics seminar} course for 3rd year BSc students in Theoretical and Experimental Physics at the Faculty of Physics, University of Belgrade (April 2019)
    \item Talk \textit{Orbital dynamics under the influence of time-periodic perturbations} at the Mathematical Institute of Serbian Academy of Arts and Sciences in Belgrade as a part of Seminar in mechanics and Seminar in theory of relativity and cosmological models (May 2019). This talk was a part of the Annual  Award   of  the  Mathematical  Institute  of  the  Serbian  Academy  of  Sciences  and  Arts  in  the  field  of  mathematics  and mechanics  for  BSc  students selection process, that I have obtained.
\end{itemize}

\vskip 4pt
\noindent
Members of the commission for the MSc thesis were Prof. Vitor Cardoso (University of Lisbon), Prof. Marija Dimitrijević-Ćirić (University of Belgrade) and Prof. Voja Radovanović (University of Belgrade). Thesis has been successfully defended on 27.6. with the 10 (of 10) mark. This version of the thesis has several typos fixed.

\newpage


\tableofcontents

\newpage
\pagenumbering{arabic}
\setcounter{page}{1} 

\part{Introduction} \label{ch:introduction}
In this work we will present a theoretical discussion of two phenomenological avenues of axions/axion-like particles fingerprints. In this introductory part, we will set the stage by giving a brief motivation for axions in general and gravitating axion configuration in particular, as well as a brief overview on present constraints on axions. Note on terminology - as we will discuss in Section \ref{sec:strong_CP}, QCD axions are postulated in order to solve the strong CP problem. There are several models of these particles that are currently viable. In addition, there are a class of particles with similar properties (light, pseudo-scalars, pseudo-Nambu-Goldstone bosons) originating from string theory. These particles are usually labeled as axion-like particles (ALP) or ultra-light axions (ULA) and in the context of Dark Matter  (DM; Section \ref{sec:DM}) also ultra-light dark matter (ULDM) (although the last term can also refer to a broader class of light bosons). We will use these terms interchangeably except where further specification is needed, for instance in Section \ref{sec:strong_CP}. Some details on point-particle action and Kerr spacetime are left for Appendix \ref{AppPointPart} and Appendix \ref{AppKerr}, respectively.

In Part \ref{ch:structure} we will describe the structure of self-gravitating axion configurations (in the spherically symmetric approximation) and the configurations gravitating around Black Holes (BHs). We will also comment on astrophysical and cosmological channels of formation of such objects. Up until and including this Part the thesis reviews the previous results with the exception of the aspects of the discussion on the dynamics of self-gravitating axion configuration that first appeared in \cite{Boskovic:2018rub} (Sections  \ref{sec:ROscillatons} and \ref{sec:NOscillatons}) and the production of axions in strong magnetic fields around BHs from \cite{Boskovic:2018lkj} (Section \ref{sec:electroWaldPert}). We focus on non-self-interacting axions with a brief discussion of self-interacting self-gravitating configurations in Appendix \ref{AppGPP}.

Part \ref{ch:orbital} is concerned with the first example, i.e. the motion of particles and light in a time-periodic background. This problem has been examined in several papers at the fully relativistic and weak-field level but with focus on particular applications to DM physics and other areas of astrophysics. Here we will, for the first time, delve into a more general discussion, with the main result being the parametric resonance mechanism behind this dynamics. We also point to a potential application to the ULA phenomenology in the context of motion of stars around the supermassive BH (SMBH) at the center of our Galaxy. This part is mostly from \cite{Boskovic:2018rub}. A review of parametric resonances is given in Appendix \ref{AppParRes}.

Finally, in Part \ref{ch:ax-photon} we present some of the results of \cite{Boskovic:2018lkj}. There the coupling between axions and scalars  with the Maxwell sector has been investigated at the classical field level in the context of Minkowski, Reissner–Nordström and Kerr backgrounds and with particular focus on instabilities. Here and in Appendix \ref{app:sc-photon} we describe the quenching of superradiant clouds through axion- and scalar-photon parametric resonances, respectively.

To start, let us fix our global variables and conventions; we will adopt geometric units ($G=c=1$) throughout, and a ``mostly plus'' signature $(-+++)$. Greek indices $(\alpha, \beta...)$ denote spacetime components and run from $0$ to $3$. Latin indices $(i,j,...)$ label the spatial components. As we adopt the geometric units, we will often use the mass parameter $\mu= m/\hbar$ with geometric-units dimension of $L^{-1}$. Primes stand for radial derivatives while $\partial_t\equiv \partial/\partial t$ and dot for proper time derivatives.

\newpage

\section{Why axions?}


\subsection{Three paradigms of fundamental physics}

Our fundamental understanding of the physical Universe is at present described by three paradigms - General Theory of Relativity, Standard Model of Elementary Particles and $\Lambda$CDM Cosmological Model + Inflation. The first one describes the spacetime arena in which the matter content of the Universe resides. This matter content and its fundamental interactions are described by the second paradigm. The third one describes the evolution of the Universe. Although these three paradigms are hugely successful they have both internal and mutual inconsistencies. First, we will give a brief overview of these three paradigms and then we will describe several of their problems and outline how the existence of new light $(10^{-33} \leqsim \mu_{\text a}[\text{eV}] \leqsim 1 )$ pseudo-scalar elementary particle could help in solving some of them or point to the nature of the solution.


\subsubsection{General Theory of Relativity}

General Theory of Relativity\footnote{In this text we mostly refer to textbooks by Weinberg \cite{bookWeinberg:1972} and Zee \cite{Zee:2013dea} in general and by Poisson and Will \cite{poissonwillbook} for relativistic astrophysics applications.} (GR) is a classical field theory that describes the dynamics of the gravitational field (real symmetric rank-2 tensor) $g_{\mu \nu}$. However, GR is not only a theory of gravity, but also a theory of spacetime. The foundational principle of GR is the Equivalence Principle (EP) which states that the effects of gravity can locally disappear with a suitable choice of coordinates. Thus, one can interpret $g_{\mu \nu}$ as the metric tensor that encodes the geometric properties of the spacetime \cite{bookWeinberg:1972}.

The dynamics of spacetime follows from the Einstein-Hilbert action which respects the EP and gives the field equations which reduce to Newtonian gravity (Poisson equation) in the weak-field limit:
\be \label{eq:action_gravity}
S_{\text{EH}}=\frac{c^4}{16\pi G}\int  d^4x \sqrt{-g} \Big(g^{\mu \nu} R_{\mu \nu} - 2 \Lambda \Big).
\ee
Here,
\beq
R_{\mu \nu}=\partial_{\sigma} \Gamma^{\sigma}_{\mu \nu}+\Gamma^{\sigma}_{\kappa \sigma} \Gamma^{\sigma}_{\mu \nu} - (\partial_{\nu} \Gamma^{\sigma}_{\mu \sigma}+\Gamma^{\sigma}_{\kappa \nu} \Gamma^{\kappa}_{\mu \sigma})
\eeq
is the Ricci tensor, whose contraction gives Ricci scalar $R=g^{\mu \nu}R_{\mu \nu}$ and
\begin{equation} \label{eq:Christoffel}
\Gamma^{\mu}_{\beta\alpha} \equiv \frac{g^{\mu \nu}}{2}\left(\partial_{\beta}g_{\nu\alpha} + \partial_{\alpha}g_{\nu\beta} - \partial_{\nu}g_{\alpha\beta}\right)\,
\end{equation}
is a Christoffel symbol. We have restored the constants in the action in order to comment on their values in various unit systems. $G$ is the Newtonian gravitational constant whose measured value is $G = 6.67 \cdot 10^{-11} \rm{m}^3/(\rm{kg} \, \rm{s}^2) $. The estimate of what we now call $G$ was already done by Newton, while the modern measurements started with the work of Cavendish in 1798. On the other hand, the value of the cosmological constant $\Lambda$ was measured\footnote{The sign and the upper value of this constant where estimated by Weinberg from anthropic arguments about a decade earlier.} for the first time in 1998.

$S_{\text{g}}/\hbar$ is dimensionless so $\Lambda/(8\pi G)$ has units of energy density in natural units system $(\hbar \equiv c \equiv 1)$ and the value of $\Lambda/(8\pi G) \sim (10^{-3} \text{eV})^4$, while in Planck units $(\hbar \equiv c \equiv G \equiv 1)$ it has the value $\Lambda \simeq 10^{-120}$. With the quantum aspects of gravity in mind one usually defines the Planck mass $M^2_{\text{Pl}}=\hbar c/(8 \pi G)$. In Planck units $M_{\text{Pl}} \simeq 0.2$, while in natural units it has dimensions of energy $M_{\text{Pl}} = 2.4 \cdot 10^{18} \text{GeV}$.

Varying the action \eqref{eq:action_gravity} we obtain the Einstein field equations in the absence of matter
\be
R_{\mu \nu} - \frac{1}{2}g_{\mu \nu}R+\Lambda g_{\mu \nu}=0,
\ee
where $G_{\mu \nu} \equiv R_{\mu \nu} - \frac{1}{2}g_{\mu \nu}R$ is often denoted as Einstein's tensor. When coupled to the matter sector $S_{\text{m}}$ Einstein field equations become
\be \label{eq:Einstein}
R_{\mu \nu} - \frac{1}{2}g_{\mu \nu}R+\Lambda g_{\mu \nu}=\kappa T_{\mu \nu},
\ee
where the matter stress-energy tensor is
\be \label{eq:e_m_tensor}
T^{\mu \nu} \equiv \frac{2}{\sqrt{-g}} \frac{\delta S_{\text{m}}}{\delta g_{\mu \nu}}
\ee
and $\kappa=8\pi G/c^4$. One can transfer $\Lambda$ to the ``matter'' side of the equation and interpret $\Lambda$ as part of the matter sector.


\subsubsection{Standard Model of Elementary Particles} \label{sec:SM_SSB}

The standard model (SM) is a quantum field theory (QFT) obtained by the qunatization of several interacting fermionic and bosonic fields. Interactions in the SM are described by $U(1) \times SU(2) \times SU(3)$ gauge fields, where $U(1) \times SU(2)$ is the electroweak sector and $SU(3)$ is the strong sector. Gauge bosons, living in the adjoint representation of the gauge group, are force carriers while the fermions (quarks and leptons) live in various representations of the gauge groups.

More importantly, SM is not just a set of quantum fields, it has a dynamical explanation of various low-energy fermion masses and the fact that electrodynamics and the weak force behave differently at low energies. These explanations are based on the mechanism of spontaneous \footnote{Spontaneous symmetry breaking refers to a scenario when the ground state (vacuum) of the theory has a lower symmetry compared to the lagrangian. In contrast, in explicit symmetry breaking the symmetry-breaking term is explicitly introduced in the lagrangian. There is also anomalous symmetry breaking where quantum effects break the classical symmetry. The origin of the last type can be seen at the level of the path integral $\int \mathcal{D}\mathcal{F} \exp{(iS[\mathcal{F}]/\hbar)}$, where $\mathcal{F}$ is some field. If the symmetry is present at the classical level, the action will stay invariant. However, the integral measure may not and thus at the quantum level symmetry-breaking effect may become manifest. All three mechanisms will be mentioned in this Part.} symmetry breaking, which we briefly review. For further reference (Section \ref{sec:axion_PQ}) we focus on a (global) $U(1)$ symmetry of the complex scalar field lagrangian
\be
\mathcal{L}=-\partial_\mu \varphi^{\dagger} \partial^\mu \varphi-\mathcal{V}(\varphi) \,,
\ee
with
\be
\mathcal{V}(\varphi)=\lambda_\varphi \Big( \varphi^\dagger \varphi-\frac{f^2_{\rm a}}{2}  \Big)^2 \,,
\ee
where $\lambda_\varphi \,, f_{\rm a}$ are real positive constants. This potential has a continuum of minima given by $|\varphi_{\rm min}| \equiv \rho_0 = f_{\rm a}/\sqrt{2}$. We expand the field around these minima as
\beq \label{eq:SSB_pot}
\varphi=(\varrho_0+\varrho) \exp{\Big(i\frac{\Upphi}{f^2_{\rm a}} \Big)} \,.
\eeq
The lagrangian in terms of the new dynamical real fields $\{ \varrho \,, \Upphi \}$ has the form
\be \label{eq:SSB_lag}
\mathcal{L}=-\partial_\mu \varrho \partial^\mu \varrho - \frac{1}{f^2_{\rm a}}(\varrho_0+\varrho)^2 \partial_\mu \Upphi \partial^\mu \Upphi -\lambda_\varphi \big((\varrho_0+\varrho)^4- f^2_{\rm a}(\varrho_0+\varrho)^2 +\frac{1}{4}f^4_{\rm a} \big)\,.
\ee
Thus the $\Upphi$ field (Nambu-Goldstone boson) is massless and the field $\varrho$ has the mass $m_{\varrho}=2f^2\lambda_\varphi$. At the intuitive level, the $\Upphi$ field can freely cycle the potential minimum, while $\rho$ needs energy in order to ``climb the hill''. This result is a special case of the more general Goldstone theorem (e.g. \cite{Zee:2003mt}) that states that for every broken (global) symmetry generator there is an associated massless scalar or pseudo-scalar. The SM uses a related idea (Higgs mechanism) where the gauge ``symmetry'' is being broken. Then, the gauge fields becomes massive, conserving the number of physical degrees of freedom.

\subsubsection{$\Lambda$CDM cosmological model and inflation}

Observations of the Cosmic Microwave Background (CMB) indicate that the Universe is highly isotropic to a special class of comoving observers\footnote{Earth is moving through the CMB rest frame with a velocity $v = 368 \rm{km}/\rm{h}$ \cite{Baumann:Cosmo}.}. Invoking the Kopernican principle one further assumes large-scale homogeneity that together with isotropy forms the cosmological principle\footnote{Consistency of the homogeneity assumptions can be tested but not the assumption directly. Cosmological principle is equivalent to the requirement that there is isotropy around two separated points.}. Comoving observers for whom the cosmological principle is satisfied use the Friedmann–Lemaître–Robertson– Walker (FLRW) coordinates
\be
ds^2=-dt^2+a^2(t)\Tilde{g}_{ij}dx^{i}dx^{j}\label{eq:metricFRW} \,, \, \Tilde{g}_{ij}dx^{i}dx^{j}=\frac{dr^2}{1-kr^2}+r^2d\Omega^2
\ee
where $\Tilde{g}_{ij}$ denotes the metric of a maximally symmetric 3-space, $k$ is the Ricci scalar of this 3-space and $a(t)$ is a scale factor. Inserting this metric in \eqref{eq:Einstein} and considering an ideal fluid
\be \label{eq:energy_momentum_ideal_fluid}
T^{\mu \nu}=(\rho + P)u^{\mu}u^{\nu}+P g^{\mu \nu}
\ee
one finds the Friedmann equations
\begin{subequations}
\begin{align}
\label{eq:Friedmann_Eq_1}
&\frac{\partial^2_{t}a}{a}=-\frac{\kappa}{6} (\rho  + 3P) \,,\\
\label{eq:Friedmann_Eq_2}
&\Big(\frac{\partial_{t}a}{a}\Big)^2=\frac{\kappa}{3}\rho-\frac{k}{a^2} \,,
\end{align}
\end{subequations}
that describe the evolution of the scale factor, dictated by the matter content of the Universe and values of $k$ and $\Lambda$. The Hubble parameter is defined as $H \equiv \partial_{t}a/a$ and is at the order-of-magnitude level inversely proportional to the age of the Universe at the evaluated time. The equation of state of the cosmological matter is usually parametrized as
\beq \label{eq:density/pressure}
P=w\rho \,,
\eeq
with $w=0$ for dust-like collisionless non-relativistic matter and $w=1/3$ for massless particles. If one interprets the Cosmological constant as a matter component (dark energy), one must have $w=-1$. It is customary to introduce the dimensionless energy density parameter
\beq \label{eq:cosmo_energy_density}
\Omega_{i}=\frac{\rho_i(t_0)}{\rho_c} \,,\, \rho_c=\frac{3H^2_0}{\kappa} \,,
\eeq
with $\rho_c$ being the critical density of the (flat) Universe today and $H_0$ represents the present-day value of the Hubble parameter.

Performing various cosmological inferences such as luminosity-distance relation of distant Supernovae, power spectrum of CMB and large scale galaxy clustering one can obtain parameters that describe the  $\Lambda$CDM cosmological model\footnote{We don't give all significant figures and uncertainties - they can be found in \cite{Aghanim:2018eyx}. Curvature density is defined  as $\Omega_k=-k/H^2_0$.} \cite{Aghanim:2018eyx}:
\beq
\Omega_{\rm b} =0.05 \,,\,\, \Omega_{\rm DM} =0.26 \,,\,\, \Omega_{\Lambda}=0.68  \,,\,\, H_0 = 67.4 \, \text{km}/{(\text{s} \, \text{Mpc})} \,, \,\, \Omega_k = 0.0007 \,.
\eeq
It is useful to express the present value of the Hubble parameter in various representations $H_{0} \sim 10^{-33} \text{eV} \sim (14 \text{Myr})^{-1} \sim$ $70 \text{km}/{(\text{s} \, \text{Mpc})}$. Note also that the above results are consistent with the spatially flat Universe so we take $k=0$.

$\Lambda$CDM model on its own is not enough to describe the initial stages in the evolution of the Universe. Particularly, there are several fine tuning problems such as the great degree of correlation between CMB points that in the conventional picture haven't had time to be in causal contact. These considerations have lead to proposing an initial highly accelerated regime - inflation. During this phase all initial inhomogeneity has been diluted. As we will show in Section \ref{sec:cosmo_axion} a scalar field (in this context known as an inflaton field) can drive the accelerated expansion of the Universe. During this stage quantum fluctuations of the inflaton field lead to the initial inhomogeneities - the seeds of future structure formation. Although more direct tests of this theory are still needed in order for it to be treated on the same footing as the $\Lambda$CDM, it has successfully predicted zero-curvature Universe with initial almost scale-invariant inhomogeneity power spectrum \cite{Guth:2013sya}.

\subsubsection{Tensions between paradigmes}

There are several tensions between the briefly described paradigms, some evident even from their descriptions - GR is a classical field theory, whose full quantum description is still lacking, while SM is a quantum field theory; $\Lambda$CDM cosmological model requires dark matter that consist of particles not found in the SM etc.

An empirically minded classification of the severity of the problems would rang the fine-tuning problems, where the accepted values of parameters have not met naturalness theoretical arguments, as least concern. Problems of this nature are almost exclusively contained in the SM - hierarchy problem, flavour problem, strong CP problem etc., along with the dark energy/cosmological constant problem. Then there are problems of theoretical formalism. Singularities in GR and incapability of making a consistent quantum theory of gravity would be positioned here as well as baryogenesis. In other words, there are well defined areas of the parameter space where the theory gives unphysical descriptions or there is an inability to extend the theory. Most alarmingly, there are empirical problems where the theories are in direct contrast with experiments and observations - we know that neutrinos are massive but can't accommodate them in the SM; there must be some additional matter beyond SM to explain the cosmological dark matter etc. In the next few sections we will briefly describe a few of these problems in more detail in order to motivate axions.

\newpage

\subsection{Strong CP problem in Quantum chromodynamics} \label{sec:strong_CP}

\subsubsection{Classical order-of-magnitude formulation}

\begin{figure}
\hspace*{\fill}%
\begin{minipage}[c]{0.44\textwidth}
\centering
\vspace{0pt}
\includegraphics[width=0.75\columnwidth]{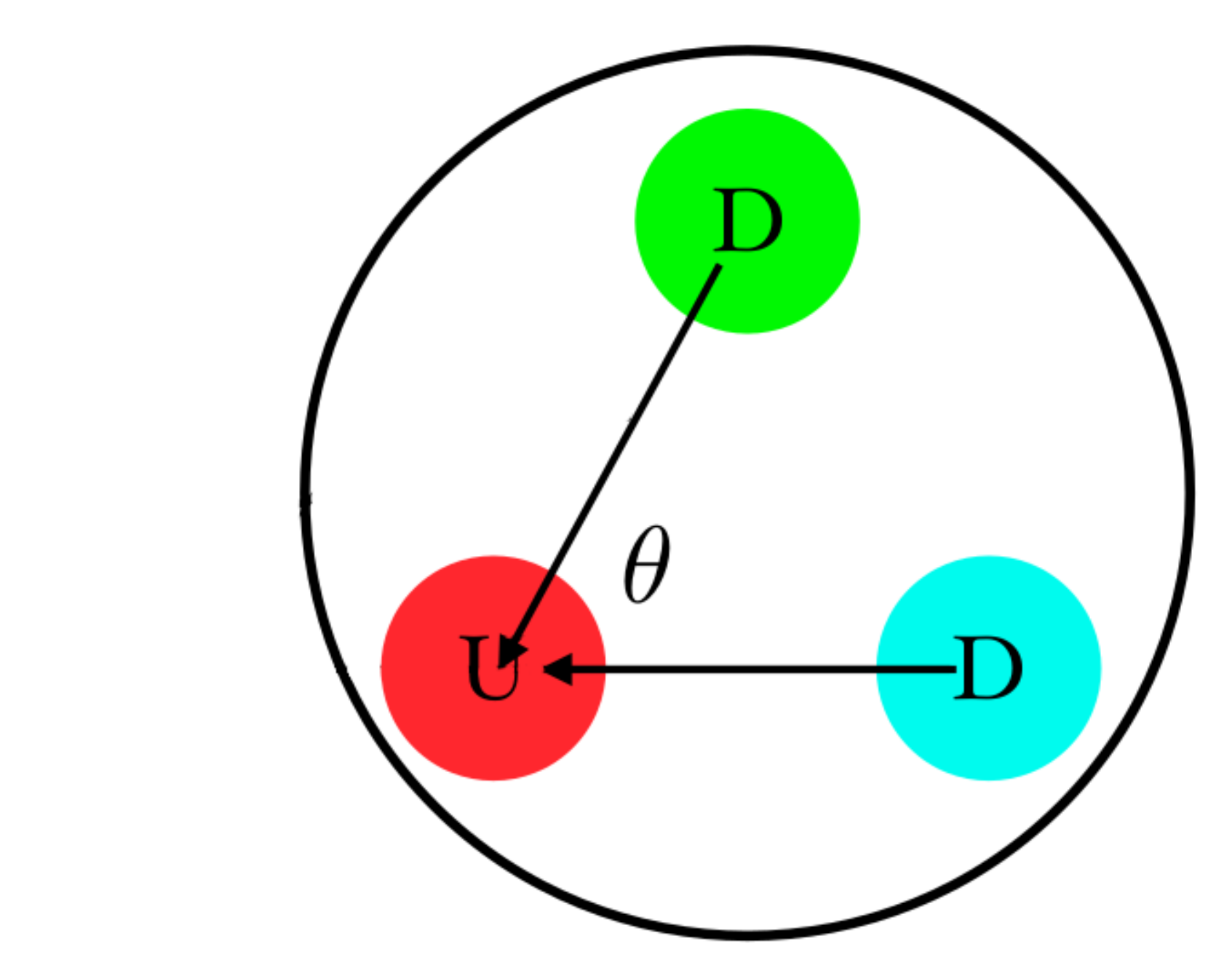}
\caption{ \fontsize{9}{12} Classical picture of the neutron prima facie.
Figure credits (adapted): \cite{Hook:2018dlk}
\label{fig:neutron}}
\end{minipage}%
\hfill
\begin{minipage}[c]{0.51\textwidth}
\vspace{0pt}
\centering
\includegraphics[width=0.5\columnwidth]{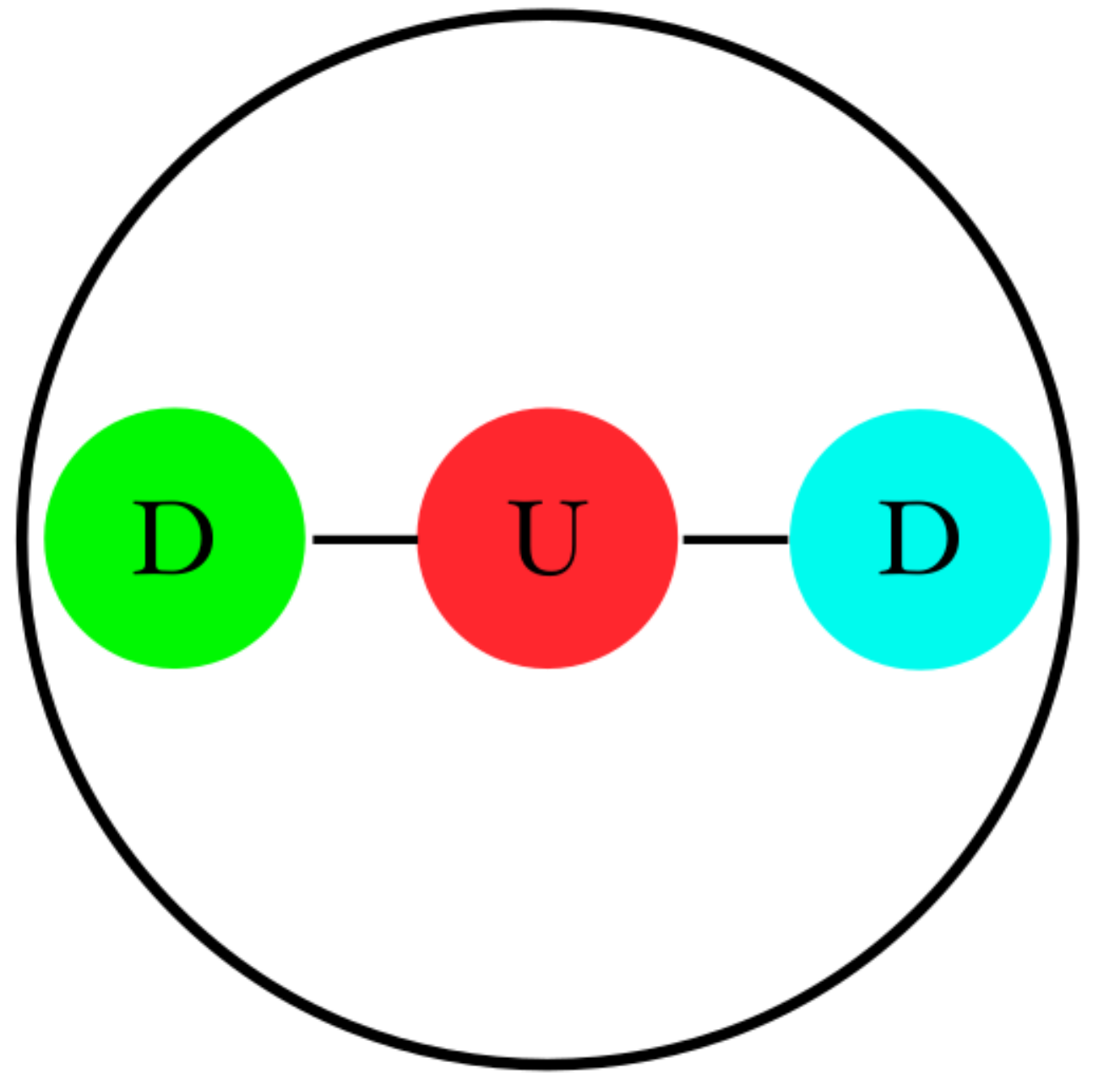}
\caption{\fontsize{9}{12} Classical picture of the neutron according to the nEDM value. Figure credits: \cite{Hook:2018dlk}
\label{fig:neutron_nEDM}}
\end{minipage}%
\hspace*{\fill}
\end{figure}

We discuss this problem in two stages. In the first stage, we discuss the classical order-of-magnitude formulation of the strong CP problem\footnote{We follow the approach of Ref. \cite{Hook:2018dlk}.}. The neutron is an electrically neutral particle, of size $\sim \hbar/(m_{\rm{n}} c) =  r_n $ that can be imagined from the classical standpoint as a collection of two $d$ and one $u$ quarks interacting with neutral gluons. These quarks are charged ($q_{d}=-1/3e,\,q_{u}=2/3e$) and their distribution in the neutron induces an electric dipole moment (nEDM). We can parametrize the value of this moment by the angle between $d$ and $u$ as in Figure \ref{fig:neutron}. EDM estimate gives
\beq
p_n \sim q_u r_n \sqrt{1-\cos{\theta}} \sim 10^{-14} e \,  \rm{cm} \sqrt{1-\cos{\theta}}.
\eeq
Present experimental probes of nEDM give an upper constraint of $p_n \leq 10^{-26} e \,  \rm{cm}$ \cite{Tanabashi:2018oca}. This means that the angle is highly tuned $\theta \sim 0$ and the quark configuration in the neutron is more like the one on Figure  \ref{fig:neutron_nEDM}.

One way to understand this fine-tuned number is to impose the symmetry that makes the nEDM go to zero as experiments suggest that the value is consistent with zero.  At the classical level, a neutron with a non-zero EDM $\bm{d}$ has spin $\bm{s}=\int \bm{r} \times \bm{p}$. With the exception of the EDM direction $\hat{d}$ there is no preferred direction in space, so $\hat{d}=\hat{s}$.  Under parity
\beq \label{eq:parity}
\mathcal{P}:\bm{x} \to -\bm{x},
\eeq
EDM transforms as $P: \bm{d} \to - \bm{d}$ and the spin $P: \bm{s} \to \bm{s}$. If we'd impose that parity is a symmetry, we would need $\bm{d}=\bm{s}=0$. However, parity is not a good symmetry of nature as it is broken by weak interactions. Similarly, one can try the solution with time inversions
\beq \label{eq:time_inv}
\mathcal{T}:t\to -t.
\eeq
under which EDM and spin transform as $T: \bm{d} \to  \bm{d}$ and $\bm{s} \to - \bm{s}$. If time inversion is a symmetry of nature, we would again need $\bm{d}=\bm{s}=0$. Yet again, $T$ is not good symmetry. The last solution, at the classical level, is that there is some dynamical mechanism that would reduce $\theta \to 0$ and this is the axion solution.

\subsubsection{Axion electrodynamics}

In order to understand at the field-theoretic level the strong CP problem we will start from the partially analogous problem in electrodynamics (we consider flat spacetime in this section). The Lagrangian for Maxwell electrodynamics is
\beq
\label{eq:lag_MaxFlat}
{\cal L}_{\rm MX} = - \frac{1}{4} F^{\mu\nu} F_{\mu\nu}  \,.
\eeq
This however is not the only Lorentz and gauge invariant term quadratic in the potential derivatives. We could also add the term of the form
\beq
\label{eq:CP_MaxFlat}
{\cal L}_{\rm top} = c \, ^{\ast}F^{\mu\nu} F_{\mu\nu}  \,,
\eeq
where $c$ is some constant and $\,^{\ast}F^{\mu\nu} \equiv \frac{1}{2}E^{\mu\nu\rho\si}F_{\rho\si}$ is Maxwell tensor dual, where $E^{\mu\nu\rho\si}$ is the totally antisymmetric Levi-Civita symbol with $E^{0123}=1$. This term is topological in nature as it doesn't depend on the metric (Box on page \pageref{box:Hodge}). Moreover, this term is a total derivative and does not contribute to the classical equations of motion. To show this we consider the action term from \eqref{eq:CP_MaxFlat} in the differential form language\footnote{We use differential forms here for pragmatic purposes in the spirit of \cite{Zee:2003mt}. Differential forms stem from the geometric formulation of gauge theories, based on fibre bundles e.g. \cite{Baez:1995sj} .}
\beq \label{eq:CP_Max_axion}
S_{\rm top} \propto \int F \wedge F.
\eeq
In Minkowski spacetime $F=dA$, so we can write the above term in the surface-term form
\beq \label{eq:Max_surface}
S_{\rm top} \propto \int d(A \wedge dA)
\eeq
as claimed.


\begin{framed}
\noindent
{\small {\it Hodge dual operator}\\ ~\label{box:Hodge}

In \eqref{eq:CP_MaxFlat} the Levi-Civitta symbol is used to contract the indices. In contrast, in \eqref{eq:lag_MaxFlat} the metric tensor is used for contraction. In the differential form representation the action from \eqref{eq:lag_MaxFlat} is of the form
\beq
S_{\rm MX} = \frac{1}{2} \int \ast F \wedge F \,,
\eeq
with $\ast$ being the metric-dependent Hodge dual operator. Hodge dual acts on a $k$-form $\omega$ in a $n$-dimensional spacetime as
\beq
\ast \omega=\frac{1}{k!(n-k)!}\frac{\omega_{\mu_1 \, ... \, \mu_k}}{\sqrt{-g}} E^{\mu_1 \, ... \, \mu_k \rho_{k+1} \, ... \, \rho_n} g_{\rho_{k+1} \sigma_{k+1}} \, ... \, g_{\rho_{n} \sigma_{n}} dx^{\sigma_{k+1}} \wedge \, ... \, \wedge dx^{\sigma_{n}} \,,
\eeq
and the components of this object are dual tensor fields.

Levi-Civitta tensor is defined as
\beq
\epsilon^{\mu\nu\rho\si}\equiv
\frac{1}{\sqrt{-g}}E^{\mu\nu\rho\si} \,.
\eeq

}

\end{framed}


Let us imagine however that the constant $c$ is actually spacetime varying field $c(x)=e^2/(16\pi^2\hbar)\theta(x)$. At the moment we consider $\theta$ as prescribed, not having a dynamics of it's own. Then the dynamical Maxwell equation will have the form
%
\beq
\label{eq:CP_max_1}
\bm{\nabla} \cdot \bm{E}&=& - \frac{\alpha c}{\pi} \bm{\nabla}\theta \cdot \bm{B} \,,\\
\label{eq:CP_max_3}
-\frac{1}{c^2} \partial_t \bm{E} + \bm{\nabla} \times \bm{B} &=& \frac{\alpha}{\pi c} (\bm{B} \partial_t \theta  +\bm{\nabla}\theta \times \bm{E}) \,,
\eeq
along with the unchanged other two (from $dF=0$)
%
\beq
\label{eq:CP_max_2}
\bm{\nabla} \cdot \bm{B}&=&0 \,,\\
\label{eq:CP_max_4}
\partial_t \bm{B} + \bm{\nabla} \times \bm{E} &=& 0.
\eeq
There is a class of materials (topological insulators) where the effective Maxwell equations are described by the variable $\theta$  from $\theta=\pi$ inside the material to $\theta=0$ outside of it (in vacuum)  and this modified (axion) electrodynamics is appropriate to use in determining the boundary conditions (e.g. \cite{Tong:GaugeTh}).

Let us note also that the added term $F \wedge F$ for generic $\theta$ is not a scalar with respect to the proper orthochronous Lorentz group (see Box on page \pageref{box:discr_sym_EL}). Starting from the field strength decomposition $F=B+E \wedge dt$ we find ${\cal L_{\rm a}} \sim \bm{E} \cdot \bm{B}$ i.e. it breaks both parity and time inversion. We note that as the term is invariant under charge conjugation there is an overall $\mathcal{C} \mathcal{P} \mathcal{T}=I$ invariance as it should be ($\mathcal{C} \mathcal{P} \mathcal{T}$ theorem in QFT).


\begin{framed}
\noindent
{\small {\it Discrete symmetries of electrodynamics}\\ ~\label{box:discr_sym_EL}

The Lorenz group $O(1,3)$ has four disjoint cosets such that each element can be written as a product of the proper orthochronous Lorentz subgroup $SO^{\uparrow}_+(1,3)$ (continuously connected to $I$) that preserves time inversion and parity and $\{I,\mathcal{P},\mathcal{T},\mathcal{P}\mathcal{T}\}$. The Lorentz scalar is a scalar under $SO^{\uparrow}_+(1,3)$, while the Lorentz pseudo-scalar is scalar under $\mathcal{P}SO^{\uparrow}_+(1,3)$.

In order to understand the transformation of the electromagnetic field with respect to the discrete symmetries $(\mathcal{P},\mathcal{T},\mathcal{C})$, we start from the II Newton law of the charged particle acted upon by the EM field (Lorentz force):
\beq \label{eq:CP_P_T_ED_IINL}
m\ddot{\bm{x}}=q(\bm{E}+\bm{v} \times \bm{B}).
\eeq
Parity operator acts as \eqref{eq:parity}, while the time inversion operator acts as \eqref{eq:time_inv}. Acting on the LHS of \eqref{eq:CP_P_T_ED_IINL} with $\mathcal{P}$ and $\mathcal{T}$ we get
\beq
\mathcal{P}:m\ddot{\bm{x}} \to - m\ddot{\bm{x}} \,, \\\
\mathcal{T}:m\ddot{\bm{x}} \to m\ddot{\bm{x}} \,.
\eeq
In order for the RHS to be consistent we must have
\beq
\mathcal{P}&:&\bm{E} \to - \bm{E} \,,\,\bm{B} \to \bm{B} \\
\mathcal{T}&:&\bm{E} \to  \bm{E} \,,\,\bm{B} \to - \bm{B} .
\eeq
Finally charge conjugation acts as:
\beq
\mathcal{C}:q\to -q
\eeq
so in order for the LHS to be invariant we must have
\beq
\mathcal{C}&:&\bm{E} \to - \bm{E} \,,\,\bm{B} \to - \bm{B} \,.
\eeq

}

\end{framed}

\subsection{Theta term in quantum chromodynamics}

Quantum electrodynamics is obtained via quantisation of Maxwell electrodynamics, an $U(1)$ gauge theory. On the other hand, the theory of the strong interaction quantum chromodynamics (QCD) is obtained by quantasing Yang-Mills $SU(3)$ theory
\beq
\label{eq:CP_QCD_gluon}
{\cal L}_{\rm YM}= - \frac{1}{2g^2} \Tr F^{\mu\nu} F_{\mu\nu}  \,,
\eeq
where the field strength carries three indices and $g$ is the Yang-Mills coupling. This object is an element of the gauge group Lie algebra and connected to the potential as
\begin{equation}
F^a_{\mu \nu}=\partial_\mu A^a_{\nu}- \partial_\nu A^a_{\mu} +f^a_{\, bc} A^b_{\mu} A^c_{\nu}.
\end{equation}
In the last expression $f^a_{\, bc}$ are Lie algebra structure constants, defined as $[T_b,T_c]=if^a_{\, bc}T_a$.
Hiding internal indices with
\beq \label{eq:MathPhysPotential}
\mathcal{A}_{\mu} \equiv -iA^a_{\mu}T_a  \,, \nonumber \\
\mathcal{F}_{\mu \nu} \equiv -iF^a_{\mu \nu}T_a  \,
\eeq
we obtain
\begin{equation} \label{eq:tenzor_krivine}
\mathcal{F}_{\mu \nu}=\partial_\mu \mathcal{A}_{\nu}- \partial_\nu \mathcal{A}_{\mu} + [\mathcal{A}_{\mu}, \mathcal{A}_{\nu}],
\end{equation}
where one should interpret $\mathcal{A}_{\mu}\mathcal{A}_{\nu}$ as a matrix product $ \mathcal{A}^j_{\mu p} \mathcal{A}^p_{\nu l} e^l \otimes e_j=(\mathcal{A}_{\mu}\mathcal{A}_{\nu})^j_{\,l} e^l \otimes e_j$, where $e_i$ is the basis in the internal space (Lie algebra) and $e^i$ it's dual. Expression \eqref{eq:tenzor_krivine} motivates the introduction of the following field strength $2$-form
\begin{equation} \label{eq:krivina_forma}
\mathcal{F}=\frac{1}{2}\mathcal{F}_{\mu \nu} dx^\mu \wedge dx^\nu,
\end{equation}
connected with the vector potential as
\begin{equation}
\mathcal{F}=d\mathcal{A}+\mathcal{A} \wedge \mathcal{A},
\end{equation}
which can be checked by substitution in \eqref{eq:krivina_forma}. In the case of Abelian groups, as is $U(1)$, corresponding Lie algebras are also Abelian (Baker-Hausdorf lemma) so that the field strength form is exact $\mathcal{F}=d\mathcal{A}$. In general however $\mathcal{A} \wedge \mathcal{A} \neq 0$. Note also that the field strength is not a gauge invariant object for non-Abelian algebras.

Analogously to the electrodynamics case we can contemplate a CP violating action term of the form \eqref{eq:CP_Max_axion}
\beq
\label{eq:CP_QCD_gluon_action}
S_{\rm a} = -\frac{\theta}{8\pi^2} \int \Tr (\mathcal{F} \wedge \mathcal{F})  \,.
\eeq
This term can also be rearranged as a total derivative (derivation is in the Box on page \pageref{box:YM_theta}):
\beq
\label{eq:CP_QCD_gluon_surface}
S_{\rm a} = \frac{\theta}{8\pi^2}  \int d^4x \epsilon^{\alpha \beta \gamma \delta}\partial_\alpha (A_{\beta}\partial_{\gamma}A_{\delta}-\frac{2}{3}iA_{\beta}A_{\gamma}A_{\delta})\,.
\eeq
The second term in the differential form representation is $A^3$, equal to zero in an Abelian gauge theory such as Maxwell electrodynamics \eqref{eq:Max_surface}. This term is responsible for non-trivial contribution of the Yang-Mills theta term on the observable aspects of the quantum theory as it can change the state spectrum even for constant $\theta$ \cite{Tong:GaugeTh}. The nontrivial role of the theta can be seen at the semi-classical level, through the instanton solutions \cite{Zee:2003mt,Tong:GaugeTh}. One of the consequences of this fact is the non-zero nEDM \cite{Hook:2018dlk}.

\begin{framed}
\noindent
{\small {\it Theta term as a total derivative - derivation}\\ ~\label{box:YM_theta}

Observe that $\Tr (\mathcal{F} \wedge \mathcal{F})$ is a $4$-form on a $4$-dimensional Minkowski spacetime. Thus, it's external derivative must be zero $d\Tr (\mathcal{F} \wedge \mathcal{F})=\Tr \, d( \mathcal{F} \wedge \mathcal{F})=0$ (it's a closed form). According to the Poincaré lema, $\Tr (\mathcal{F} \wedge \mathcal{F})$ must be also an exact form $\Tr (\mathcal{F} \wedge \mathcal{F})=\Tr (d X)$. As $X$ is a 3-form it can only be formed from $\mathcal{A}^3 \equiv \mathcal{A} \wedge \mathcal{A} \wedge \mathcal{A}$ and $\mathcal{A} \wedge d\mathcal{A}$. We can then write the following equality
\beq \label{eq:YM_theta_exact_inter_2}
\Tr(\mathcal{F} \wedge \mathcal{F})=d\Tr(c_1 \mathcal{A}^3 + c_2 \mathcal{A} \wedge dA),
\eeq
where $c_1 \,, c_2$ are coefficients to be determined. Expanding LHS we obtain
\beq
\Tr(\mathcal{F} \wedge \mathcal{F})=\Tr(d\mathcal{A} \wedge d\mathcal{A}+ 2 d\mathcal{A} \wedge \mathcal{A} \wedge \mathcal{A}),
\eeq
where we used the graded cyclicity property of the trace $\Tr(\omega \wedge \mu)=(-1)^{pq}\Tr(\mu \wedge \omega) $ and $\omega$ and $\mu$ are Lie algebra-valued $p$ and $q$ forms, respectively. This property can be easily proved, expanding the forms into their components and using linearity of trace for basic forms $dx^\mu$. Using the cyclicity of the trace again, RHS of the \eqref{eq:YM_theta_exact_inter_2} is
\beq
d\Tr(X)=\Tr(3 c_1 \, d\mathcal{A} \wedge \mathcal{A}^2 + c_2 d\mathcal{A} \wedge d\mathcal{A}),
\eeq
from where we conclude $c_1=\frac{2}{3}, \, c_2=1$. Using \eqref{eq:MathPhysPotential} we obtain \eqref{eq:CP_QCD_gluon_surface}.
}

\end{framed}

There is also the contribution of the same form as \eqref{eq:CP_QCD_gluon} from the electroweak sector, depending on the quark mass matrices \cite{Marsh:2017hbv}. In order for the effective $\theta$ to be (consistent with) zero, there must be a fine tuning between the strong and the electroweak sector. This fine tuning is the core of the CP problem.

\subsubsection{Axion solution} \label{sec:axion_PQ}

Axions are dynamical solution to the Strong CP problem. The starting point for the generation of axions is Percei-Quinn $U(1)$ symmetry. Toy model for this problem is the one discussed in Section \ref{sec:SM_SSB}. As we have seen in the reexpressed lagrangian \eqref{eq:SSB_lag} axion $\Upphi$ enjoys the shift symmetry $\Upphi \to \Upphi + c$. In reality, axion is not massless but massive particle. Because of the quantum non-perturbative effects, the shift symmetry is anomalus and thus broken to a discrete symmetry $\Upphi \to \Upphi + 2\pi n \,, n \in \mathbb{N}$ (as $\Upphi$ is an angular variable). This can be modeled by explicitly breaking the lagrangian in \eqref{eq:SSB_lag} by adding a small term to the potential  \eqref{eq:SSB_pot} (``tilting the sombrero potential'') \cite{Helfer:2017a}
\be
\mathcal{V}(\varphi) \to \mathcal{V}(\varphi)-\epsilon \varphi_1 = \mathcal{V}(\varphi)-\epsilon (\rho+\rho_0)\cos{\Big(\frac{\Upphi}{f_{\rm a}} \Big)} \,, \epsilon_{\rm sym} \ll f^3_{\rm a}
\ee
where $\epsilon_{\rm sym}$ is the strength of the symmetry breaking. The potential for the axion is now
\be \label{eq:axion_potential}
V_{\rm a}(\Upphi)=\mu^2_{\rm a}f^2_{\rm a} \Big[ 1-\cos{\Big(\frac{\Upphi}{f_{\rm a}} \Big)} \Big] \,,
\ee
with $\epsilon \equiv \sqrt{2}\mu^2_{\rm a}f_{\rm a}$, $\mu_{\rm a}$ is the axion mass and the constant term is added in order to normalize the potential. Parameter $f_{\rm a}$ is known as the decay constant. Integrating out more massive radial field \cite{Helfer:2017a}, the axion lagrangian is
\be \label{eq:axion_lag_flat}
{\cal L}_{\rm a} = - \frac{1}{2}  \p^{\mu}\Upphi\p_{\nu} \Upphi-V_{\rm a}(\Upphi) \,.
\ee

The axion couples to the QCD as
\beq
{\cal L}_{\rm aQCD} = c \frac{1}{16\pi^2} \frac{\Upphi}{f_{\rm a}} \Tr ^\ast F^{\mu \nu} F_{\mu \nu} \,
\eeq
where $c$ is a model-dependent constant, in such a way that the vacuum expectation value of the field subtracts the $\theta_{\rm eff}$.

\subsection{Axiverse scenario in String Theory}

String theory is by far the most popular candidate for the quantum theory of gravity. Among many aspects, string-theoretic models share the need to inhabit Universes with a number of dimensions higher than ours. In order for these models to meet reality, these dimensions need to compactified. This compactification typically manifest itself in low energies with a plethora of new particles and in particular ultra-light pseudo-scalar particles, called axion-like particles (ALPs) \cite{Arvanitaki:2009fg, Marsh:2017hbv}. This is the so-called axiverse scenario. Such particles could make a fraction or whole of the dark matter \cite{Hui:2016ltb}.  Thus, detection or observational imprint of such particles could be very strong hint in favour of string theory or at least extra dimensions.

\subsection{Nature of Dark Matter}\label{sec:DM}

\subsubsection{Dark Matter problem} \label{sec:DM_axions}

The idea of dark matter (DM) has a long history in physics and astronomy. Observational arguments for the necessity of the presence of cosmological DM started in the thirties in the 20th century and eventually became an established idea in the seventies \cite{Bertone:2016nfn, deSwart:2017heh}. In the early nineties it became evident that this DM can't be made from the SM particles. There are now several very strong arguments in favour of DM on various scales. We will here illustrate the simplest one (but not the strongest). To a first and very rough approximation one can imagine all the matter in the galaxy in the homogeneous sphere of radius $R$. Radial velocities of particles are then given by
\beq
v(r)=\sqrt{\frac{GM(r)}{r}} \propto \begin{cases} r  & \,,\, r<R \\  \frac{1}{\sqrt{r}} & \,,\, r>R \end{cases} \,.
\eeq
However observation of distant stars away from the concentration of the visible matter suggest
\beq
v(r)=\sqrt{\frac{GM(r)}{r}} \propto \begin{cases} r  & \,,\, r<R \\  \rm{const.} & \,,\, r>R \,.\end{cases}
\eeq
This result then leads one to conclude that there must be an additional presence of ``invisible'' matter that scales as $\rho \propto r^{-2} \,,r \gg R$. Alternatively, one can assume that the gravitational theory is modified in the regime of very low accelerations of the motion of analyzed objects\footnote{Analogous situation happened in 19th century. Leverie (successfully) proposed the existence of the new planet in the Solar System (Neptune), based on the comparison of the Uranus trajectory observations and celestial mechanics predictions from the gravitational influence of the Sun and other known planets. The same method lead him later, on the basis of Mercury's motion, to propose a new planet between the Sun and Mercury. After almost half a century of the unsuccessful search for this planet (or alternative ``DM'' models), GR gave a satisfying quantitative description of its motion \cite{Hanson:1962, Wells:2011st}.}. From the data one can construct phenomenological law describing modified gravity, so-called Modified Newtonian Dynamics (MOND). However, all the relativistic generalizations of MOND have been strongly constrained through the advances of GW astronomy \cite{Chesler:2017khz} and have been unsuccessful in explaining large scale structure power spectrum \cite{Dodelson:2011qv}.

While the alternatives to non-baryonic DM are on a weak footing, most popular DM candidate Weakly Interacting Massive Particles (WIMP), part of the supersymmetric extension of the SM, have also faced strong constraints (e.g. \cite{Lisanti:2016jxe}). In addition, current experiments haven't detected any supersymmetric particles. As a consequence, the astroparticle community has widened the scope of DM searches \cite{Bertone:2018xtm}. One of these candidates that have recently gained an increased attention are axions and ALPs. QCD axions have been recognised as the potential DM candidate from the very beginning (early 80s) and their present constraints are not nearly as stringent as are the constraints for WIMPs. In addition, recently ALPs in the range  $10^{-23} < \mu_{\text a}[\text{eV}] < 10^{-21}$ (fuzzy DM - FDM) have gained interest (see also Section \ref{sec:small_scale_chalenges}) \cite{Hu:2000ke, Marsh:2016rep, Hui:2016ltb}.

\subsubsection{Axions in an expanding Universe} \label{sec:cosmo_axion}

In order to understand axionic behaviour in a cosmological context, we will start with the Klein-Gordon equation \eqref{eq:KG} in the FLRW spacetime. Intuitively, flat space-time KG will be amended by an additional frictional term $\propto \dot{\Upphi}$, as (at least for ordinary matter) we expect that expansion of space-time will lead to matter dilution. In order to be dimensionally consistent we need a cosmological object with units $L^{-1}$ and the natural choice is the (Hubble) rate of the Universe expansion. Formally, calculating the covariant derivative for FLRW metric \eqref{eq:metricFRW} we obtain\footnote{Because of the cosmological principle, $\Upphi$ can only depend on $t$.}
\beq \label{eq:KG_FLRW}
\Ddot{\mathcal{\Upphi}}+3H\dot{\Upphi}+\mu^2_{\text{a}}\Upphi=0
\eeq

There are two asymptotic regimes of this equation.  When $\mu_{\text{a}} \ll H$ the field is exponentially suppressed to a constant value (dumped regime)
\beq \label{eq:KG_FLRW_damped}
\Upphi_{\mu=0}=\Upphi(0)+\dot{\Upphi}(0)\int^t_0 dt' \exp{\Big(3\int^{t'}_0 H(t'')dt''\Big)} \,,
\eeq
so that $\Upphi_{\mu=0} \to \rm{const.}$ quickly. In the other regime, $\mu_{\text{a}} \gg H$ we have linear harmonic oscillator (LHO)
\beq
\Upphi_{H=0}=\Upphi(0) \cos{(\mu_{\rm a}t)}+\frac{\dot{\Upphi}(0)}{\mu_{\rm a}} \sin{(\mu_{\rm a}t)}\,.
\eeq
Amplitude in the oscillation limit can be obtained by the WKB-styled analysis. We assume the ansatz \cite{Marsh:2016rep}
\beq
\Upphi = \mathcal{A}(t) \cos{(\mu_{\rm a}t+\Upsilon)} \,,\, \dot{\mathcal{A}} \sim \epsilon
\eeq
with $\epsilon/\mu_{\rm a} \sim H/\mu_{\rm a} \ll 1$. Inserting the ansatz in \eqref{eq:metricFRW} we obtain
\beq \label{eq:cosmo_axion_ansatz}
\mathcal{A}(t)=a^{-3/2}(t)+\mathcal{O}(\eps^2) \,.
\eeq

Identifying the stress-energy tensor \eqref{eq:e_m_tensor} obtained from the scalar field action with the one for an ideal fluid \eqref{eq:energy_momentum_ideal_fluid} we find [as comoving observers are static $u^{\mu}=(1,0,0,0)$] the field density
\be
\rho_{\Upphi}=\frac{1}{2}(\partial_{t}{\Upphi})^2+\frac{1}{2}\mu^2_{\rm a}\Upphi^2 \,,
\ee
and pressure
\be
P_{\Upphi}=\frac{1}{2}(\partial_{t}{\Upphi})^2-\frac{1}{2}\mu^2_{\rm a}\Upphi^2 \,.
\ee
In the overdamped limit, density-pressure ratio \eqref{eq:density/pressure} quickly becomes
\be
w_{\Upphi}=-1 \,.
\ee
That is, axions behave like a DE (or inflaton field) for $\mu_{\rm a} < H_0$. In the other limit ($H_0 \ll \mu_{\rm a}$) from \eqref{eq:cosmo_axion_ansatz} we find
\be
\rho_{\Upphi} \propto a^3 \,,\, \langle  w_{\Upphi} \rangle =0
\ee
and the axions behave like a CDM on large scales.

Previous discussion is valid for ALP, while for QCD axions, formed at comparatively lower energies, one should also include mass dependence on temperature and hence time. For $T \ll 200 \rm{MeV}$ (QCD phase transition scale), QCD axion mass settles to the zero-temperature value and the above analysis is applicable \cite{Marsh:2016rep}.
\subsubsection{Cosmological production of axion DM} \label{sec:cosmo_axion_production}

After inflation, the Universe was in a hot and dense state. As the Universe expanded, this thermal bath of particles cooled and various particles decoupled from it. Most of these products (thermal relics) have had mostly unchanged populations from the frezzout. There are two time scales that basically dictate this dynamics - (microscopic) particle interaction timescale $\Gamma=n\sigma v$ and the (macroscopic) Hubble rate $H$, where $\sigma$ is an interaction cross section that can be found from QFT, $v$ is average velocity and $n$ is number density. When $\Gamma \gg H$, particles are thermalized. In the other limit they decouple from the thermal bath. In particular, this is the mechanism for the production of WIMPs and in order for them to make a DM population, $\sqrt{\langle \sigma v \rangle} \sim 0.1 \sqrt{G_{\rm F}}$, where $G_{\rm F}$ is the Fermi constant \cite{Baumann:Cosmo}. This result is known as WIMP miracle as the cosmological conditions ``require'' new particles at the $\rm{TeV}$ scale. Axions can be also produced thermaly, however this mechanism is largely constrained \cite{Marsh:2016rep} and the most attractive mechanism for axion production is non-thermal.

Most popular axion DM production mechanism is misalignment or vacuum realignment. In this scenario the number of axions are produced from the breaking of the PQ symmetry, discussed in Section \ref{sec:axion_PQ}, with initial conditions
\be
\Upphi(t_{i})=f_{\rm a}\theta_i \,,\, \dot{\Upphi}(t_{i})=0
\ee
and $\theta_i=\mathcal{O}(1)$. Initially axion density is misaligned from the vacuum \eqref{eq:KG_FLRW_damped} and then it dynamically realignes \eqref{eq:cosmo_axion_ansatz} as it rolls down the potential well. In order to be a priori relevant DM candidate it must have begun oscillating around the potential minimum at latest at the matter-radiation equality $H(a_{\rm eq})=10^{-28} \rm{eV}$, since after that we see DM imprints in the CMB. If ALP satisfies this condition then it's density is given by [from Eq. \eqref{eq:cosmo_energy_density}]
\beq
\Omega_{\rm a}(a) =\frac{8\pi G}{3H^2_0} \rho_{\rm a}(a_{\rm osc}) \Big( \frac{a_{\rm osc}}{a} \Big)^3  \,.
\eeq
As in this period expansion is radiation dominated we find $a_{\rm osc}$ from\footnote{In the radiation dominated expansion, from \eqref{eq:Friedmann_Eq_1} and \eqref{eq:Friedmann_Eq_2}, $a \propto \sqrt{t}$ and $2H=1/t$.}
\beq
a_{\rm osc}=\frac{a_{\rm eq}}{\sqrt{2t_{\rm eq}H_{\rm osc}}} \,,
\eeq
where $a_{\rm eq} = 3 \times 10^{-4}$, $t_{\rm eq} = 60 \text{kyr}$ and $H_{\rm osc} = \mu_{\rm a}$, while $\rho_{\rm a}(a_{\rm osc})$ is roughly set by the initial conditions
\be
\rho_{\rm a}(a_{\rm osc}) \approx \frac{1}{2}\mu^2_{\rm a}\Upphi_i^2 \,.
\ee
Thus,
\be
\Omega_{\rm a} \approx 0.1  \Big(\frac{\mu_{\rm a}}{10^{-22} \rm{eV}}\Big)^{1/2}  \Big(\frac{f_{\rm a}}{10^{17} \rm{GeV}}\Big)^2 \Big( \frac{\theta_i}{1} \Big)^2 \,.
\ee
In order for $f_{\rm a}$ to be between the GUT $(10^{16} \rm{GeV})$ and the Planck scale $(10^{18} \rm{GeV})$ axion mass has to be in the FDM range, which is similar numerical coincidence as in the WIMP miracle \cite{Hui:2016ltb}.

\subsection{Small scale challenges of $\Lambda$CDM cosmology} \label{sec:small_scale_chalenges}

In order to describe the behaviour of matter in the evolving Universe one must consider deviations from the cosmological principle. In the early stages of the evolution of the Universe, this can be done semi-analytically using cosmological perturbation theory (e.g. \cite{Baumann:Cosmo}). Modern picture of the structure formation is hierarchical in the sense that first DM halos (virializes self-gravitating structure), formed from the inflatory seeds of inhomogeneities, merge between themselves to form larger structures. Galaxy formation and evolution is in the highly non-linear regime. As CDM model assumes that DM particles behave as a collisionless gas, one can use $N$-body simulation that track many-body gravitational interactions between DM halos\footnote{In practice artificial particles that represent them.}. Modern simulation are hydrodynamical in nature and can describe also baryonic effects. The holy grail of the join effort of cosmological perturbation theory and cosmological simulation is to reconstruct present picture of the Universe from the initial conditions generated from inflation. There are several problems with state-of-the-art comparison between the simulation outcome and the observational infered structure of the Universe in low redshifts, collectively labeled as a small scale challenges of $\Lambda$CDM (for a recent review see \cite{Bullock:2017xww}). We will present here only so-called cusp-core problem.

Pure CDM simulation point toward ``cuspy'' center of the DM halo, given by Navaro-Frank-White (NFW) profile
\begin{equation} \label{eq:NFW}
\rho_{\text{NFW}}(r)=\frac{\rho_s}{\frac{r}{r_s}\big(1 + \frac{r}{r_s} \big)^2}\,.
\end{equation}
Here, $\rho_s$ is related to the density of the Universe at the moment the halo collapsed and $r_s$ is NFW scale radius. As $r \to 0$, $\rho_{\text{NFW}} \propto r^{-1}$. Observational inferences of DM densities (from velocity curves) however point towards more ``cored'' DM profiles with $\rho_{\rm{DM}} \propto r^{\gamma} \,, \gamma \approx 0-0.5$ \cite{Bullock:2017xww}. Baryons are missing from the DM only simulations and full hydrodynamical simulations point toward more cored profiles, through various mechanisms. For example, supernovae in the galaxy centres can disperse DM from the centers and lead to cored profiles. This mechanism becomes stretched for DM-rich dwarf galaxies. Furthermore, hydrodynamical simulation require tuning of large numbers of astrophysical parameters whose variation leads to different outcomes. Similarly to the other small scale challenges, hydrodynamical simulations are at the moment not robust enough to point whether these problems come from our understanding of complex baryonic astrophysics or fundamental properties of DM \cite{Bullock:2017xww}. As strongest arguments for ``cold'' in CDM come from linear regime, one can therefore ask whether properties of DM itself, e.g. self-interaction, are in part responsible for the structure of evolved galaxies.

If DM is actually FDM $(10^{-23} < \mu_{\text a}[\text{eV}] < 10^{-21})$ De Broglie wavelength of axions is comparable to the galactic scales
\beq \label{eq:FDM_de_broglie}
\lambda_{\text{dB}}\sim 0.1 \text{kpc} \Big(\frac{v_\text{vir}}{10^{-3}c}\Big)^{-1} \Big(\frac{\mu}{\mu_{22}}\Big)^{-1} \,,
\eeq
where $v \sim 10^{-3}c$ are typical virialized volicites in a Milky Way and $\mu_{22}=10^{-22} \text{eV}$. On these scales one can expect that the wavelike (``fuzzy'') nature of axions can contribute to the small scale behaviour of DM and in that way alleviate some of the small scale challenges $\Lambda$CDM (see \cite{Marsh:2015wka, Hui:2016ltb} and Section \ref{sec:darkhalo_descr}).

\subsection{Nature of dark, compact objects}

A special type of astrophysical objects are \textit{compact objects}. Traditionally, this is the name of the set that contains neutron stars (NS), black holes (BH) and sometimes white dwarfs (WD) (e.g. \cite{shapiro}). All of the three objects are relativistic (WDs mildly) and their structure is significantly different from regular stars - WDs and NSs are supported by degeneracy pressure, while BHs are spacetime structures left over from matter collapsing. As Chandrasekhar famously put it \textit{the only elements in their construction are our concepts of space and time}. From a purely phenomenological bottom-up approach one can ask if there are other similar objects in the Universe. These hypothetical objects are called exotic compact objects (ECOs) (for a review see \cite{Cardoso:2019rvt}). The era of gravitational wave astronomy provides a way to answer this question and constrain the models that describe ECOs.

Furthermore, astrophysical BHs are to a first approximation (in the sense of neglecting surrounding matter in accretion disk and the interstellar space) described by a Kerr spacetime (Appendix \ref{AppKerr}). Famously, no hair theorems established that we need only two numbers to describe such objects at the classical level, their mass $M$ and angular momentum $J$  (for a review see \cite{Cardoso:2016ryw}). There is a sharp divide between BH interiors and exteriors in the form of the horizon which acts as a one-way membrane. While the exteriors are regular, BH interiors are pathological and it is expected from a quantum theory of gravity to resolve the problem of spacetime singularities. Preliminary steps in using quantum mechanics to understand BHs lead however to problems, such as BH information paradox with no accepted solution at the moment (e.g. \cite{Compere:2019ssx}).

In principle Kerr BHs (and others) could exist without the horizon if they saturate the Kerr bound $a \geq M$ [naked singularities, see \eqref{eq:Kerr_horizon}]. However, it is commonly believed that these singularities are hidden from observers (Penrose's cosmic censorship hypothesis). While the topic is still under discussion in generic spacetimes (e.g. \cite{Cardoso:2017soq}), for astrophysical BHs it looks as if the hypothesis holds \cite{Sperhake:2009jz}. From a more theoretical standpoint, BH vanishing tidal Love number\footnote{This object described the deformation (rigidity) of the self-gravitating structure from tidal deformations of the companion, e.g. \cite{poissonwillbook}.} can be phrased as a problem in the naturalness perspective \cite{Porto:2016zng}.   Thus, one can from a more top-down theoretical approach ask, given all of these fascinating properties and problems  - classical and quantum, if BHs actually exist or is the Universe instead populated by a plethora of ECOs that just look like BHs and evade their problems (BH mimickers). At this moment, it is hard not to answer negatively to this question in general but in some parts of the parameter space alternatives are still not ruled out. It is also a sensible effort to develop a phenomenological paradigm to quantify the existence of BHs or constrain the alternatives \cite{Cardoso:2019rvt}. In order to do this, one has to theorize various ECOs and consider their observational imprints.

There are three reality checks that one has to perform on the models
\begin{itemize}
    \item Are these objects stable? If not, how do instability timescales compare with the relevant astrophysical and cosmological scales?
    \item Can these objects \textit{in principle} form?
    \item Is there a viable astrophysical and/or cosmological formation channel for these objects?
\end{itemize}
Compact axion stars are the most conservative models of ECOs and fairly easily pass all of the above reality checks, as we will argue in Section \ref{sec:axion_stars}. On the other hand, they can't be so compact in order to serve as BH mimickers, but their existence is still valid from the perspective of the question of the existence of various ECOs.

\newpage

\section{Why axion gravitating structures?}

In this work we will focus on axion phenomenology through gravitating structures. Axions are expected to interact very weakly with the SM, so their pile-up can help with detecting them. As axions are bosons so large number configurations are in principle possible. In Part \ref{ch:structure} we will discuss mechanisms for these configuration to be produced.

Owing to their low masses, these configurations could have very high population numbers and allow for classical field description. This description can be of help in making clear predictions regarding their phenomenology, which allows us to constrain the model or detect the particle signature. Let us, for example, imagine that axions constitute DM in galactic haloes and estimate their occupancy number
\begin{equation} \label{eq:boson_DM_occupancy}
\mathcal{N} \approx \frac{\Delta N}{\Delta V (\Delta \Pi)^3} \sim 10^7 \Big( \frac{\rho_{\text DM}}{1 \text{GeV}/\text{cm}^3} \Big) \Big( \frac{\mu}{1 \text{eV}}\Big)^{-4}  \Big( \frac{v_\text{vir}}{100 \text{km}/s}\Big)^{-3} ,
\end{equation}
where $\Pi=p/\hbar$ and $\rho_{\text DM}$ is dynamically estimated DM density in the Solar neighborhood \cite{SalucciNesti:2013}. For typical QCD axion mass $\mu_{\text{QCDa}} \sim 10^{-5} \text{eV}$ we find $\mathcal{N} \sim 10^{27}$ and for typical ALP mass $\mu_{\text{ALP}} \sim 10^{-22} \text{eV}$ we find $\mathcal{N} \sim 10^{55}$. These are huge occupancy number that allow for classical field description.

The axion lagrangian  \eqref{eq:axion_lag_flat} in curved spacetime (using the principle of general covariance) is
\be \label{eq:Lagrangian_scalar}
{\cal L} \supset - \frac{1}{2} g^{\mu\nu} \p_{\mu}\Upphi\p_{\nu} \Upphi-V_{\rm a}(\Upphi) \,.
\ee
Expanding the potential for small value of the axion field first two contributions are
\begin{equation}
V_{\rm a}(\Upphi) = \frac{1}{2} \mu^2_{\rm a} \Upphi^2 -\frac{1}{4!}\lambda_{\rm a}\Upphi^4 + \mathcal{O}\Big(\Big(\frac{\Upphi}{f_{\rm a}}\Big)^6\Big)\,,
\end{equation}
with $\lambda_a=\mu^2_{\rm a}/f^2_{\rm a}$. Unless stated otherwise, we focus on non-self-interacting scenario with the scalar potential containing only the mass term. Minimizing the action containing only the above terms in the prescribed spacetime and neglecting back-reaction we obtain the Klein-Gordon equation
\begin{equation} \label{eq:KG}
\left(\nabla^{\mu}\nabla_{\mu} - \mu^2_{\rm a} \right) \Upphi =0\,.
\end{equation}
In the Minkowski spacetime $\nabla^{\mu}\nabla_{\mu}=\partial^{\mu}\partial_{\mu}$. 

Let us discuss the weak-field regime of self-gravitating solution \footnote{We here mostly follow \cite{Coleman:1985rnk, Nastase:2019pzh}.}, where we should effectively add the gravitational potential. Due to Derrick theorem (Box on page \pageref{box:Derrick}), in $D \geqslant 3 $ static solutions are not possible. There are ways for one to circumvent Derrick theorem
\begin{itemize}
    \item Solutions can be time dependent. However, this can be realized either through the breaking of the Lorenz invariance (if not, we could boost to the rest frame) or through dissipation. If dissipation is large with comparison to relevant astrophysical or cosmological scales, one can have quasi-stable configuration. In the axion context, this solution is known as the oscillaton (Section \ref{sec:axion_stars}).
    \item  Through coupling to other fields, e.g. Lee's model of a coupled real and complex scalar \cite{Coleman:1985rnk}.
    \item Higher derivatives in the lagrangian (problem with the renormalizability).
    \item If we consider complex scalar instead of the real, configuration could be protected by a charge, as for Q-balls and boson stars (footnote \ref{fn:q_ball}).
\end{itemize}

Aforementioned arguments lead us to the conclusion that we need to consider time-dependent and (as we will see) time-periodic axionic background. In such backgrounds possibility of resonances between the background and the visible matter inside it occurs. These resonances could lead to enhancement of the background fingerprint and point to its existence or help in constraining the models that describe it.


\begin{framed}
\noindent
{\small {\it Derrick theorem}\\ ~\label{box:Derrick}

We consider the Lagrangian of the form \eqref{eq:Lagrangian_scalar} for $(1,D)$ Minkowski spacetime and generic potential $U(\Phi)$. Energy of the configuration is $E=T+V$ with $V=V_1+V_2$ and
\beq
T \equiv \frac{1}{2} \int d^D  \bm{x} \,(\partial_t \Phi)^2 \,,\, V_1 \equiv \frac{1}{2}\int d^D \bm{x} \, (\bm{\nabla}\Phi)^2 \,,\, V_2 \equiv \int d^D \bm{x} \, U(\Phi).
\eeq
As $T\,,V_1 \geqslant 0$, energy bounded (necessary for the system stability) from below $E \geqslant E_{\rm min}$ implies $\exists U_{\rm min} \,, U \geqslant U_{\rm min}$. Thus, we can always redefine the potential so that $U_{\rm min}=U \big( \Phi_0)=0$ for the ground state $ \Phi_0 $. Note that $V_{1,2} \geqslant 0$ and are simulationlsy zero for the ground state.

Now, let us analyze the family of solutions of the form $\Phi_\lambda ( \bm{x} ) = \Phi_0 (\lambda \bm{x}) \,, \, \lambda \in \mathbb{R}_{+}$. The energy of these solutions is given by
\beq
E_\lambda=\lambda^{2-D} V_1^{(0)} + \lambda^{-D} V_2^{(0)},
\eeq
where the label $^{(0)}$ implies that $V_{1,2}$ are evaluated for the ground state. If $\lambda=1$ is the ground state, then it must have the minimal energy  $\partial_\lambda E_{\lambda}|_{\lambda=1} = 0$ so that we obtain
\beq
(D-2)V_1+DV_2=0.
\eeq
From the sign of $V_{1,2}$ we conclude that for $D \geqslant 2$, $\Phi=0$. In lower dimensions one can have static configurations (solitons), e.g. sine-Gordon or Higgs model \cite{Coleman:1985rnk, Nastase:2019pzh}. Recently, there have been steps generalize the Derrick theorem to curved spacetime \cite{Carloni:2019cyo}.

}

\end{framed}

\newpage


\section{Axion coupling to the visible matter}

Axionic observational fingerprints can arise from their gravitational and other interactions with the ``visible matter''. In this work we are interested in the two types of interactions - motion of point particles in the axionic background and the interaction between axions and the Maxwell sector. We give general framework for these two types of interactions in turn. Note that all the spacetimes considered in this work are asymptotically flat $(\Lambda=0)$.

\subsection{Point particle in the axion background}

Motion of particles in the axion background can be described through a particle-metric coupling, where the metric depends on the distribution of the axion field. Parameterizing the spacetime trajectory with an affine parameter, one obtains the geodesic equation for both massive and massless particles
\begin{equation} \label{eq:geodesic_main}
\ddot{x}^\mu+\Gamma^{\mu}_{\hphantom{\mu}\alpha \beta} \dot{x}^\alpha \dot{x}^\beta=0\,.
\end{equation}
Details can be found in Appendix \ref{AppPointPart}.

\subsection{Coupling to the Maxwell sector} \label{sec:axMaxframe}

Lagrangian field density for the Maxwell sector and its coupling to the axion field $ \Phi$ in a general-relativistic curved spacetime contains
\beq
\label{eq:MFaction}
{\cal L} \supset - \frac{1}{4} F^{\mu\nu} F_{\mu\nu} - \frac{k_{\rm a}}{2} \Phi \,^{\ast}F^{\mu\nu} F_{\mu\nu} \,,
\eeq
where $F_{\mu\nu} \equiv \na_{\mu}A_{\nu} - \na_{\nu} A_{\mu}$ is the Maxwell tensor and we refer to Box on page \pageref{box:Hodge} on the dual tensor. Note that under parity $\Phi \to - \Phi$ (axion is a pseudo-scalar\footnote{In the main text we will sometimes refer to axion as a scalar, keeping in mind that it has odd parity and is really a pseudo-scalar. In the Appendix \ref{app:sc-photon} we briefly consider interactions of (even-parity) scalars with the Maxwell sector. }) and $\,^{\ast}F^{\mu\nu} F_{\mu\nu} \sim \bm{E} \cdot \bm{B} \to - \bm{E} \cdot \bm{B}$, so that the Lagrangian is Lorentz scalar.

If $\Phi$ is the QCD axion ($\alpha_{\rm EM}$ is the EM fine structure constant),
\be \label{eq:ax_ph_k_K}
k_{\rm a}=\frac{\alpha_{\rm EM} K}{4\pi f_{\rm a}} \,
\ee
with~\footnote{Model names \cite{Marsh:2016rep}: Kim-Shifman-Vainshtein-Zakharov (KSVZ) and Dine-Fischler-Srednicki-Zhitnitsky (DFSZ).} \cite{Tanabashi:2018oca}
\be
K=\Big(\frac{E}{N}-1.92 \Big) \,,\, \frac{E}{N}=\begin{cases} \frac{8}{3} & \,,\, \rm{DFSZ} \\ 0 & \,,\, \rm{benchmark \, KSVZ}\end{cases}
\ee
leading to
\be
\frac{\sqrt{\hbar}}{k_{\rm a}}=\frac{10^{16}}{0.203\frac{E}{N}-0.39}\left(\frac{10^{-5}\,{\rm eV}}{\mu_{\rm a}}\right)\,{\rm GeV}\,. \label{eq:k_ax_ph_values}
\ee
In some alternative models $K$ could be as high as $\sim 10^2$ or higher allowing for coupling to hidden sector photons \cite{Hertzberg:2018zte}. Thus, we consider arbitrary coupling constant to keep the discussion as general as possible.

We get the following equations of motion for the Lagrangian above (other terms are in \eqref{eq:Lagrangian_scalar} and we consider only mass term in the potential):
%
\beq
\label{eq:MFEoMScalar}
\left(\nabla^{\mu}\nabla_{\mu} - \mu^{2}_{\rm a} \right) \Phi &=&\frac{k_{\rm a}}{2} \,^{\ast}F^{\mu\nu} F_{\mu\nu}\,,\\
\label{eq:MFEoMVector}
\nabla^{\nu} F_{\mu\nu} &=& - 2 k_{\rm a} \,^{\ast}F_{\mu\nu} \nabla^{\nu} \Phi\,,\\
T_{\mu\nu} &=& F_{\mu}{}^{\rho} F_{\nu\rho}- \frac{1}{4} g_{\mu\nu} F^{\rho\sigma} F_{\rho\sigma} +  \nabla_{\mu}\Phi \nabla_{\nu} \Phi \nonumber  \,\\
&& - \frac{1}{2} g_{\mu\nu} \left( \nabla^{\rho} \Phi \nabla_{\rho}\Phi + \mu^{2}_{\rm a}\Phi\Phi \right) - \frac{k_{\rm a}}{2} \Phi g_{\mu\nu} \,^{\ast} F^{\rho\sigma} F_{\rho\sigma} \,. \label{eq:MFEoMTensor}
\eeq
These are (inhomogeneous) Klein-Gordon and Maxwell equations, respectively, followed by the stress-energy tensor for the Einstein equation \eqref{eq:Einstein}.

\newpage

\section{Present constraints}

\begin{figure}
\hspace*{\fill}%
\begin{minipage}[c]{0.44\textwidth}
\centering
\vspace{0pt}
\includegraphics[width=1\columnwidth]{Figures/axion_constraints_18.png}
\caption{ \fontsize{9}{12} Exclusion plot of axion-photon coupling with respect to the axion mass. Gray zones and lines represent experimental constraints while the coloured ones represent experiments. Some of the constraints assume that axions dominantly contribute to the DM abduance (telescopes, haloscopes), while the other relay on the axion production through the Primakoff process $\gamma+Ze \to Ze + a$.
Figure credits: \cite{Tanabashi:2018oca}
\label{fig:QCD_constraints}}
\end{minipage}%
\hfill
\begin{minipage}[c]{0.51\textwidth}
\vspace{0pt}
\centering
\includegraphics[width=1\columnwidth]{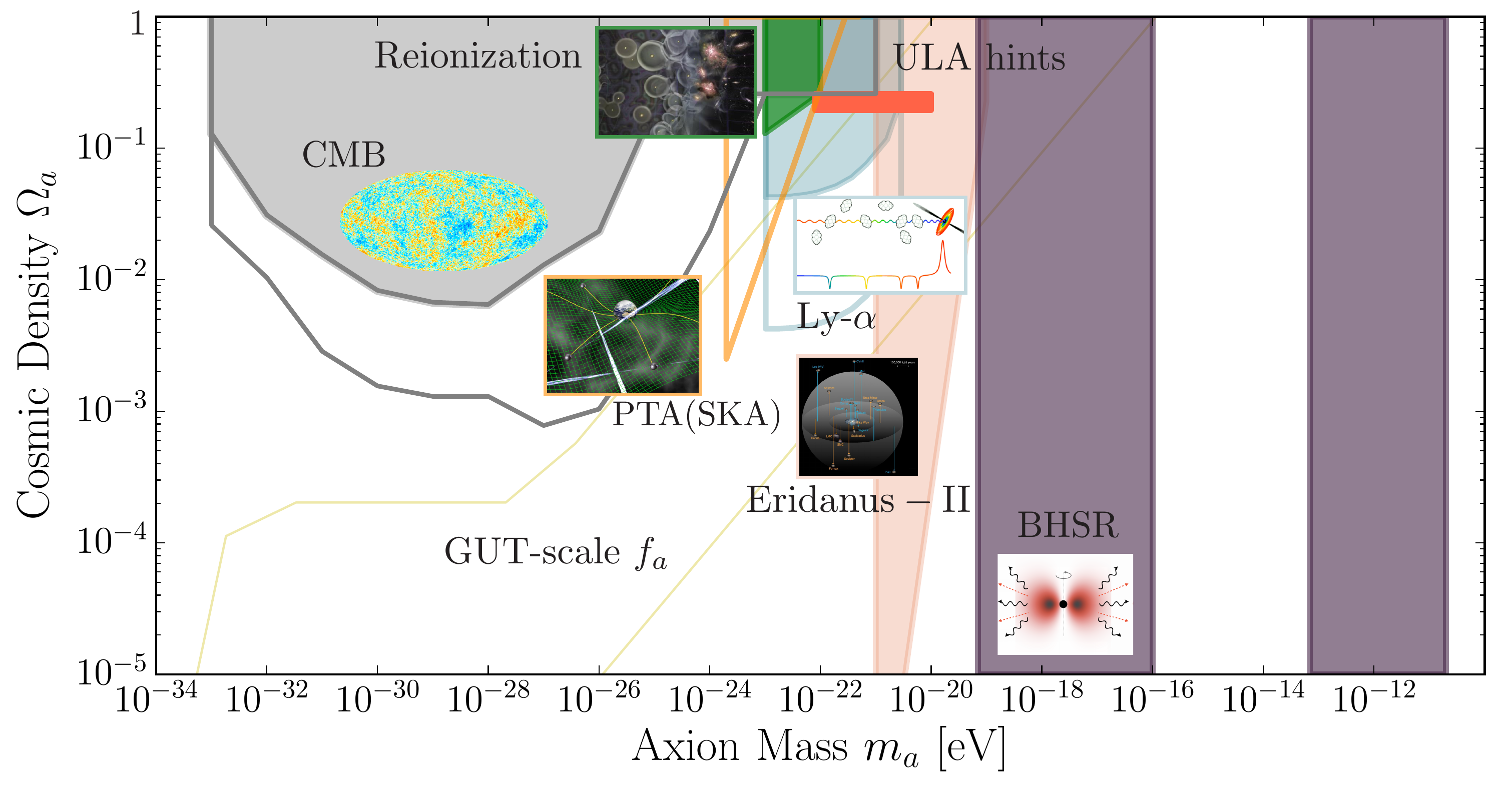}
\caption{\fontsize{9}{12} Exclusion plot of ALP energy density with respect to the mass. Shaded zones are present constraints, while the thick lines represent planned programs. All constraints are indirect - cosmological (CMB, Reionization, Ly-$\alpha$) and astrophysical (others), and assume only gravitational interaction. Figure credits: \cite{Grin:2019mub}
\label{fig:ULA_constraints}}
\end{minipage}%
\hspace*{\fill}
\end{figure}

Present constraints on axions and ALPs are shown of Figs. \ref{fig:QCD_constraints}, \ref{fig:ULA_constraints}, combining various direct detection experiments and indirect astrophysical and cosmological constraints. Fig. \ref{fig:QCD_constraints} is concentrated on heavier axions, such as QCD axions, and the processes that rely on electromagnetic coupling. KSVZ and DFSZ  are the most popular QCD axion models, which are largely unconstrained. Fig.  \ref{fig:ULA_constraints} focuses on lighter axions and indirect astrophysical and cosmological probes. Shaded zones give present constraints, while the thick lines present planned empirical programs (see \cite{Grin:2019mub}). Recently, ALP range has also began to be probed through laboratory experiments (e.g. \cite{Abel:2017a}), while several other experiments are planned.

In this work we will focus on FDM mass range (Section \ref{sec:darkhalo_descr}) and the mass range applicable to BH superradiance (Section \ref{sec:ax_grav_BH}). Although we investigate the impact of axion-photon coupling on present constraints based on BH superradiance, dominant physics of both approaches relies only on the EP. This is precisely the opportunity of gravitational probes of axions/ALP - however weakly they interact with the rest of the matter, they must gravitate and if they are localised, strength of the gravitational field could be significant.

\newpage

\part{Gravitating axion configurations} \label{ch:structure}
In this Part we will describe self-gravitating axion configurations as well as gravitating configurations around BHs. We will neglect both the axion self-interaction and the coupling to the Maxwell sector in the Lagrangian \eqref{eq:MFaction}, while we treat the later perturbatively in Part \ref{ch:ax-photon}.

\section{Self-gravitating axion configurations} \label{sec:axion_stars}

We consider time-dependent, spherically symmetric, real\footnote{Complex scalar field counterparts to oscillatons are known as boson stars, whose metric is stationary but have a harmonic boson~\cite{Liebling:2012fv}.
These configurations have $\text{U}(1)$ symmetry and are hence protected by a charge i.e. evade Derrick theorem. Corresponding flat spacetime object is known as Q-ball \cite{Nastase:2019pzh}. \label{fn:q_ball}} scalar field solutions $\Psi(t,r)$ of the coupled Einstein-Klein-Gordon (EKG) equations \eqref{eq:Lagrangian_scalar}. These solutions are known as oscillatons~\footnote{Object is regarded as an axion star if one takes self-interactions or even full axion potential \eqref{eq:axion_potential} into account~\cite{Helfer:2017a, Visinelli:2017ooc, Braaten:2018nag}.} for the first time constructed in~\cite{Seidel:1991zh}. The dynamics and stability of these objects were studied in Refs.~\cite{Seidel:1991zh,Alcubierre:2003sx,Okawa:2013jba,Brito:2015yfh}, where a set of stable ground states were found (excited states are unstable and we do not discuss them here). These solutions actually have a small radiating tail, as they can't be solitons (see Box on page \pageref{box:Derrick}), but the mass-loss rate is for much of the parameter space larger than a Hubble time\footnote{The reason why they are sometimes called pseudo-solitons.} ~\cite{Page:2003rd,Fodor:2009kg,Grandclement:2011wz}. Such solutions can be (in principle) formed through gravitational collapse and cooling mechanisms~\cite{Seidel:1993zk,Guzman:2004wj,Guzman:2006yc,Okawa:2013jba}. We analyse cosmological channels for their formation in Section \ref{sec:axion_conf_cosmo_astro}.

Most general spherically symmetric spacetime in radial coordinates $(t,r,\theta,\varphi)$ has the form
\begin{equation}
ds^2=g_{\mu\nu}d x^\mu dx^\nu=-Adt^2+Bdr^2+r^2d\Omega^2\,,\label{eq:SSmetric}
\end{equation}
where $A\equiv A(t,r)>0,\, B\equiv B(t,r)>0$ and $d\Omega^2=d\theta^2+\sin^2(\theta)d\varphi^2$ is the metric on the two-sphere. Note that we consider asymptotically flat spacetime. For computational convenience, this metric is rewritten as
\begin{equation}
 ds^2 = B(t,r) \left(- \frac{1}{C(t,r)}dt^2 +  dr^2 \right) + r^2 d\Omega^2 \label{eq:metric2}\,.
\end{equation}
In this section we will set units such that $\mu_{\rm a} = 1$ (unless explicitly stated otherwise) and redefine the scalar through
\be
\Phi = \frac{\Psi}{\sqrt{4 \pi}}\,.
\ee
With these definitions, Eqs.~\eqref{eq:MFEoMScalar} and \eqref{eq:MFEoMTensor} lead to\footnote{Non-zero Cristoffel symbols and Ricci tensor components for the time-dependent spherically-symmetric metric can be found in Ref. \cite{bookWeinberg:1972} - Chapter 11, Section 7. See also comment in the Appendix of Chapter V.4. in \cite{Zee:2013dea} on the missing factor in \cite{bookWeinberg:1972}.}:
\begin{align}
  & -\frac{B'}{r B}+B \left(\Psi^2-\frac{1}{r^2}\right)+C \partial_t{\Psi}^2+\Psi'^2+\frac{1}{r^2} = 0\,, \label{eq:1of4}\\
  & 2 \Psi' \partial_t{\Psi}-\frac{\dot{B}}{r B} = 0\,, \label{eq:2of4}\\
  & \frac{B'}{B}+B \left(r \Psi^2-\frac{1}{r}\right)+\frac{1}{r} = \frac{C'}{C}+r C\partial_t{\Psi}^2+r \Psi'^2\,, \label{eq:3of4}\\
  & r B \Psi+\frac{r C'}{2C} \Psi'+\frac{r \partial_t{C}}{2} \partial_t{\Psi}-2\Psi'-r \Psi''+r C \partial^2_t{\Psi} = 0\,. \label{eq:4of4}
\end{align}
%





\subsection{Basic physical picture} \label{sec:OOMbs}

Let us first understand at the order-of-magnitude level characteristics of these configuration. In self-gravitating configurations made up from fermions without the energy source, such as thermonuclear reactions in ordinary stars, e.g. WD and NS, degenerate pressure (originating from Pauli principle) opposes gravitational collapse. In the bosonic case, collapse can be halted only because of Heisenberg's uncertainty principle. Let $R$ be a characteristic size of our configuration (boson star or oscillaton) and $v_\text{vir}$ is virialized velocity. From the uncertainty principle
\beq
R \mu_{\rm a} v_\text{vir} \geqsim 1.
\eeq
As $v_\text{vir} \sim \sqrt{M/R}$, we obtain the mass-radius relation
\beq
\mu_{\rm a}  M \sim \frac{1}{\mu_{\rm a} R}.
\eeq
We can also estimate the maximum mass of these configurations. We expect (e.g. from the hoop conjecture) that the minimal radius of the object is the Schwarzschild radius $R_\text{s}=2M$, so the maximall mass is (restoring SI units)
\beq
M_{\rm max} \sim \frac{\hbar c}{G m_{\rm a}} \sim \frac{M^2_{\rm pl}}{m_{\rm a}}.
\eeq

%

\subsection{Fully relativistic results} \label{sec:ROscillatons}
Oscillaton geometries can be obtained through the expansion of the metric coefficients and the field:
\begin{align}
  & B(t,r) = \sum_{j=0}^{\infty} b_{j}(r) \cos(2 j \omega t)\,, \label{EKGmetricB} \\
  & C(t,r) =  \sum_{j=0}^{\infty} c_{j}(r) \cos(2 j \omega t)\,, \label{EKGmetricC} \\
  & \Phi(t,r) =  \sum_{j=0}^{\infty} \Phi_{j+1}(r) \cos([2 j+1] \omega t)\,, \label{EKGmetricPhi}
\end{align}
truncated at a finite $j$, which depends on the accuracy necessary (for our study $j=1$ is sufficiently accurate).
Accordingly, we will also use reference spacetimes for which,
\begin{align}
  & B(t,r) =  b_{0}(r) + b_{1}(r) \cos(2 \omega t)\,,\label{eq:expB}\\
  & C(t,r) = c_{0}(r) + c_{1}(r) \cos(2 \omega t)\,,\label{eq:expC}\\
  & \Phi(t,r) = \Phi_{1}(r) \cos(\omega t) + \Phi_{2}(r) \cos(3 \omega t)\,.\label{eq:expPhi}
\end{align}

\begin{figure}
\hspace*{\fill}%
\begin{minipage}[t]{0.45\textwidth}
\centering
\vspace{0pt}
\includegraphics[width=\textwidth]{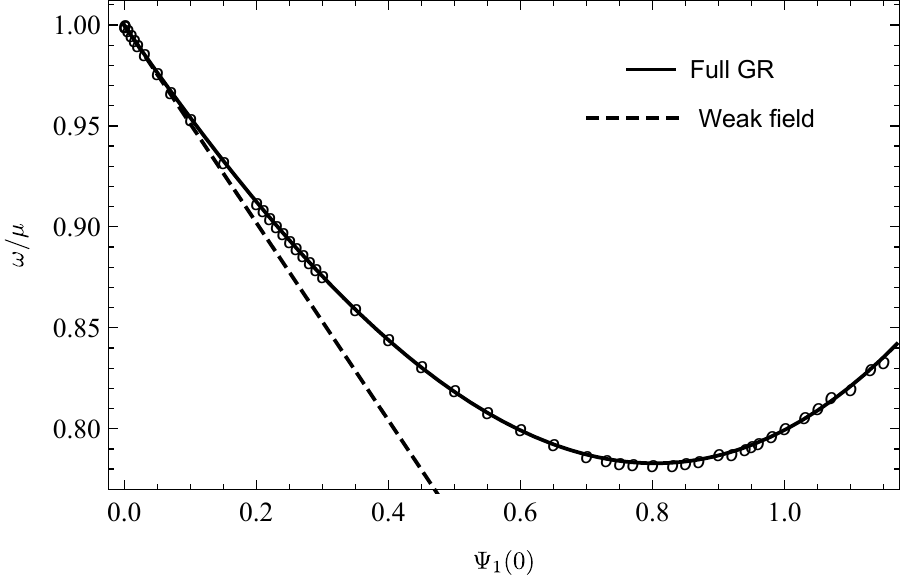}
\caption{\fontsize{9}{12} The fundamental frequency $\omega$ of a scalar oscillaton, as a function of the central value of the scalar field. The minimum value of the frequency is given by $\omega/\mu \sim 0.782$.
The dashed line is the weak-field prediction, discussed in the next section, and agrees well with the full relativistic results at low compactness. Figure credit: \cite{Boskovic:2018rub}.}\label{fig:OvsPhi0}
\end{minipage}%
\hfill
\begin{minipage}[t]{0.45\textwidth}
\centering
\vspace{0pt}
\includegraphics[width=\textwidth]{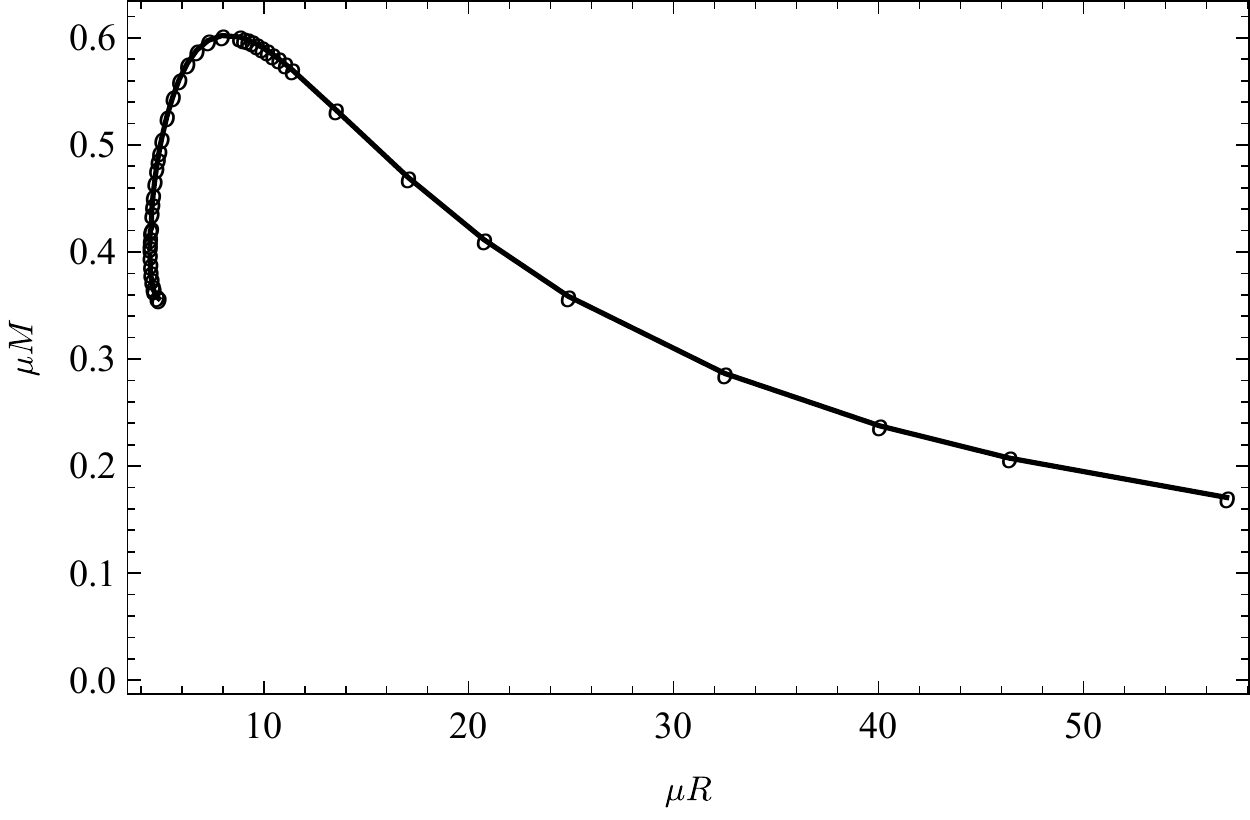}
\caption{\fontsize{9}{12} Mass of the scalar oscillatons as a function of its radius. The maximum mass of stable configurations is $\mu M_{\text{max}} \sim 0.6$ and corresponds to a compactness $\mathcal{C} \sim 0.07$. These values mark the boundary between stable and unstable oscillaton configurations \cite{UrenaLopez:2002gx}: oscillatons with larger radii are stable, and are unstable for smaller radii. It is in the unstable branch of this plot that the maximum compactness is attained, $\mathcal{C}_{\text{max}} \sim 0.1$. Figure credit: \cite{Boskovic:2018rub}.}\label{fig:MvsR}
\end{minipage}%
\hspace*{\fill}
\end{figure}

The coefficients $b_{0},\,b_{1}...$ can be obtained by inserting the expansion above in the equations of motion, and requiring the vanishing of each harmonic term, order by order.
In this particular case, one finds six ODEs for the variables ${b_0,b_1,c_0,c_1,\Phi_1,\Phi_2}$. These equations can be solved numerically using a shooting method \cite{Brito:2015yfh, Boskovic:2018rub}, giving  the profiles of all the components of the metric and the scalar field as well as a value for the fundamental frequency $\omega$ of the oscillaton -- see Fig.~\ref{fig:OvsPhi0}.
%

Notice that since $A(t,r) = B(t,r)/C(t,r)$ [see Eqs.~\eqref{eq:SSmetric} and \eqref{eq:metric2}] , the coefficients of $A$ are obtained like this
\beq
a_0&=& \frac{2 b_0 c_0- b_1 c_1}{2 c_0^2+c_1^2}\,,\\
a_1&=& \frac{2 b_1 c_0 - 2 b_0 c_1}{2c_0^2 - c_1^2}\,,
\eeq
such that $A$ is written as
\be
 A(t,r)\equiv \frac{B}{C} = a_0(r) + a_{1}(r) \cos(2 \omega t)\,.\label{EKGmetricA}
\ee
Given that the solutions are spherically symmetric and asymptotically flat, the effective mass of these configurations is given by the following expression (recovering $\mu_{\rm a}$)
\be
M=\frac{1}{\mu_{\rm a}} \lim_{r \to \infty}\frac{r}{2}\left(1 - \frac{1}{b_{0}(r)}\right)\,.\label{eq:OscillatonMass}
\ee
We (arbitrarily) define the radius of the oscillaton as the location at which $98\%$ of the total mass is contained. This results, obtained in Ref. \cite{Boskovic:2018rub}, are in a good agreement with previous works on the subject -- see Figs.~\ref{fig:OvsPhi0}-\ref{fig:MvsR} and compare with Refs.~\cite{Seidel:1991zh,UrenaLopez:2001tw,Brito:2015yfh}.

The dynamical oscillaton spacetime can be characterized by comparing the magnitude of its time-dependent to its time-independent components.
These quantities depend on the compactness $\mathcal{C}$ of the spacetime,
\be
\mathcal{C} = \frac{M}{R}\,.
\ee
Numerical results indicate that at small $\mathcal{C}$, and restoring the mass $\mu_{\rm a}$, one has
\be
\mu_{\rm a} R \approx \frac{9.8697}{\mu_{\rm a} M}\,,\label{eq_C_num}
\ee
scaling expected from the order-of-magnitude arguments from Section \ref{sec:OOMbs}.

\begin{figure}
\centering
\includegraphics[width=0.5\textwidth]{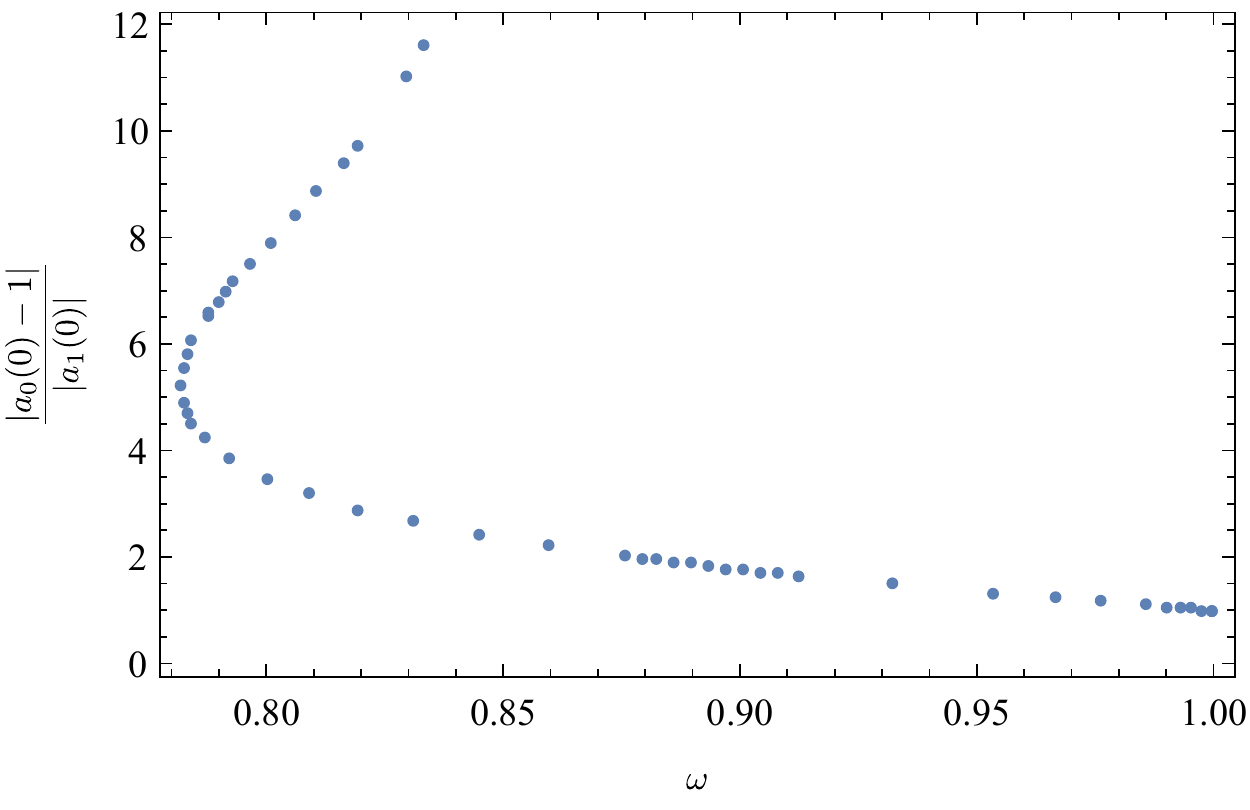}
\caption{Ratio between the effective static and the dynamic part of the metric coefficient for Oscillaton configuration. $\omega$ is given in units of $\mu$.}
\label{fig:osc_static_dynamic}
\end{figure}

At large distances, the scalar profile decays exponentially and the spacetime is described by the Schwarzschild geometry. We thus focus on the metric components close to the origin, $r\ll 1/(M\mu_{\rm a}^2)$. Our numerical results, for $\mathcal{C} < 0.07$, are described by:
\beq
\frac{a_1(0)}{a_0(0)}       &\sim&  6.2 \mathcal{C}+21.8 \mathcal{C}^2 -126 \mathcal{C}^3 + 6160.2 \mathcal{C}^4\,,\\
\frac{|b_1(0.5)|}{b_0(0.5)} &\sim& -0.0003\mathcal{C}+ 0.08 \mathcal{C}^2 - 6.3 \mathcal{C}^3 +325.8\mathcal{C}^4  \,.
\eeq
The error associated is of order $0.3\%$ for $a_1/a_0$ and $2\%$ for $b_1/b_0$ (at the level of accuracy with which we work). From these fits, we see that the time-dependent part of the $g_{tt}$ component isn't always subdominant with respect to the corresponding static part. Unlike the time-dependent part of $g_{rr}$, which remains subdominant for all oscillatons, we see that for $g_{tt}$ the time-dependent part grows such that its magnitude becomes comparable, and even dominant, to the magnitude of the static part. In order to appreciate the dynamical aspect of the spacetime we have subtracted constant asymptotic term from the static part of the metric  $|a_0-1|$ on Fig. \ref{fig:osc_static_dynamic}. The metric itself is not observable object (because of diffeomorphism invariance) and especially in the Newtonian regime only derivatives of the metric coefficients will be relevant (Section \ref{sec:SphSymTimePeriodWeak}) so that this representation clearly underlines the highly dynamical nature of the weak field regime of the $g_{tt}$.

One can take a closer look at the way in which compactness influences the spacetime metric by observing that its components can be written, for $r\mu_{\rm a} < 1$ and $\mathcal{C} < 0.01$, as:
\beq
\label{eq:EKGmetricCRexpansion}
  a_0(r) &=& f_{a0}(\mathcal{C}) + g_{a0}(\mathcal{C})\mu_{\rm a}^2 r^2\\
  a_1(r) &=& f_{a1}(\mathcal{C}) + g_{a1}(\mathcal{C})\mu_{\rm a}^2 r^2 \\
  b_0(r) &=& f_{b0}(\mathcal{C}) + g_{b0}(\mathcal{C})\mu_{\rm a}^2 r^2 \\
  b_1(r) &=& f_{b1}(\mathcal{C}) + g_{b1}(\mathcal{C}) \mu_{\rm a}^2 r^2
\eeq
where the coefficients depend only on the compactness and are given in Table~\ref{tab:funcs}. We have also restored the mass $\mu$ for clarity.
The errors on the corresponding functions, in this range of values, are at most $(0.6,3.4,0.05,3.0)\%$ for $a_0,\,a_1,\,b_0,\,b_1$ respectively.

\begin{table}[]
\centering
\caption{The behavior of oscillaton spacetimes at small radii, as described by \eqref{eq:EKGmetricCRexpansion}, where $\mathcal{C} \equiv M/R$. These results were obtained for $\mathcal{C} < 0.01$.}
\label{tab:funcs}
\begin{tabular}{ll}
$(f_{a0},\,f_{a1})=$ & \!\!\!\!\!\!\!($1 - 6.456 \mathcal{C} - 1673.7 \mathcal{C}^3\,,\quad6.163 \mathcal{C} - 1400.5 \mathcal{C}^3$)\\
$(g_{a0},\,g_{a1})=$ & \!\!\!\!\!\!\!($1.808 \mathcal{C}^2 + 77.162 \mathcal{C}^3\,,\quad -5.486 \mathcal{C}^2 - 2.820 \mathcal{C}^3$)\\
\,\,\,\,\,\qquad$f_{b0}=$ &  \!\!\!\!\!$1 - 0.819 \mathcal{C}^3 + 156.20 \mathcal{C}^4 - 6107.0 \mathcal{C}^5$\\
\,\,\,\,\,\qquad$g_{b0}=$ &  \!\!\!\!\!$1434.94 \mathcal{C}^3 - 163338 \mathcal{C}^4 + 5.816 \mathcal{C}^5$\\
\,\,\,\,\,\qquad$f_{b1}=$ &  \!\!\!\!\!$0.013 \mathcal{C}^3 - 2.34 \mathcal{C}^4 + 64.604 \mathcal{C}^5$\\
\,\,\,\,\,\qquad$g_{b1}=$ &  \!\!\!\!\!$-5.56 \mathcal{C}^3 - 63.71 \mathcal{C}^4 - 976.58 \mathcal{C}^5$
\end{tabular}
\end{table}

%

\subsection{Weak field regime} \label{sec:NOscillatons}

The small compactness regime of oscillatons and boson stars corresponds to the Newtonian-like limit: velocities are small, and the gravitational potential is everywhere weak. We are expanding the frequency of the scalar field around its mass so that, up to the second order in the group velocity $v=p/\mu_{\rm a}$ ($p$ is wave number), we can write
\begin{equation} \label{eq:NomegaSR}
\omega = \mu_{\rm a} + \frac{p^2}{2\mu_{\rm a}} + {\cal O}(p^4)\,.
\end{equation}
As the field is ``trapped'' by self-gravity, $p^{2}<0$ and we expect for the long-range behaviour to be of the form $\psi \sim e^{i p r} \sim e^{-|p| r}$.

\subsubsection{Weak field limit of the Einstein-Klein-Gordon equations for the real scalars}

We review the weak field expansion of EKG, following ~\cite{Seidel:1990jh,UrenaLopez:2002gx,Guzman:2004wj}. First, we write the truncated metric coefficients corresponding to the EKG background, \eqref{EKGmetricB} and \eqref{EKGmetricA}, as slightly perturbed away from the Minkowski metric:
\beq
A(t,r) &=& 1 + 2V(r)+2V_{1}(r) \cos(2 \omega t)\,,\label{eq:NexpA2}\\
B(t,r) &=& 1 + 2W(r)+ 2W_{1}(r) \cos(2 \omega t)\,.\label{eq:NexpB2}
\eeq
The ansatz for the field \eqref{EKGmetricPhi} is
\be  
\Phi(t,r) = \phi(r) \cos(\omega t) \,.\label{eq:NexpPhi2}
\ee

The order of magnitude of the various derivatives in the weak field regime can be estimated using $\phi_{1}(r) =e^{ikr}$. Then:
\beq
&&\partial_t \Phi(t,r) \sim - \mu \phi(r) \sin(\omega  t)+{\cal O}(v^2) \label{eq:NtderivativePhi} \,,\\
&& \Phi'(t,r) \sim {\cal O}(v)\,.  \label{eq:NrderivativePhi}
\eeq
In the Newtonian limit of the Einstein's equations we expect that $V \sim 1/r \sim {\cal O}(v^2) $ and $W \sim {\cal O}(v^4) $ and mutatis mutandis for $V_{1}$ and $W_{1}$. Differentiating \eqref{eq:NexpA2} with respect to the time and radial coordinates we obtain:
\beq
&& A'(t,r) \sim {\cal O}(v^4) \,, \label{eq:Nrderivativea} \\
&&\partial_t A(t,r) =- 2\mu V_1(r)\sin(2 \omega t) + {\cal O}(v^4) \,. \label{eq:NtderivativeA}
\eeq
%

Applying this ansatz to \eqref{eq:4of4} we obtain
\be \label{eq:NSchrodinger}
e\phi=-\frac{1}{2\mu^2 r}(r\phi)''+V\phi\,,
\ee
with $e=k^2/(2\mu)<0$. 

In order to get Poisson equation, we will follow the approach of obtaining weak-field limit of a relativistic star \cite{bookWeinberg:1972}. We will introduce notation $\nu=2V(r)+2V_{1}(r)\cos(2 \omega t)$ and $\sigma=2W(r)+2W_{1}(r)\cos(2 \omega t)$. Einstein tensor components $G_{tt}$ and $G_{rr}$ are:
\beq
G_{tt}&=&\frac{A}{r^2} \Big( \frac{1}{B} \Big( \frac{B'}{B}r-1   \Big)+1\Big)\,,\\
G_{rr}&=&\frac{B}{r^2} \Big( \frac{1}{B} \Big( \frac{A'}{A}r+1   \Big)-1\Big)\,.
\eeq
Using \eqref{eq:NexpA2} and \eqref{eq:NexpB2} and expanding up to $v^4$ we get:
\beq
G_{tt}r^2&=&\sigma+ r \sigma'\,, \\
G_{rr}r^2&=&r\nu' -\sigma\,.
\eeq
Equating last two expressions with the corresponding component of the stress-energy tensor \eqref{eq:MFEoMTensor} we obtain:
\beq
(\sigma r)'&=& 8\pi r^2T_{tt} \,,\label{eq:NEinsteintt}\\
\nu' &=& 8\pi rT_{rr}+\frac{\sigma r}{r^2} \,.  \label{eq:NEinsteinrr}
\eeq
Differentiating equation \eqref{eq:NEinsteinrr} and combining it with \eqref{eq:NEinsteintt}
\be
\nu''+\frac{2\nu'}{r}=8\pi ( r T'_{rr}+T_{rr}+T_{tt})\,.\label{eq:NalmostPoisson}
\ee
If we reinstate $c$ and remember that $G_{\mu \nu}=(8\pi/c^4)T_{\mu \nu}$, it stands out that the term in brackets on the right hand side of the last equation should be treated up to $c^2$ order i.e.
\be
rT'_{rr}+T_{rr}+T_{tt}=\frac{1}{2}\phi^2\mu^2-\cos(2\omega t)\left(\frac{1}{2}\phi^2\mu^2+r\mu^2\phi \phi' \right)\,.
\ee
If we redefine the scalar, $\phi = \psi/\sqrt{8 \pi}$, and equate terms in front of the $\cos(0)$ and $\cos(2\omega t)$ on both sides of the equation we find:
\beq
&&V''+\frac{2V'}{r}=\frac{1}{2} \psi^2\mu^2 \,,\label{eq:NPoissonApp}\\
&&V''_{1}+\frac{2V'_{1}}{r}=-\frac{1}{2} \psi^2\mu^2-r\mu^2\psi \psi'\,.\label{eq:NTimePoisson}
\eeq
Equation \eqref{eq:NPoissonApp} is nothing but Poisson equation. We rewrite \eqref{eq:NTimePoisson} as,
\be
V_{1}'= -\frac{1}{2} r\psi^2\mu^2+\frac{1}{2r}\int^{r}_{0} r^2\psi^2\mu^2 dr\,.\label{eq:NTimePoisson2}
\ee
We see from \eqref{eq:NEinsteinrr} that $\sigma \sim \partial_{r}{\nu} \sim {\cal O}(v^4)$ as claimed. Finally, using \eqref{eq:NEinsteintt} we see that the second term on the r.h.s of \eqref{eq:NTimePoisson2} is of order $o(v^6)$. Thus, the second term (mass) on the r.h.s. of \eqref{eq:NTimePoisson2} is smaller than the first and can be neglected.
Therefore, setting $\mu_{\rm a}=1$, we obtain:
\beq
(rV)''&=&\frac{1}{2} r\psi^2\,, \label{eq:NPoisson} \\
V'_{1}&=&-\frac{1}{2} r\psi^2\,. \label{eq:NV2}
\eeq
%

\subsubsection{Perturbative description of the Newtonian oscillatons profile}

In the previous subsection we showed that, up to second order in $v$, EKG system reduces to Eqs. \eqref{eq:NSchrodinger}, \eqref{eq:NPoisson} i \eqref{eq:NV2}. We refer to $V(r)$ as the Newtonian potential and to $V_{1}(r)\cos(2\omega t)$ as the time-dependent potential. Note that the equations \eqref{eq:NSchrodinger} and \eqref{eq:NPoisson} are decoupled from \eqref{eq:NV2} and form the Schrödinger-Poisson (SP) system~\cite{Hui:2016ltb, Guzman:2004wj}. When the additional self-interacting potential is present this system is called Gross-Pitaevskii-Poisson  system \cite{Chavanis:2011} (see Appendix \ref{AppGPP}).
Equation \eqref{eq:NV2} is present for oscillatons (and not for boson stars where the field is complex and harmonic) and is responsible for the time-dependence of the $A(t,r)$ metric coefficient. As we have chosen a normalization of the wavefunction in the form $\int dV |\psi(r)|^2=N$, where $N$ is the number of the particles in the system, we can find the mass of the Newtonian oscillaton as $M=\int_{0}^{\infty} d V \rho(r)$, where $\rho(r) \equiv T_{tt}=\psi^2/(8\pi)$, and see that by definition it does not depend on the function $b_0(r)$ as is the case in general (\ref{eq:OscillatonMass}) and as expected from fully relativistic analysis (see Figure 1 in Ref.~\cite{Brito:2015yfh}).

Analytical solutions for these systems in general do not exist but there is a high precision approximate analytical solution
in the case of the non-self-interacting fields~\cite{KlingRajaraman:2017}, which is the focus of this work.
Non-self-interacting oscillatons exhibit a Yukawa-like behavior at large distances. Thus, there is no well-defined notion of surface, even at a Newtonian level. The radius of this kind of object is defined as we did in the fully relativistic case.

As the SP system admits scale symmetry, solutions corresponding to different masses can be obtained from a unique solution by rescaling~\cite{Guzman:2004wj,KlingRajaraman:2017}.
The scaling that leaves SP system invariant for various parameters is given by,
\beq
r    &\to& \frac{r}{\lambda}\,,\,e \to \lambda^2 e\,,\nonumber\\
\psi &\to& \lambda^2 \psi\,,\,V \to \lambda^2 V\,,\, V_{1} \to \lambda^2 V_{1}\,,\, M  \to \lambda M\,,\label{eq:NSPscale}
\eeq
where $\lambda$ is the scale factor. We will fix this factor as in Ref.~\cite{KlingRajaraman:2017} by identifying $2\lambda^2=-e$. A scale-independent field is found by expanding field around zero value of the radial coordinate and at infinity and matching these solutions. The free parameters are found by fitting this solution onto the numerical solution of the scale invariant SP system. These parameters are proportional to the scale invariant value of the central $(s_{0})$ and long-range $(\alpha)$ field expansion, the scale invariant mass $(\beta)$ and linearly related to the central value of the scale-invariant Newtonian potential $(v_{0})$. Technical details are collected in the Box on the page~\pageref{subsec:NOpert}. The numerical values of these parameters, along with the scale invariant radius $Z$, are:
\beq
s_{0} &=&  1.022\,,\quad v_{0} =  0.938 \nonumber\,,\\
\alpha &=& 3.495\,,\quad \beta =  1.753\nonumber\,,\\
Z  &= &  5.172\,.
\eeq
From Eq.~\eqref{eq:NSPscale}, it is obvious that the scaling between mass and radius is of the form
\begin{equation} \label{eq:R_M_N_Osc}
R = \frac{Z\beta}{M}=\frac{9.065}{M}
\end{equation}
and $\lambda=\sqrt{\mathcal{C} Z/\beta}$. Notice the excellent agreement with the low compactness full numerical result,
Eq.~\eqref{eq_C_num}. From the scaling relations, we can find the dependence of the field frequency \eqref{eq:Nomega} on the central value of the field
\begin{equation}
\omega=1-\frac{\psi(0)}{2s_{0}}.
\end{equation}
The plot of this function is superposed on the relativistic $\omega-\Psi(0)$ plot  (Fig. \ref{fig:OvsPhi0}). We can see that the agreement for small values of $\Psi(0)$ is very good.


\begin{framed}
\noindent
{\small {\it Analytical profile of Newtonian oscillatons}.\\ ~\label{subsec:NOpert}


From scaling symmetry \eqref{eq:NSPscale}, one can define the scale-invariant field $s=\psi /\lambda^2 $, where $\lambda=\sqrt{\mathcal{C} Z/\beta}$ (scale factor), is found by expanding around zero value of the radial coordinate and at infinity and matching these solutions. Once the field is known, density can be found as $\rho=\mu_{\rm a}^2\psi^2/(8\pi)= \Lambda (
\mu_{\rm a} s)^2$ and $\Lambda=\lambda^4/(8\pi)$. Expansion of the scale-invariant field around the center is given by
\beq
s_{<}=\sum^{\infty}_{n=0}s_{n}z^n, \label{eq_appB_small_r_exp}
\eeq
where $z$ is the scale-invariant radial coordinate $z=\lambda \mu_{\rm a} r$. This expansion is not convergent after $z>4$ \cite{KlingRajaraman:2017}. At large radius adequate expansion is of the form
\beq
s_{>}=\sum^{\infty,\infty}_{n,m=0,0} s^n_{m}\Big(\frac{e^{-z}}{z^\sigma} \Big)^n z^{-m}. \label{eq_appB_large_r_exp}
\eeq
The series in $m$ is only asymptotic to the $s$, for large $z$, and in Ref. \cite{KlingRajaraman:2017} optimal asymptotic approximation (see \cite{benderbook}) is performed by truncating the series with the adequate $m_\star$. From this expansion we see that the long-range behaviour of the density is
\be
\rho(r) \sim \Lambda \mu_{\rm a}^2 \alpha^2 (\lambda \mu_{\rm a} r)^{2\sigma} e^{-2 \lambda \mu_{\rm a}  r} \,,
\ee
where $\alpha=s^1_0$, $\sigma=1+\beta$.

Object linearly related to the scale-invariant Newtonian gravitational potential is defined as
\be
2\Big (\frac{e}{\mu_{\rm a}} - V \Big )=\lambda^2 v \,.
\ee
Expansions for $v$ have the same form as for $s$:
\beq
v_{<}=\sum^{\infty}_{n=0}v_{n}z^n \,,\, v_{>}=\sum^{\infty,\infty}_{n,m=0,0} v^n_{m}\Big(\frac{e^{-z}}{z^\sigma} \Big)^n z^{-m}. \label{eq_appB_v_exp}
\eeq
Series coefficients can be found by inserting expansions for $s$ and $v$ into SP system \cite{KlingRajaraman:2017}. Then, the expansions are matched at the matching point $2.5<z_{\star}<3.5$ and free parameters $s_{0}$, $\alpha$, $\beta$ and $v_{0}$ are found by fitting onto numerically obtained solutions. For the parameter values, we used one given in Eq. 31 in Ref. \cite{KlingRajaraman:2017} and reconstructed terms up to $n_\star =50$ for $s_{(n_\star)} \approx  s_{<}$ and $n_\star  = 3 , m_\star = 6 $ for $s^{(n_\star)}_{(m_\star)} \approx s_{>}$, where $n_\star$ and $m_\star$ refer to orders of series truncation, with $z_{\star}=3$. Value of the scale-invariant radius $Z$ is found by inverting $m(Z)=0.98 M$, where $M$ is the total mass and $m(Z)$ is the Newtonian mass function.

}

\end{framed}


We will now provide comparison between small radius metric coefficients expansion in terms of compactness $\mathcal{C}$ obtained in fully relativistic analysis summarized in \eqref{eq:EKGmetricCRexpansion} and Table \ref{tab:funcs} and in Newtonian limit. The small $r$ behaviour of Newtonian oscillaton density is (see Box on page ~\pageref{subsec:NOpert})
\begin{equation} \label{eq:Ndensitysmallr}
\rho(r) = \Lambda (a+b(\lambda r)^2) + {\cal O}(r^4)\,,
\end{equation}
where $\Lambda=\lambda^4/8\pi $, $a=s_{0}^2,\,b=-s_{0}^2 v_{0}/3$.

Newtonian oscillatons do not have defined surface and the normalisation procedure for the Newtonian potential is not the same as for the sphere in Newtonian gravity. We have
\begin{equation}
V(r) = -\int^{\infty}_{0}\frac{dr}{r^2}m(r)+\int^{r}_{0}\frac{d\Tilde{r}}{\Tilde{r}^2}m(\Tilde{r})\,,
\end{equation}
where $m(r)$ is the Newtonian mass function.
The first term -- proportional to $\mathcal{C}$ (as can be seen from a dimensional analysis) -- is integrated using the full expansion described in the box. The second term reduces to $2\pi \Lambda a r^2/3$ at ${\cal O}(r^3)$. Similarly
\begin{equation}
V_{1}(r) =4\pi \int^{\infty}_{0}dr r\rho (r)-4\pi \int^{r}_{0}d\Tilde{r}  \Tilde{r} \rho(\Tilde{r}).
\end{equation}
The small-$r$ expansion for the second term gives us $-2\pi\Lambda a r^2$  at ${\cal O}(r^3)$. The first, of the order $\mathcal{C}$, is integrated using the full expansion. We get the following results for the parameters defined in \eqref{eq:EKGmetricCRexpansion},
\beq \label{eq:Nmetricexpansion}
f_{a0}&=& 1 - 5.720 \mathcal{C}\,,\quad f_{a1}= 5.720 \mathcal{C}\,,\\
g_{a0}&=& 1.514 \mathcal{C}^2\,,\quad g_{a1}= - 4.543 \mathcal{C}^2\,,\\
f_{b0} &=& 1\,,\\
g_{b0} &=& g_{b1}=f_{b1}=0\,.
\eeq
in very good agreement with respect to fully relativistic expansion from Table~\ref{tab:funcs} (notice that the fully relativistic expansion is restricted to only mildly Newtonian oscillatons).

For small $r$, $V_{1}$ is larger in magnitude than the Newtonian potential. This seemingly odd result was recognized in Ref.~\cite{UrenaLopez:2002gx}. The physical origin of this property can be traced to the scalar pressure, which is of the same order of magnitude as the energy density. Calculating the stress-energy tensor in a spherically-symmetric spacetime \eqref{eq:SSmetric}
\be
T_{tt}=\frac{1}{2}(\partial_t \Phi)^2 + \frac{1}{2}\frac{A(t,r)}{B(t,r)}(\Phi')^2+\frac{1}{2}A\mu^2\Phi^2 \,,
\ee
and using the weak-field ansatz \eqref{eq:NexpPhi2} we obtain
\beq
\rho=T_{tt}=\frac{1}{2}\phi^2 \mu^2_{\rm a}+\mathcal{O}(v^2)\,, \\
T_{rr}=-\frac{1}{2}\phi^2 \mu^2_{\rm a} \cos{(2\omega t)}+\mathcal{O}(v^2) \,.
\eeq
As the weak-field limit is dynamical we are in a weak-field but Newtonian-like limit. The gradient of the Newtonian potential is dominated~\footnote{This fact, that there is no \textit{weakly dynamical} approximation of the oscillatons was mistakenly interpreted as an absence of the \textit{weak field} limit~\cite{Seidel:1991zh}. As this time-dependent potential is not important to the exploration of the oscillaton structure, owing to the fact that the \eqref{eq:NV2} is decoupled from the SP system, its existence was not explored further in the most of the literature. On the other hand, this pressure is important for understanding the values of the metric coefficients, and ipso facto for understanding the motion of test particles in this background as recognized in \cite{Khmelnitsky:2013lxt} and upon we will further comment in Part \ref{ch:orbital}.} by the magnitude of the gradient of the time-dependent potential for $\lambda r \leqsim  0.57Z$ and becomes an order of magnitude larger at $\lambda r \approx 1.06Z$. \\

\subsection{Cosmological production}\label{sec:axion_conf_cosmo_astro}

\subsubsection{Structure formation with Fuzzy DM} \label{sec:darkhalo_descr}

As elaborated in Section \ref{sec:DM_axions} and Section \ref{sec:small_scale_chalenges}, axion DM particles with masses around $\mu_{\rm a} \sim 10^{-23} \rm{eV}$ (FDM) could have interesting consequences for the galaxy structure and dynamics. Notably, in their cores could form Newtonian oscillatons of the $\sim \rm{kpc}$ scales [Eq. \eqref{eq:FDM_de_broglie}]. The connection between Newtonian oscillatons and DM halos is not straightforward. It is theoretically expected that a dark halo consists of a nearly homogeneous core surrounded by particles which are behaving like CDM \cite{Marsh:2015wka}. The density profile of such effectively cold, DM region is described by the NFW profile~\eqref{eq:NFW}. Both cosmological and galaxy formation simulations of fuzzy DM of several groups \cite{Schive:2014dra, Veltmaat:2016rxo,Mocz:2017wlg}  have confirmed such a picture and revealed non-local scaling relations between the parameters that describe the soliton and the whole halo ~\footnote{There is a different type of this scaling between Refs. \cite{Schive:2014hza,Veltmaat:2016rxo} and Ref. \cite{Mocz:2017wlg}. In Ref. \cite{Bar:2018acw} it was argued that the mismatch between these two scalings is a consequence of the unnatural choice of the initial conditions in the Ref. \cite{Mocz:2017wlg}}.

We can describe approximate density profile for the whole halo as~\cite{Marsh:2015wka}
\begin{equation}  \label{eq:halodensity}
\rho(r)=\rho_{\text{sol}}(r)\theta(r_{\epsilon}-r)+\rho_{\text{NFW}}(r)\theta(r-r_{\epsilon})\,,
\end{equation}
where
\begin{equation} \label{eq:solitondensity}
\rho_{\text{sol}}(r)=\frac{\rho_c}{(1 + 0.091(r/r_c)^2)^8} \,,
\end{equation}
is an oscillaton density. In the previous equations $\theta$ is the Heaviside function, $\rho_c$ is the central density of the soliton, $r_c$ (the core radius) is the point at which the density falls off to half of its central value, $r_{\epsilon}$ is soliton-NFW transition radius. Demanding continuity of the soliton and NFW densities at the transition (and optionally their first derivative), we are left with only four (three) free parameters which can be found by fitting galactic rotation curves. The soliton density function~\eqref{eq:solitondensity} was found by fitting onto results of galaxy formation simulations~\cite{Schive:2014dra}. The fitted density distribution \eqref{eq:solitondensity} for the soliton is in excellent agreement with our approximate analytical solution of Section~\ref{sec:NOscillatons}. One of the two soliton parameters $(\rho_c, r_c)$ can be replaced instead by the axion particle mass. This is a global parameter independent of the galactic details. From the definition of $r_c$ and the scaling in Eq. \eqref{eq:NSPscale},
\begin{equation} \label{eq:fuzzy_cdensity}
\rho_c=1.94 \times 10^{-2} \Big ( \frac{r_c}{1\text{kpc}} \Big )^{-4} \Big ( \frac{\mu}{\mu_{22}} \Big )^{-2}  \frac{M_{\odot}}{\text{pc}^3}.
\end{equation}
In the last we used our analytical profile to obtain the numerical prefactor. Simulations indicate that the transition radius usually corresponds to $r_{\epsilon} \approx 3.5 r_c$ \cite{Mocz:2017wlg}. The scale-invariant radius of that point is $Z_s \equiv \lambda \mu r_{\epsilon} \approx 1.035 Z$. Profile \eqref{eq:halodensity} was used for fitting galactic rotation curves~\cite{Marsh:2015wka,GonzalezMoralesMarsh:2017, Bernaletal:2017a}. We will use reference parameters for the Milky Way (MW), estimated in Refs.~\cite{Schive:2014dra, DeMartinoBroadhurst:2017}: $m=0.8m_{22}$ and $r_c=120 \text{pc}$, for which $\rho_c=146 M_{\odot}/\text{pc}^3$.

Oscillaton profile is cored so in this way FDM could solve the small scale cusp-core problem \cite{Marsh:2015wka}. Recently, this picture has been contested - further analysis showed that not only the oscillaton profile is not adequate to match the observationaly found cores \cite{Deng:2018jjz} but the existence of the oscillaton predicts rotation curve artefacts not found in the observations \cite{Bar:2018acw, Bar:2019bqz}. These artefacts allowed for constraining $\mu_{\rm a} \leqsim 10^{-21} \rm{eV}$. These constraints match the cosmological ones from Ly-$\alpha$ forest (see Figure \ref{fig:ULA_constraints} and \cite{Marsh:2016rep}).

More recent core zoom-in simulations have also found exited quasi-normal modes of the FDM halo cores\footnote{For the systematic investigation of quasi-normal modes of Newtonian oscillatons see Ref. \cite{Guzman:2018bmo}.} \cite{Veltmaat:2018dfz}. These oscillatons can be considered as the De Broglie scale oscillations, compared to the Compton scale ones analysed in Section \ref{sec:NOscillatons}. Long-term effect of such oscillations on the old stellar cluster in ultrafaint dwarf galaxy Eridanus II have put constraints for FDM $\mu_{\rm a} \leqsim 0.8 \cdot 10^{-21} \rm{eV}$ with the potential of further analysis to probe $\mu_{\rm a} \leqsim \cdot 10^{-19} \rm{eV}$. \cite{Marsh:2018zyw}. Constraints from the Compton scale oscillatons will be discussed in Section \ref{sec:motion_pheno}.

There are still reasonable caveats that allow for further investigation of FDM and reevaluation of most of the mentioned constraints and we mention some of them:
\begin{itemize}
    \item Cosmological production of oscillatons has not been confirmed in cosmological simulations in the whole range of FDM masses; this assumption is at the core of the some of the above arguments;
    \item Proper investigation of the baryon effects on the FDM structure has not been performed in the simulations;
    \item It has been argued at the order-of-magnitude level that strong self-interactions can alter the structure formation \cite{Desjacques:2017fmf}, this yet has to be both confirmed in cosmological simulations and the consequences for the existence and structure of the oscillatons have to be investigated.
\end{itemize}

%
%

\subsubsection{Axion DM clumps and relativistic axion stars}

Besides large cores in the FDM range, there are also arguments for the existence of smaller objects, both dilute (DM clumps) \cite{Guth:2014hsa,Levkov:2018kau} and compact (relativistic axion stars) \cite{Helfer:2017a}. One way for these objects could form is through enhanced axion  power spectrum on small scales \cite{Widdicombe:2018oeo}. These objects could reveal themselves through GW signals from mergers \cite{Helfer:2018vtq,Clough:2018exo} but also in other window, notably axion-photon resonances (Part \ref{ch:ax-photon}) \cite{Tkachev:2014dpa,Hertzberg:2018zte}.

\pagebreak

\section{Axion configurations gravitating around Black Holes} \label{sec:ax_grav_BH}

Besides cosmological production of axions (and gravitating axionic configurations), there are other channels related to the instabilities of BH spacetimes. One of the most explored ones is related to superradiant instability of Kerr BHs. The consequences of this instability could be probed through electromagnetic and gravitational astronomy.

As reviewed in Appendix \ref{AppKerr} Kerr spacetime admits ergoregions, where there are no stationary observers. An ergoregion allows for extracting energy and angular momentum from Kerr BHs. At the fundamental level, this extraction can be realized through particle and fluid or field processes. In the first sense this extraction process is labelled as a Penrose process and in the second superradiance. Here we sketch the basic physics behind this process, while a detailed overview including historical references, generalizations and applications can be found in \cite{Brito:2015oca}.

\subsection{Superradiance instability of Kerr Black Holes}

Let us imagine an incoming scalar wave into the ergoregion
\beq
\Psi \sim \text{Re} [e^{-i\omega t + i m \varphi}f(r,\theta)] \,.
\eeq
From the time-like Killing vector we can form the covariantly conserved\footnote{Proof: $\nabla_\mu J^\mu = (\nabla_\mu T^{\mu \nu}) (\xi_t)_\nu + T^{\mu \nu} \xi^\sigma_t \nabla_\mu g_{\sigma \mu}=0.$} energy current of the field
\beq
J_\mu = - T_{\mu \nu} \xi^\nu_t \,,
\eeq
while from the stress-energy tensor we find
\beq
J_\mu= g_{\mu t} \mathcal{L}+\partial_t \Psi \partial_\mu \Psi \,.
\eeq

\begin{figure}
\centering
\includegraphics[width=0.5\textwidth]{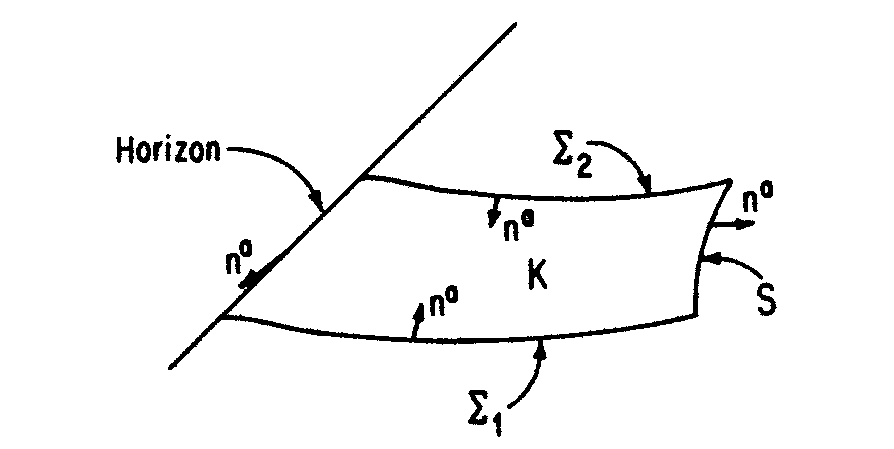}
\caption{With the description of the Kerr BH superradiant instability. Figure credits: \cite{waldbook}}
\label{fig:SR}
\end{figure}

The region where this current is conserved is bordered by the (outer) horizon, spatial infinity and two spacelike hypersurfaces at constant initial $t_i$ and final $t_f$ time through which we evaluate the energy flux (see Fig. \ref{fig:SR}). From the Stokes theorem (see Box on page \pageref{}) we find
\beq
E(t_f)-E(t_i)=-\Delta E_{\rm BH} \,
\eeq
with change of the energy at the BH horizon (corresponding to the \textit{ingoing} flux) given by
\beq
\Delta E_{\rm BH}=\int^{t_f}_{t_i} dt \int dS_{\rm BH}|_{r_+} \, J_\mu l^\mu \,,
\eeq
where $\int dS_{\rm BH}|_{r_+}$ is the angular integral evaluated at the BH horizon and $l^\mu$ is the $4$-vector that defines the Kerr BH horizon \eqref{eq:horizon_null_surface}. Evaluating the last result we obtain
\beq \label{eq:energy_extraction_BH}
\Delta E_{\rm BH}=\int^{t_f}_{t_i} dt \int dS_{\rm BH}|_{r_+} \, \omega (\omega - m \Tilde{\Omega}_+)\sin^2(\omega t - m\varphi)|f(r,\theta)|^2 \,,
\eeq
and $\Tilde{\Omega}_+$ angular velocity at the horizon (Appendix \ref{AppKerr}). From the signs of the terms we see that if the superradiant condition
\beq \label{eq:SR_condition}
\frac{\omega}{m}<\Tilde{\Omega}_+ \,
\eeq
is met, the wave can drain the energy ($E(t_f)>E(t_i)$) (and the angular momentum as one can show similarly) from the BH. This phenomenon is a wave phenomenon analogue to the Penrose process for particles\footnote{Superradiance does not occur for fermionic fields \cite{Brito:2015oca} and the precise relation between the Penrose process and superradiance is not entirely clear \cite{Vicente:2018mxl}.}.

Press and Teukolsky imagined a BH bomb scenario \cite{Press:1972zz} where the BH is surrounded by a reflection mirror and the superradiant condition is met. Then, due to the avalanche-like  process (the wave is amplified and then reflected back from the mirror iteratively) instability occurs (BH bomb). Massive particles are natural ``mirrors'' as their mass confines them around the BH. In such way extended configurations (scalar clouds) could be produced. Similar phenomena occurs also for vector and tensor fields \cite{Brito:2015oca}.

\begin{framed}
\noindent
{\small {\it Stokes' theorem}\\

Stokes' theorem is a differential-geometric generalization of several theorems of multivariable and vector calculus
\beq
\int_M d\omega=\int_{\partial M} \omega
\eeq
with $M$ being the $n$-dimensional compact orientable manifold with the boundary $\partial M$ and $\omega$ is an $n-1$ form on $M$. Coordinate representation of this theorem in the form useful in GR is \cite{carrollbook}
\beq
\int_M d^n x \sqrt{-g}\nabla_\mu V^\mu=\int_{\partial M} d^{n-1} y \sqrt{-\gamma}n_\mu V^\mu
\eeq
with $\gamma_{\mu \nu}$ being the induced metric on a submanifold $\partial M$ and $\omega=\star V$.

}
\end{framed}

\subsection{Scalar clouds around Kerr Black Holes}

Imagine now that the superradiant instability produces an extended configuration (scalar cloud). Here we will describe the properties of these objects.  We neglect the backreaction of the scalar field onto the geometry, an approximation which is justified both perturbatively and numerically~\cite{Brito:2014wla} as we will elaborate more in the Section \ref{sec:sr_rates}.

\subsubsection{Basic physical picture}  \label{sec:grav_atom_oom}

For most of the parameter space region, the cloud spreads over a large volume around the BH  and the system is in the weak field regime. Similarly to discussion in Section \ref{sec:NOscillatons} we have, up to the second order in the group velocity $p/\mu_{\rm a}$,
\begin{equation} \label{eq:Nomega}
\omega=\mu_{\rm a}+\frac{p^2}{2\mu_{\rm a}}+{\cal O}(p^4)\,.
\end{equation}

The length scale associated with the particle's momentum is De Broglie wavelength $\lambda_{\text{D}} =2\pi/ |p|$ and in the near-horizon limit the important scale is the gravitational radius $r_g \sim M$. We will use the virial theorem\footnote{The leading order behaviour of the weak-field gravitational potential is $\propto 1/r$.} $(2T \simeq V)$ to understand the dependence of the average size of the cloud $r_c$ (where the typical particle is located) on $\mu_{\rm a}$ and $M$:
\begin{equation} \label{eq:cloud_radius}
\frac{|p^2|}{\mu_{\rm a}} \sim \frac{M \mu_{\rm a}}{r_c}.
\end{equation}
The de Broglie wavelength of the wave on radius $r_c$ depends on the number of modes excited as $n\lambda_{\text{D}}=2r_c \pi$. We find (Bohr radius)
\begin{equation}
r_c \simeq \frac{n^2}{\mu_{\rm a} \alpha},
\end{equation}
where $\alpha$ is the fine structure constant
\begin{equation}
\alpha = \frac{r_g}{\lambda_{c}} =\mu_{\rm a}M,
\end{equation}
and $\lambda_{c}=1/\mu_{\rm a}$ is the (reduced) Compton wavelength. Finally, we see that the behaviour of the real part of the spectrum is the same as for the hydrogen atom, mutatis mutandis:
\begin{equation} \label{eq:Nomega2}
\omega_n=\mu_{\rm a} \Big(1- \frac{\alpha^2}{2n^2} \Big).
\end{equation}
%

\subsubsection{Weak field approximation of Klein-Gordon equation on Kerr BH background}  \label{}

Neglecting backreaction, let us work in the perturbative framework at the first order. There is no known exact analytical solution to the Klein-Gordon equation in Kerr spacetime, although analytical progress can be made. Teukolsky showed that the symmetries of the spacetime allow for separation of variables \cite{Teukolsky:1973ha}:
\beq \label{eqn: Psi decomposition}
\Psi=\text{Re}\Big[ \int \frac{d\omega}{2\pi}\sum^{}_{l,m}e^{-i\omega t + i m \varphi}S_{lm}(\theta)R_{\omega lm}(r) \Big],
\eeq
where $S_{lm}$ is the spin-weighted spheroidal harmonic and $R_{lm}$ is radial function. The spheroidicity parameter $c^2=a^2(\omega^2-\mu_{\rm a}^2)$ goes to zero when $\omega \sim \mu_{\rm a}$  and $\lim_{c \to 0}e^{i m \varphi}S_{lm}(\theta) = Y_{lm}(\theta,\varphi) $, where $Y_{lm}$ are spherical harmonics \cite{Berti:2005gp}. As first shown in \cite{Detweiler:1980uk} in the $\alpha \ll 1 $ limit, the radial function can be analytically solved in the near-horizon and long-range regions and asymptotically matched. We will here focus on the long-range behaviour and expand the Lagrangian for the scalar field in terms of $\alpha$.

We assume the following ansatz for the field
\beq
\Psi =  \frac{1}{\sqrt{2\mu_{\rm a}}} \left[ \psi \, e^{-i \mu_{\rm a} t} + \psi^\ast  \, e^{+ i \mu_{\rm a} t}  \right]  , \label{eqn: real scalar field 2}
\eeq
working in the regime $\alpha \ll 1$ and $r \sim r_c \simeq (\mu_{\rm a} \alpha)^{-1}$ \eqref{eq:cloud_radius}. Compare and contrast the approach with one for the weak field limit of self-gravitating structures in Section \ref{sec:NOscillatons}. If we insisted that $\psi=\psi^\ast$, we would obtain the same ansatz as \eqref{eq:NexpPhi2}. In essence, there we considered a spherically-symmetric background that admits time-inversion so we eliminated the $\sin$ term in the weak-field expansion. In the Kerr background we can't do this (Appendix \ref{AppKerr}). The action for $\psi$ reads
\beq
S = \int d^4 x \sqrt{-g} \left( - \frac{1}{2 \mu_{\rm a}} \left[  \nabla_\alpha \psi^\ast \nabla^\alpha \psi + i \mu_{\rm a} g^{t \alpha}  \left( \psi^\ast \nabla_\alpha \psi -  \psi \nabla_\alpha \psi^\ast \right) + \mu_{\rm a}^2 ( g^{tt} + 1) \psi^\ast \psi \right]  \right)  , \label{eqn: psi action}
\eeq
We illustrate here the scaling arguments, based on order-of-magnitude estimates from Section \ref{sec:grav_atom_oom}
\beq
\frac{M}{r} \sim  M\mu_{\rm a} \alpha &\sim& \alpha^2 \,, \\
\Big( \frac{a}{r} \Big)^k=\frac{\Tilde{a}^k M^k}{r^k} &\sim& \Tilde{a}^k \alpha^k \,, \\
\frac{aM}{r^3} &\sim& \tilde{\alpha} \mu_{\rm a} \alpha^5
\eeq
Note that the dominant and subdominant frequency contributions are separated so
\beq
\psi \sim e^{ikr}e^{-ip^2/(2\mu_{\rm a})t} \,,
\eeq
Using similar arguments as in \eqref{eq:NtderivativePhi} and
\eqref{eq:NrderivativePhi} we estimate the derivatives scaling
\beq
\psi' \sim \mu_{\rm a}\alpha \, \\
\partial_t \psi \sim \mu_{\rm a} \alpha^2 \,.
\eeq

Varying the Lagrangian at the lowest order (see Box on page \pageref{box:scalar_cloud_expansuon} for the details of the expansion) we obtain the Schrodinger equation for the hydrogen atom
\begin{align}
i \frac{\partial }{\partial t} \psi(t, \textbf{r}) = \left[ - \frac{1}{2\mu_{\rm a}} \nabla^2 - \frac{\alpha}{r} \right] \psi(t, \textbf{r}) \, .  \label{eqn: Schrodinger like with gravity}
\end{align}
We further expand $\psi$ in terms of the (quasi-)stationary eigenstates\footnote{We follow \cite{Baumann:2018vus, Berti:2019wnn} with the notation for the principal number. In the superradiance literature more commonly $\tilde{n}=n+l+1$ is used instead of $n$.}
\beq
\psi(t, \textbf{r}) = \int \frac{d\omega}{2\pi}  \sum_{l m}  \psi_{n l m}(t, \textbf{r}) \, ,
\eeq
and using~\eqref{eqn: Psi decomposition}, we can identify (in the small spheroidicity approximation)
\begin{align}
\psi_{n \ell m} (t, r, \theta, \varphi)  \simeq  e^{-i (\omega - \mu_{\rm a})t}  \bar{R}_{n \ell}(r) Y_{\ell m}( \theta, \varphi)  \, , \label{eqn: Bohr radial eigenfunction}
\end{align}
with $\bar{R}_{n \ell}(r) \equiv \sqrt{\mu_{\rm a}/2}  R_{\omega  \ell m}(r)$. As the action \eqref{eqn: psi action} is expanded at large distances, the boundary condition at the BH event horizon is replaced by a regular boundary condition at the origin. Consequently, the solutions $\bar{R}_{n \ell}(r)$ take the form of the  radial functions of the hydrogen atom. Higher-order corrections (fine and hyperfine spectrum) can be found in \cite{Baumann:2018vus}. We use Dirac-styled notation  $|n \, l \, m  \rangle$ for quasi-eigenstates.


\begin{framed}
\noindent
{\small {\it Weak field expansion of the Klein-Gordon equation in the Kerr background - details}\\ ~\label{box:scalar_cloud_expansuon}

Let us use these estimates to expand the lagrangian term-by-term. For example
\beq
-\frac{\mu_{\rm a}^2}{2\mu_{\rm a}}(g^{tt}+1)\psi^\ast \psi &=& \alpha \psi \psi^\ast \frac{r }{r^2\left(1+\frac{a^2-2 Mr}{r^2}\right) } \frac{\left(1+\frac{a^2}{r^2}\right)}{\left(1+ \frac{a^2}{r^2} \cos^2 \theta  \right)} \, , \\
&=& \alpha  \frac{\psi \psi^\ast}{r} + \mathcal{O}(\alpha^4) \,.
\eeq
A similar procedure gives the following expansion for other terms
\beq
-\frac{i\mu_{\rm a}}{2\mu_{\rm a}}g^{tt}(\psi^\ast \partial_t \psi - \psi \partial_t \psi^\ast) &=& \frac{i}{2} (\psi^\ast \partial_t \psi - \psi \partial_t \psi^\ast) (1+\mathcal{O}(\alpha^2)) \,, \\
-\frac{i\mu_{\rm a}}{2\mu_{\rm a}}g^{t  \varphi}(\psi^\ast \partial_\varphi \psi - \psi \partial_ \varphi \psi^\ast) &=& \frac{iMa}{r^3}(\psi^\ast \partial_\varphi \psi - \psi \partial_ \varphi \psi^\ast)  (1+\mathcal{O}(\alpha^2)) \,, \\
-\frac{1}{2\mu_{\rm a}} g^{tt} \partial_t \psi^\ast \partial_t \psi  &=& \frac{1}{2\mu_{\rm a}}  \partial_t \psi^\ast \partial_t \psi (1+\mathcal{O}(\alpha^2))\\
-\frac{1}{2\mu_{\rm a}} g^{rr} \partial_r \psi^\ast \partial_r \psi  &=& -\frac{1}{2\mu_{\rm a}}  \partial_r \psi^\ast \partial_r \psi    (1+\mathcal{O}(\alpha^2)) \\
-\frac{1}{2\mu_{\rm a}} g^{\theta \theta} \partial_\theta \psi^\ast \partial_\theta \psi  &=&  -\frac{1}{2\mu_{\rm a} r^2}  \partial_\theta \psi^\ast \partial_\theta \psi (1+\mathcal{O}(\alpha^2))  \label{eq:super_Kerr_theta_exp} \\
-\frac{1}{2\mu_{\rm a}} g^{\varphi \varphi} \partial_\varphi \psi^\ast \partial_\varphi \psi  &=&  -\frac{1}{2\mu_{\rm a} r^2 \sin^2 \theta}  \partial_\varphi \psi^\ast \partial_\varphi \psi(1+\mathcal{O}(\alpha^2))   \label{eq:super_Kerr_theta_phi} \,.
\eeq
We also need to expand the metric determinant
\beq \label{eq:super_Kerr_det_Exp}
\sqrt{-g}=r^2 \sin^2 \theta \,(1+\mathcal{O}(\alpha^2)) \,.
\eeq
Compare this results with Ref. \cite{Baumann:2018vus} - it looks as if they are missing $\mathcal{O}(\alpha^4)$ contributions from \eqref{eq:super_Kerr_theta_exp}, \eqref{eq:super_Kerr_theta_phi} and \eqref{eq:super_Kerr_det_Exp}.

Thus, the first few contributions to the weak-field expansion of the lagrangian are of the form
\beq
\mathcal{L}=\mathcal{L}_2+\mathcal{L}_4+\mathcal{L}_5 \,
\eeq
with $\mathcal{L}_n \sim \alpha^n$.
}

\end{framed}

\subsubsection{Superradiant rates} \label{sec:sr_rates}

Superradiance is a process and a cloud doesn't suddenly pop out of the BH. In other words, calculated eigenstates are truly not stationary and also contain the imaginary component
\beq
\omega \to \omega + i \Gamma \,.
\eeq
Let us go back to \eqref{eq:energy_extraction_BH} and consider the infinitesimal change in energy i.e. $\Delta t=t_f - t_i \to 0$. Then, we can find the superradiant rate as
\beq
 \Gamma \simeq \frac{\partial_t E_{\rm BH}}{E_{\rm c}},
\eeq
where $E_{\rm c}$ is the energy of the scalar cloud. In practice, for the weak-field regime the decay width was analytically calculated \cite{Detweiler:1980uk} by matching the Klein-Gordon solutions in the near-horizon regime and the weak-field regime. There was a missing factor of two in the original derivation and the correct expression \eqref{eqn: Detweiler approximation} can be found in e.g. \cite{Brito:2015oca, Baumann:2018vus}.

Here we give just the rough estimate in order to understand the scaling argument. In order for the angular momentum to be conserved we need quasi-stationary states with $l \neq 0$. Hydrogen radial wavefunction scale as 
\beq
\psi \propto \left(\frac{r}{r_{\rm c}}\right)^l
\eeq
so that the rate estimate is (the sign correspond to the energy going \textit{into} the cloud)
\beq
\Gamma &\sim & \frac{\mu_{\rm a}( m \Omega_+ - \omega)r^2_+ \left(\frac{r_+}{r_{\rm c}}\right)^l}{\mu_{\rm a}^2 \int^r_{\rm c} dr r^2  \left(\frac{r}{r_{\rm c}}\right)^l} \,, \nonumber \\
&\sim & ( m \Omega_+ - \omega) \alpha^{4l+5} \,. \label{eq:SR_rate_estimate}
\eeq
where we used $\eqref{eq:energy_extraction_BH}$, $E_{\rm c}=\mu_{\rm a}^2 \int dV \psi^\dagger \psi$ and $r_+ \sim M$. As $\alpha \ll 1$ rate is strongly suppressed for large $l$. Thus, for $m$ that satisfies \eqref{eq:SR_condition} we need minimal $l=m=1$.

The full result for the rate in the weak-field approximation is
\beq
 \Gamma_{n \ell m} = \frac{2 r_+}{M} C_{ n \ell m }(\alpha)  \left( m \Tilde{\Omega}_+ - \omega \right) \alpha^{4\ell + 5}\, ,  \label{eqn: Detweiler approximation}
\eeq
with
\beq
C_{n \ell m}\left( \alpha \right) \equiv \frac{2^{4\ell +1} (n + \ell)!}{n^{2\ell + 4} (n - \ell- 1)!} \left[ \frac{\ell !}{(2\ell)! (2\ell + 1)!}\right]^2 \prod_{j=1}^{\ell} \left[ j^2 \left( 1 - \tilde{a}^2 \right)  + \left(\tilde{a} m - 2  \frac{ r_+}{M}+ \alpha  \right)^2 \right] \,. \label{eqn: Detweiler coeff}
\eeq
The dominant growth mode is $|2 \, 1 \, 1  \rangle$ \cite{Detweiler:1980uk,Dolan:2007mj}. When the superradiance shuts down $\omega_{|2 \, 1 \, 1  \rangle}=\Tilde{\Omega}_+$ all the higher states $|n \, 1 \, 1  \rangle \,,n>2$ violate the superradiant condition as $\omega_n > \omega_2$ and start decaying into BH. In the dominant state the field is described by
\begin{equation} \label{eq:211psi}
\Psi=A_0 r M \mu_{\rm a}^2 \exp{\Big(-\frac{1}{2}rM\mu_{\rm a}^2\Big)}\cos{(\varphi-\omega t)}\sin{\theta} \,,
\end{equation}
while the superradiant rate is approximated by \cite{Detweiler:1980uk}
\beq \label{eq:SR_inst_time}
\tau_{\rm SR} \approx 0.28 \text{yr}\,  \Big(\frac{\alpha}{0.07} \Big)^{-9} \frac{M}{3 M_\odot} \Tilde{a}^{-1} \,.
\eeq
Amplitude $A_{0}$ of the scalar field \eqref{eq:211psi} is normalized in a way that the integrated density gives the cloud mass $M_{\rm c}$ \cite{Brito:2014wla}
\beq \label{eq:cloud_mass}
A^2_0 \approx \frac{1}{32\pi} \frac{M_{\rm c}}{M}\alpha^{-4} \,.
\eeq

In the previous discussion we have assumed $\alpha \ll 1$ regime. It has been shown both analytically (e.g. \cite{Arvanitaki:2010sy}) and numerically \cite{Dolan:2007mj} that for $\alpha \leqsim 0.5$ superradiance instability is most efficient. We will now argue that this condition will be consistent with the weak-field description of the cloud. There are two more important reasons - firstly, there is an irreducible mass of the Black Hole that can't be extracted (Appendix \ref{sec:Penrose}). Secondly, while the strength of the BH-axion interaction is dictated by the Compton scale, the structure of the configuration is dictated by the de Broglie scale and $r_{\rm c} \simeq \lambda_{\rm dB} \gg \lambda_{\rm c}$. In conclusion
\beq
\frac{M_{\rm c}}{r_{\rm c}} < M \mu_{\rm a} \simeq \alpha \ll 1 \,,
\eeq
where $M_{\rm c}/r_{\rm c}$ is a measure of the cloud self-gravity.

\subsubsection{Phenomenological implications} \label{}

Relevant values of axion mass are between \cite{Baumann:2018vus}
\begin{equation} \label{eq:ax_mass_range}
\frac{\alpha^{\rm (min)}}{0.07} \Big(\frac{M}{10M_{\odot}}\Big)^{-1}< \frac{\mu_{\rm a}}{10^{-12} \text{eV}} < \frac{\alpha^{\rm (max)}}{0.07} \Big(\frac{M}{10M_{\odot}}\Big)^{-1}\,,
\end{equation}
with
\begin{equation}
\alpha^{\rm (min)}=0.006 \Big(\frac{M}{10M_{\odot}}\Big)^{\frac{1}{9}}\,,
\end{equation}
and $\alpha^{\rm (max)}$ depending on $\Tilde{a}$. For example, $\alpha^{\rm (max)}=0.42$ for $\Tilde{a}=0.7$ and $\alpha^{\rm (max)}=0.19$ for $\Tilde{a}=0.8$ (see Ref.~\cite{Dolan:2007mj}). Physically, the lower limit arises from the condition that the significant growth of the cloud occurs during the age of the Universe, while the upper limit is numerically estimated from the growth rate function. For primordial BHs\footnote{Primordial BHs are hypothetical objects and a potential DM candidate. Primordial BH-axion mixed dark matter scenario was considered in Ref. \cite{Rosa:2017ury}.} $M/M_{\odot} \in (10^{-10},10^{-4})$ we find $\mu_{\rm a}/(10^{-12}\text{eV}) \in (10^3,10^{12})$, while for stellar ($M/M_{\odot} \in (10^{0},10^{2})$) and supermassive ($M/M_{\odot} \in (10^{6},10^{10})$) BHs we find $\mu_{\rm a}/(10^{-12}\text{eV}) \in  (10^{-2},10^{2})$ and $\mu_{\rm a}/(10^{-12}\text{eV}) \in (10^{-9},10^{-5})$, respectively.

Previous sections imagined a scenario where the superradiant instability produces a scalar cloud and then shuts down leaving the time-independent cloud. However, as the scalar is real there can be no solitonic configurations and the cloud will lose energy by GW emission with the estimated power \cite{Brito:2014wla}
\beq
\begin{aligned}
\frac{d E_{\rm GW}}{dt} \approx 0.01 \left(\frac{M_{\rm c}}{M} \right)^2 \alpha^{14} \, . \label{eqn: Schwarzschild strain 2}
\end{aligned}
\eeq
From $\partial_t M_{\rm c}=-\partial_t E_{\rm GW}$ one obtains
\beq
M_{\rm c}(t) = \frac{M_{\rm c}(0)}{1 + t/\tau_{c}} \,,  \label{eqn: cloud evolution}
\eeq
where $M_{\rm c}(0)$ is the initial mass of the cloud (when the superradiance stops) and $\tau_{\rm c}$ the lifetime of the cloud, given by \cite{Baumann:2018vus}
\beq
\begin{aligned}
\label{equ:tauc}
\tau_{\rm c} &\,\simeq\,  10^7 \text{yr} \left( \frac{M}{3M_\odot} \right) \left( \frac{\alpha}{0.07}\right)^{-15}\,,\\
&\,\simeq\,  
10^9 \text{yr} \left( \frac{M}{10^5 M_\odot} \right) \left( \frac{\alpha}{0.1}\right)^{-15}   \, .
\end{aligned}
\eeq

In order for this process to be astrophysically realistic one has to include accretion from the surrounding medium (accretion disks, interstellar medium) that can supply the BH with lost mass and angular momentum on superradiance or even trigger the instability by increasing the BH rotation \cite{Brito:2014wla}. There are other influences on the evolution of the superradiant clouds - presence of the binary \cite{Baumann:2018vus,Berti:2019wnn}, several ALPs triggering superradiance \cite{Stott:2018opm}, strong axion self-interaction \cite{Yoshino:2012kn,Yoshino:2015nsa} and the interaction with the electromagnetic field (produced e.g. by the plasma in the accretion disk) \cite{Rosa:2017ury,Ikeda:2019fvj,Boskovic:2018lkj}. We will comment more on the last topic in Section \ref{sec:ax_ph_clouds}.

Superradiant instability of axions (or real scalars/massive vectors) can leave observable effects in essentially three ways:
\begin{itemize}
    \item Superradiance tends to drive the BH to lower rotation rates and thus there could be holes in Regge $a - M$ Plane \cite{Arvanitaki:2009fg,Arvanitaki:2010sy,Arvanitaki:2014wva,Arvanitaki:2016qwi,Cardoso:2018tly,Zhang:2018kib,Stott:2018opm}
    \item GW signal could be detectable either through resolvable events or in the form of the stochastic signal \cite{Arvanitaki:2009fg,Arvanitaki:2010sy,Arvanitaki:2014wva,Arvanitaki:2016qwi,Brito:2017wnc,Brito:2017zvb}
    \item Change in the spacetime structure around BHs through because of the cloud can leave the specific imprint in the GW signal of Extreme Mass Ratio Inspirals (EMRI) \cite{Cardoso:2011xi,Ferreira:2017pth,Zhang:2018kib,Hannuksela:2018izj}. Specifically note that time-periodic potential, originating from the pressure, will be present as in the Newtonian oscillatons (Section \ref{sec:NOscillatons}) \cite{Ferreira:2017pth}.
\end{itemize}
Present analysis stems from the first two points lead to constraints of the following range of masses of ALPs
\beq
6\cdot 10^{-13} \rm{eV} \leqsim \mu_{\rm a} \leqsim 2 \cdot 10^{-11} \rm{eV} \,,\, 10^{-18} \rm{eV} \leqsim \mu_{\rm a} \leqsim 10^{-16} \rm{eV} \,.
\eeq
%

\subsection{Gravitating axion configurations around compact objects - other mechanisms} \label{sec:electroWaldPert}

Besides BH superradiance, there are other avenues in which gravitating axionic configurations could be produces - superradiance through axion-plasma coupling in pulsar magnetospheres \cite{Day:2019bbh} or production of axions in strong EM fields \cite{Garbrecht:2018akc, Boskovic:2018lkj}. We breifly comment on the second scenario (in the BH context), while refering for details to mentioned references.

It has been shown by Wald~\cite{Wald:1974np} that, neglecting backreaction, Kerr BHs immersed in a homogeneous magnetic field $B$ aligned with the BH axis of symmetry allows for an exact analytical solution of Maxwell's equations (electro-vacuum solution):
\beq
A_{\mu}=\frac{1}{2}B \big((\xi_\varphi)_{\mu}+2a (\xi_t)_{\mu} \big) \,.
\eeq
This field would lead to the BH accreting surrounding charge in the accretion disk and the interstellar medium. Therefore BHs would acquire a charge in those enviroments and be described by a Kerr-Newman spacetime, with a total vector potential given by~\cite{Wald:1974np}
\beq
\tilde{A}_{\mu}=\frac{1}{2}B \big(((\xi_\varphi)_{\mu}+2a (\xi_t)_{\mu} \big)-\frac{1}{2}q (\xi_t)_{\mu},
\eeq
with $q=Q/M$ and $Q$ is the accumulated BH charge. At equilibrium, the  BH charge-to-mass-ratio is given by $q=2B a$. We can therefore  analyse two different cases: (i) the BH is uncharged, and there is a net flow of charge from the surrounding medium, (ii) the BH is charged, but there is no net flow of charge from the surrounding medium. Recent estimates for supermassive BH in the Galactic center suggest that rotationally-induced charge is stable with respect to the discharging processes from the surroundings of an astrophysical plasma \cite{Zajacek:2018ycb}. Let us then focus on the second (equilibrium) case and estimate the importance of the induced charge on the background geometry. Using the limit for a maximal astrophysically realistic magnetic field\footnote{In natural units, the strength of a magnetic field around a source of mass $M$ can be measured defining the characteristic magnetic field $B_M=1/M$ associated to a spacetime curvature of the same order of the horizon curvature. In physical units this is given by $B_M\sim 2.4\times 10^{19} \left(M_{\odot}/M\right) {\rm Gauss}$. For astrophysical BHs, a reference value for the largest magnetic field that can be supported in an accretion disk is given by $B\sim 4\times 10^8 \left(M/M_\odot\right)^{-1/2}{\rm Gauss}$~\cite{ReesAGN} so that the approximation $B\ll B_M$ is well justified.}, we find $q \leq  10^{-11} a/M$,  i.e. the geometry is still well described by the Kerr metric.

Hence, we here consider a Kerr spacetime with the vector potential of the form
\beq
A^{\rm Wald}_{\mu} = \frac{1}{2} B g_{\mu \nu} \xi^{\nu}_\varphi =
\frac{B\sin^{2}\theta}{2\Sigma} (-2 a M r, 0, 0, \mathcal{F} )\,,\label{eq:WaldSol}
\eeq
where $\mathcal{F}$ is a metric function given in Eq.~\eqref{eq:KerrBLfcts}.

Let us now consider, instead of Maxwell's equations, the generalized axionic equations~\eqref{eq:MFEoMVector}. For $k_{\rm a}=0$, Wald's solution is a solution to the problem, together with a vanishing scalar field. Thus, we are interested in a first-order\footnote{We note that when expanding in $k_{\rm a}$ with a background EM field, one is effectively considering expansions of the form $k_{\rm a} \langle A \rangle$, where $ \langle A \rangle$  is a characteristic, dimensionless and Lorentz-invariant measure of the EM field strength (e.g. $ \langle A \rangle =Q^2/M^2$ for a charged BH). In other words, strong EM fields can compensate for a ``small'' value of $k_{\rm a}$ and produce observable consequences. A similar approach was recently considered in the context of pulsar magnetospheres~\cite{Garbrecht:2018akc}.} (in $k_{\rm a} B^2M^2$) production of axions, as a consequence of the EM background. The dominant term describing
the axionic field is the equation 
\beq 
\left(\nabla^{\mu}\nabla_{\mu} - \mu^{2}_{\rm S} \right) \Psi = &
          \frac{1}{2}k_{\rm a} \,g^{\alpha \mu} g^{\beta \nu} ~ \,^{\ast}F^{(0)}_{\mu \nu} F^{(0)}_{\alpha \beta},\label{eq:KG_Wald}
\eeq       
where $F^{(0)}_{\mu\nu}$ denotes the Maxwell tensor corresponding to Wald's solution. 
Using Eq. \eqref{eq:WaldSol} we find, to fifth order in the spin $\tilde{a}=a/M$,
\beq
g^{\alpha \mu} g^{\beta \nu} ~ \,^{\ast}F^{(0)}_{\mu \nu} F^{(0)}_{\alpha \beta}&=&-\frac{12aB^2M\cos\theta\sin^2\theta}{r^2}+\frac{4a^3B^2M\cos\theta\sin^2\theta\left(2r-M+\cos(2\theta)(M+5r)\right)}{r^5}\nonumber\\
&-&\frac{2a^5B^2M\cos^3\theta\sin^2\theta\left(-10M+r+\cos2\theta(10M+21r)\right)}{r^7}\,.
\eeq

One can now expand the left-hand side of Eq. \eqref{eq:KG_Wald} order by order in the spin, with $\Psi=\Phi_1 \tilde{a}+\Phi_2 \tilde{a}^2+\ldots$. To first order in rotation (and for $\mu_{\rm a}=0$) one gets 
\beq
&&\frac{\partial}{\partial\theta}\left(\sin\theta \frac{\partial\Phi_1}{\partial\theta}\right)+\sin\theta\frac{\partial}{\partial r}\left((r^2-2Mr+a^2) \frac{\partial\Phi_1}{\partial r}\right) =\nonumber\\
&&=-6k_{\rm a}B^2M^2\cos\theta\sin^3\theta\,,
\eeq
and similar equations for higher order terms, each of which can be solved with an expansion in spherical harmonics. 
Finally, to first order in $k_{\rm a} M^2 B^2$ and fifth order in the spin for massless ``axions'' we find
\beq
\Psi=k_{\rm a}B^2M \cos\theta\left(\frac{3a}{2}+\frac{a^3}{2r^2}\right)-\cos^3\theta\left(\frac{a}{2}+\frac{a^3}{r^2}+\frac{a^5}{2r^4}\right)
+\cos^5\theta\left(\frac{a^3}{2r^2}+\frac{a^5}{r^4}\right)-\cos^7\theta\frac{a^5}{2r^4}\,.
\eeq
This field will in turn contribute to the background EM field, via Eq.~\eqref{eq:MFEoMVector}, but as a second order 
(in $k_{a}$) effect. Further properties and phenomenological consequences of this solution should be subject of further study. 

We note that this results is related to the fact that a background electromagnetic field induces an axionic instability\footnote{This is in some sense inverse problem than the one in Part \ref{ch:ax-photon} where we discuss instabilities from electromagnetic fluctuations on the background axionic configuration.} in flat space, for electric fields above a certain threshold value \cite{Ooguri:2011aa,Boskovic:2018lkj}. When carried over to curved spacetime, this phenomena translates into generic instabilities of charged black holes. In the presence of charge, black hole uniqueness results are lost and one can find solutions which are small deformations of the Kerr-Newman geometry (of which the preceding discussion is an example) {\it and}
hairy stationary solutions without angular momentum but which are ``dragged'' by the axion \cite{Boskovic:2018lkj}.

\newpage

\part{Orbital dynamics in the axion configurations background} \label{ch:orbital}
The problem of motion in General Relativity is a fundamental one.
It is the motion of objects and of light that allows for precise
tests of the theory, by connecting it to observations. Conversely, the way that objects move allows one to infer, study and map the amount of matter contributing to the motion. When the object (a star, a planet, etc...) is idealized as point-like, it moves
-- to first approximation -- along geodesics of the spacetime ``generated'' by the rest of the universe.

Static, spherically symmetric objects in otherwise empty spacetime give rise to a Schwarzschild geometry. Geodesic motion around a Schwarzschild background has been studied for decades. The symmetries of the spacetime allow for three constants of motion, which simplify considerably the analysis and make the problem integrable. Nonetheless, spherical symmetry does not necessarily imply staticity when matter pervades the geometry. For example, radially oscillating stars produce an effective geometry which is time-dependent in their interior. Birkhoff's theorem guarantees that the spacetime outside such a configuration is described by a Schwarzschild geometry \cite{bookWeinberg:1972}. 

Motivated by spacetimes discussed in Part \ref{ch:structure} we wish to consider the full problem of geodesic motion, in what looks like a classic problem in Newtonian physics and General Relativity: how do particles move in a time-dependent and periodic gravitational potential? 

Some aspects of this question were considered previously within a very specific context -- that of oscillating bosonic DM -- and within some approximations~\cite{Khmelnitsky:2013lxt,Blas:2016ddr,Ferreira:2017pth, AokiSoda:2017a}. We will comment upon significane of such topics for axion phenomenology in Section \ref{sec:motion_pheno}. 

\newpage

\section{Particles in spherically symmetric and time-periodic background: relativistic results} \label{sec:SphSymTimePeriodRel}

\subsection{Geodesics in the time-dependent geometry} \label{sec:motionfullgeo}

Let us start from the most general spherically symmetric and time-periodic background of the form \eqref{eq:SSmetric}. For the Lagrangean~\eqref{Lagrangean} corresponding to this metric,  $\varphi$ is a cyclic coordinate. Thus, its conjugate momentum, the angular momentum along the $z-$axis $r^2\sin(\theta)\dot{\varphi}=J$, is a conserved quantity. Due to the spherical symmetry of the metric, the geodesics will always be planar. Therefore, without loss of generality, we set $\theta = \frac{\pi}{2}$. The geodesic equations are  reduced to two nontrivial coupled equations,
\beq
&&\ddot{t}+\frac{1}{2A}\left(\del_t A\dot{t}^2+ 2A'\dot{r}\dot{t}+\del_t B\dot{r}^2\right)=0 \,,\label{geodesics1}\\
&&\ddot{r}+\frac{1}{2B}\left(B'\dot{r}^2+A'\dot{t}^2- 2 r\dot{\varphi}^2 + 2\del_t B\dot{r}\dot{t}\right)=0 \,. \label{geodesics2}
\eeq

Two simplest examples of trajectories concern circular and radial motion, for which 
\beq
r(\tau)&=& r_0\,,\, \dot{r}=0\,,\, \ddot{r}=0\\
r(\tau=0)&=&r_{\rm init}\,,\,\dot{\varphi}=0\,,
\eeq
respectively.

Consider first circular motion in our coordinates. The substitution $\dot{\varphi} = \Omega$ in equation \eqref{geodesics2} yields,
\be
\frac {A'}{2B}\dot{t}^2-\frac{r_0\Omega^2}{B}=0\,.
\label{eq:dott0}
\ee
There is a solution  if $A'>0$. Solving Eq.~\eqref{eq:dott0} for $\dot{t}$ and differentiating to find $\ddot{t}$ we may rewrite equation~\eqref{geodesics1} as
\begin{equation}
\frac{\del_{t} A'}{A'}-\frac{\del_t A}{A}=0.
\end{equation}
Any non-null separable function $A(t,r)=a_t(t)a_r(r)$ satisfies this condition, making it sufficient for circular motion to be allowed. This condition reduces to
\be
\Omega=\frac{1}{\sqrt{\frac{2r_0a_r(r_0)}{a^{'}_r(r_0)}-r^{2}_0}},\label{Angular_freq}
\ee
implying that $2a_r/a'_r>r_0$ at $r_0$. Note that the {\it coordinate} angular velocity is,
\be
\tilde{\Omega}=\frac{d\varphi}{dt}=\frac{d\varphi}{d\tau}\frac{d\tau}{dt}=\frac{\Omega}{\dot{t}}\,.
\label{eq:OmtotildeOm}
\ee
From \eqref{eq:dott0},
\be
\dot{t}=\Omega\sqrt{\frac{2r_0}{a'_r\left(r_0\right)a_t\left(r_0\right)}}\,,
\ee
thus we find
\be
\tilde{\Omega}=\sqrt{\frac{a'_r\left(r_0\right)a_t\left(r_0\right)}{2r_0}}\,, \label{eq:coordinatangfreq}
\ee
in agreement with standard results for $a_t(t)=1$ e.g.~\cite{Teukolsky:1973ha}.

\subsection{Geodesics in weakly dynamic spacetimes}

Up to now, our results are generic and valid for any geometry of the form \eqref{eq:SSmetric}. In most of the applications in mind, the spacetime is to a good approximation static, and the time-dependence is but a small deviation away from staticity. Thus, we find it convenient to expand the metric around a static reference spacetime $g^{(0)}_{\mu\nu}$
\be
g_{\mu\nu}=g^{(0)}_{\mu\nu}+\epsilon g^{(1)}_{\mu\nu}\,,\label{metric_expansion}
\ee
with $\epsilon$ beeing a small dimensionless book-keaping parameter.

Let $x^{\mu}_0(\tau)$ be the solution of the equations of motion derived from the Lagrangian 
$L_0(x^\mu,\dot{x}^\mu)=g^{(0)}_{\alpha\beta}\dot{x}^\alpha\dot{x}^\beta$. Let us consider what happens to the solution when the metric is slightly perturbed as in Eq.~\eqref{metric_expansion}, such that it gives rise to a new Lagrangian 
$L(x,\dot{x})=L_0(x^\mu,\dot{x}^\mu)+\epsilon L_1(x^\mu,\dot{x}^\mu)$. The associated geodesics will have a different solution, which should lie sufficiently close to $x^\mu_0(\tau)$. Lets find it, by expanding around $x_0^\mu$: 
\beq
&&x^\mu(\tau)=x_0^\mu(\tau)+\epsilon\,\eta^\mu(\tau)\,,\\
&&L(\eta,\dot{\eta},\tau)=L_0(x_0)+\epsilon L_1(x_0) + \epsilon \frac{\partial L_0}{\partial x^\alpha} \eta^\alpha+\epsilon \frac{\partial L_0}{\partial \dot{x}^\alpha} \dot{\eta}^\alpha\nonumber\\
&+&  \epsilon^2 \left(\frac{1}{2}\frac{\partial^2L_0}{\partial x^\alpha x^\beta}\eta^\alpha \eta^\beta+\frac{1}{2}\frac{\partial^2L_0}{\partial \dot{x}^\alpha \dot{x}^\beta}\dot{\eta}^\alpha \dot{\eta}^\beta+\frac{\partial^2L_0}{\partial {x}^\alpha \dot{x}^\beta}\eta^\alpha \dot{\eta}^\beta \right)\nonumber\\
&+& \epsilon^2 \frac{\partial L_1}{\partial x^\alpha} \eta^\alpha +\epsilon^2 \frac{\partial L_1}{\partial \dot{x}^\alpha}\eta^\alpha+{\cal O}(\epsilon^3)\,.
\label{expansion}
\eeq
Although not explicitly stated, each partial derivative of the Lagrangean is to be evaluated at $x_0(\tau)$, here and in the following.

Since we assume that $x_0(\tau)$ is known as it solves the $0^{th}$ order equations of motion, the above expansion only depends on $\eta$, $\dot{\eta}$ and $\tau$. Thus, to extremize the action of this Lagrangean, the variation must be done in $\eta$. The Euler-Lagrange equations of ~\eqref{expansion} yield, up to second order in $\epsilon$, the following nontrivial result
%
\begin{equation}
\frac{d}{d \tau}\left(\frac{\partial \zeta}{\partial\dot{x}^i}  \right)-\frac{\partial \zeta}{\partial x^i}  = \frac{\partial L_1}{\partial x^i}  - \frac{d}{d\tau}\left(\frac{\partial L_1}{\partial \dot{x}^i} \right)\,,
\label{initgeoress}
\end{equation}
where we defined
\begin{equation}
\zeta =\frac{\partial L_0}{\partial x^j}\eta^j + \frac{\partial L_0}{\partial \dot{x}^j}\dot{\eta}^j\,.
\end{equation}
Expressing $L_0$ as a function of the metric, equation \eqref{initgeoress} can be further simplified into
%
\beq
&&\ddot{\eta}^\gamma+2\Gamma^{\gamma}_{(0)\alpha\beta}\dot{x}_0^\alpha\dot{\eta}^\beta+\left(\partial_\delta \Gamma^{\gamma}_{(0)\alpha\beta}\right)\dot{x}_0^\alpha\dot{x}_0^\beta \eta^\delta\nonumber\\
&&=-\frac{1}{2}g^{(0)\, \gamma \beta}\left(\frac{d}{d\tau}\left(\frac{\partial L_1}{\partial \dot{x}^\beta} \right) -\frac{\partial L_1}{\partial x^\beta} \right)\,.
\label{georessonance}
\eeq
%

If the LHS was equated to 0, one would have the geodesic deviation equation, describing the evolution of a perturbation on the geodesic itself. However, there is a ``force-term'' presented on the RHS illustrating how the metric's perturbations disturb the geodesics of the background static spacetime. This result can also be obtained by firstly applying Euler-Lagrange equations to $L(x,\dot{x})$ and only then considering the expansion around $x_0$.

By inspection of equation \eqref{initgeoress}, if $x_0^\gamma(\tau)$ is cyclic in $L_0$, the equation may be integrated on both sides, leading to a first order equation
%
\beq
\frac{\partial \zeta}{\partial\dot{x}^\gamma} &=& \frac{\partial^2L_0}{\partial \dot{x}^\gamma \dot{x}^\beta} \dot{\eta}^\beta + \frac{\partial^2L_0}{\partial \dot{x}^\gamma x^\beta} \eta^\beta\nonumber\\
&=&2g^{(0)}_{\gamma\beta}\dot{\eta}^\beta+2\partial_\beta (g^{(0)}_{\gamma\alpha})\dot{x}_0^\alpha\eta^\beta\nonumber\\
&=&\int_{\tau_0}^{\tau}\left(-\frac{d}{dy}\left(\frac{\partial L_1}{\partial \dot{x}^\gamma} \right) +\frac{\partial L_1}{\partial x^\gamma} \right)dy+C_\gamma,
\label{cyclicgeoress}
\eeq
%
where $C_\gamma$ is a real constant depending on the initial conditions.

\subsubsection{Motion in weakly-dynamic and time-periodic spacetimes} \label{sec:periodic_spacetimes}

We will now specialize the discussion to weakly-dynamic \textit{and} time-periodic geometries. Having in mind applications in Section \ref{sec:axion_stars}, we consider spacetimes for which
\beq
A(t,r)&=&a_0(r)+\epsilon a_1(r)\cos\left(2\omega t\right)\,,\label{eq:weakly1}\\
B(t,r)&=&b_0(r)+\epsilon b_1(r)\cos\left(2\omega t\right)\,,\label{eq:weakly2}
\eeq
thus describing time-periodic geometries with period $T=\pi/\omega$. For the background metric, both $t$ and $\varphi$ are cyclic coordinates in $L_0$, allowing the corresponding two equations in system \eqref{georessonance} to be rewritten as first order equations using \eqref{cyclicgeoress}:
\beq
\dot{\eta}^\varphi&=&-\frac{2}{r_0}\dot{\varphi}_0\eta^r+\frac{C_\varphi}{2r_0^2} +\frac{F_{\varphi}}{2r_0^2}\,,\label{cyclivarsphi}\\
\dot{\eta}^t&=&-\frac{a_0'}{a_0}\dot{t}_0\eta^r-\frac{C_t}{2a_0}-\frac{F_t}{2a_0}\,,
\label{cyclivarst}
\eeq
where
\beq
EL_\gamma &=& \left(-\frac{d}{d\tau}\left(\frac{\partial L_1}{\partial \dot{x}^\gamma} \right) +\frac{\partial L_1}{\partial x^\gamma} \right)\,,\\
{\cal F}_t&=&\int_{\tau_0}^{\tau} EL_t(y) \, dy\,,\quad {\cal F}_\varphi=\int_{\tau_0}^{\tau} EL_\varphi(y) \, dy\,,
\eeq
and $a_0(r_0) = g^{(0)}_{tt}(r_0)$, $b_0(r_0) = g^{(0)}_{rr}(r_0)$.
As we discussed previously, geodesics on the background metric $g^{(0)}$ defined in Eq.~\eqref{metric_expansion} are planar, thus allowing the choice $\theta_0(\tau)=\pi/2$. 
Replacing the relations \eqref{cyclivarsphi} and \eqref{cyclivarst} on the system \eqref{georessonance}, we are left with a system of two decoupled second-order equations for $\eta^r$ and $\eta^\theta$:
%
\beq
&&2 r_0   \left(\ddot{\eta}^\theta+\eta^\theta \dot{\varphi}_0^2\right)+4\dot{\eta}^\theta  \dot{r}_0=\frac{EL_\theta }{r_0 }\label{2ordertheta}\,, \\
&& \eta^r  \left(\dot{r}_{0}^2 b_0''+\dot{t}_{0}^2 a_0''+6 \dot{\varphi}^2\right)+2 \dot{\eta}^r  \dot{r}_0  b_0'+2 b_0 \ddot{\eta}^r\nonumber\\
&&- \frac{\eta^r  b_0'\left(\dot{r}_0^2 b'_0+\dot{t}_0^2 a_0'-2 r_0  \dot{\varphi}^2\right)}{b_0} = \frac{\dot{t}_0  a_0' \left({\cal F}_t+C_t+2 \eta^r  \dot{t}_0  a_0'\right)}{a_0}\nonumber\\
&&+\frac{2 \dot{\varphi}  \left({\cal F}_\varphi+C_\varphi\right)}{r_0}+EL_r  \,\label{2orderr}\,.
\eeq
%

\subsection{Symmetries of motion} \label{sec:symmetries}

Since the metric coefficients of a dynamic spacetime are time dependent, time homogeneity -- valid in the  time-independent spacetime -- no longer holds generically and time is not a cyclic coordinate. Specifically, in the case of time-periodic spacetimes, symmetry is not completely lost but reduced to a discrete subgroup, akin to (space) translation symmetry in crystals~\footnote{It would be tempting to label this system as a time crystal, but this label is technically not applicable as in the time crystals symmetry is spontaneously broken~\cite{ShapereWilczek}. For a cosmological example of a similar system, see Ref.~\cite{ArkaniHamed:2003uy}.}.

Breaking of the time homogeneity has interesting consequences for the study of the test particle initial value problem; take for example the case of circular background motion ($\dot{r}_0=0, \ddot{r}_0=0$). In the weakly-dynamic spacetime, the time-dependent part of the metric is treated as a perturbation. Assume therefore a metric expansion of the form \eqref{eq:weakly1} and \eqref{eq:weakly2}, where the background metric 
$g^{(0)}_{\alpha\beta}$ is static and spherically symmetric. If we want our ``initial time'' $t_{i}$ to correspond to a vanishing perturbation, we should set it to $\pi/(4\omega)$ or to $3\pi/(4\omega)$. The solution to the problem depends on the specific choice one makes, because of the different signs of the cosine derivatives in the time-dependent part of the metric. 
Note that this situation is consistent with the spherical symmetry of the problem: for initial times different from zero, the relation $\varphi \propto t$ is not valid anymore, instead $\varphi \propto (t-t_{i})$. In other words, it is irrelevant at what point on the initial orbit (at fixed times) our particle is when the perturbation is turned on; however, it is important at what point in time that happened. Time-periodic perturbations break the time homogeneity and the same initial conditions, but different $t_i$, don't lead to the solutions which can be related by the time translation in $t_i$.

In the quantum treatment of the electron motion in crystals, Bloch's theorem implies that there is a conservation of the crystal momentum (e.g. \cite{Tong:CondMat}). However, as the symmetry is discrete, Noether theorem is not applicable and the conservation law is a consequence of the linearity of Quantum Mechanics (Schrödinger equation is subject of the Floquet theory for the symmetry in question - see Appendix \ref{AppParRes}). As General Relativity is highly non-linear, we should not expect that point particle energy will be periodic in $\pi/\omega$, in analogy with electrons in crystals, in general. We can calculate the change of the particle energy function $E(t)=-\partial L/\partial \dot{t}$, at the order ${\cal O}(\epsilon^2)$, between two arbitrary moments in time
\beq 
\label{eq:energy_periodicity_general}
&& E(t_2)-E(t_1)= \nonumber\\
&& 2\epsilon \omega  \int^{\tau(t_2)}_{\tau(t_1)}\left(b_1(r_{0})\dot{r}_{0}^2-a_1(r_{0})\dot{t}_{0}^2\right)\sin{(2\omega t_0(y))} dy\,.
\eeq
The integrand is not necessarily periodic in $\pi/\omega$, so we can't claim $E(t)=E(t+n\pi/\omega),n \in \mathbb{N}$. However, this conclusion will change when the equations of motion become linear, as in Section \ref{sec:circularmotionexample}.

We should also note that spacetimes with metric expansion as in the example do admit time-inversion symmetry. This symmetry is broken when friction is present, as in Section~\ref{sec:toy_model_dissipation}.

\subsection{Circular and radial background motion - linear regime \label{sec:circularmotionexample}} 
%

\subsubsection{Circular background motion}

We start by considering the equations of motion from Section \ref{sec:periodic_spacetimes} in the context of background circular geodesics. Imposing the circularity condition ($\dot{r}_0=0, \ddot{r}_0=0$) in Eq.~\eqref{geodesics2} one finds
\be
r_0(\tau)=r_0\,,\quad   \varphi_0(\tau)=\Omega\tau\,,\quad t_0(\tau)=\frac{\Omega}{\tilde{\Omega}}\\
\tau+t_i\,,\label{Circular Solution}
\ee
where $\tilde{\Omega}$ is given by equation \eqref{eq:coordinatangfreq} and $t_i$ is a constant. The requirement that the motion is timelike is equivalent to requiring that $\Omega$ be given by Eq.~\eqref{Angular_freq}. Furthermore, $EL_\theta$ and $EL_\varphi$ vanish; replacing $EL_t$ and $EL_r$ by their corresponding expressions, an analytical solution is found for $\eta^r(\tau)$  through \eqref{2orderr}:
\be
\eta^r(\tau) = \mathcal{D}(\omega) \cos \left(2 \omega t_0(\tau)\right)+C_1 \cos (\Theta \tau + C_2) + C_3\,.
\label{circularsol}
\ee
where $C_1$ and $C_2$ are real constants dependent on the initial conditions of $\eta^r$, $C_3$ is related with the initial conditions of $\eta^t$ and $\eta^\varphi$, and:
\beq
\mathcal{D}(\omega) &=& \frac{r_0 \left(a_0 a_1'-a_1 a_0'\right) }{8 r_0 \omega^2 b_0 a_0+2 r_0 (a'_0)^2-a_0 \left(r_0 a_0''+3 a_0'\right)}\,, \nonumber\\
\Theta &=& \Omega  \sqrt{\frac{r_0 a_0 a_0''-2 r_0 a_0'^2+3 a_0 a_0'}{b_0a_0a_0'}}\,.\label{Theta}
\eeq
It is clear that there may exist {\it resonances} in the motion, when the amplitude $\mathcal{D}(\omega)$ diverges. In order to understand physical behaviour in this case, we need higher-order terms of the particle Lagrangian in $\epsilon$.  Resonance occurs at frequencies
$\omega=\omega_{\rm res}$ for which the denominator of $\mathcal{D}(\omega)$ above vanishes:
{\begin{equation} \label{eq:circresonant}
\omega_{\rm res}=\pm\frac{\Theta}{2\dot{t}_0}\,.
\end{equation}}

The frequency $\Theta$ corresponds to the proper radial epicyclic frequency for this static, axially symmetric spacetime \cite{RonaldoVieira:2017}. 
To obtain the radial epicyclic frequency in coordinate time we need to divide $\Theta$ by $\dot{t}_0$. Note that the frequency of the metric perturbation in \eqref{eq:weakly1} and \eqref{eq:weakly2} is in fact $2\omega$. Then, the effective frequency of resonance corresponds to $2\omega_{res}=\Omega/\dot{t}_0$ which, as stated, is the radial epicyclic frequency in coordinate time.
Thus, our system behaves as a classic, driven harmonic oscillator: when the ``forcing'' frequency equals the natural (epicyclic) frequency, a resonance occurs. 
If $\Theta$ differs from $\Omega$, the geodesics will precess. The above is a very generic prediction of a smoking-gun of time-periodic spacetimes.

Finally, let us apply \eqref{eq:energy_periodicity_general} to this specific background motion. As $r_0$ and $\dot{t}_0$ do not depend on the proper time, the integral reduces to zero when $t \rightarrow t+n\pi/\omega,n \in \mathbb{N}$. This conclusion is not valid during the resonant motion when the higher order terms become important. 

\subsubsection{Radial background motion}

We now  focus on perturbations on radial geodesics. Imposing $\dot \varphi_0(\tau)=0$ and $\ddot \varphi_0(\tau)=0$, an explicit analytic solution for $\eta^r$ is not possible for general radial geodesics. Thus we specialize to motion of small amplitude around the geometric center of our spacetime. Expanding $L_0$ to first order around $r=0$ and $\dot{r}=0$, the geodesics following from $L_0(x,\dot x)$ admit the following solution:

\beq 
r_0(\tau)&=&\tilde{r}_0\cos\left(\Omega_0 \tau\right)+\frac{\dot{ \tilde{r}}_0}{\Omega_0}\sin\left(\Omega_0\tau\right)\,,\\
t_0(\tau)&=&\alpha\tau+t_i\,,\\
\Omega_0&=&\alpha\sqrt{\frac{a_0''}{2b_0}}\,,\quad\alpha=\dot{t}_0(\tau)\,,\label{Radial Solution}
\eeq
where $\tilde{r}_0$ and $\dot{\tilde{r}}_0$ are initial conditions. Although not explicitly stated, all quantities are to be evaluated at $r=0$. In the preceding derivation, we used the fact that the parity of $a_0$, $b_0$, $a_1$ and $b_1$ implies that these functions and their odd radial derivatives vanish at the origin, for regular spacetimes.

Expanding \eqref{2orderr} on $r$ and $\dot{r}$ around 0, using the appropriate expressions for $EL_\varphi$, $EL_t$ and $EL_r$, we obtain the following analytic solution,
\beq
\eta^r(\tau)&=&-\mathcal{Q}(\omega)\frac{\dot{r}_0(\tau ) \cos (2 \omega \alpha  \tau  ) 
}{(\alpha  \omega ) \left(4 b_0 a_0\left(2 b_0\omega ^2-a_0''\right)\right)}\nonumber\\
&-& \mathcal{G}(\omega)\frac{r_0(\tau) \sin (2 \omega \alpha  \tau   ) 
}{4 b_0a_0 \left(2 b_0 \omega^2-a_0''\right)}\nonumber\\
&+&C_1 \cos (\omega_0 \tau  +C_2)-C_t\frac{\dot{r}_0(\tau )}{2 \alpha  a_0} \tau\,, \label{nrradial}
\eeq
$C_1$ and $C_2$ are constants depending on the initial conditions on $\eta^r$. $C_t$ is the integration constant in \eqref{cyclivarst} and
\beq
\mathcal{G}(\omega)&=&b_0\left(a_0a_1''-a_1a_0''\right)+b_1a_0 a_0''\,,\\
\mathcal{Q}(\omega)&=&4 b_0b_1 a_0\omega^2+\mathcal{G}(\omega)\,.
\eeq
The solution grows linearly in time unless $C_t=0$.  This is an expected result as, by inspection of equation \eqref{cyclivarst}, one may conclude that, for a non vanishing $C_t$, $\dot{\eta}^t$ is given by a sum of a constant with the integral of a periodic function, inducing a linear growth of $\eta^t$. As all our equations depend on a small and stable evolution of the motion, this result may be alarming. Nevertheless one may always choose $C_t$ to be null, as it is only related to the initial conditions of $\dot{\eta}^t$ which are decoupled of the initial perturbation on the radial direction. For these reasons, we will use $C_t=0$. 
 
Again, there is a frequency $\omega$ for which the two denominators on \eqref{nrradial} vanish, corresponding to a resonance. This frequency is
\begin{equation}\label{eq:freqrad}
\omega_{\rm res}=\sqrt{\frac{a_0''(0)}{2b_0(0)}}=\frac{\Omega_0}{\alpha}\,.
\end{equation}
Now, resonance occurs when the perturbation frequency $(2\omega)$ is two times the "natural" frequency by which $r_0(\tau)$ oscillates. This result is extremely intuitive, if one imagines the spacetime pushing the object away from the centre, while the latter is also going away from it, and pulling inwards when the object starts moving towards the centre. Then, in each half period of the small oscillation in $r_0$, the metric's perturbation must complete a full period. Similarly to the circular case, \eqref{eq:energy_periodicity_general} implies energy periodicity in $\pi/\omega$ for the radial motion, when we concentrate on small amplitude deviations ${\cal O}(\tilde{r}_0^2)$.

\newpage

\section{Particles in spherically symmetric and time-periodic background: non-relativistic regime} \label{sec:SphSymTimePeriodWeak}

\subsection{Non-relativistic motion in the weak-field regime} \label{sec:N_dynamics}

In order to precisely establish the analogy with the driven harmonic oscillator from the end of the previous section, and to understand dynamical aspects beyond the linear regime, we consider non-relativistic particle motion in the weak field limit. The Lagrangian for the test particle non-relativistic equatorial motion in a weak and asymptotically flat spherically-symmetric spacetime, like the one given by \eqref{eq:NexpA2} and \eqref{eq:NexpB2} is described by
\begin{equation} 
2L=-(1+\nu)\dot{t}^2+\dot{r}^2+r^2\dot{\varphi}^2.
\end{equation}
Here $\nu(t,r)=2V(r)+\epsilon 2V_{1}(r)\cos(2 \omega t)$. $V(r)$ is the Newtonian gravitational potential and $V_1(r)$ originates from time-dependent part of the $A(t,r)$ metric coefficient. 
These quantities are defined in Eq.~\eqref{eq:NexpA2} and are related to the metric coefficients in \eqref{eq:weakly1}, \eqref{eq:weakly2} as $1+2V=a_0$ and $2V_1=a_1$.

The Euler-Lagrange equations reduce to:
\begin{align} 
  & -(1+\nu)\ddot{t}=\frac{1}{2} \partial_t \nu\dot{t}^2+ \nu' \dot{t}\dot{r},\\
  & \partial^2_t r+\partial_t r\frac{\ddot{t}}{\dot{t}^2}=-\frac{1}{2} \nu'+r\tilde{\Omega}^2. \label{eq:EL_weak_field_radial_0}
\end{align}
As we are interested in the non-relativistic motion $\dot{r}<<\dot{t}$
\begin{align} 
  & \partial^2_t r-\frac{\tilde{J}^2}{r^3}=-\frac{1}{2}\nu', \label{eq:ELweakfieldradial}
\end{align} 
where we have introduced the coordinate angular momentum $\tilde{J}=r^2\tilde{\Omega}$. We see that the second term on the l.h.s. of \eqref{eq:EL_weak_field_radial_0} is of order $\sim v/c^2$, when we restore $c$. These equations are valid even for the highly-dynamical spacetime, as we will further elaborate in the Section~\ref{sec:Mathieu_rapid}.

Now we focus on the weakly-dynamical spacetime $(\epsilon<<1)$  and weak orbital perturbations from background circular motion $r=r_0+\epsilon \eta^r  $ as in \eqref{Circular Solution}. Equation \eqref{eq:ELweakfieldradial} then reduces to:
\begin{align} 
& \tilde{\Omega}=\sqrt{\frac{1}{r_0}V'}, \\
& \partial^2_t\eta^r+(V'' +3\tilde{\Omega}^2 )\eta^r=-V_1'\cos{(2\omega t)} \label{eq:ELweakfieldcircpert}. 
\end{align}
In the last equation and until the end of this and the next section both potentials and their derivatives, with respect to $r$,  are  to  be  evaluated  at $r_0$. Last equation represents equation of motion for the driven linear harmonic oscillator, as claimed. Resonance occurs when
\begin{equation} \label{eq:N_res_freq}
\omega_{\rm res}=\frac{1}{2}\sqrt{V''+3 \tilde{\Omega}^2 }\,.
\end{equation}
This result agree with the appropriate limit of Eq.~\eqref{Theta}.

For radial motion $(\tilde{J}=0)$, equation \eqref{eq:ELweakfieldradial} yields:
\begin{align} 
& \partial^2_t r_0=-V', \label{eq:EL_weak_field_radial_pert_0th} \\
& \partial^2_t\eta^r+V'' \eta^r=-V_1'\cos{(2\omega t)} \label{eq:EL_weak_field_radial_pert_1st}. 
\end{align}
The solution of these equations depends on the form of the potential. However (see also Section~\ref{sec:circularmotionexample}), we can expand around the initial state $\{ r_0=0, \partial_t r_0=0 \}$ and for small amplitudes $\tilde{r}_0$ obtain
\begin{equation} 
r_0(t) = \tilde{r}_0\cos\left(\tilde{\Omega}_0 t \right)+\frac{\dot{\tilde{r}}_0}{\tilde{\Omega}_0}\sin\left(\tilde{\Omega}_0 t \right)\,,
\end{equation}
with $\tilde{\Omega}_0=\sqrt{V''(0)}$. Expanding $V_1(0)$ to the first non-zero term (as $V'(0)=V_1'(0)=0$, see Section~\ref{sec:circularmotionexample}), \eqref{eq:EL_weak_field_radial_pert_1st} becomes
\begin{equation} 
\partial^2_t\eta^r+\tilde{\Omega}^2_0 \eta^r=-V_1''(0)\cos{(2\omega t)r_0(t)}\,.\label{eqeta2}
\end{equation}
When we solve this equation, we see that resonance occurs when
\begin{equation} \label{eq:N_res_freq_radial}
\omega_{\rm res}=\tilde{\Omega}_0\,.
\end{equation}
This result coincides with the relativistic one~\eqref{eq:freqrad}.

\subsection{Higher order corrections: background circular orbits} \label{sec:Mathieu}

The motion described by \eqref{eq:ELweakfieldradial} is formally the same as that of a point particle (in Newtonian gravity) around a spherical body whose luminosity changes~\cite{Saslaw:1978}~\footnote{Such scenario is relevant for the analysis of dynamics of dust or small planetary systems' bodies around variable stars, where the time-dependent radiation pressure, acting as a perturbation, influences orbital dynamics.}. In the following, we take a similar approach in order to assess the dynamics beyond the linear regime.

In the previous sections, we used a linear expansion 
$x^{\mu}=x^{\mu}_0+\epsilon \eta^{\mu} $ in the small parameter $\epsilon$ to understand the evolution of the perturbation. This expansion is in fact a truncated version
of the correct full series $x^{\mu}=x^{\mu}_0+\sum^{\infty}_{n=1}\eta^{\mu}_{n} \epsilon^n$. To understand what new features can arise in the full theory,
we now expand in the radial coordinate, still at the linear level, but with different parameter strength $\lambda < \epsilon$ - This will allow us to ``effectively'' probe the higher-order terms. Using this expansion, the equation of motion for the $\eta^r$ [Eqs.~\eqref{eq:EL_weak_field_radial_pert_1st} and \eqref{eq:ELweakfieldcircpert}] now  reduces to
\beq 
&& \lambda \partial^2_t\eta^r+\lambda \Big((2\omega_{\rm res})^2+\epsilon V_1 '' \cos{(2\omega t)} \Big)\eta^r \nonumber\\
&=& -\epsilon F\cos{(2\omega t)} +\mathcal{O}(\lambda^2)\,, \label{eq:ELweakfieldcircpertMathieu}
\eeq
where $F=V_1 '$ for circular orbits and $F=V_1''r_0(t)$ for radial motion is the forcing term. The corresponding resonance frequencies $\omega_{\rm res}$ are given by Eqs.~\eqref{eq:N_res_freq} and \eqref{eq:N_res_freq_radial} for circular and radial motion respectively. For $\lambda=\epsilon$, we recover the previous results at linear order in these parameters. The $\epsilon\lambda$ term in \eqref{eq:ELweakfieldcircpertMathieu} impacts the equations of motion of $\eta^{r}_2$, thus explaining our claim that we are ``probing'' higher order behaviour. From now on, we absorb $\epsilon$ in $V_1$ and $\lambda$ in $\eta^r$. Equation \eqref{eq:ELweakfieldcircpertMathieu} is known as the inhomogeneous Mathieu equation.

We can classify the motion described by \eqref{eq:ELweakfieldcircpertMathieu}, by comparing the driving and natural frequencies $\omega$ and $\omega_{\rm res}$ respectively. The motion can then be in {\it adiabatic} $(\omega_{\rm res} \gg \omega)$, nearly {\it parametric-resonant} $(\omega_{\rm res} \sim \omega)$ or {\it rapidly oscillating background} $(\omega_{\rm res} \ll \omega)$ regime. In the next few subsections we will focus on the circular background motion, but the analysis is easily generalized.

\subsubsection{Adiabatic regime} \label{sec:Mathieu_adiabatic}

Here it is natural to use an adiabatic approximation \cite{LLbookmechanics, binney2011galactic} in order to understand the motion of particles. For time scales of the order of $1/\omega_{\rm res}$, Eq.~\eqref{eq:ELweakfieldcircpertMathieu} describes an harmonic oscillator with a constant driving force. This type of external force does not deform the phase portrait of the oscillator, but only shifts it by a constant amount $\eta^r_{c,0}=-V_1 '/W(0)^2$ (think of the mass attached to the vertical elastic spring), where  
\be
W(t)=\sqrt{(2\omega_{\rm res})^2+V_1 '' \cos{(2\omega t)}}\,.
\ee
The time dependence will, on the one hand, modify the center of the phase-space trajectory [non-homogeneous term in \eqref{eq:ELweakfieldcircpertMathieu}] as $\eta^r_c \approx (W(0)/W)^2\eta^r_{c,0}\cos{(2\omega t)}$. On the other hand, the phase-space trajectory will be itself deformed because of the time dependence of the effective frequency $W$ [homogeneous part of \eqref{eq:ELweakfieldcircpertMathieu}]. The Hamiltonian that effectively describes the $\{\eta^r,\partial_t\eta^r \}$ motion is
\begin{equation} 
H=\frac{1}{2}(\partial_t\eta^r)^2+\frac{1}{2}W^2(\eta^r-\eta^r_{c})^2=IW\,,\label{eq:hamiltonian_adiabatic}
\end{equation}
where we introduced the action-angle variables $\{ I,\theta \}$ as $\eta^r-\eta^r_c=\sqrt{2I/W} \sin{\theta}$, $\partial_t\eta^r=\sqrt{2IW} \sin{\theta}$. As the action is approximately preserved during the adiabatic process we can calculate it at the initial time $I(0)=I_0$ and find approximate analytical solution to the \eqref{eq:ELweakfieldcircpertMathieu} in this regime
\begin{equation} \label{eq:adiabatic_solution}
\eta^r(t) \approx \eta^r_c(t)+\sqrt{\frac{2I_0}{W}} \sin{\theta(t)},
\end{equation}
where $\theta(t) \approx  (2\omega_{\rm res})t$ and we used Hamiltonian equations of motion $\partial_t \theta \equiv \partial H / \partial I= W$. From \eqref{eq:hamiltonian_adiabatic} we can also see the leading behavior of the energy function - it will be periodic with the period of $\pi/\omega$.


\subsubsection{Parametric resonances} \label{sec:Mathieu_parametric}
When the driving and natural frequencies are similar, parametric resonance can occur. This result is natural to understand - time-periodic potential effectively modulates the gravitational constant. This is most dramatically seen when $V_1 \propto r^{-2}\cos{2\omega t}$, as then at the full non-linear level $G_{\rm eff}-G \propto \cos{2\omega t}$ ~\cite{Saslaw:1978}. In general, this manifests itself at the perturbative level i.e.  for the epicyclic frequency. 

We will introduce dimensionless time and rescale parameters in \eqref{eq:ELweakfieldcircpertMathieu} as  $T = 2 \omega t$, $a=(\omega_{\rm res}/\omega)^2$, $2\epsilon=V_1 ''/(2 \omega)^2$ and $f=-V_1 '/(2 \omega)^2$:
\begin{equation} \label{eq:inhomogen_Mathie}
\partial^2_T\eta^r+(a+2\epsilon \cos{T})\eta^r=f\cos{T}.
\end{equation}
Stability analysis discussed in Appendix \ref{AppParRes} predicts instabilities when 
\beq 
 \omega&=&2\omega_{\rm res}/n \nonumber\\
&=& \{ 2\omega_{\rm res},\omega_{\rm res}, 2\omega_{\rm res}/3, \omega_{\rm res}/2, 2 \omega_{\rm res}/5, \rm{...} \}. \label{eq:N_omega_instable}
\eeq
%


\subsubsection{Rapidly oscillating background}
\label{sec:Mathieu_rapid}

Motion in a rapidly oscillating background is effectively dictated by the static background, because the perturber acts so rapidly that the system doesn't have time to adapt, similarly to the sudden approximation in Quantum Mechanics. This is the case, as we shall see, even when the ``perturbing'' force is the same order of magnitude as the ``non-perturbing'' force. Because of this,  we will be general and start the discussion by rewriting  \eqref{eq:ELweakfieldradial} as
\begin{equation} \label{eq:NII_law_osc}
\partial^2_t r = - U'(r)+F(r)\cos{(2\omega t)},
\end{equation}
where $U'(r)=V'(r)-\tilde{J}^2/r^3$ and $F(r)=-V_1'(r)$. Let the radial coordinate be decomposed as $r=r_{s}+\xi^r$, where $r_{s}$ and $\xi^r$ are slowly and rapidly varying parts, respectively. After this decomposition, the slow and rapid parts of the equation of motion \eqref{eq:NII_law_osc} must be separately satisfied \cite{LLbookmechanics}~\footnote{Discussion of the initial conditions when there is a non-zero phase can be found in \cite{RidingerDavidson:2007}.}. Rapid part will have the form
\begin{equation}
\partial^2_t\xi^r=F(r_s)\cos{(2\omega t)}+\mathcal{O}(\xi^r),
\end{equation}
where we assumed that $\xi^r$ terms are small. This equation can be easily integrated
\begin{equation} \label{eq:osc_motion_eta}
\xi^r(t)=-\frac{1}{(2\omega)^2} F(r_s)\cos (2 \omega t)
\end{equation}
and we can see that $\xi^r$ is indeed small, because of $1/\omega^2$ suppression, and that our assumption that the coordinate can be perturbatively decomposed irrespective of the fact that the ``perturbative'' force is bigger than the ``non-perturbative'' is correct. 
The slowly varying part of Eq.~\eqref{eq:NII_law_osc}, after averaging, has the form
\begin{equation}
\partial^2_t r_{s}=-U'(r_s)-\frac{1}{2(2\omega)^2} F(r_s)F'(r_s),
\end{equation}
As claimed, motion is governed by the time-independent effective potential and time-varying part is suppressed by $1/\omega^2$.

\subsection{Numerical evolution in the homogeneous background} \label{sec:toy_model}
%
\begin{figure*}[th]
\begin{tabular}{cc}
\includegraphics[width=0.45\textwidth,clip]{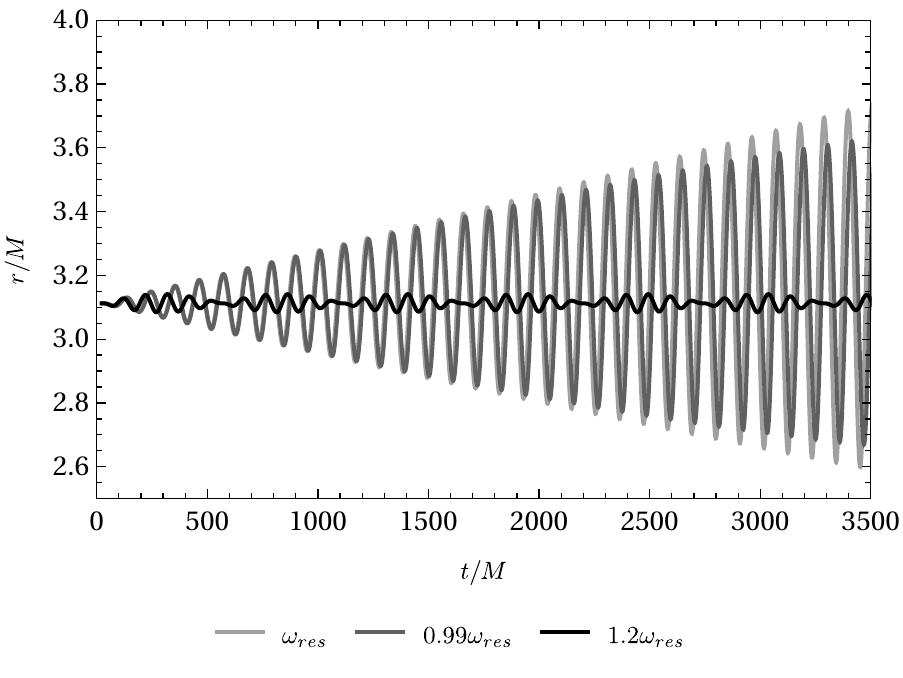}&
\includegraphics[width=0.45\textwidth,clip]{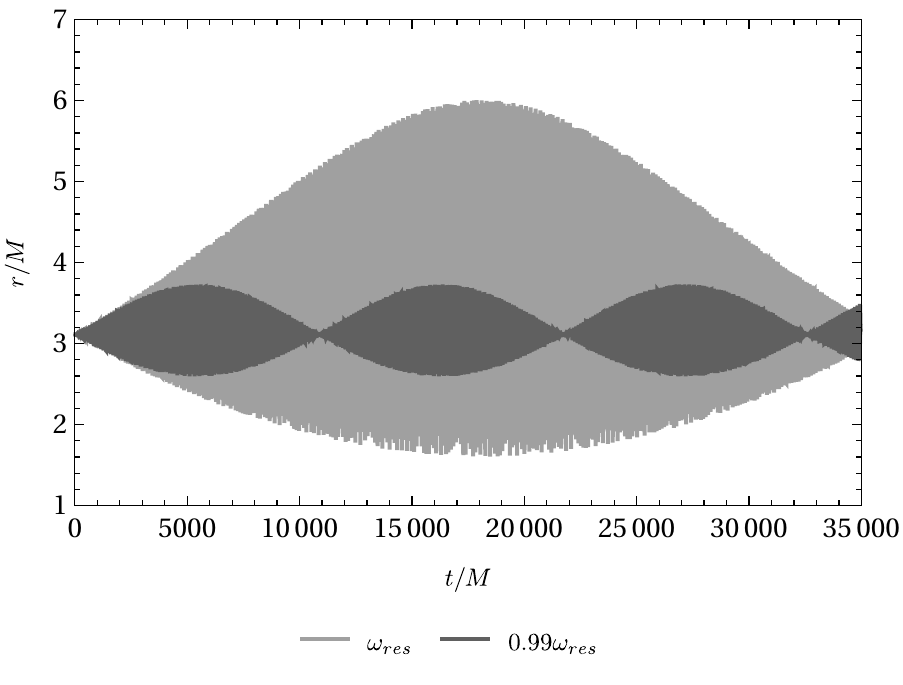}
\end{tabular}
\caption{Evolution of an initially circular geodesic in the spacetime of a time-periodic geometry~\eqref{eq:toyA}-\eqref{eq:toyB}, for different spacetime frequency $\omega$. The geodesic was circular in the static geometry of a constant density star with radius ${\mathcal C}=0.1$, placed at an initial radius $r=3.11 M$. Because the full geometry is now time-dependent with $\epsilon=10^{-3}$, the motion is not perfectly circular nor closed. For this example, there is a resonance at $M\omega=M\omega_{\rm res}=34.7\times 10^{-3}$. It is apparent that as $\omega$ is tuned in close to resonance the motion differs wildly from its unperturbed circular trajectory.}
	\label{fig:GeoNumCirc}
\end{figure*}
\begin{figure*}[th]
\centering
\includegraphics[scale=1.3]{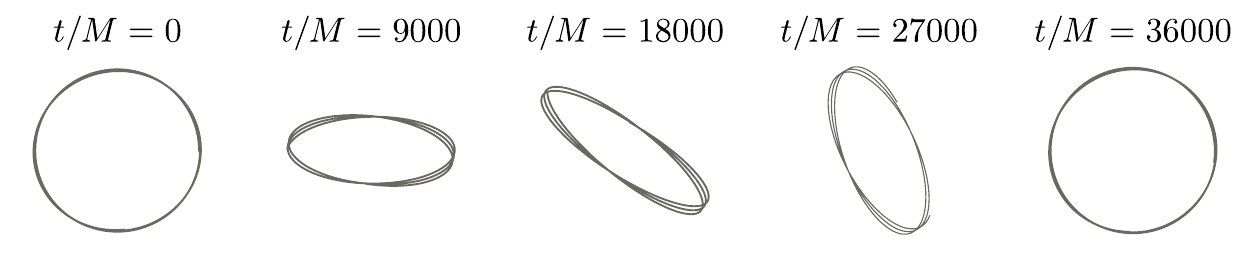}
\caption{Parametric representation of the geodesic corresponding to resonance; the central object is the same as that in Fig.~\ref{fig:GeoNumCirc}. The values of $t/M$ are the initial instant for which the geodesic is being represented, during $\delta t/M=500$.  As stated, the motion is not perfectly circular nor closed. The geodesic oscillates periodically from the initial orbit to an ellipse of large eccentricity, returning to the unperturbed motion after a beating period.
}
\label{fig:ParametricCirc}
\end{figure*}

To explore these results in a specific setup, we solved the unperturbed equations of motion numerically in an artificial, toy-model:
a constant density star spacetime, on top of which a time-periodic fluctuation was superposed. To be specific, we set the metric components:
\beq
A&=&A_{\rm star}(r)+\frac{M^2\,A_{\rm star}(r)}{M^2+r^2}\epsilon \cos(2\omega t)\,, \label{eq:toyA} \\
B&=&B_{\rm star}(r)+\frac{M^2\,B_{\rm star}(r)}{M^2+r^2}\epsilon \cos(2\omega t)\,,\label{eq:toyB}
\eeq
Here, $A_{\rm star}(r), B_{\rm star}(r)$ correspond to the geometry of a constant density star, of mass $M$ and radius $R$ in General Relativity~\cite{bookWeinberg:1972, Macedo:2013jja}. This rather arbitrary choice could mimic for example radially oscillating stars or other geometries. For us here, it is merely a toy arena where we can test of previous results.
We assume that there is no coupling between the fluid in the star and the orbiting object, and that therefore the object follows a geodesic. A straightforward analysis shows that
for $\epsilon=0$ there are {\it stable} circular timelike geodesics for any $r^2<R^3/(2M)$ and that they are all stable if the compactness $\mathcal{C}<23/54$.

In the weak field limit of our toy model \eqref{eq:toyA}, $A_{\rm star}(r)=1+2V_{\rm star}(r)$ and
\begin{equation} \label{eq:N_star_potential}
V_{\rm star}=-\mathcal{C}\Big(\frac{3}{2}- \frac{r^2}{2R^2}\Big)
\end{equation}
corresponds to the potential of a homogeneous sphere in the Newtonian gravity (or spherical harmonic oscillator). From \eqref{eq:N_res_freq} we obtain $\omega_{\rm res}=\tilde{\Omega}$. This result is consistent with the evaluation of \eqref{eq:circresonant} for dilute relativistic stars described by \eqref{eq:toyA}, when $\Theta \approx 2 \Omega$. Note that in this setup the homogeneous part of \eqref{eq:ELweakfieldcircpert} is the result expected from Newtonian gravity - precession occurs for the radial perturbations of circular motion inside the homogeneous sphere with the epicyclic frequency $2\tilde{\Omega}$ \cite{binney2011galactic}.

For the relativistic numerical investigation we took a star with compactness $\mathcal{C}=0.1$ and $\epsilon=10^{-3}$. Using the above, the functions $a_0\,,\,a_1\,,\,b_0\,,\,b_1$ [defined in \eqref{eq:weakly1}-\eqref{eq:weakly2}] are trivially known, and the geodesics can be numerically solved without approximations, using the full metric. We imposed initial conditions corresponding to fully unperturbed circular geodesics,
and monitored the position $r(t)/M$. The trajectory is shown in Fig.~\ref{fig:GeoNumCirc} for three different ``driving'' frequencies $\omega$. 
Since the equations of motion are accurate up to order ${\cal O}(\epsilon^2)$, an absolute resonance is not featured in our fully numerical solution. Instead, we find a beating pattern due to the interference of the two sinusoidal signals in \eqref{circularsol}. We applied a numerical Fourier analysis to these solutions to understand the spectrum of frequencies present in the data.
Our results show a clear, discrete spectrum of two frequencies for each example. These match, to within an error of $~0.1\%$, to the ones given by $\Theta(\omega)$ in \eqref{Theta} and $\omega$,  confirming the validity of our perturbative analytic results. The beating frequency is defined by the half difference between the frequencies of the two signals in \eqref{circularsol}
\beq \label{eq_beating_freq}
\omega_{\rm beat}=\frac{\omega-\omega_{\rm res}}{2}\,.
\eeq
Therefore, the beating period becomes larger as $\omega$ approaches $\omega_{\rm res}$ , given by Eq. \eqref{eq:circresonant}, as seen in Fig.~\ref{fig:GeoNumCirc}. Likewise, the position $r(t)$ grows to larger 
amplitudes as $\omega$ reaches $\omega_{\rm res}$, indicating that this is, in fact, a resonance.

To understand whether instabilities, as predicted by the analysis of 
Sections \ref{sec:Mathieu_parametric}, were possible, we numerically evolved the orbits for ``driving'' frequencies $\omega$ given by \eqref{eq:N_omega_instable} and explored the parameter space spanned by $\{ \mathcal{C},\,\epsilon,\,r_0,\, t_i\}$. The motion is confined to within the ``star'' at all times, i.e. $r(t)<R$. For all the parameter values that we explored, we found {\it resonant-like} behaviour for $\omega=2 \omega_{\rm res}$. For $\omega= \omega_{\rm res}/2$, the behaviour depends on the
value of the parameters. When $\mathcal{C} \sim 0.1$, small $r_0$ and large $\epsilon$ we found resonant-like behavior. However, for other points in the parameter space, the envelope of the $r(t)$ is seemingly linearly and very slowly growing. This growth may be tamed at some proper time, but we haven't observed this in all cases. Regarding the phase $t_i$, the existence of resonances seems
to be independent on it.

We have not found any resonant or unbounded motion for $\omega=2\omega_{\rm res}/3$, $\omega=2\omega_{\rm res}/5$ or $\omega=\omega_{\rm res}/3$. As this conclusions haven't changed for dilute backgrounds (i.e. weak fields) where $\mathcal{C} \sim 10^{-3}$ we can conclude that higher order terms of the expansion are responsible for the taming of the  instabilities predicted by the Mathieu equation. It should be noted that we numerically evolved trajectories until $\tau/M \sim 10^{9}$ and that resonant or unbounded motion may become apparent at later times in the cases where it was not found. 
The orbital motion in the adiabatic and rapidly-oscillating background regime is in very good agreement with the behaviour described by Eq.~\eqref{eq:adiabatic_solution} and in Section \ref{sec:Mathieu_rapid}, respectively, for small and qualitatively even for large compactness~\footnote{In these two regimes, the equations of motion are stiff: they contain two time scales with big gaps between them and one should use appropriate integrators \cite{PressRecipes:2007}.}.
\begin{figure}
\centering
\includegraphics[width=0.5\columnwidth]{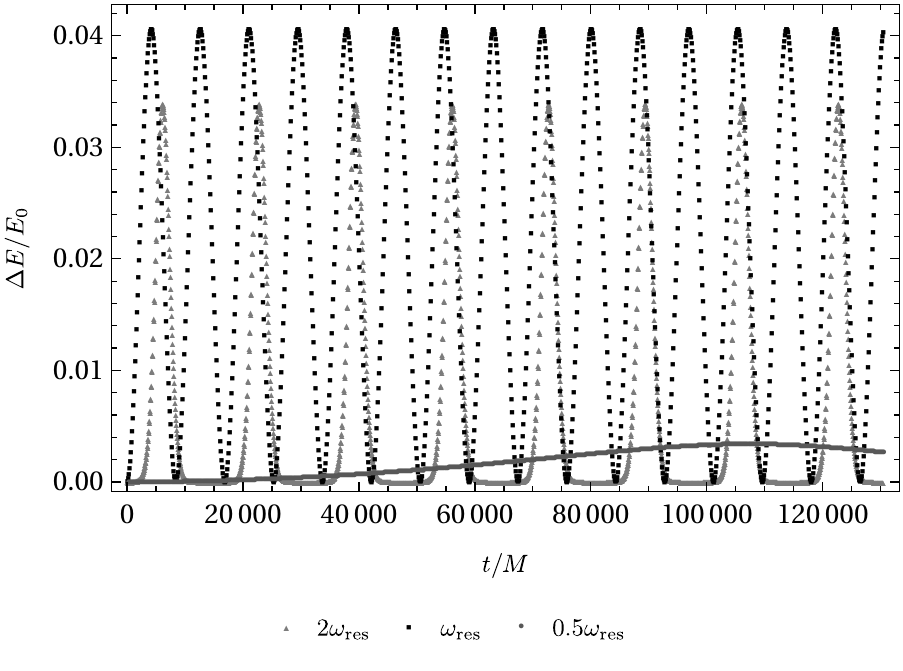}
\caption{Time evolution of the variation of the orbit's energy ($E=-\partial L/\partial t$), weighted by its initial value. The particle has a background circular motion around the toy model under the same conditions that were simulated in fig. \ref{fig:GeoNumCirc}, with $\epsilon = 1/100$. The perturbative term has frequencies, $2\omega_{\rm res}$, $\omega_{\rm res}$ and $.5\omega_{\rm res}$, for which the evolution of the energy is periodic with the frequency of the envelope on resonant motion. Figure credits: \cite{Boskovic:2018rub}}
\label{fig:EnergyvTime_circular}
\end{figure}

The time evolution of the energy, $E=-\partial L/\partial t$, is as expected from our earlier general considerations. In particular, focus on the behaviour of $\Delta E=|E(n\pi/\omega)-E_0|$, $n \in \mathbb{N}$ and $E_0=E(0)$. When the driving frequency is not given by $\omega=\{\omega_{\rm res}/2,\omega_{\rm res},2\omega_{\rm res} \}$, we find $\Delta E/E_0 \ll 1$. On the other hand, when $\omega$ equals one of these resonant frequencies, the relative change is periodic, as seen in Fig.~\ref{fig:EnergyvTime_circular}. We find that the period is the same as the envelope of $r(t)$ in Fig.~\ref{fig:GeoNumCirc}.

Finally, the parametric representation of the geodesic in Cartesian coordinates, shown in Fig.~\ref{fig:ParametricCirc}, features the predicted precession of the geodesic. The initial circular geodesic is deformed into an ellipse whose eccentricity peaks when the deviation is also maximum, before returning to circular after a beating period.

We have also studied radial motion in this background. The features of the motion strongly depend on the parameters. In general, if the time-varying component of the metric is strong enough
(i.e. for large enough $\epsilon$), resonances can be excited for {\it any} initial conditions {\it and} for any background frequency. We observed such behavior for sufficiently dilute configurations.
The natural frequency of small-amplitude oscillations in general differs from the background's. However, the spacetime drives the object to frequencies
which seem to be a multiple of background's.
This drift in frequency is confirmed by numerical Fourier analysis and the phenomena is observed both for ``driving'' frequencies much larger and smaller than the natural frequency (we find a drift even when the driving frequency is two order of magnitude larger or smaller). The behaviour of the solution in this scenario departs strongly from the one described in Section~\ref{sec:circularmotionexample}; however, all the resonances are tamed at some point in proper time. This behavior seems to be strongly model-dependent. In the small $\epsilon$ regime, our results
are similar to the circular case.

\subsubsection{Impact of dissipation} \label{sec:toy_model_dissipation}

The above analysis neglects dissipative effects. It is, in principle, possible that the resonances do not leave any observable imprint
in realistic situations: gravitational drag, along with gravitational radiation losses could, for instance, drive the inspiralling body inwards without
even being affected by resonances. To test this, we have added a dissipative force $F$ to the motion of the body, of initial mass $\mu_p(\tau=0)=0$ and radius $R_p$. The equations of motion are given by
\beq
\mu_p\left(\ddot{x}^{\gamma} + \Gamma^{\gamma}_{\alpha\beta} \dot{x}^{\alpha} \dot{x}^{\beta}\right) = F^{\gamma}.
\label{eq:dragmotion}
\eeq 
The force can describe several effects, such as gravitational radiation reaction, accretion of gravitational drag~\cite{Macedo:2013qea,Barausse:2014tra}. Regarding accretion, for a small compact object (radius $R_p$ much smaller than the mean free path) its mass growth is determined by
\beq
\dot{\mu}_p = \frac{\pi \rho R_p^2}{v},
 \label{eq:accretion}
\eeq 
where $\rho$ is the density of the compact object ``generating'' the dynamical spacetime configuration and $v$ the relative velocity between the orbiting body and the compact star. The gravitational drag force may be modelled by dynamical friction on a constant-density medium~\cite{Ostriker:1998fa,Macedo:2013qea,Barausse:2014tra}
\beq
F_{DF}=- \frac{4 \pi \mu_p ^2 \rho}{v^2} I_v\,,
\eeq
with, 
\beq
I_v = 
\begin{cases}
\frac{1}{2}\log \left(\frac{1 + v/c_s}{1 - v/ c_s}\right) - v/c_ s , \quad &v < c_s \\
\frac{1}{2}\log \left(1-\frac{c_s^2}{v^2}\right) + \log \left(\frac{v t}{r_{\rm min}}\right) , \quad &v > c_s \\
\end{cases}
\eeq
where $c_s$ is the velocity of sound in this medium and $r_{\rm min}\sim R_p/v$~\cite{Macedo:2013qea}.
In our simulations, we always used subsonic motion.

Choosing once again $\theta = \pi/2$ and taking into account the effect of accretion and gravitational drag, the dissipation force may be modelled in a Newtonian way:
\beq
F_D^t &=& 0\,,\\
F_D^r &=& -\dot{\mu}_p\dot{r} + F_{DF} \frac{\dot{r}}{v}\,,\\
F_D^\varphi &=& -r\dot{\mu}_p\dot{\varphi} + F_{DF} \frac{r\dot{\varphi}}{v}\,.\label{eq:dissipationequation}
\eeq
The system \eqref{eq:dragmotion} and \eqref{eq:accretion} determines the motion of an object through this perturbed spacetime.

\begin{figure}
\centering
\includegraphics[width=0.5\textwidth]{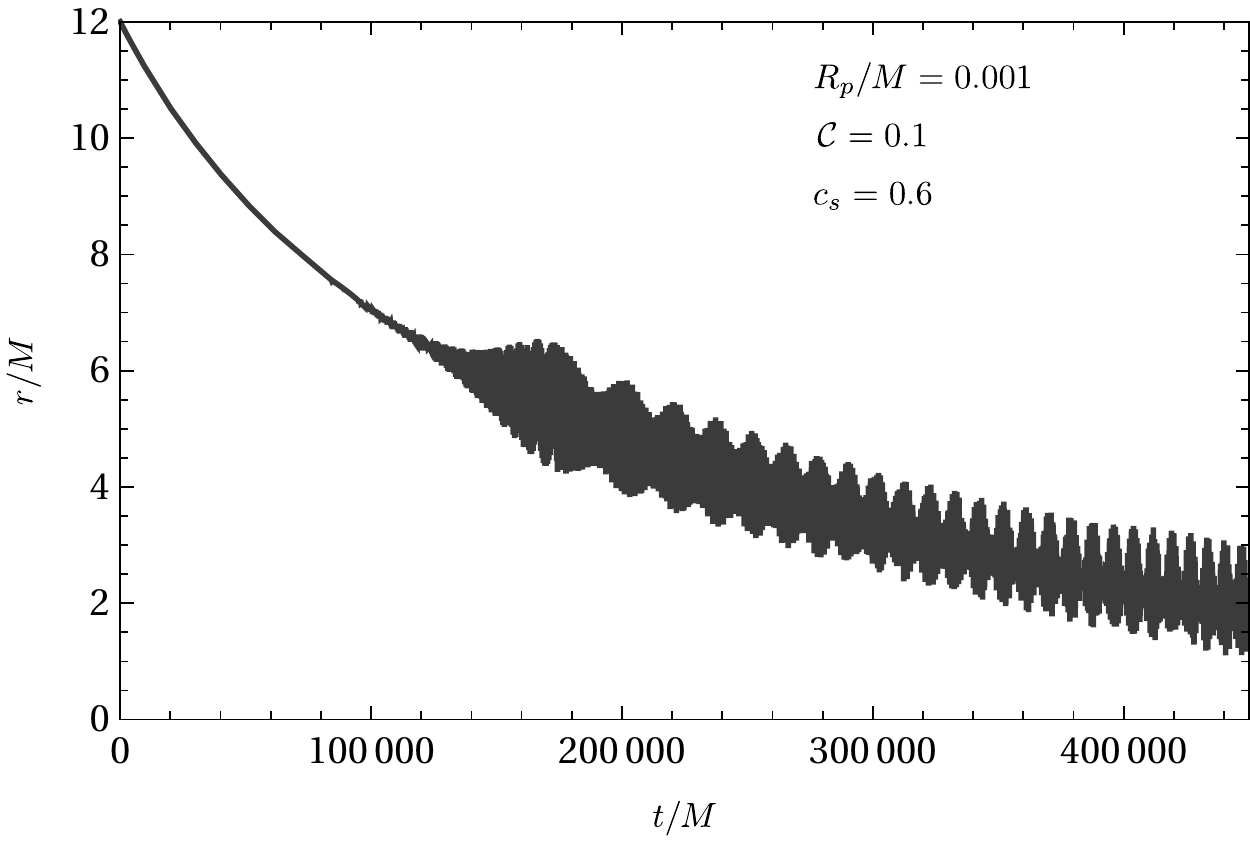}
\caption{Radial motion of an object subject to very weak dissipation on the time-periodic geometry \eqref{eq:toyA}-\eqref{eq:toyB}, where resonance occurs for $r/M = 6$. The star has $\mathcal{C}=0.1$, the inspiralling object has $R_p = 10\, \mu_{p_0} =10^{-3}M$. The sound speed was taken to be $c_s=0.6 c$. Figure credits: \cite{Boskovic:2018rub} \label{fig:DragGeoNumCirc}}
\end{figure}

Fig.~\ref{fig:DragGeoNumCirc} represents the radial evolution of an originally circular motion of a very small object in the previous homogeneous toy model, undergoing subsonic dissipation. As in the Section \ref{sec:toy_model}, the compactness of the star is taken to be $\mathcal{C}=0.1$, the density of the medium $M^2\rho=3M^3/(4 \pi R^3)\simeq2.38732 \cdot 10 ^{-4}$ and the speed of sound ($c_s$) was chosen to be $0.6 c$. Regarding the orbiting object, its initial mass and radius was chosen to be very small on the geometry ($R_p = 10 \,\mu_{p_0} = 10^{-3}M$), such that the effect of the drag would be small. The oscillating frequency of this toy model was chosen such that resonance would occur at $r/M = 6$. It is clear from Fig.~\ref{fig:DragGeoNumCirc} that the object undergoes a very slow inspiralling motion until it reaches $r/M \sim 6$. Then, the eccentricity of the orbit lowers, and the trajectory becomes similar to the resonant behaviour studied in the previous section. This implies that an object captured by the gravity of this periodic structures will undergo a resonance (if possible) when reaching the correspondent radius. 

Our results for larger damping indicate that the drag hastens the decay of the object to the center of the star. For large enough damping, the resonance (and forcing) has little impact on the motion, as expected. Consequently, if the friction is too large, resonance might be unobservable.

\newpage

\section{Massless particles in time-periodic backgrounds}

Geodesic equation for the 4-momentum of the photon
\begin{equation}
\frac{dP^{\mu}}{d\lambda} + \Gamma^{\mu}_{\alpha \beta} P^{\alpha} P^{\beta} = 0,
\end{equation}
where the Christoffel symbols $\Gamma^{\mu}_{\alpha \beta}$ are calculated using the full time dependent metric of Eq.~\eqref{metric_expansion}.

Here we will discuss gravitational redshift in time-periodig geometries. Observers measure different proper-time intervals depending on the properties of the underlying metric; thus, two observers may measure different frequencies for the same pulse of radiation 
if they are at different points in spacetime. This frequency difference can be quantified by the quantity
\begin{equation}
\frac{\omega_i - \omega_j}{\omega_j}\,,
\end{equation}
where $\omega_i$ and $\omega_j$ are frequency measurements (of the same radiation signal) by two different observers. This quantity is called the gravitational redshift.

Given the umbilical relation between the spacetime and the gravitational redshift, for a time-dependent metric the redshift also varies in time. 
This possibility has been investigated, in a non-cosmological context, for light propagating through a gravitational wave~\cite{kaufmann1970redshift,faraoni1991frequency,smarr1983gravitational} and in the context of quantum fluctuations of spacetime \cite{thompson2006spectral}. For simplicity, but without loss of generality, focus on radiation emitted by a ``star'' at rest at the origin $r=R_e=0$ and received at a distance $r=R_r$, in an oscillaton spacetime.
Considering that the emitter and the receiver are at rest, one can write their 4-velocities as
\begin{equation}
u^{\mu} = \left(\frac{1}{\sqrt{A(t,r)}},0,0,0\right),
\end{equation}
such that the 4-momentum of a radial photon is
\begin{equation}
P^{\mu} = \left(\frac{E_r}{\sqrt{A(t,r)}},\frac{E_r}{\sqrt{B(t,r)}} ,0,0\right)\,,
\end{equation}
where $r=R_e,\,R_r$ is the radial coordinate of emission or of reception of the signal.

The energy of the photon is given by $E_r = \hbar \omega_r$, where $\omega_r$ is the associated frequency; using this definition together with the explicit form of its 4-momentum, one can write the redshift between the emitter and receiver as
\begin{equation}
\label{eq:red-general}
Z\equiv 1 + z \equiv \frac{\omega_e}{\omega_r} = \sqrt{\frac{A(t_e,R_e)}{A(t_r,R_r)}}\frac{P^t(R_e)}{P^t(R_r)}\,,
\end{equation}
where $P^t$ stands for the time component of the 4-momentum of the photon and $R_e$ and $R_r$ correspond to the radial coordinates of the emitter and the receiver of the photon, respectively.

For static spacetimes, there's an exact time-like Killing vector which guarantees that $P^t(R_r) a_0(R_r) = P^t(R_e) a_0(R_e)$. Using this result in Eq.~\eqref{eq:red-general} it follows that
for static spacetimes,
\begin{equation}
\label{eq:static-redshift}
Z_{\rm static} = \sqrt{\frac{a_0(R_e)}{a_0(R_r)}} \frac{a_0(R_r)}{a_0(R_e)} = \sqrt{\frac{a_0(R_r)}{a_0(R_e)}}\,,
\end{equation}
a well-known result analysed in the context of static Boson stars~\cite{schunck1997gravitational}.

In Ref. \cite{Boskovic:2018rub} oscilalting redshift was considered in the context of oscilaton spacetimes. Weak-field limit of this problem was considered in \cite{Khmelnitsky:2013lxt} (see also \ref{sec:orbital_other}).

\newpage

\section{Phenomenological implications} \label{sec:motion_pheno}

\subsection{Motion in oscillatons} \label{sec:applications_oscillatons}

We now briefly discuss how the previous results apply to oscillaton spacetimes, studied in Section~\ref{sec:axion_stars}.
The motion in the spacetime describing oscillatons has been studied numerically for very specific conditions~\cite{2006GReGr..38..633B}, the results obtained agree qualitatively with the conclusions drawn from Section~\ref{sec:toy_model} for homogeneous backgrounds, and with our own simulations of motion in oscillatons. In Ref.~\cite{2006GReGr..38..633B}, background circular geodesics were perturbed and numerically solved for an oscillaton, with initial conditions corresponding to a turning point. For different values of initial radius and angular momentum, bound orbits with elliptic-like behavior were observed, similar to the ones in our toy model example of Fig.~\ref{fig:ParametricCirc}. Both the period and amplitude of the oscillation were sensible to these conditions, as seen in both our analytical and numerical results. Ref.~\cite{2006GReGr..38..633B} also found that there are initial conditions for which the amplitude of oscillations are negligible.
Our results agree with these findings.

An obvious question concerns the existence of resonances for these objects. We used our numerical results from Section~\ref{sec:ROscillatons} to compute the ratio between the resonant and the oscillaton's frequency, for both circular, Eq.~\eqref{eq:circresonant}, and radial, Eq.~\eqref{eq:freqrad} motion. We paid also special attention to multiples of such ratio, for which the Mathieu equation predicts instabilities - Eq. \eqref{eq:N_omega_instable}. Such ratio, by virtue of being dimensionless, can only depend on the product of oscillaton $M$ and the scalar $\mu$ mass, and the compactness $\mathcal{C}$ is a suitable choice of dimensionless combination. For the most compact oscillatons $(\mathcal{C} \sim 0.07)$, the difference between $\omega$ and $2\omega_{res}$ was a factor of two too large, and the gap widens as $\mathcal{C}$ decreases.
In summary, our results indicate that neither on circular nor radial motion is able to excite resonances in oscillaton spacetimes.

We can use the analytical solution in the Newtonian regime (Section~\ref{sec:NOscillatons}) to confirm these findings for dilute oscillatons. Taking the expansion~\eqref{eq:Nmetricexpansion} and using it in Eq.~\eqref{eq:freqrad}, we find that, for radial motion near the origin the resonance frequency is
\beq
\omega_{\rm res}\left(\mathcal{C}\right)=1.33791 \, \mathcal{C}\,.
\eeq
Therefore, the resonance frequencies are bounded from above by the maximum allowed compactness. For the maximum compactness for which the Newtonian analysis is valid, $\mathcal{C}\approx 0.01$, one finds $\omega_{\rm max}=0.013379$. This upper bound is considerably {\it lower} than the oscillatons's frequency of Fig.~\ref{fig:OvsPhi0} (even in the relativistic scenario).
Thus no resonances are excited by radial motion.

For circular motion, using Eq.~\eqref{eq:circresonant} we find,
\beq
\omega_{\rm res}(\mathcal{C})=\frac{\mathcal{C}}{2\sqrt{2}}\sqrt{\frac{14.32 - 88.6408 \, \mathcal{C}}{1 + \mathcal{C} \,(-6.19 + 1.79\, r^2  \, \mathcal{C})}}\,.
\eeq
Thus, we find again an upper bound
\beq
\omega_{\rm res}<\frac{\mathcal{C}}{2\sqrt{2}}\sqrt{\frac{14.32}{1 -6.19\, \mathcal{C}}}\,.\label{eq:freqcircmaj}
\eeq
The r.h.s grows with compactness between 0$<\mathcal{C}<0.161551$. This means that for the Newtonian regime ($\mathcal{C}<0.01$) the frequency of resonance in circular background motion is bounded by the r.h.s of \eqref{eq:freqcircmaj} evaluated at $\mathcal{C}=0.01$, which is $\omega_{\rm max}= 0.0138134$. Once again, this value does not get near the frequencies of oscillatons corresponding to the Newtonian regime in Figure \ref{fig:OvsPhi0}. We conclude that the motion in Newtonian oscillatons is in the rapidly oscillating background regime $(\omega \gg \omega_{\text{res}})$. It is, thus, appropriate to use the formalism developed in Section \ref{sec:Mathieu_rapid}, to understand motion in these oscillatons, applicable irrespective of the fact that spacetime is highly-dynamical or not.

We can focus the discussion on homogeneous oscillatons where, via \eqref{eq:NPoisson}, \eqref{eq:NV2}  and \eqref{eq:N_star_potential},
\beq
V'(r)=-\frac{1}{3}V_{1}'(r)=\frac{M}{R^3}r.
\eeq
This description is, by \eqref{eq:Ndensitysmallr}, valid for motion near the center. The amplitude of the rapidly varying part of the radial coordinate  \eqref{eq:osc_motion_eta} is proportional to $\tilde{\Omega}^2/ \omega^2$, where $\tilde{\Omega}^{-1}=\sqrt{R^3/M}$ is the dynamical time scale of the slowly-varying radial component. If the object has $\mathcal{C}=10^{-2}$, on the verge of the weak-field limit validity, then (see also Fig. \ref{fig:res_self_int_osc}) $\tilde{\Omega}^2/\omega^2 \sim 10^{-4}$. Thus, the motion is always well described by a smooth transition between background (circular or radial) motion on which small perturbations are superposed.

\subsubsection{Impact of self-interaction} \label{AppGPP_self_int_motion}

We will now find the ratio of the orbital frequency and the frequency of oscillatons in the weak coupling and in the Thomas-Fermi regime. The qualitative picture is simple: attractive self-interactions make the object more compact. This causes $\tilde{\Omega}(z)/\omega$ to increase, with respect to the non-self-interacting value. On the other hand, the time-dependent potential will be overtaken by the Newtonian gravitational potential at smaller distances. Repulsive self-interaction will make an opposite effect. From scaling relations (and $\omega \approx \mu$), we find
\begin{equation} \label{eq:osc_self_int_omega_ratio}
\frac{\tilde{\Omega}(z)}{\omega} = \mathcal{C}\frac{Z(\gamma)}{\sqrt{2}\beta(\gamma)}\sqrt{\frac{1}{z^3}\int^{z}_{0}y^2(s(y;\gamma))^2 dy}.
\end{equation}
The ratio is proportional to $\mathcal{C}$, with a coefficient of order one, see Fig. \ref{fig:res_self_int_osc} (where we used the profile description given in Appendix \ref{AppGPP} to compute the integral). As $\mathcal{C} \leqsim 0.01$ in the weak-field regime, even with self-interactions included, motion in the oscillatons will be in the rapidly oscilating background regime. 
\begin{figure}
\centering
\includegraphics[width=0.5\textwidth]{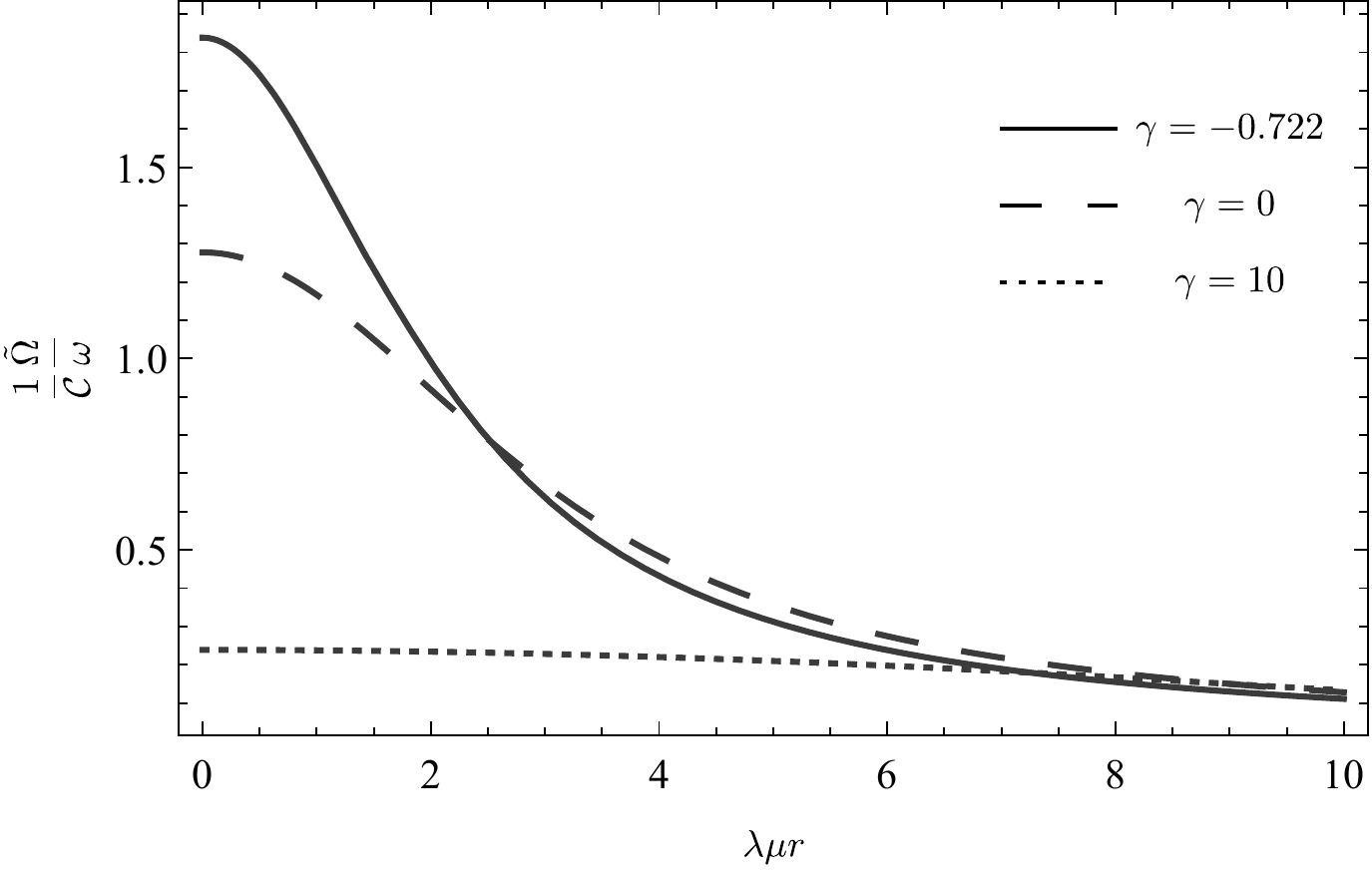}
\caption{Ratio of Keplerian and oscillaton frequency for different signs and values of the scalar self-interacting coupling. Figure credits: \cite{Boskovic:2018rub}}
\label{fig:res_self_int_osc}
\end{figure}
%

\subsection{Fuzzy DM halo spacetime dynamics and general features of motion} \label{sec:darkhalo_motion}

In Section \ref{sec:NOscillatons} we showed that the spacetime describing the soliton is highly dynamical, in the sense that the gradient of the time-dependent potential (time-dependent force) $|V'_{1}|$ is of the same order of magnitude as or larger than the Newtonian gravitational force $V'$. Regarding the halo ``atmosphere'' (outer layer), as $r_s \sim r_\epsilon$ \cite{Bernaletal:2017a}, we are interested in the $r\gg r_s$ limit of the NFW profile $\rho_{\text{NFW}}(r) \sim \rho_s ( r/r_s )^{-3}$. In this limit, the time-dependent force becomes {\it logarithmically} smaller than the Newtonian force $V'/|V'_{1}| \sim \ln(r/r_s)$. The dark halo radius, taken as $r_{200}$, the point when the mean halo density is $200$ times bigger than the cosmological critical density, is usually two orders of magnitude larger than $r_s$. Thus, even at the halo radius the dynamical component is of the same order of magnitude as the static component. We can conclude that the whole halo is highly dynamical.

As we saw in Section~\ref{sec:applications_oscillatons}, the motion in solitons is in a rapidly oscillating background regime, as $\tilde{\Omega} \ll \omega$, where $1/\tilde{\Omega}$ is timescale associated with the motion dictated by Newtonian force. The ratio of the celestial object's Keplerian orbital and the oscillaton frequency depends on the core radius and the particle's mass near the soliton center\footnote{A useful number to keep in mind is that the period of oscillation for ULA with $m=m_{22}$ is $T \sim 1 \text{yr}$.} [from Eq. \eqref{eq:fuzzy_cdensity}]
\begin{equation} \label{eq:fuzzy_omega_ratio}
\frac{\tilde{\Omega}}{\omega} \approx 4 \times 10^{-9}  \Big ( \frac{r_c}{1\text{kpc}} \Big )^{-2} \Big ( \frac{m}{m_{22}} \Big )^{-2}\,.
\end{equation}
For reference MW parameters from Section \ref{sec:axion_conf_cosmo_astro}, we find $\tilde{\Omega}/\omega \sim 4.34 \times 10^{-7}$. As rotation curves in outer regions of the halo ``atmosphere'' develop plateaus, the orbital frequencies must go further down at large distances. The effect of the time-dependent force on the orbital motion in this regime is, as explained in Sections~\ref{sec:Mathieu_rapid} and \ref{sec:applications_oscillatons}, suppressed by $(\tilde{\Omega}/\omega)^2$. In the absence of other matter sources, such suppression is extremely large in the galactic context (for the above mentioned estimates for MW, this is equal to $10^{-13}$) irrespective of the highly dynamical nature of the spacetime.

\subsection{Constraining ULA density at the Galactic center} \label{sec:darkhalo_res}
We now show that the motion of bright S stars can be used to constraint ULA DM densities at the center of our galaxy. 
Data from the last 20 years was used to probe Yukawa-like modifications of gravity \cite{HeesPRL:2017}, and new data is expected to be of further help in this endeavour~\cite{HeesELT:2017}. This year the S0-2 star will be at its closest distance from the SMBH, and a redshift measurement is expected~\cite{HeesProc:2017}.

The behaviour of matter in the sub-parsec region is dominated by the SMBH gravity. We should stress that our understanding of DM behaviour in the presence of the SMBH and during the galactic evolution timescales is still in its infancy and mostly focused on CDM~\cite{ GenzelEisenhauer:2001, MerrittD:2010} (but see also Refs.~\cite{Hui:2016ltb,Ferreira:2017pth, Helfer:2017a, Bar:2018acw, Bar:2019pnz}). The core density estimate from Section~\ref{sec:darkhalo_descr} may not apply in the sub-parsec region, since the SMBH can make the region denser, e.g. by adiabatic growth \cite{MerrittD:2010}. Thus, we use constraints from previous analysis of the orbit of S stars as a rough upper limit on the extended background and treat fraction of ULA component as a free parameter. Present constraints allow for $1\sigma$ upper limit of $M_{\text{ext}}=10^{-2}  M_{\text{SMBH}}$, where $ M_{\text{SMBH}}=4.02 \times 10^6 M_{\odot}$ and background radius cutoff was fixed at $R=11 \text{mpc}$ in order to encompass whole of S0-2 star orbit~\cite{BoehleGhez:2016}. Some CDM estimates in this region are of order $\sim 10^3 M_\odot$ \cite{GhezSalim:2008}. This background consists of faint stars, compact objects and DM. We arbitrarily take the maximum contribution of ULA, $\lambda_{\text{ULA}}$, to be $30\%$. In this approach we don't have prior restrictions on the axion mass range that we probe but orbital timescales focus the range on FDM and MDM.

To estimate the impact of ULA time-dependent force, consider a simplified model where stellar orbits are influenced only by the (non-rotating) SMBH at the Post-Newtonian (PN) level and all other matter components are incorporated in the homogeneous spherically-symmetric extended background $\rho_{\text{ext}}$. The stellar equations of motion are, from Sections \ref{sec:NOscillatons} and \ref{sec:N_dynamics} and Refs.~\cite{RubilarEckart:2001, poissonwillbook}
\beq 
\partial^2_t \vec{r} &=&-\frac{1}{r^3}\Big[\Big(1+4\frac{1}{rc^2}+\frac{v^2}{c^2}\Big)\vec{r}-4\vec{v}\frac{\vec{v} \cdot\vec{r}}{c^2}\Big]\nonumber\\
&-&\frac{M_{\text{ext}}(r)}{r^3}\vec{r}+4\pi  \lambda_{\text{ULA}} \rho_{\text{ext}}\vec{r}\cos{(2\omega t+2\Upsilon)}\,.\label{eq_motion_smbh}
\eeq
In the above we use dimensionless quantities,
\be \label{eq_variables}
t =\frac{t}{\tau_{\text{dyn}}}\,,\quad r =\frac{r}{10\text{mpc}}\,,\quad M =\frac{M}{M_{\text{SMBH}}}\,,
\ee
$\tau_\text{dyn}=\sqrt{GM_{\text{SMBH}}/r_{I}^3}^{-1}=7.4\text{yr}$ is the dynamical timescale associated with the gravity of SMBH at $r_{I}=10\text{mpc}$ and $\Upsilon$ represents phase difference. 
In the context of our results from Sections \ref{sec:circularmotionexample}-\ref{sec:toy_model} for the circular example, the presence of the additional SMBH potential does not change the picture significantly, as it can be incorporated in $V(r_0)$. 

Most studied S stars around Sgr $\text{A}^{\star}$ are on a highly eccentric orbits, and the detailed treatment of such motion is outside the scope of this work. For such orbit, Eq. \eqref{eq:N_res_freq} is not applicable as the resonant frequency can change by as much as one order of magnitude along the orbit. We will onward focus only on S0-2 star. Its (initial) orbital parameters are $a_{0}=4.878(8)\text{mpc}$, $e_{0}=0.892(2)$ and $T_{0}=15.92(4)\text{yr}$ \cite{BoehleGhez:2016}. 
Present constraints on periastron precession are: $|\dot{w}_{0}|<1.7\times10^{-3}\,\text{rad}/\text{yr}$~\cite{HeesPRL:2017,HeesProc:2017}.
We have numerically solved the equations of motion, for different extended mass $M_{\text{ext}}$ and ULA abundance $\lambda_{\text{ULA}}$ as well as phase difference $\Upsilon$.
The system starts from the apocenter, and we monitor the secular (osculating) orbital parameters, averaging over one orbit. The motion is contained within a fixed orbital plane, even at the PN level \cite{poissonwillbook}, and, ipso facto, orbital inclination and longitude of the ascending node are fixed, which leaves us with semi-major axis, eccentricity, orbital period and periastron precession. From this set of orbital elements, only periastron precession is affected by PN effects \cite{poissonwillbook} and the homogeneous background \cite{RubilarEckart:2001, ZakharovNucita:2007, JiangLin:1985}, when a time-dependent ULA force is neglected. 

\begin{figure}
\hspace*{\fill}%
\begin{minipage}[c]{0.45\textwidth}
\centering
\vspace{-1pt}
\includegraphics[width=1\textwidth]{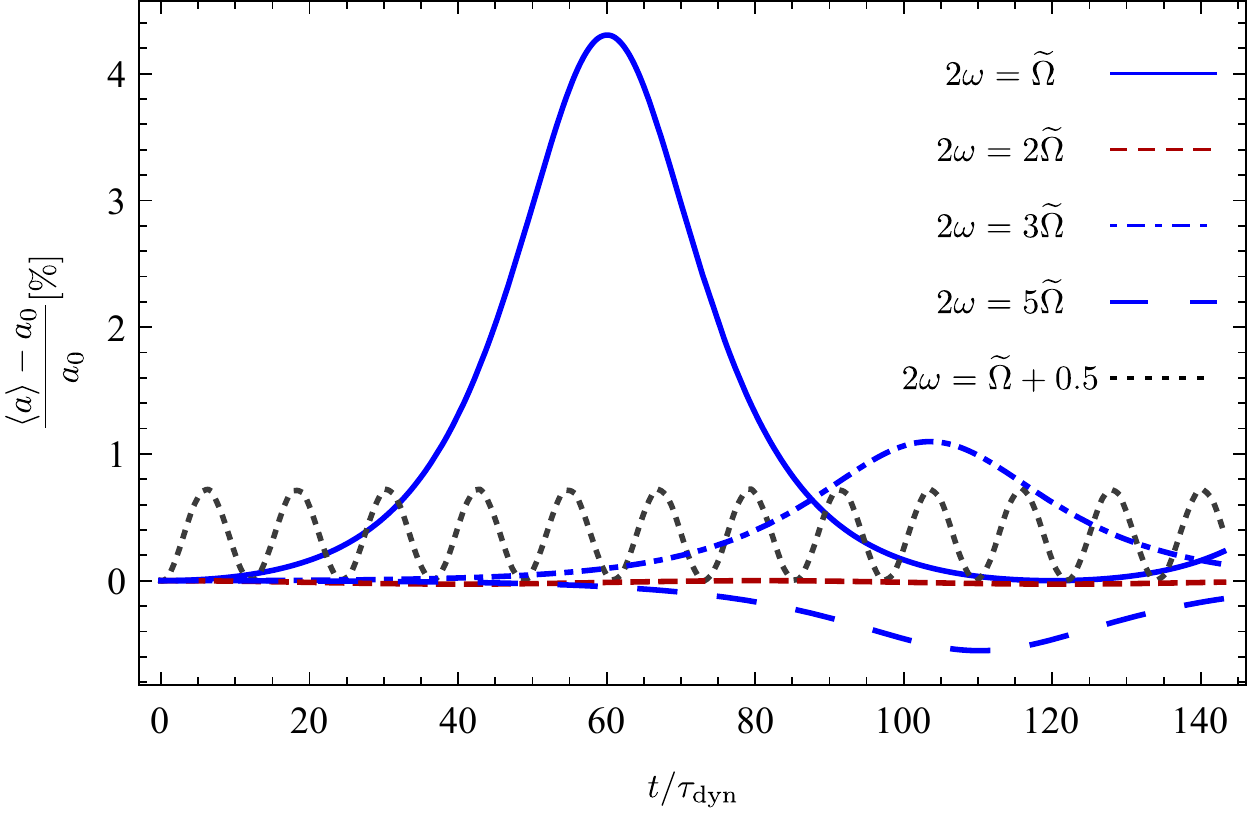}
\caption{ \fontsize{9}{12}
Relative change of secular semi-major axis $a$ with respect to the value without present time-dependent force $a_0$ for three dominant resonant (blue; solid, dashed and dashed-dotted lines), one non-resonant (dashed red line) and one near-resonant (dashed black line) frequencies. The background is, for illustrative purposes, taken as the least conservative one: $M_{\text{ext}}=10^{-2}M_{\text{SMBH}}$, $\lambda_{\text{ULA}}=0.3$ and we take $\Upsilon=0$. The axion particle masses correspond to multiples of mean motion and some of the values can be found in Tables \ref{tab:S2_prim_res} and \ref{tab:S2_sec_res}.  Figure credits: \cite{Boskovic:2018rub}.
\label{fig:SMBH_semi_major}}
\end{minipage}%
\hfill
\begin{minipage}[c]{0.47\textwidth}
\centering
\vspace{0pt}
\includegraphics[width=\textwidth]{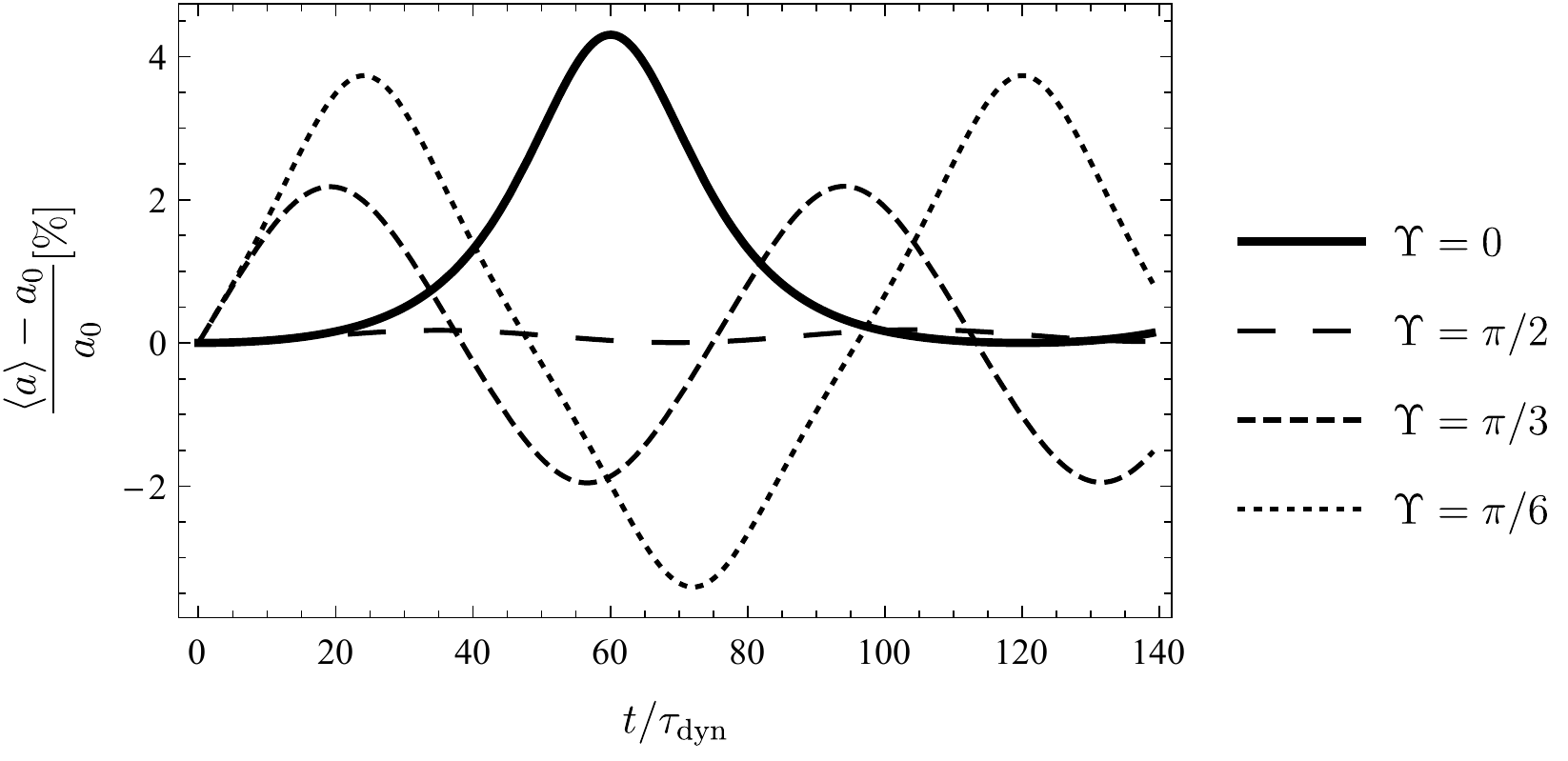}
\caption{\fontsize{9}{12} Relative change of secular semi-major axis $a$, with respect to the reference one, for dominant (odd) resonant frequency and different values of phase differences $\Upsilon$. Background is the same as in Figure \ref{fig:SMBH_semi_major}. Figure credits: \cite{Boskovic:2018rub}.
\label{fig:SMBH_semi_major_phase}}
\end{minipage}%
\hspace*{\fill}
\end{figure}

General orbital behaviour is similar to the one described in Section \ref{sec:toy_model}. Numerical calculations show, for the cases that we examined, that (anti-)resonant behaviour of the radial coordinate can be found both when $2\omega
=(2n+1)\tilde{\Omega} $ (odd resonances) and $2\omega
=2n\tilde{\Omega}$ (even resonances), where $\tilde{\Omega}=2\pi/T$ is orbital mean motion, $T$ is orbital period and $n \in \mathbb{N}$ \,\,\footnote{Similar ratios are known from resonant phenomena (mean-motion resonances) in celestial mechanics and galactic astronomy \cite{book_murraydermott, binney2011galactic}. This observations demands further investigation.}. Resonances become less pronounced, in absolute terms, with increasing $n$. When $\Upsilon=0$ only odd resonances occur and their shape is Gaussian-like. The ``sign'' of these resonances, i.e. whether they lead to increase or decrease of the oribtal radius as well as the timescales involved, depend on the environment. We also find, as in Section \ref{sec:toy_model}, (non-symmetric) window around dominant resonance inside which oscillations are slightly amplified with respect to the other driving frequencies. These cases can be analytically understood with the help of first-order perturbation theory (Section \ref{AppEliptical}). In Fig. \ref{fig:SMBH_semi_major} we show the semi-major axis secular evolution for the first three resonant frequencies as well as one non-resonant and one close to the dominant resonance. Qualitative behaviour of the $e$ and $T$ is similar. In Tables \ref{tab:S2_prim_res} and \ref{tab:S2_sec_res} we list amplitudes of secular changes of $a$ for two dominant resonant frequencies, as well as their sign. Notice that the current observational precision $0.16\%$ is enough to probe almost all the scenarios that we examined during the resonant amplification.

For $\Upsilon=\pi/2$ only even resonances occur with the same phenomenology as their odd counterparts. For all other phase differences, of the form $\Upsilon=\pi/m$, $m \in \mathbb{N} \setminus \{1,2\}$, the resonance evolution is akin to a sinusoidal shape (see Fig.~\ref{fig:SMBH_semi_major_phase}). For a given resonance, the amplitude of orbital parameters are mildly dependent on $\Upsilon$: at most (for $a$) they were lowered by $50 \%$ (for $m=3$), compared to the $m=1$ or $m=2$ case, but notice that now $a$ periodically becomes both larger and smaller than its undisturbed value (in absence of time-dependent force). This qualitative dependence on the relative phase is consequence of absence of time-translation symmetry as discussed in Section \ref{sec:symmetries}.

\begin{figure}
\hspace*{\fill}%
\begin{minipage}[c]{0.45\textwidth}
\centering
\vspace{-1pt}
\includegraphics[width=1\textwidth]{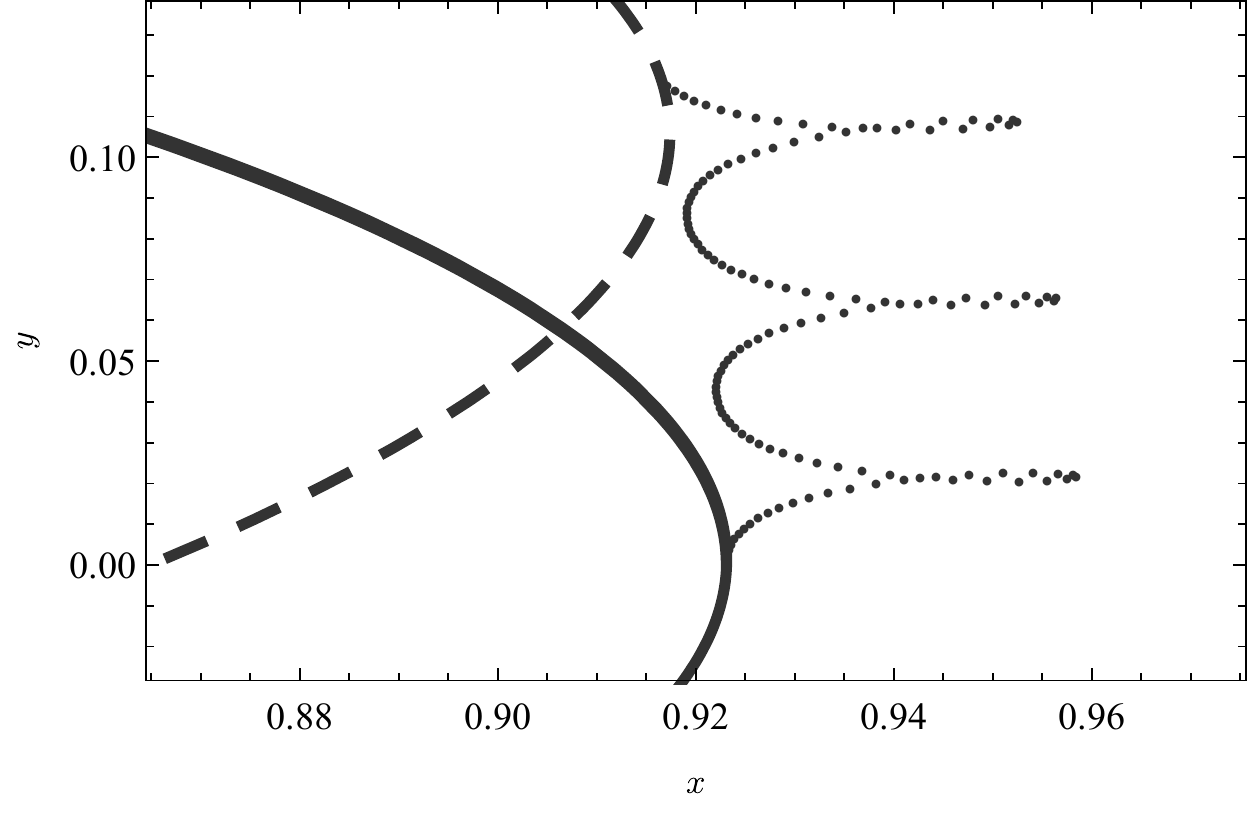}
\caption{ \fontsize{9}{12}
Apoastron shift of S0-2 star $(e=0.892)$ during $t=300\tau_{\text{dyn}}$ in orbital plane for $M_\text{ext}=8 \times 10^{-3}M_{\text{SMBH}}$, $\lambda_{\text{ULA}}=0.3$, $\Upsilon=0$ and resonant motion $(2\omega=\tilde{\Omega})$. The orbits are presented at the beginning (full line) and at the end of the interval (dashed). Black dots correspond to the apoastron position during this interval. Notice that SMBH dominates background in determining periastron shift direction (retrograde), but during resonant motion short change of direction of apoastron shift occurs. Orbital coordinates correspond to $x=r\cos{\varphi}$ and $y=r\sin{\varphi}$ and are given in the units of $10 \text{mpc}$.  Figure credits: \cite{Boskovic:2018rub}.
\label{fig:SMBH_apoastron}}
\end{minipage}%
\hfill
\begin{minipage}[c]{0.48\textwidth}
\centering
\vspace{0pt}
\includegraphics[width=\textwidth]{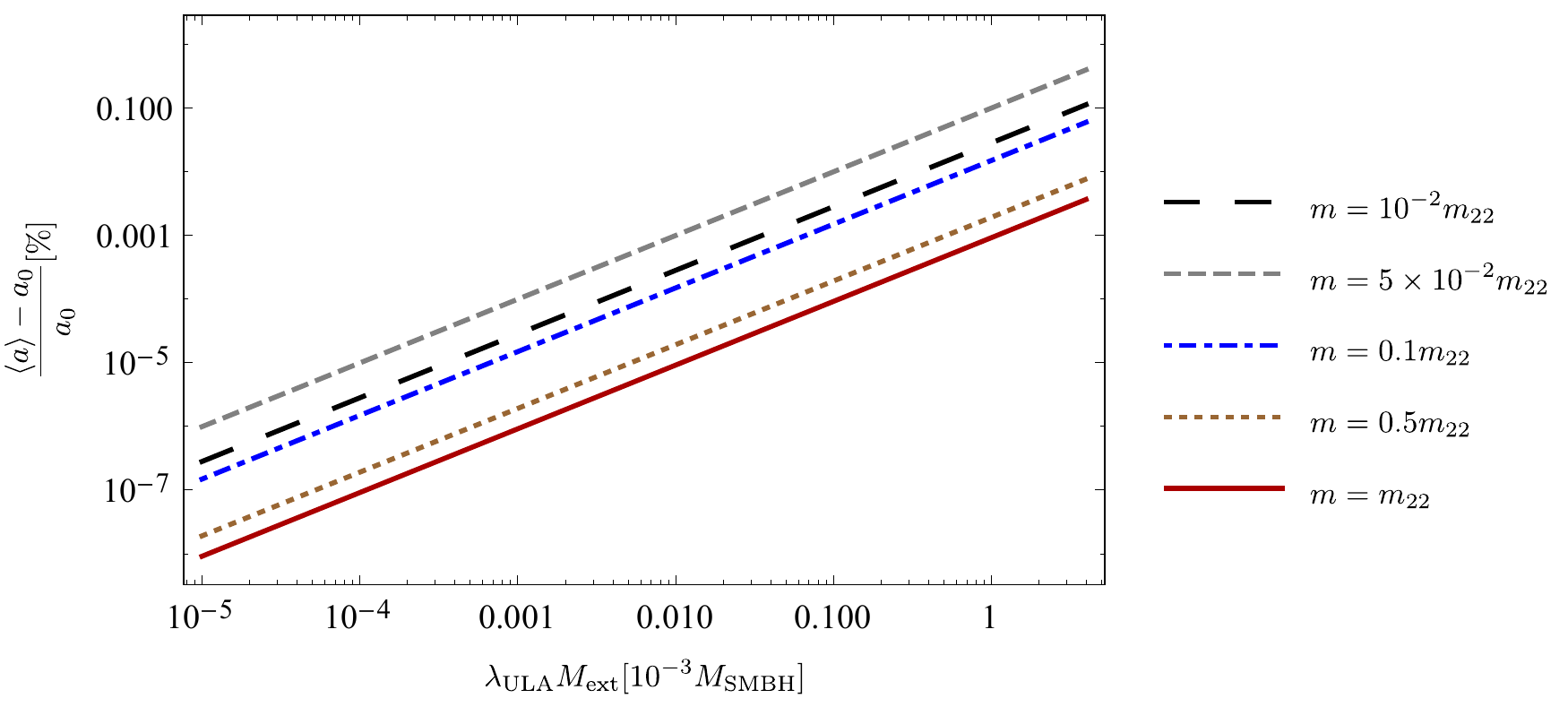}
\caption{\fontsize{9}{12} Relative change of secular semi-major axis $a$ amplitude, with respect to the reference one for different particle masses and ULA abundances. Results are obtained using first-order pertrubation theory (see Appendix \ref{AppEliptical}) for primary near-resonant window. We have neglected dependence of the phase difference as is it is, in this case, part of the argument of a harmonic function and influences only the time at which  the semi-major axis develops a maximum. Note that the maximum values are obtained for $m=5 \times 10^{-2} m_{22}$ contour, as the primary resonance corresponds to $m=4.1 \times 10^{-2} m_{22}$ (see Table \ref{tab:S2_prim_res}). Figure credits: \cite{Boskovic:2018rub}.}
\label{fig:contur_plot_S2_A}
\end{minipage}%
\hspace*{\fill}
\end{figure}

\begin{table}[]
\centering
\caption{Secular amplitudes of S0-2 star semi-major axis $a$, that correspond to primary resonance, and whose sign is denoted by $\sigma$: $+$ for amplification and $-$ for depletion. In each case the homogeneous background mass $M_{\text{ext}}$ (in units of $[0.01 M_{\text{SMBH}}]$) and ULA abundance $\lambda_{\text{ULA}}$ were varied.  We fix phase difference to $\Upsilon=0$. 
Note that the resonance is located at frequencies $\omega=m/\hbar$.}
\label{tab:S2_prim_res}
\begin{tabular}{|c  c | c | c c |}
\hline \hline
$M_{\text{ext}} $  & $m\,[m_{22}]$ & $\lambda_{\text{ULA}}$ & $\sigma$ & $\langle a \rangle \,[\text{mpc}]$   \\
\hline

$1$  & $0.0412$ & $0.3$ & $+$  & $5.089$    \\

$$   &  $$ & $0.05$ & $+$  & $4.963$    \\

$$  &  $$ & $0.005$  & $-$  & $4.854$   \\

\hline

$0.1$  & $0.0413$ & $0.3$  & $+$ & $4.931$ \\

$$ & $$  & $0.05$ & $+$ &  $4.895$  \\

\hline

$0$ & $/$ & $0$ & $/$ & $4.878$ \\
\hline
\end{tabular}
\end{table}
\begin{table}[]
\centering
\caption{Same as Table \ref{tab:S2_prim_res} for secondary resonances.}
\label{tab:S2_sec_res}
%
\begin{tabular}{|c  c | c | c c |}
\hline \hline
$M_{\text{ext}}$  & $m\,[m_{22}]$ & $\lambda_{\text{ULA}}$ & $\sigma$ & $\langle a \rangle \,[\text{mpc}]$   \\
\hline
 
$1$ & $0.124$  & $0.3$ &  $+$  & $4.932$    \\

$$ & $$  & $0.05$  &  $-$  & $4.859$   \\

$0.1$ & $0.124$  & $0.3$  & $+$ & $4.887$  \\
\hline
\end{tabular}
\end{table}
%
The extended background leads to a retrograde periastron shift of stellar orbits, as reviewed in Section \ref{sec:toy_model}. On the other hand, relativistic effects of the strong SMBH gravity lead to a prograde periastron shift, potentially masking the previous effect. Periastron precession was found by identifying successive radial maximums (in order to evade numerical difficulties explained in Ref. \cite{RubilarEckart:2001}). Sign of $\dot{w}$ direction depends on the background mass $M_{\text{ext}}$ i.e. whether it is dominated by the SMBH PN or background Newtonian contribution. The shape of periastron precession with respect to time is similar to the other orbital parameters. For resonant motion, base value of $\dot{w}$ tends to be lower (in relative terms), as compared to non-resonant one, and $\dot{w}$ increases when amplification occurs. Depending on the values of  $M_{\text{ext}}$ and $\lambda_{\text{ULA}}$ this can lead to change of sign and, as a consequence, direction of periastron precession. This manifests in apoastron developing some kind of helix trajectory as in Fig.~\ref{fig:SMBH_apoastron}. When the oscillating frequency is similar to the resonant one, range of $\dot{w}$ values is similar to the one that correspond to resonant frequency.
For all cases that we considered, value of $\dot{w}$ was inside present constraints.

Inference of ULA mass and abundance from semi-major axis is highly degenerate, as the phase difference, type of resonance and timescales over which the resonance develops significantly contribute to the problem\footnote{Whether inference of other orbital elements could significantly lower the degeneracy of the problem should be subject of further studies.}. A rough picture can be obtained by focusing on a near-resonant window and the first-order perturbation theory, as described in Fig.  \ref{fig:contur_plot_S2_A}. Long-term and precise observations of this and other S stars (and comparison with other constraints) will allow for constraining ULA densities for axion masses that correspond to the resonant frequencies.  Stars with longer orbital periods (or equivalently smaller axion masses), cannot be probed in this way as dynamical timescales become large. However, this type of resonant phenomena is known, in celestial mechanics and galactic astronomy \cite{binney2011galactic, book_murraydermott}, to leave fingerprints in the orbital parameter space, something that deserves further scrutiny. One of the better known of these structures, and potentially similar to this problem, are Kirkwood gaps in the distribution of semi-major axes of the main-belt asteroid orbits \cite{book_murraydermott}. Identification of these structures could be possible with the observation of a large number of stars in the central sub-parsec and parsec scales~\cite{ValluriDebattista:2012}.

\subsubsection{Elliptical orbits in ultralight DM background: perturbation theory} \label{AppEliptical}

In order to analytically understand the influence of time-periodic backgrounds on elliptical orbits in the context of ULA DM, consider first-order perturbation theory in celestial mechanics \cite{poissonwillbook}. We here assume that some other matter components, e.g. SMBH as in Section \ref{sec:darkhalo_res}, are dominant and consider DM background as a homogeneous one. This approach is equivalent to the one in Ref. \cite{Blas:2016ddr}, but stated in a different language. For example, taking secular equation for the change of semi-major axis $a$ and working at the first order we obtain
\begin{equation}
\langle \dot{a} \rangle_{\text{ULA}}(t) =-4\rho_{\text{ULA}}T a \frac{J_n (ne)}{n}f(t),
\end{equation}
where we used notation from Section \ref{sec:darkhalo_res}, $J_n$ is Bessel function, $f(t)=\sin(\delta \omega t+ 2\omega t_{0}+2\Upsilon)$, $\delta \omega=2\omega-n\tilde{\Omega}$ and $t_0$ is the time of the first periastron passage\footnote{In order to obtain this equation, one should change, when averaging, from integrating with respect to time to integrating with respect to the eccentric anomaly and use Kepler's equation. Also,  the term $\delta \omega t$ is approximately constant on the orbital timescales.}. Prediction of this equation is in good agreement with our numerical results from Section \ref{sec:darkhalo_res} for the driving frequencies close to the resonante one. Note that as we work at the first order, we can add different contribution to orbital element secular change. In the context of Section \ref{sec:darkhalo_res}:  $\langle a \rangle=\langle a \rangle_{\text{ULA}}$, as $\langle a \rangle_{\text{PN}}=0$.

This result is \textit{not} applicable when the true resonance occurs  i.e. $\delta \omega$ is sufficiently small as it would imply linearly diverging secular evolution of $\langle a \rangle$. Secular evolution is tamed at some point in time, as observed in Section \ref{sec:darkhalo_res}, and this phenomenon is the generic one as discussed in Sections \ref{sec:circularmotionexample}-\ref{sec:toy_model}. 

\subsection{Motion in stellar and planetary systems inside fuzzy DM halo} \label{AppBinary}

Consider now how time-periodic forces influence the motion of objects within a stellar or planetary system, which itself moves around a halo. For simplicity we focus on binary systems, like those studied recently~\cite{Blas:2016ddr,Rozner:2019gba}. Objects in such systems experience acceleration (neglecting all other contributions except their mutual gravitational interaction and time-dependent force)
\begin{equation}
m_{i}\vec{r}_{i}=\pm\frac{m_{1}m_{2}}{r^3}\vec{r}+\vec{F}^{\text{TD}}_{i}\,,
\end{equation}
where $\vec{F}^{\text{TD}}_{i}=4\pi \rho(\vec{r}_{i}) m_{i} \vec{r}_i \cos{(2\omega t+2\Upsilon(\vec{r}_{i}))}$ is the time-dependent force in the context of ULA DM and $\vec{r}=\vec{r}_2-\vec{r}_1$. We can transform perspective to that of dynamics of the relative particle $m_{\text{red}}=m_1 m_2 /M$, where $M=m_1+m_2$, and the center-of-mass (CM) $\vec{R}=(m_1 \vec{r}_1 + m_2 \vec{r}_2)/M$. Note that $\rho(\vec{r}_i) \approx \rho(\vec{R})$ and $\Upsilon(\vec{r}_i) \approx \Upsilon(\vec{R})$. Thus, time-dependent force decomposes to center-of-mass and relative component. Relative component is just a small perturbation with respect to the other forces that dictate dynamics in stellar and planetary systems, but it can lead to potentially observable consequences if resonance occurs. For example, Ref.~\cite{Blas:2016ddr} studied how it influences the secular change of binary pulsar periods and whether such effect is measurable. As noted in Ref.~\cite{Pani:2015qhr} there are two more contributions to the period change, that of the CM motion along the line-of-sight and the one induced by variation of orbital inclinations. The last effect depends on the binary orbit orientation. Furthermore, Solar System barycenter, with respect to whom pulsar timing are measured, also oscillates. 

Let us estimate ULA DM consequence on relative motion in the Solar System. Using first order perturbation theory (see Section \ref{AppEliptical} and Ref. \cite{Blas:2016ddr})
\beq
\langle \dot{a} \rangle_{\text{ULA}}(t) \approx -3 \times 10^{-5} \frac{\text{m}}{\text{yr}} \Big ( \frac{\rho_{\text{ULA}}}{1.13 \times 10^{-2} \frac{M_\odot}{\text{pc}^3}}  \Big ) \Big ( \frac{T}{1\text{yr}} \Big ) \Big ( \frac{a}{1\text{AU}} \Big ) \frac{J_n (ne)}{n}f(t)\,.
\eeq
We used values for the local DM density from Ref.~\cite{SalucciNesti:2013}. Present accuracy in Solar System observations is $\sim 10^{-1} \text{m}/\text{yr}$ for Mars and $\sim 10^{-2} \text{m}/\text{yr}$  for Moon \cite{Minazzoli:2017vhs}. Note also that Solar System planets have low eccentricites and $J_n$ additionally suppresses this ratio: e.g. for Mars $e=0.0934$ and $J_1(e)\approx4\times10^{-2}$. We stress that the truly resonant case should be properly studied numerically.


\subsection{Other topics} \label{sec:orbital_other}

Original work that initiated search for axionic imprint from time-periodic potential concerned itself with pulsar timing \cite{Khmelnitsky:2013lxt}. Pulsars are very precise rotators and measurement of their signals over time can give very precise  pulses times of arrival. If the space between the pulsar and Earth is immersed in ULA halo the spacetime will oscillate and hence there will be oscillating gravitational redshift superposed on pulsar EM signal \cite{Khmelnitsky:2013lxt}. If pulsar is inside halo core (oscillaton), the signal will be additionally enhanced \cite{DeMartinoBroadhurst:2017}. This approach already produced ome constraints on FDM local density $\rho < 6 \text{GeV/cm}^3$ ($\mu_{\text a} \leq 10^{-23}$ eV) from The Parkes Pulsar Timing Array (PPTA) \cite{Porayko:2018sfa}. These results are not particularly significant at the moment, as we dynamically know that local DM is  $\rho_{\text DM} \approx 0.4 \text{GeV/cm}^3$ \cite{SalucciNesti:2013}. However, this is remarkable proof-of-principle and future missions, such as Square Kilometer Array (SKA), could be used to probe ULA thoroughly. In addition, FDM can resonantly induce change of the orbital period to binary pulsars \cite{Blas:2016ddr,Blas:2018aok,Rozner:2019gba} that can also be detected through pulsar timing.

In the context of scalar cloud, LISA could probe stellar mass objects EMRI in clouds \cite{Hannuksela:2018izj}. As inspiraling objects loses energy on GW it scans through a range of frequency (as in Section \ref{sec:toy_model_dissipation}) and in that way can reach the resonant ones. In addition, absence of spherical symmetry leads to new phenomena: Lindblad and co-rotating resonances \cite{Ferreira:2017pth}.

\newpage

\part{Axion-Photon resonances} \label{ch:ax-photon}
The photon-axion mixing in the presence of an external magnetic or electric field can be used to impose strong constraints on axion-like particles due to intergalactic magnetic fields~(see e.g. \cite{Mirizzi:2006zy} for a review) or can lead to a detectable signature in the spectra of high-energy gamma ray sources~\cite{Hooper:2007bq}. 

In Section \ref{sec:ax_grav_BH} we have argued that around massive, spinning BHs superradiant instabilities can be triggered, through which the axion field grows and ``condensates'' . In an astrophysically relevant situation, BHs are often surrounded by a plasma in an accretion disk, which generates its own EM field. In addition, galactic magnetic fields and background EM radiation is present. The presence of magnetic fields in regions where gravity is strong may give rise to new phenomena. It has been argued recently that the coupling of superradiant axion clouds with photons can lead to bursts of radiation which in the quantum version resemble laser-like emission~\cite{Rosa:2017ury,Sen:2018cjt}. Thus, the evolution of superradiant instabilities would produce a periodic emission of light. These arguments are order-of-magnitude, highly approximate and partially inconsistent, but have very recently been put on a firmer ground through the full numerical solution of the relevant equations~\cite{Ikeda:2019fvj}. More generally, the study of axion electrodynamics in curved spacetimes has been the topic of a few studies, with some results in the Schwarzschild background in the context of Pulsar magnetospheres \cite{Garbrecht:2018akc} and polarization of EM waves passing through the scalar clouds around BHs~\cite{Plascencia:2017kca}. Here we will study the coupling between axions and the Maxwell sector in flat spacetime and in the context of scalar clouds. Scalar-Maxwell coupling, as opposed to the pseudo-scalar axionic one, is discussed in Appendix \ref{app:sc-photon}.

\newpage

\section{Background axion field in flat space-time}\label{sec:mink_axion}

\subsection{Homogeneous configuration} \label{sec:mink_axion_hom}

First we consider a constant background axion field $\Phi$ with a harmonic time dependence (in the non-relativistic approximation, set by the boson mass, see Part \ref{ch:structure}) $\Phi \sim e^{\pm i \mu_{\rm a} t}$. The analysis has the simplest form if we work in the Coulomb gauge\footnote{Calculation in the Lorenz gauge can be found in the Section III B in \cite{Boskovic:2018lkj}.} $\bm{\nabla} \bm{A}=0$. In general, the space component of Maxwell's equations (with sources) reduces to \cite{JacksonED}
\beq
\nabla^2\bm{A}-\partial^2_t \bm{A}-\nabla (\partial_t A_0)=-\bm{j} \, .
\eeq
Using Helmholtz theorem we can decompose $\bm{j}=\bm{j}_l+\bm{j}_t$, with $\nabla \times \bm{j}_l=0$ (longitudinal component) and $\nabla\bm{j}_t=0$ (transverse component). Finally, the time component of Maxwell's equations gives $\partial^2_t \bm{A}-\nabla^2\bm{A}=\bm{j}_t$. Notice that the effective current sourced by non-relativistic axions (where $|\nabla \Phi| \ll |\partial_t \Phi| $) is irrotational. Applying this to Eq. \eqref{eq:MFEoMVector} we obtain
\begin{equation} \label{eq:max_ax_ed_Mink}
\partial^2_t\bm{A}-\bm{\nabla}^2 \bm{A}+2k_{\rm a} \partial_t \Phi \bm{\nabla} \times \bm{A} =0\,.
\end{equation}
The momentum space representation of the previous equation shows that the fluctuations of the Fourier-transformed vector potential $\bm{A}_{\bm{p}}$ are described by
\begin{equation} \label{eq:Mathieu_mod}
\partial^2_t\bm{A}_{p}+p^2\bm{A}_{p}+ik_{\rm a}\bm{p} \times \int\frac{d^3\bm{p}'}{(2\pi)^3}\partial_t{\Phi}_{\bm{p}-\bm{p}'}\bm{A}_{\bm{p}'}=0\,.
\end{equation}

Consider the homogeneous axion field $\Phi=\Phi_0 \cos{(\mu_{\rm a} t)}$. As shown in Ref.~\cite{Hertzberg:2018zte}, in a circular polarization representation  $\bm{A}_{\bm{p}}=\sum_{\lambda}y_{\bm{p}}\bm{\xi}^{(\lambda)}_{\bm{p}}+\text{c.c.}\,$ the vectors ${\xi}^{(\lambda)}_{\bm{p}}$ decouple and after the variable change $\mu_{\rm a} t=T+\pi/2$, we are left with
\begin{equation} \label{eq:Mathieu0}
\partial^2_T x_p+\Big(\frac{\omega^2}{\mu^2_{\rm a}}-2\Phi_0 k_{\rm a}\frac{p}{\mu_{\rm a}} \cos{T} \Big)x_p=0\,.
\end{equation}
In other words, we find that our problem is completely reduced to the well-known Mathieu equation with\footnote{In the Appendix \ref{AppParRes} we used $a$ instead of $\Upsilon$, but not here in order not to confuse this with the BH rotation parameter $a$.} $\Upsilon=\omega^2/\mu^2_{\rm a}$ and $\epsilon=-\Phi_0 k_{\rm a}p/\mu_{\rm a}$. Applying \eqref{eq:hom_k_instable} to this problem demonstrates that the dominant rate of the instability (for the small effective coupling $\Phi_0 k_{\rm a}$) is given by
\beq \label{eq:axion_rate_hom}
\lambda_\ast=|\Phi_0 k_{\rm a}(\omega_{\ast}/\mu_{\rm a})|\mu_{\rm a}=\frac{1}{4}|\Phi_0 k_{\rm a}|\mu_{\rm a}.
\eeq

Compare and contrast this problem with the one in Part \ref{ch:orbital}. There, parametric resonances manifested themselves at the quasi-linear level in perturbation theory and then the natural question is whether non-linear terms will quench the instability. In contrast, as the Maxwell-Klein-Gordon system is linear, there are no non-linear terms to suppress the instability. Furthermore, in  Part \ref{ch:orbital} orbital parameters need to be finely tuned in order for the resonance to happen, while here resonances will always happen as the Fourier transformation to the real space will scan through \textit{all} wave numbers including the ones that trigger the instability.

\subsection{Inhomogeneous configuration} \label{sec:mink_axion_inhom}

\subsubsection{Analytical estimates} \label{sec:mink_axion_inhom_an}

Compared to the flat space analysis from Section \ref{sec:mink_axion_hom}, the localized axion configuration introduces one more timescale in the problem, that of the time $d$ needed for photons to leave the axion configuration, where $d$ is a measure of the configuration size. Thus, there is another rate in the problem,
\begin{equation}
\lambda_\gamma \sim \frac{1}{d}\,.\label{eq:inst_rate2}
\end{equation}
If $\lambda_\gamma>\lambda_\ast$, with $\lambda_\ast$ being the estimate of the EM field instability rate for the homogeneous condensate, photons leave the configuration before the instability ensues and the effective rate of the instability is zero. In the other extreme, we can approximate the rate of the dominant instability by
\begin{equation}\label{eq:inst_rate3}
\lambda \approx \lambda_\ast\,[\langle \Phi \rangle]-\lambda_\gamma,
\end{equation}
where $\langle \Phi \rangle$ is some estimate of the average value of the axion field, to be implemented in the expression obtained for the homogeneous case [see Eq.~\eqref{eq:axion_rate_hom}]:
\begin{equation}
\lambda_\ast\,[\langle \Phi \rangle]    \approx \frac{1}{2}k_{\rm a}\langle \Phi \rangle\mu_{\rm a}\,.
\end{equation}
This estimate was conjectured previously \cite{Hertzberg:2010yz,Hertzberg:2018zte} and compared to numerical results for setups different from the one that we will now consider. Sketch of the more formal analytical understanding of this problem are outlined in Box on page \pageref{subsec:inhoman}.

Keeping the scalar clouds around Kerr BHs (Section \ref{sec:ax_ph_clouds}) in mind, let us consider the dominant mode in the gravitational atom \eqref{eq:211psi}, ``frozen'' and embedded in Minkowski spacetime, and estimate the instability rate. For the measure of $d$ we use the full-width-at-half-maximum (FWHM) of the function \eqref{eq:211psi}
\begin{equation} \label{eq:sc_cloud_size}
d \approx \frac{4.893}{M\mu_{\rm a}^2}.
\end{equation}
For the estimate of the field value we take the radial mean of the field on the FWHM and maximal contribution from the harmonic part of the function
\begin{equation} \label{eq:sc_cloud_av_field}
\langle \Phi \rangle \approx (1/d)\int_{\text{FWHM}}|\Phi(r)|dr \approx 0.592 A_0\,.
\end{equation}
%


\begin{framed}
\noindent
{\small {\it Analytical description of Maxwell sector instabilities for inhomogeneous axion configurations - Sketch}\\ ~\label{subsec:inhoman}

Maxwell sector instabilities in localized axion configurations have been previously considered at the order-of-magnitude and numerical (Floquet analysis) level and in the context of weak-field self-gravitating axion configurations (axion clumps) in Refs. \cite{Hertzberg:2010yz,Hertzberg:2018zte}. We would like here to provide a more analytical understanding of a general problem, besides the full time-domain numerical analysis in Section \ref{sec:mink_axion_num}. We will consider a 1+1 toy model of our problem
\begin{equation} \label{eq:Mathieu_mod2}
\partial^2_t f_{p} + p^2 f_{p} + \Big( \alpha  p  \int^{\infty}_{-\infty} dp' K_{p-p'} f_{p'} \Big)\, \cos{t}=0,
\end{equation}
where $K_{p-p'}$ is a Gaussian kernel of width $\sigma \sim 1/d$, $d$ is the characteristic size of our configuration, $\alpha$ mops up various constants, including the coupling constant and $k$ is in the units of $\omega$.

Let us first comment on the validity of our toy model. First, we effectively reduced a $2+1$ problem to a $1+1$ problem. In the case of spherically-symmetric axion clumps we indeed expect effective one-dimensional differential equation and this equation reduces to the one that has the same structure as our problem in the case of $l=1,m=0$ excitations \cite{Hertzberg:2018zte}. However, $\{ l,m \}$ mode coupling is generally not captured by our toy model. Secondly, we approximated $\partial_t \Phi_{\bm{p}-\bm{p}'}$ with $K_{p-p'}$. Fourier transform of the axion configuration wavefunction in Eq. \eqref{eq:211psi} is known in the context of atomic physics (e.g. \cite{bransden_atoms_molecules}). However, general asymptotic behaviour of both our axion configuration and the one used to describe spherically symmetric or rotating axion clumps can be captured by this kernel - as $d \to 0: K_{p-p'} \to 0$ and as $d \to \infty: K_{p-p'} \to \delta(p-p')$. Second limit asymptotes to the homogeneous background and the Mathieu equation and the first one reduces our configuration to the one of measure zero. In the latter case, there is no possibility for either linear or parametric resonances and we are left with the vacuum solution of Maxwell equations. These limits correspond to the supercritical and subcritical regime, respectively. Besides that, convolution with the Gaussian kernel (also known as the generalized Gauss-Weierstrass transformation) has a nice property that its eigenfunctions are $e^{(a+bi)x}$, with the eigenvalues $e^{(1/2)\sigma^2(a+bi)^2}$.

The symmetries of our toy model will put constraints on the form of its solution. The problem is linear with a time-periodic coefficient and respects time-inversion symmetry, so the Floquet theorem demands solutions of the form \cite{benderbook}
\begin{equation}
f_p=e^{\mu_p t}\phi(t)+e^{-\mu_p t}\phi^{\ast}(t),
\end{equation}
where $\phi(t)$ is a $2\pi$-periodic function and $\mu_p$ is Lyapunov exponent. The system is stable, in the sense of Lyapunov, iff $\text{Re}\{ \mu_p \} = 0$.

For non-zero $\sigma$, differential equation governing $f^{(1)}_{1/2}$
(see Appendix \ref{app:RG_time_scales_1st} for the notation) has the form
\begin{equation}
\partial^2_t f^{(1)}_{1/2}+\Big(\frac{1}{2} \Big)^2 f^{(1)}_{1/2}= - \frac{1}{4} e^{-\sigma^2 t^2/2} \Big(
 A^{\ast}  e^{i(1/2)t} +  A  e^{i(3/2)t} \Big) +\text{c.c.}
\end{equation}
It is left for future work to see whether the criterion \eqref{eq:inst_rate3} can arise from the perturbative framework.

}

\end{framed}

\subsubsection{Numerical results} \label{sec:mink_axion_num}

\begin{figure}
\hspace*{\fill}%
\begin{minipage}[c]{0.44\textwidth}
\centering
\vspace{-1pt}
\includegraphics[width=\textwidth]{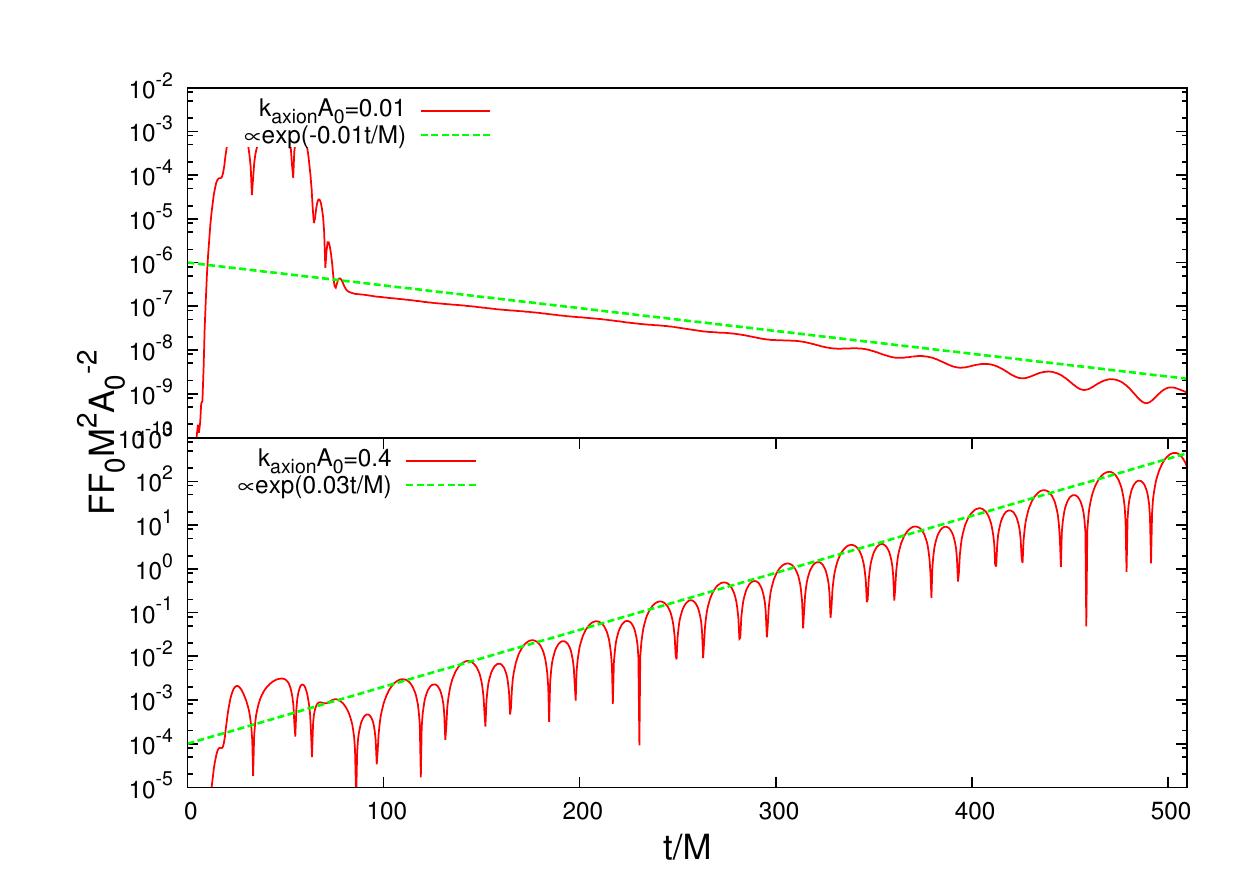}
\caption{ \fontsize{9}{12}
Time evolution of the monopole part of the EM scalar $F_{\mu\nu}F^{\mu\nu}$ for a coupling $M\mu=0.2$ at $r=20M$ when $k_{\rm a}A_{0}=0.01$ (upper panel) and $k_{\rm a}A_{0}=0.4$ (lower panel), in a Minkowski background. The scalar field is kept fixed and described by \eqref{eq:211psi}. The initial profile is described by $(E_{0}/A_{0},w/M,r_{0}/M)=(0.001,5.0,40.0)$, but the qualitative features of the evolution are independent on these. For small couplings $k_{\rm a}A_0$ there is no instability and the initial EM fluctuation decays exponentially. For large couplings, on the other hand, an exponential growth ensues. Figure credits: \cite{Ikeda:2019fvj}.
\label{graph_fixed_scalar}}
\end{minipage}%
\hfill
\begin{minipage}[c]{0.48\textwidth}
\centering
\vspace{0pt}
\includegraphics[width=\textwidth]{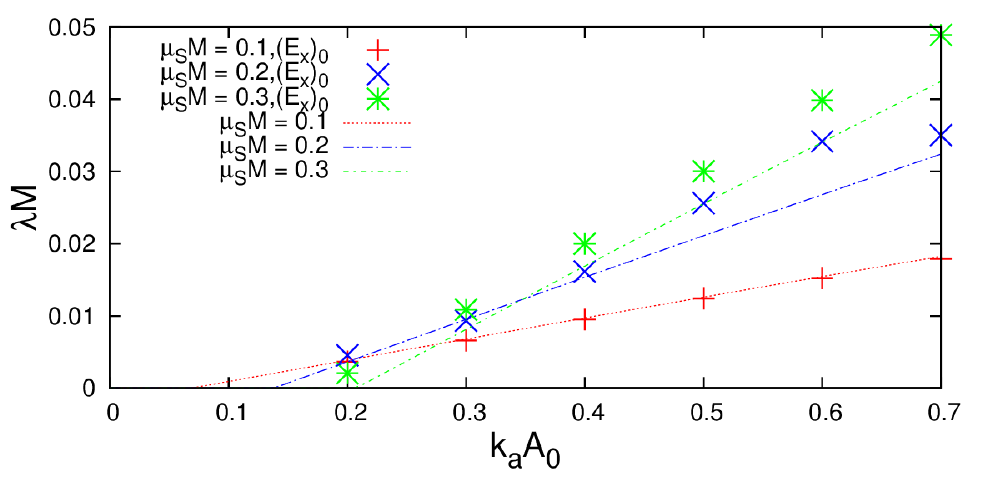}
\caption{\fontsize{9}{12} Instability rates for axionic couplings, in the presence of a background axion described by the cloud \eqref{eq:211psi}, for a Minkowski background. The analytical estimates for the instability rate $\lambda$ for axionic couplings, as given to first order by Eq.~\eqref{eq:inst_rate3} (dashed lines, full expansion is described in Appendix \ref{app:RG_time_scales_1st}), are compared with the numerical results of Ref.~\cite{Ikeda:2019fvj} (crosses). We find good agreement between our analytical estimates and numerical data for small mass ($M\mu_{\rm a}\sim0.1$) and small axion couplings. Figure credits: \cite{Boskovic:2018lkj}.
\label{fig:Mink_axion_rates}}
\end{minipage}%
\hspace*{\fill}
\end{figure}

In order to check whether the simple qualitative description from the previous section is realistic we will compare it with the numerical results. The numerical setup, in which the dynamics was treated as a Cauchy problem in full $3+1$ numerical relativity, was developed  in~\cite{Ikeda:2019fvj, Boskovic:2018lkj}. We refer the reader to these works for the details on the numerical procedure and only summarize the initial data and the analysis tools in Box on the page \pageref{subsec:numset}.

In the setup, initial data were described by \eqref{Eq.electromagnetic field initial data}, while the axion configuration was ``freezed'' - the Klein Gordon equation is not evolved, and the axion is described by Eq.~\eqref{eq:211psi} at all times. Time evolution was calculated for several different $k_{\rm a}A_{0}$ and $\mu M=0.1,\,0.2,\,0.3$, where $M$ is the BH mass that supports the solution~\footnote{In the Minkowski scenario this should be seen just as the configuration profile parameter.}~\eqref{eq:211psi}. The behavior of the EM fields is shown in Fig.~\ref{graph_fixed_scalar}. The shown results are qualitatively representative for the investigated parameter space \cite{Ikeda:2019fvj}. We see that when the coupling $k_{\rm a} A_{0}$ is small, the initial EM fluctuation dissipates and seems to vanish exponentially (a zero EM field is an exact solution of the field equations). On the other hand, when the coupling $k_{\rm a} A_{0}$ is larger than a certain critical value $k_{\rm a}^{\rm c} A_{0}$ (supercritical regime), the EM field grows exponentially  with time $\sim e^{\lambda t}$, as is apparent from the bottom panel. This behaviour is in contrast with the order-of-magnitude statements in~\cite{Rosa:2017ury} but consistent with arguments in Section \ref{sec:mink_axion_inhom_an}.


The exponential growth rate $\lambda$ of the electric field was estimated using best-fits to the local maxima at $r=60M$.  In Fig.~\ref{fig:Mink_axion_rates} the analytical estimate~\eqref{eq:inst_rate3} is superposed on the numerical results obtained in ~\cite{Ikeda:2019fvj}. It is clear that these estimates are in good qualitative agreement with the numerical results. In addition, a cutoff coupling below which no instability arises shows up naturally. In summary, a very simple and elegant analytic formula explains most of the results that we observe numerically. It should be noted that in ~\cite{Boskovic:2018lkj} a comparison between the estimate~\eqref{eq:inst_rate3} and the \textit{full} numerical results was performed for the first time. Previously, \cite{Hertzberg:2018zte}
used Floquet analysis (see Appendix \ref{sec:Floquet} for the description) in the $p$-space, decomposing the electromagnetic field into vector spherical harmonics with the focus only on the channel $\{l,m\}=\{1,0\}$.


\begin{framed}
\noindent
{\small {\it Numerical setup (axion coupling): Initial data and analysis tools}\\ ~\label{subsec:numset}

In order to construct the initial data, one must solve the constraint equations, described in \cite{Ikeda:2019fvj, Boskovic:2018lkj}. For the axion cloud profile \eqref{eq:211psi}, the initial data that was used in \cite{Ikeda:2019fvj, Boskovic:2018lkj} is given by
\begin{eqnarray}
E^{r}&=&E^{\theta}=0,\\
E^{\varphi}&=&E_{0}(r,\theta),
\end{eqnarray}
where $E_{0}(r,\theta)$ is an arbitrary function of $r$ and $\theta$. For $E_{0}(r,\theta)$ a Gaussian profile was used
\begin{eqnarray} \label{Eq.electromagnetic field initial data}
E_{0}(r,\theta)=E_{0}e^{-\left(\frac{r-r_{0}}{w}\right)^{2}}\Theta(\theta),
\end{eqnarray}
where $E_{0}$, $r_{0}$, and $w$ are the amplitude, the peak radius and the width of the initial electric field, respectively. Two types of functions $\Theta(\theta)$ have been considered. The ``extended profile'' is a simple constant value,
\be
\Theta(\theta)=1\,,\label{ID_extended}
\ee
called that way as it is direction-independent. The``localized profile'' is
\begin{equation}\label{ID_localized}
\Theta(\theta)=
\left\{\begin{array}{cc}
\sin^{4} 4\theta&~{\rm for}~0<\theta<\frac{\pi}{4}\\
0&~{\rm for}~\frac{\pi}{4}<\theta<\pi \,,
\end{array}\right.
\end{equation}
since it is sharply peaked along some directions only. In~\cite{Ikeda:2019fvj} it was shown that the numerical results between the ``localized'' and the ``extended'' profile do not change at a qualitative level. These profiles were used in both Minkowski and Kerr scenarios.

To gain information about the time development,
the physical quantities extracted from the numerical simulation are the multipolar components of the physical quantities $Z_i=\{\Phi_{i},FF_{i},  (T^{\rm EM}_{tr})_{i} \}$, with $FF=F_{\mu\nu}F^{\mu\nu}$ and
\begin{eqnarray}
Z_{0}(t,r)&:=&\int d\Omega Z(t,r,\theta,\phi) Y_{00}(\theta,\phi)\,,\\
Z_{1}(t,r)&:=&\int d\Omega Z(t,r,\theta,\phi) Y_{R}(\theta,\phi)\,.
\end{eqnarray}
In the previous definitions $Y_{R}=\frac{1}{2}\left(Y_{1,1}+Y_{1,-1}\right)$ and $Y_{\ell m}(\theta,\phi)$ are spherical harmonics.

}

\end{framed}

\subsection{Interaction with plasma} \label{sec:Mink_plasma}

Thus far, the system was assumed to evolve in a vacuum environment, when in reality the universe is filled with matter. Particularly, we will consider interaction with plasma.  The influence of plasma on axion-photon conversion has been discussed for superradiant axions~\cite{ Rosa:2017ury,Sen:2018cjt,Day:2019bbh}, but also in other contexts~\cite{Hertzberg:2018zte, Carlson:1994yqa}. EM wave propagation through plasma is described by the modified dispersion relation \cite{JacksonED}
\begin{equation}
p^2=\omega^2-\frac{\omega^2_{\rm plasma} \omega}{\omega+i \nu},
\end{equation}
where $\nu$ is the collision frequency and
\begin{equation}
\omega_{\rm plasma}=\frac{4 \pi e^2 n_{e}}{m_{e}},
\end{equation}
is the plasma frequency; $m_e$, $e$ and $n_e$ are the mass, charge and the concentration of free electrons, respectively. Conceptually, it is helpful to consider two limiting cases - collisional ($\omega \ll \nu$; appropriate in the context of plasma in the accretion disks) and collisionless ($\omega \gg \nu$; in the context of interstellar matter or a thin accretion disk).

EM waves in the collisionless limit that we discuss here have a modified dispersion relation that is equivalent to providing a photon with a mass\footnote{Limitations of such approach are briefly discussed in \cite{Baryakhtar:2017ngi}. See also \cite{Day:2019bbh}.} $\mu_{\rm V}=\omega_{\rm plasma}$. For high $\mu_{\rm V} \geq (1/2)\mu_{\rm a}$, decay processes become kinematically forbidden\footnote{The decay process is $a\to \gamma +\gamma$ (for a Lagrangian with a $\Phi F^2$ term), so if the photon has an (effective) mass $\mu_{\rm V}$, in order for the decay to be energetically favourable, we should have $\mu_{\rm a} \geq 2\mu_{\rm V}$.}.
For interstellar matter \cite{Carlson:1994yqa}
\begin{equation} \label{eq:plasma_ISM}
\omega_{\rm plasma}=\sqrt{\frac{n_e}{0.03\text{cm}^{-3}}}(6.4  \,\times\, 10^{-12} \rm{eV}),
\end{equation}
the plasma frequency is below  the range of the QCD axion mass and some of the ULA.

We will now concisely reproduce some of the results of Ref.~\cite{Sen:2018cjt} (and expand them to the scalar coupling case in Appendix \ref{app:scalar_plasma}). Consider \eqref{eq:Mathieu0} now with
\begin{equation}  
\Upsilon \to \Upsilon + \left( \frac{\mu_{\rm V}}{\mu_{\rm a}} \right)^2 \,.
\end{equation}
Critical stability curves on the $\Upsilon-\epsilon$ diagram are given by (see Appendix \ref{sec:Mathieu_zone_border})
\begin{equation}
\label{eq:AvsEps}
\Upsilon=\frac{1}{4} + \eps+\mathcal{O}(\eps^2)\,.
\end{equation}
Inserting the appropriate $\Upsilon$ and $\eps$, we can find the values of the parameters for which~\eqref{eq:AvsEps},  a quadratic equation in $p$, has real solutions. These are the critical plasma frequency
\begin{equation}
\omega^{\rm crit}_{\rm plasma}=\frac{1}{2}\mu_{\rm a} \sqrt{1+(\Phi_0 k_{\rm a})^2}\,,
\end{equation}
in agreement with Ref. \cite{Sen:2018cjt}. One can also straightforwardly find corrections to the instability rate, induced by the effective mass
\begin{equation} \label{eq:ax_rate_plasma}
\lambda_{\rm a}=\frac{1}{2}\Phi_0 k_{\rm a}\mu_{\rm a} \sqrt{1-4\left(\frac{\omega_{\rm plasma}}{\mu_{\rm a}}\right)^2}+\mathcal{O}(k^2_{\rm a}).
\end{equation}

\newpage

\section{Phenomenological implications}

\subsection{Axion DM clumps}

In \cite{Hertzberg:2018zte} axion-photon resonances have been considered in the context of DM clumps. Results indicate that these resonances can occur only for non-standard couplings, repulsive self-interaction or high angular momenta of DM clumps. If realized they could lead to pileup of critically massive clumps and in such way dramaticaly shape the population of such objects. 

\subsection{Axion clouds} \label{sec:ax_ph_clouds}

We have discussed the exponential growth of an EM field around a ``frozen'' axion cloud in a flat space background in Section \ref{sec:mink_axion}. These results pose the question whether such dynamics occurs also in curved spacetime, in particular, in the context of scalar clouds around Kerr BHs. From the flat-space discussion one can conjecture that when the effective coupling is larger than a threshold value, the EM field may grow exponentially -- fed by the axionic cloud, which itself grew through superradiance and extracted its energy from the spinning BHs. Similar arguments at the semi-classical and order-of-magnitude level can be found in~\cite{Rosa:2017ury}. These conjectures were shown (numerically) to be true for the evolution of Maxwell's field equations coupled to an axion field in a (fixed) Kerr background. In particular it was shown that for critical values of the coupling $k_{\rm a}\Phi_0$, EM fields are spontaneously excited in such environments, even at the classical level~\cite{Ikeda:2019fvj}. These instabilities can be indeed completely  understood in the context of classical field theory, owing to the bosonic nature of axions and photons, that allows buildup of macroscopic numbers of particles. Here we give an overview of the numerical results, while we refer the reader to \cite{Ikeda:2019fvj,Boskovic:2018lkj}  on technical details.

\begin{figure}
\hspace*{\fill}%
\begin{minipage}[c]{0.45\textwidth}
\centering
\vspace{-1pt}
\includegraphics[width=\textwidth]{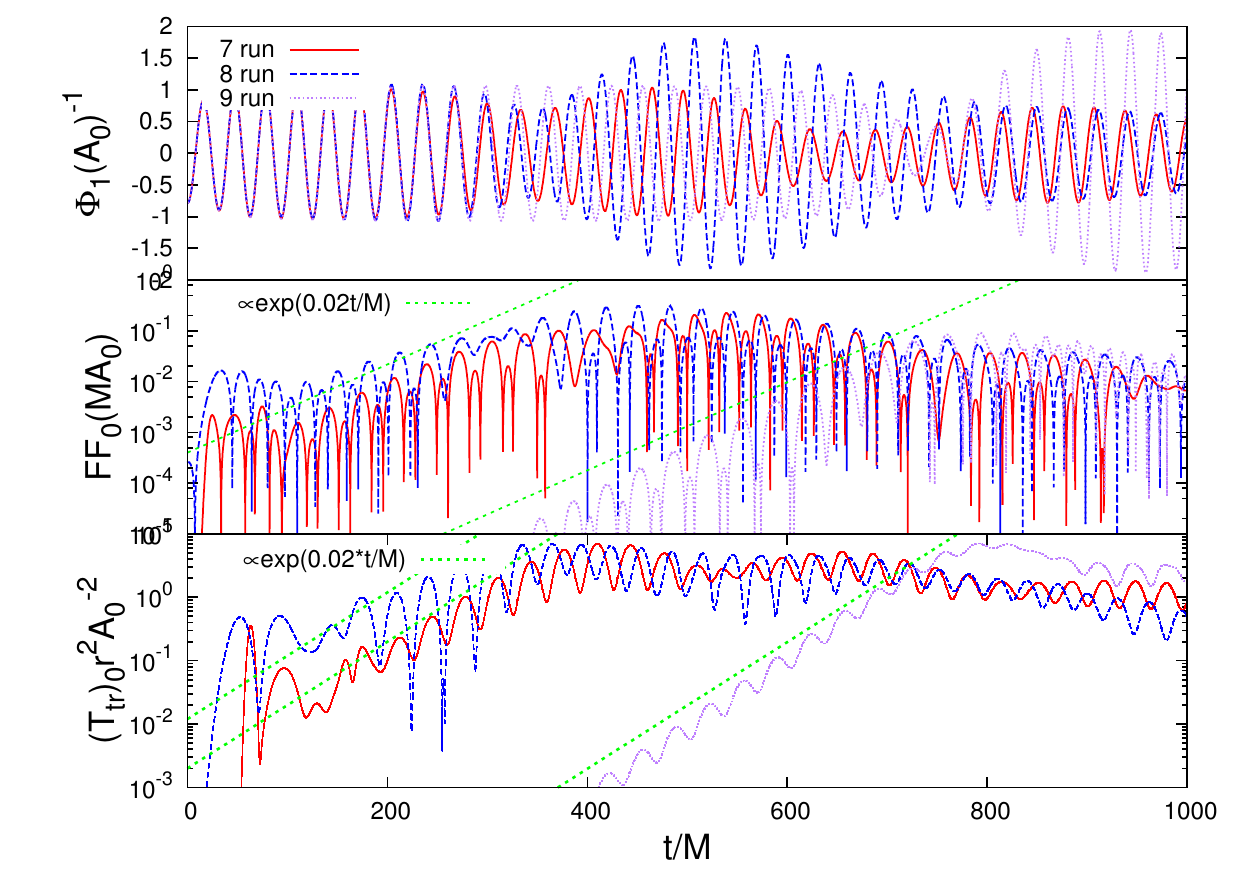}
\caption{ \fontsize{9}{12}Time evolution of $\Phi_{1}$ (upper panel), $FF_{0}$ (middle panel) at $r=20M$ and the energy flux (bottom panel) for an axion with mass
$M\mu=0.2$ around a BH with $a=0.5M$. The coupling constant is super-critical with $k_{\rm a}A_0=0.3$.
The initial EM profile is described by $(E_0/A_0,w/M)=(10^{-3},5),(10^{-3},20),(10^{-4},5)$ for run $7, 8, 9$ respectively, and $r_0=40$. The overall behavior and growth rate of the instability
at large timescales are insensitive to the initial conditions. Figure credits: \cite{Boskovic:2018lkj}
\label{graph_Kerr_sample}}
\end{minipage}%
\hfill
\begin{minipage}[c]{0.45\textwidth}
\centering
\vspace{0pt}
\includegraphics[width=\textwidth]{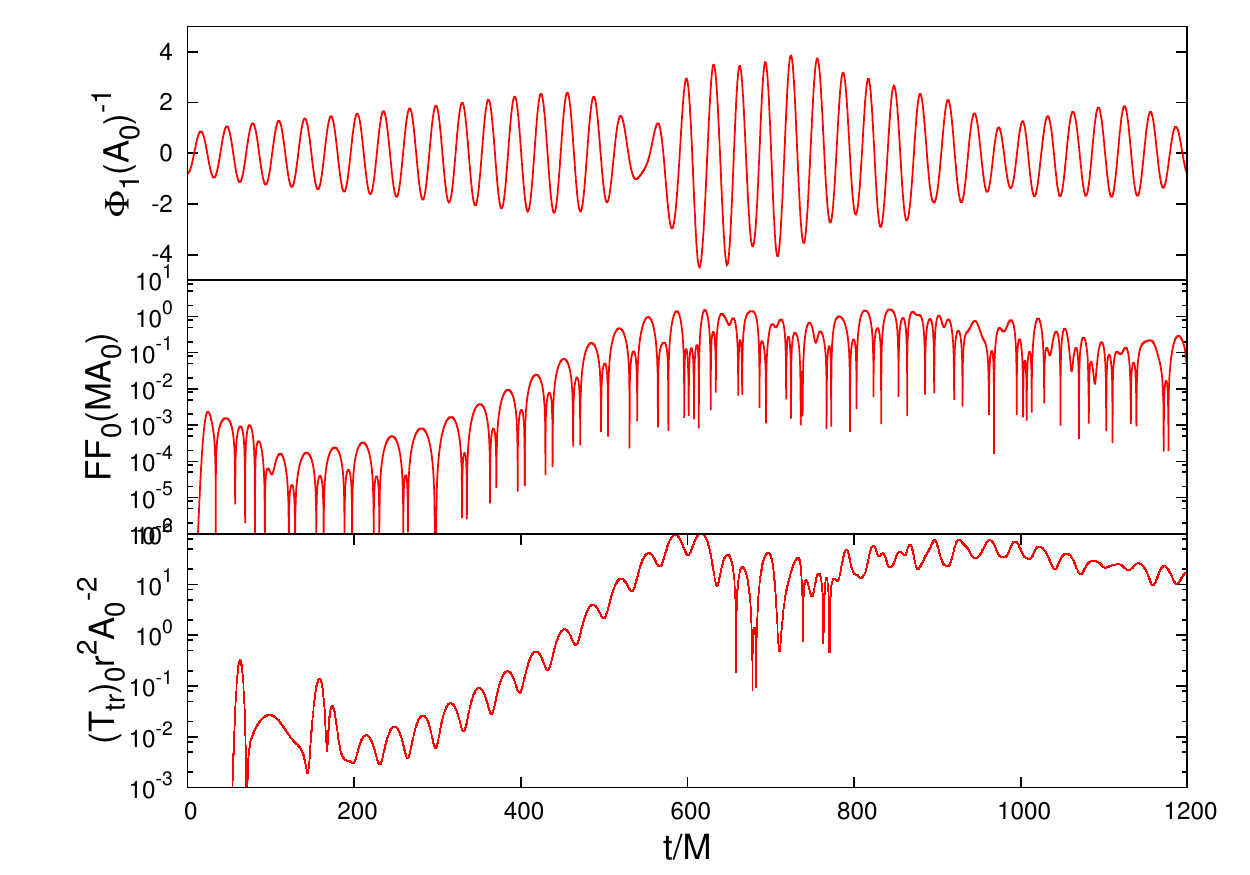}
\caption{\fontsize{9}{12} Evolution of an initially sub-critical axion, driven supercritical by a superradiant-like term. After the axion becomes super-critical, an instability sets in
which gives rise to a burst of EM radiation, leading to a depletion of the scalar, until a new superradiant growth set in. Our results are consistent with the triggering of
periodic bursts. These results describe an axion with mass $M\mu=0.2$ around a BH with $a=0.5M$. The coupling constant is sub-critical $k_{\rm a}A_0=0.15$.
The initial EM profile is described by $(E_0/A_0,w/M)=(10^{-3},5)$ and $r_0=40$. Figure credits: \cite{Ikeda:2019fvj}
\label{graph_Kerr_sample_C}}
\end{minipage}%
\hspace*{\fill}
\end{figure}

Consider a background geometry described by a Kerr BH of mass $M$ and angular momentum $Ma$, while both the axion and the EM fields are evolved.  Numerical results of ~\cite{Ikeda:2019fvj} show that, as in the Minkowski case, there still exists a critical coupling beyond which an instability arises. Analogously to the flat spacetime scenario, the initial instability growth is exponential. However, the dynamics of the axion cloud acts as a negative feedback. As the axion field is suppressed, effective coupling can drop below the critical value and the system is stable. Now, one can wait for the superradiance to grow the axion cloud again and drive it to the supercritical regime. Results are shown in Fig.~\ref{graph_Kerr_sample} through the axion field, Maxwell scalar and energy flux perspective. Compare and contrast these results with the ones for ``frozen'' axion cloud in Fig~\ref{graph_fixed_scalar}. These results are at the qualitative level independent of initial conditions and BH spin \cite{Ikeda:2019fvj}. Qualitativly similar phenomenon have been found for scalar condensates with a self-interacting potential, but in the absence of couplings to the EM sector~\cite{Yoshino:2012kn,Yoshino:2015nsa}.

Let us use \eqref{eq:inst_rate3} to estimate the critical coupling and compare it with \eqref{eq:k_ax_ph_values}. Using \eqref{eq:cloud_mass}  we find 
\beq\label{crit_kaxion}
\frac{\sqrt{\hbar}}{k_{\rm c}}  \simeq   1.8 \cdot 10^{18} \text{GeV} \, \alpha \sqrt{\frac{M_{\rm c}}{M}} \,.
\eeq
Note the different scaling with $\alpha$ with respect to~\cite{Ikeda:2019fvj}. There the expression was formed based on the numerical results that the critical value for the instability is  $k_{\rm a}A_{0}\sim 0.2-0.3$ always. This assumption is not physically satisfying and the analytical estimate \eqref{eq:k_ax_ph_values} is more sensible (note also that the probed $\alpha$ range in \cite{Ikeda:2019fvj} was $0.1 \,-\, 0.3$). Comparing \eqref{crit_kaxion} with \eqref{eq:k_ax_ph_values} 
\beq
\frac{k_{\rm a}}{k_{\rm c}} \simeq 1.3 \cdot 10^{-6} \, (0.203\frac{E}{N}-0.39) \left(\frac{\mu_{\rm a}}{10^{-12}\,{\rm eV}}\right)  \left( \frac{\alpha}{0.07}\right) \sqrt{\frac{M_{\rm c}}{M}}\,
\eeq
we see that either one needs an unconventionally high coupling in order for the critical one to be reached (as mentioned, this could happen if there is an axionic coupling to the hidden sector \cite{Hertzberg:2018zte}) or one needs to consider superradiance from primordial BHs (PBH) that can be triggered by axions with higher mass, compared to the axions needed for stellar and SMBHs superradiance. The second scenario was considered in \cite{Rosa:2017ury}, in the context of mixed QCD axion-primordial BH DM. There it was further speculated whether this scenario could lead to an explanation of fast radio bursts\footnote{These are $\sim \rm{ms}$ radio signals originating from some high energy astrophysical process not yet identified.}. Their estimates suggest that one would need QCD axion of $\mu_{\rm a} \sim 10^{-5} \rm{eV}$ with\footnote{See \eqref{eq:ax_ph_k_K} for the definition of $K$.} $K \sim 1$ and PBH\footnote{For reference $M_{\leftmoon} \sim 7.4 \cdot 10^{22} \, \rm{kg} \sim 3.6 \cdot 10^{-8} M_{\odot}$.} with $M \sim 10^{23} \,-\, 10^{24} \rm{kg}$. This PBH mass range is marginally acceptable as a subdominant DM component (see Fig. 4 in \cite{Murgia:2019duy}), while the axion in the mentioned range could soon be propped in planned experiments \cite{Rosa:2017ury}.

\subsubsection{Blast vs. leakage} \label{sec:blast_leak}

From \eqref{eq:SR_inst_time}, \eqref{eq:inst_rate3} and \eqref{crit_kaxion} we can estimate the ratio of the superradiance instability $\tau_{\rm SR}$ timescale and the instability timescale associated with the axion-photon resonance $\tau_{\rm EM}$:
\beq
\frac{\tau_{\rm SR}}{\tau_{\rm EM}}  \simeq 5.8 \cdot 10^{8} \, \Tilde{a}^{-1}  \left( \frac{\alpha}{0.07}\right)^{-7} \Big( \frac{k_{\rm a}}{k_c} -1 \Big) \,.
\eeq
and similarly for the ratio between the timescale of the cloud energy loss through GW radiation \eqref{equ:tauc} and the axion-photon resonance
\beq
\frac{\tau_{\rm c}}{\tau_{\rm EM}}  \simeq 2.5 \cdot 10^{16}  \left( \frac{\alpha}{0.07}\right)^{-13} \Big( \frac{k_{\rm a}}{k_c} -1 \Big) \,.
\eeq
Thus, there is a huge separation of scales $\tau_{\rm c} \gg \tau_{\rm SR} \gg \tau_{\rm EM}$ even for coupling constants much higher than the critical one. The picture that one expects is - superradiance develops the scalar cloud until the electromagnetic instability sets in (if appropriate) and the cloud is depleted.

However, in principle it is possible that there are regimes where there is a uniform EM flux during the superradiant evolution. Because of the large time scales it is very hard to perform full numerical evolution of the superradiance, but in \cite{Ikeda:2019fvj} an artificial superradiant-like term was introduced to explore this possibility. This term, of the form $C \partial\Phi/\partial t$ was introduced to the Klein-Gordon equation~\eqref{eq:MFEoMScalar}. The modification mimics and gives rise to superradiance, and was used in Zel'dovich's pioneering work on rotational superradiance~\cite{zeldovich1,zeldovich2,Cardoso:2015zqa,Brito:2015oca}. The term is Lorentz violating as it describes the absorption on a timescale $\sim 1/C$ in a co-rotating frame with absorbing medium. The study in \cite{Ikeda:2019fvj} showed (Fig.~\ref{graph_Kerr_sample_C}) that the physical results did not change with respect to frozen-superradiance evolutions. However, while the results of Ref.~\cite{Ikeda:2019fvj} are compelling and lead to periodic bursts of EM radiation, completely ruling out an alternative scenario would require a full numerical simulation.

\subsubsection{Plasma effects}                   \label{sec:plasma_axion_Kerr}

We will reconsider the Minkowski discussion of the plasma impact on the axion-photon resonances from \ref{sec:Mink_plasma}. This topic has been considered at the order-of-magnitude level in \cite{Rosa:2017ury,Sen:2018cjt}. In \cite{Boskovic:2018lkj} this problem was approached through numerical simulation of the axion-Proca system. The time domain study is qualitatively similar to the scalar-Proca case discussed in Appendix \ref{sec:Kerr_scalar} and summarized on Fig.~\ref{graph_Proca_a05_mu02_ks05_p2_phi_FF_Sr_E1}. This confirms the flat space picture: when the (effective) mass of the vector field is larger than the axion mass, the burst is suppressed. In the context of BH superradiance from the interstellar matter estimate \eqref{eq:plasma_ISM} one can expect EM instability not to be quenched on the primordial and lower range of the stellar BHs mass spectrum.

The preceding analysis neglects the time-dependence of the plasma distribution, in particular it neglects also the backreaction by the cloud on the plasma. Although the full problem is outside the scope of this work, we note that the arguments of Ref.~\cite{Hertzberg:2018zte} suggest that non-harmonic time dependence would not jeopardize parametric resonances as long as the $\epsilon$ in \eqref{eq:Mathieu0} is much smaller than the plasma frequency. However, the time-periodic background of real scalars can drive matter resonantly in peculiar configurations oscillating with the (multiple of) scalar mass \cite{Boskovic:2018rub, Ferreira:2017pth} or possibly deplete it from the central regions~\cite{Ferreira:2017pth}. Such scenarios should be separately studied.

With regards to the collisional regime estimates from Ref. \cite{Sen:2018cjt} we find
\begin{equation}
\nu  \sim  \Big( \frac{M}{10 M_{\odot}} \Big)^{-\frac{5}{8}} 10^{-6}\text{eV}\,,
\end{equation}
for the collision frequency and
\begin{equation}
\omega_{\rm plasma} \sim \Big( \frac{M}{10 M_{\odot}} \Big)^{\frac{1}{2}} 10^{-3}\text{eV}\,,
\end{equation}
for the plasma frequency in the inner rim of the accretion disk around BHs (collisional regime). For BHs larger than $M \sim 10^{-3} M_{\odot}$, $\omega_{\rm plasma}>10^{-5} \text{eV}$ and the axion decay is forbidden in all of the parameter range interesting in a BH-superradiance context.

However, one should also consider the geometry of the problem. Accretion disks are planar structures (when thin), immersed in a ``spheroidal'' scalar cloud. The EM field enhancement can originate in the space external to the accretion disk (there is a limitation from interstellar matter there, discussed in the previous subsection). Such waves can lead to Ohmic heating of the disk or disperse it through the radiation pressure. The quantitative analysis of this would probably depend on the geometry of the initial fluctuations. We should also note that the estimates of the peak luminosity from Ref.~\cite{Rosa:2017ury} (which are even lower than the ones estimated numerically in Ref.~\cite{Ikeda:2019fvj}) indicate that the radiation pressure (if EM instability ensues) would blow away the surrounding plasma.

Besides astrophysical plasma, large electric fields can lead to Schwinger $e^{+}e^{-}$ pair production. It was argued, at the order-of-magnitude level, that such plasma can indeed be created and reach large enough densities (and consequently critical $\omega_{\rm plasma}$) to block EM bursts~\cite{Rosa:2017ury}. Subsequently, $e^{+}e^{-}$ annihilations would drive the plasma density down and restart the process again.

\newpage

\section{Discussion} \label{ch:discussion}
Axions are one key possibility to solve the strong CP problem and axion-like particles
could be one solution to the dark matter puzzle. In addition, the existence of ultra-light fields could provide empirical hints towards string theory or compactified dimensions. Extensive experimental and observational efforts are actively looking for ``axionic'' imprints. Given the nature of the coupling, and the universality of free fall, nontrivial important effects are expected in regions where gravity is strong. Rotating black holes (immersed, or not in magnetic fields) are a prime example of such regions.

If ultra light axions with $\mu_{\rm a} \sim 10^{-22} \rm{eV}$ (fuzzy DM) make up a dominant fraction of DM they can form self-gravitating cores at the centers of galactic dark halos. Such objects are of interest in solving small scale challenges of $\Lambda$CDM cosmology. In other models and mass ranges, axions could form objects smaller in size - compact objects or DM clumps. Axion and axion-like light particles -- even with negligible initial abundance -- trigger superradiant instabilities around massive, spinning BHs. The instability extracts rotational energy away from the spinning BH and deposits it into a cloud of scalars, with a spatial extent $\sim 1/(M\mu_{\rm a}^2)$. Over long timescales, when the mass of the cloud is sufficiently large, GW-emission becomes important, and leads to a secular spin-down of the cloud (and BH), and a consequent cloud decay. Such systems are a promising source of GWs, both as resolvable and as a stochastic background, that can be detected with current and future detectors.

In this work we considered possibilities of axion phenomenology from gravitational coupling with objects moving in the background of gravitating axionic configurations as well as axion-photon coupling. We have shown that both scenarios could lead to parametric resonances. In the first case, these resonances originate from the fact that the oscillating background modulates the epicyclic frequency of celestial objects. Nonlinear terms in the equations of motion tame the instabilities into resonances. Such phenomena could be observed in motion of objects around SMBH at the Galactic center, motion of binary pulsars and EMRIs in scalar clouds.

In the second case, instabilities could be tamed only through finite size effects. Namely there exist critical coupling beneath which instabilities are not triggered. This critical coupling can't be reached in the context of stellar BHs and conventional QCD axions but for primordial BHs or unconventional coupling (allowed if there exist hidden radiation channels) it could. Such instabilities could on the one hand relieve the present constraints from BH superradiance but on the other provide additional phenomenological channel for scalar clouds.

There are various questions that could be the topic of further studies - more detailed understanding of the axion-photon-plasma-BH square and the interrelations of various types of instabilities that can arise in such systems; more detailed theoretical understanding of orbital resonances, including connections with standard mean-motion resonances in celestial mechanics and the behaviour in the absence of spherical symmetry etc. On the other hand, related beyond-SM and DM candidate-particles are ultra-light vectors \cite{East:2017ovw,East:2018glu,Baryakhtar:2017ngi,Baryakhtar:2018doz,Cardoso:2018tly}. In principle one would expect that all of the above phenomena would apply also for them and in fact some progress has been made in this direction \cite{LopezNacir:2018epg}. However, there is a lot more that could be done in order to understand what fraction of these results is transferable.

\newpage

\appendix

\section{Point particle action} \label{AppPointPart}

Here we describe the action for massive and massless particles in GR. Let us start with flat spacetime. The action for the point particle, described only by its mass $m$, that satisfies the Poincaré symmetry and is dimensionally consistent is
\beq
S[\gamma]=-m \int_\gamma \sqrt{-dx^\mu dx_\mu}=-m\int^{\lambda_f}_{\lambda_i} d\lambda \sqrt{-\dot{x}^\mu \dot{x}_\mu},
\eeq
where dot stand for taking the derivative with respect to the parameter $\lambda$ and $\gamma$ is a curve on the spacetime. Of course, physics can't depend on the choice of the parameter, so it represents a gauge redundancy of our description of the particle dynamics. This action does not work for massless particles $m=0$ and the square root function is not pleasant to work with in general. Thus, one can use a different action that cures the problems of the previous. We introduce the auxiliary field $\eta$ and consider the action \cite{Zee:2013dea, Polchinski:1998rq}
\beq
S'=\frac{1}{2} \int_\gamma d\lambda (\eta^{-1}\dot{x}^\mu \dot{x}_\mu-\eta m^2).
\eeq
In order for $S'$ to be reparametrization invariant we must demand that the auxiliary field transforms as $\eta(\lambda) d\lambda = \Tilde{\eta}(\Tilde{\lambda}) d\Tilde{\lambda}$. Varying $S'$ we see that the auxiliary field is not dynamical and
\beq
\Big(\eta^{2}=-\frac{\dot{x}^\mu\dot{x}_\mu}{m^2} \wedge   m \neq 0 \Big) \vee \Big(\dot{x}^\mu\dot{x}_\mu=0 \wedge m=0 \Big).
\eeq
Using $\eta$ for $m \neq 0$ we see that $S'$ reduces to $S$. However, for $m=0$ we find
\beq \label{eq:action_photon_dynamics}
S'_{m \neq 0}[\gamma]=\frac{1}{2} \int_\gamma d\lambda (\eta^{-1}\dot{x}^\mu \dot{x}_\mu).
\eeq

These results are carried over to curved spacetime via the principle of general covariance (itself being a consequence of EP) as we use the metric $g_{\mu \nu}$ to lower the indices. Varying $S$ we obtain the geodesic equation
\begin{equation} \label{eq:geodesic}
\ddot{x}^\mu+\Gamma^{\mu}_{\hphantom{\mu}\alpha \beta} \dot{x}^\alpha \dot{x}^\beta=0\,,
\end{equation}
where the Christoffel symbols are defined by \eqref{eq:Christoffel}. It is useful to the choose proper time $\tau$
for the parameter $\lambda$, so that $\dot{x}^\mu \dot{x}_\mu = -1$.

On the other hand, varying \eqref{eq:action_photon_dynamics} we get
\begin{equation}
\frac{d}{d\lambda}(\eta^{-1}g_{\alpha \nu} \dot{x}^\nu)-\frac{1}{2}\eta^{-1}\partial_\alpha g_{\mu \nu} \dot{x}^\mu \dot{x}^\nu=0.
\end{equation}
Choosing $\lambda'=\eta \lambda$, the equation of motion has the form of the geodesic equation \eqref{eq:geodesic} and $\lambda'$ is then called the affine parameter. There is still a leftover freedom to rescale the parameter. It is often useful to choose the affine parameter in such a way that $\dot{x}^\mu=P^\mu$, where $P^\mu$ is the $4$-momentum of the photon.

Formally it is useful to consider the lagrangian $L=g_{\alpha\beta}\dot{x}^\alpha \dot{x}^\beta$, where it is assumed that we are differentiating with respect to the affine parameter. Then, the particle action breaks the reparametrization invariance
\be
S[\gamma(\tau)]= \int_{\tau_i}^{\tau_f}g_{\alpha\beta}\dot{x}^\alpha \dot{x}^\beta d\tau  \label{Lagrangean}\,,
\ee
and we can always rescale the parameter such that $L=-1,0,1$ for
timelike, null or spacelike geodesics, respectively.

\newpage

\section{Kerr spacetime} \label{AppKerr}

The Kerr spacetime describes rotating BHs in an asymptotically flat Universe~\footnote{Introductory overviews can be found in e.g.  \cite{Zee:2013dea}, \cite{shapiro}  and \cite{Teukolsky:2014vca}.}. In Boyer-Lindquist coordinates (BL) $(t,r,\theta,\varphi)$ metric is given by
\begin{align}
\label{eq:KerrBLlineelement}
\dif s^{2} = & - ( 1 - \frac{2 M r}{\Sigma} ) \dif t^{2}
        - \frac{4 M r a \sin^{2}\theta}{\Sigma} \dif t \dif \varphi
        + \frac{\Sigma}{\Delta} \dif r^{2}
        + \Sigma \dif \theta^{2}
        + \frac{\mathcal{F}}{\Sigma} \sin^{2}\theta \dif \varphi^{2}
\,,
\end{align}
where
\begin{eqnarray}
\label{eq:KerrBLfcts}
\Delta &=& r^{2} + a^{2} - 2 M r
       = (r - r_{+})(r - r_{-} )
\,,\\
\Sigma &=& r^{2} + a^{2} \cos^{2}\theta
\,,\\
\mathcal{F} &=& ( r^2 + a^2 )^{2} - \Delta a^{2} \sin^{2}\theta
\,,
\end{eqnarray}
and $a=J/M$, while $J$ and $M$ are the BH's angular momentum and mass, respectively ($r_\pm$ is defined below). There are two length scales that characterize the Kerr spacetime - Schwarzhild $r_{\rm s}=2M$ and $a$. It is also useful to define the dimensionless rotational parameter $\tilde{a}=a/M$ and note that $0<\tilde{a}<1$ (consequence of the cosmic censorship conjecture). As $a \to 0$, BL coordinates reduce to Schwarzshild coordinates of a non-rotating BH in the asymptotically flat Universe (Schwarzschild BH).

As Kerr spacetime is axi-symmetric (invariant under $\varphi$ rotations) and stationary (invariant under time translations and simultaneous $t \to -t$ and $\varphi \to -\varphi$ inversions), it possess two Killing vectors $\xi^\mu_t=(1,0,0,0)$ and $\xi^\mu_\varphi=(0,0,0,1)$. These Killing vectors correspond to two conserved quantities - energy and angular momentum (per unit mass)
\begin{equation} \label{eq:Kerr_Killing}
E \equiv -   \xi^\mu_t v_\mu \, , \, J \equiv  \xi^\mu_\varphi v_\mu \,.
\end{equation}

Singularities of the BL metric occur when
\beq
\Delta=0 \Rightarrow r_{\pm}=M \pm \sqrt{M^2-a^2}  \,,\label{eq:Kerr_horizon}
\eeq
and
\beq
\Sigma=0 \Rightarrow r_{\bullet}=0 \wedge \theta_{\bullet}=\frac{\pi}{2}  \,.
\eeq
The nature of these singularities can be explored through the analysis of the appropriate scalar. As is the case for Schwarzschild BHs, Kerr BH is a vacuum solution to the Einstein equation so the Ricci scalar is by definition $R=0$. Instead, one can use the Kretschmann scalar
\beq
K \equiv R_{\alpha \beta \gamma \delta}R^{\alpha \beta \gamma \delta} =&& -\frac{96}{(a^2 \cos (2 \theta )+a^2+2 r^2)^6} \Big( a^6 \cos (6 \theta )+10 a^6-180 a^4 r^2 +240 a^2 r^4 \nonumber  \\
&&  +6 a^4 (a^2-10 r^2) \cos (4 \theta )+15 a^2 (a^4-16 a^2 r^2+16 r^4) \cos (2 \theta ) -32 r^6 \Big) .
\eeq
Taking appropriate limits we find that $r_{\pm}$ are only coordinate singularities of the BL metric (Kretschmann scalar is regular at $r_{\pm}$), while the $(r_{\bullet},\theta_{\bullet})$ represents a physical ring-like singularity of the Kerr BH spacetime.

Besides these singularities, and in contrast with the Schwarzschild metric, Kerr spacetime admits another feature, ergoregion, where static observers can't exist. The proper velocity of static observer is, by definition, $v^\mu=(\dot{t},0,0,0)$ and $d\tau^2=-g_{tt}dt^2$. Therefore, in order for static observers to exist $-g_{tt}>0$. It is easy to see that for the Schwarzschild metric this condition is always satisfied outside of the horizon. However for Kerr spacetime it is not. From \eqref{eq:KerrBLlineelement} one finds that
\beq
g_{tt}=0 \Rightarrow r_{\rm{S}\pm}=M \pm \sqrt{M^2-a^2\cos^2{\theta}}  \,.\label{eq:Kerr_ergo}
\eeq
Surfaces defined in this way, called the infinite-redshift surfaces, match the horizons only at $\theta=0$ and $\theta=\pi$ (as $a \to 0$ these surfaces approach each other). The spacetime region between the outer horizon and the outer infinite-redshift surface is called the ergoregion. This region is located outside the outer horizon so it is observationally accessible. Note that inside the ergoregion, the time Killing vector becomes space-like as $g_{\mu \nu} \xi_t^\mu \xi_t^\nu=g_{tt}>0$.

Let us now analyse more general stationary observers i.e. ones whose proper velocity is $v^\mu=(\dot{t},0,0,\dot{\varphi})=\dot{t}(1,0,0,\Tilde{\Omega})$, with $\Tilde{\Omega}=d\varphi/dt$ being the coordinate angular velocity. In order for time-like observers to exist $v^2<0$, so the corresponding critical angular velocity is
\beq
\Tilde{\Omega}_{\circlearrowleft, \circlearrowright}=\frac{-g_{t\varphi} \pm \sqrt{\Delta} \sin{\theta}}{g_{\varphi \varphi}} \,.
\eeq
On the horizon $r_\pm$ we have $\Tilde{\Omega}_{\pm}=$ $\Tilde{\Omega}_\circlearrowleft=$ $\Tilde{\Omega}_\circlearrowright$. The angular velocity of the outer horizon is [from \eqref{eq:KerrBLlineelement}]
\beq
\Tilde{\Omega}_{+}=\frac{a}{2Mr_{+}} \,.\label{eq:horizons_angular}
\eeq
We can use the preceding analysis to define the horizon as a null hyper-surface described by the null vector
\beq \label{eq:horizon_null_surface}
l^\mu=(1,0,0,\Tilde{\Omega}_+)=\xi^\mu_t+\Tilde{\Omega}_+ \xi^\mu_\varphi
\eeq
and two space-like vector orthogonal to it $h^\mu=(0,0,1,0)$ and $k^\mu=(0,0,0,1)$.

\subsection{Penrose process} \label{sec:Penrose}

Let us imagine that some (sub-)nuclear process occurred in the ergoregion and that one of the particles will end up inside the Kerr BH horizon $(\dagger)$, while the other one will end up outside of the ergoregion $(\heartsuit)$. Both particles will have conserved energy and angular momentum \eqref{eq:Kerr_Killing} along their respective geodesics. Besides, locally we have four-momentum conservation at the time when the (sub-)nuclear process occurred $p^\mu_{\rm in}=p^\mu_{\dagger}+p^\mu_{\heartsuit}$. Let us first dot the conservation law with $(\xi_t)_\mu$:
\beq
p^\mu_{\rm in}(\xi_t)_\mu=p^\mu_{\dagger}(\xi_t)_\mu+p^\mu_{\heartsuit}(\xi_t)_\mu \,.
\eeq
Now, each of the terms is separately conserved along the geodesic so
\beq
E_{\rm in}(\infty)-E_{\heartsuit }(\infty)=-p^\mu_{\dagger}(\xi_t)_\mu  \,.
\eeq
There are no prior restrictions on the sign of the RHS so the outcome could be $E_{\rm in}(\infty) < E_{ \heartsuit }(\infty)$ i.e. energy extraction from the BH. Note that in the absence of the ergoregion $g_{tt}<0$, so that $-g_{tt} p^\mu_{\dagger} \xi^\mu_t \geq 0$ and $E_{\rm in}(\infty) \geq E_{\heartsuit}(\infty)$.

From the thermodynamics laws, we can find the irreducible mass of the BH that can't be extracted by the Penrose process or other classical mechanisms (Christodoulou mass, e.g. \cite{Zee:2013dea})
\beq
M_{\rm irr} =\frac{1}{2} (M^2+\sqrt{M^4-J^2}) \,.
\eeq

\newpage


\section{Gross-Pitaevskii-Poisson system} \label{AppGPP}

As mentioned in the Introduction, various types of scalar fields have been considered as DM models. Most of them develop cores that can be described by the Gross-Pitaevskii-Poisson (GPP) system \cite{Berezhiani:2015bqa, Hui:2016ltb, RindlerDallerShapiro:2012}. In the other extreme, when the self-gravitating scalar configurations are candidates for compact objects, GPP is a useful arena for understanding qualitative aspects of models. The GPP equation describes systems of large number of bosons at zero temperature, with the mean-field approximated self-interaction and coupled to the external potential \cite{RogelSalazar:2013}
\be
e\psi=\left(-\frac{\hbar^2}{2\mu}\nabla^2+V_{\text{ext}}+U(\psi)\right)\psi\,.
\ee
Usually, self-interactions $U(\psi)$ are described up to two-body processes $g_{2}|\psi|^2$. The external potential is gravitational, in the DM physics context, and various optical/magnetic traps in the cold atoms research. The gravitational potential $V_{\text{ext}}=\mu V$ is described by the Poisson equation. When the self-interaction is negligible the system is called Schrodinger-Poisson (SP).

FDM models presume very small attractive self-interaction that can be neglected on a galactic scale \cite{Hui:2016ltb,Desjacques:2017fmf}. On the other hand, scalar field DM models with strong repulsive\footnote{Labeled as a ``wrong sign'' potential in the context of axion stars.} self-interaction has also been considered~\cite{RindlerDallerShapiro:2012, Chavanis:2011}. In the superfluid DM model, the condensate (when the phonon excitations are neglected) is characterized by the self-interaction through primarily three-body processes, captured in the GPP picture by $g_{3}|\psi|^4$ \cite{Berezhiani:2015bqa}. Other types of self-interactions have also been studied \cite{Macedo:2013jja,UrenaLopez:2001tw}.

Gross-Pitaevskii equation can, in a hydrodynamical picture, be rewritten as Navier-Stokes equation \cite{RogelSalazar:2013, Chavanis:2011} with the present additional term $(1/m)\nabla Q$, called quantum pressure\footnote{Although the more appropriate label would be quantum potential as it is not in the form $(1/\rho) \nabla p^{(Q)}$. Term can be expressed as $(1/\rho) \del_i {p^{(Q)}_{ij}}$\cite{Mocz:2017wlg}.} $Q$ and defined as
\begin{equation}
\frac{1}{m} Q=-\frac{1}{2\mu^2} \frac{\nabla^2 \sqrt{\rho}}{\sqrt{\rho}}\,.
\end{equation}
This term originates from the kinetic term in the Hamiltonian. Pressure arising from (repulsive) self-interaction of both dominant two- and three-particle processes, has a polytropic equation of state 
\be
P=K\rho^{1+1/n}\,,
\ee
where $K=g_{2}/(2\mu^2)$ ($K=2g_{3}/(3\mu^3)$) for dominant two (three)-particle  interaction. When the flow is stationary and the quantum pressure is negligible with respect to the the pressure originating from the (repulsive) self-interaction and gravity, system is in the Thomas-Fermi regime and Navier-Stokes and Poisson equation reduce to Lane-Emden equation with polytropic indices $n=0.5$ for superfluid DM and $n=1$ for repulsive/fluid DM. 
 
In the rest of this Appendix we will describe analytical construction of the profile of the self-interacting Newtonian oscillatons. Detailed analytical and numerical analysis of the SP/GPP system can be found elsewhere, e.g. \cite{Hui:2016ltb,Chavanis:2011,Guzman:2004wj,KlingRajaraman:2017, KlingRajaraman:2017b} and references therein.

\subsection{Analytical profile of self-interacting Newtonian oscillatons} \label{AppGPP_self_int}

We focus here only on two-particle interactions,  $U(\psi)=g_{2}|\psi|^2$ in the GPP system. It is well known that the Lane-Emden equation with $n=1$ allows for analytical solution of the form $\rho=\rho_{c}\sin{(\xi)}/(\xi)$, where $\rho_{c}$ is density at the center of the polytrope, $\xi=r/\alpha$ is a standard notation for the dimensionless radius of polytropes and $\alpha^2=(n+1)K\rho_{c}^{1/n-1}/(4\pi)$ is the scale factor. Polytropes with $n<5$ do admitt well-defined surface and their radius-mass relation is given by 
\begin{equation}
R \propto M^{\frac{1-n}{3-n}},
\end{equation}
where the constant of proportionality depends on $K$ and $n$ \cite{poissonwillbook}.

Recently, the methodology of the approximate analytical solution construction for the SP system that we have used in this paper \cite{KlingRajaraman:2017}, has been expanded to include self-interactions \cite{KlingRajaraman:2017b}. For weak couplings, the expansion is perturbatively constructed around non-self-interacting solution. Intermediate and strong coupling are obtained by perturbations around the Thomas-Fermi solution, described in the previous paragraph.

We parametrise the strength of the self-interaction, as in Ref. \cite{KlingRajaraman:2017b},
\begin{equation}
\gamma=\frac{\Delta \Lambda(\gamma)}{2\mu^2}.
\end{equation}
Fixing the value of $\gamma>\gamma_{\text{min}}$ and $\mathcal{C}$ uniquely specifies the value of $\Delta=32\pi \mu g_2$, as can be seen on Fig. 4 in Ref. \cite{KlingRajaraman:2017b}. There is a minimum value of coupling for attractive self-interactions $\gamma_{\text{min}}=-0.722$, after configurations become unstable under small perturbations  \cite{KlingRajaraman:2017b}.

For weak couplings, the field and potential expansion is the same as in \eqref{eq_appB_small_r_exp}, \eqref{eq_appB_large_r_exp} and \eqref{eq_appB_v_exp}, but now free parameters depend on the strength of the self-interaction (Eq. 29 in Ref. \cite{KlingRajaraman:2017b}). As we are interested only in general aspects of this models, we used $n_\star=20$ for $s_{(n_\star)}$ and $s^{(n_\star)}_{(m_\star)}$ is identified with the Whittaker function, that well describes leading order long-range behaviour \cite{KlingRajaraman:2017,KlingRajaraman:2017b}. The matching point was estimated as a point where difference between $s_{(n_\star)}$ and $s^{(n_\star)}_{(m_\star)}$ is the smallest.
For the strong coupling we used Thomas-Fermi solution, matched with the Whittaker function (for the long-range behaviour). The matching point can be found as the point where the self-interacting and the quantum pressure have the same value \cite{KlingRajaraman:2017b}.

\newpage

\section{Parametric resonance} \label{AppParRes}

\subsection{Linear and parametric resonances} \label{}

Small oscillations are a cornerstone of almost all theoretical modelling in physics in both classical and quantum context. Some oscillating systems can develop resonance for the particular choice of the parameters that describe them. In this case the oscillation amplitude becomes significantly larger with respect to the non-resonant one and the system evolution may even become unbounded. There are various ways in which the resonance can be excited - by means of the driving force (regular resonances), nonlinearities or because the parameters of the system are themselves periodically varying (parametric resonances).

\begin{figure}
\hspace*{\fill}%
\begin{minipage}[c]{0.46\textwidth}
\centering
\vspace{0pt}
\includegraphics[width=1\columnwidth]{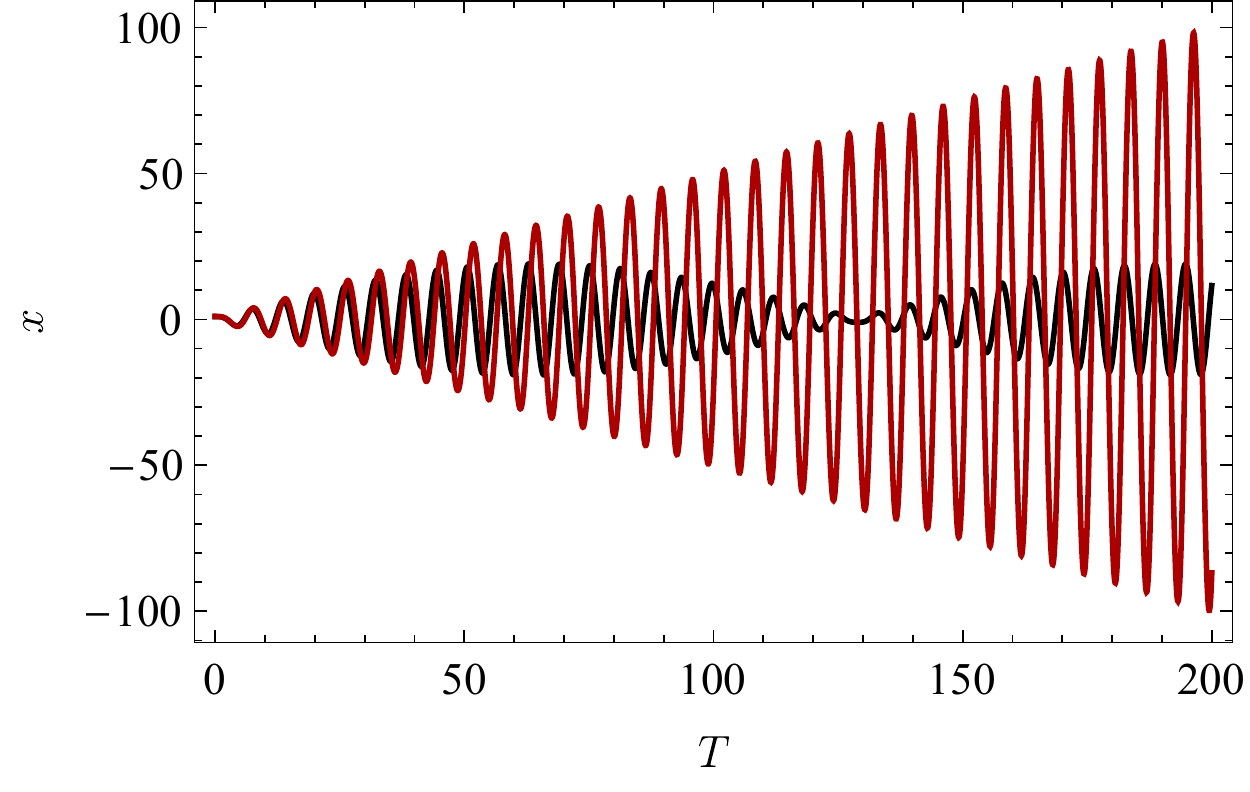}
\caption{ \fontsize{9}{12}
Time evolution of the driven LHO with $f_0=1$ and $\epsilon=0$ (red curve) and $\epsilon=0.1$ (black curve), with $\epsilon=(\omega/\omega_0)^2$.
\label{fig:LHO}}
\end{minipage}%
\hfill
\begin{minipage}[c]{0.49\textwidth}
\vspace{0pt}
\centering
\includegraphics[width=1\columnwidth]{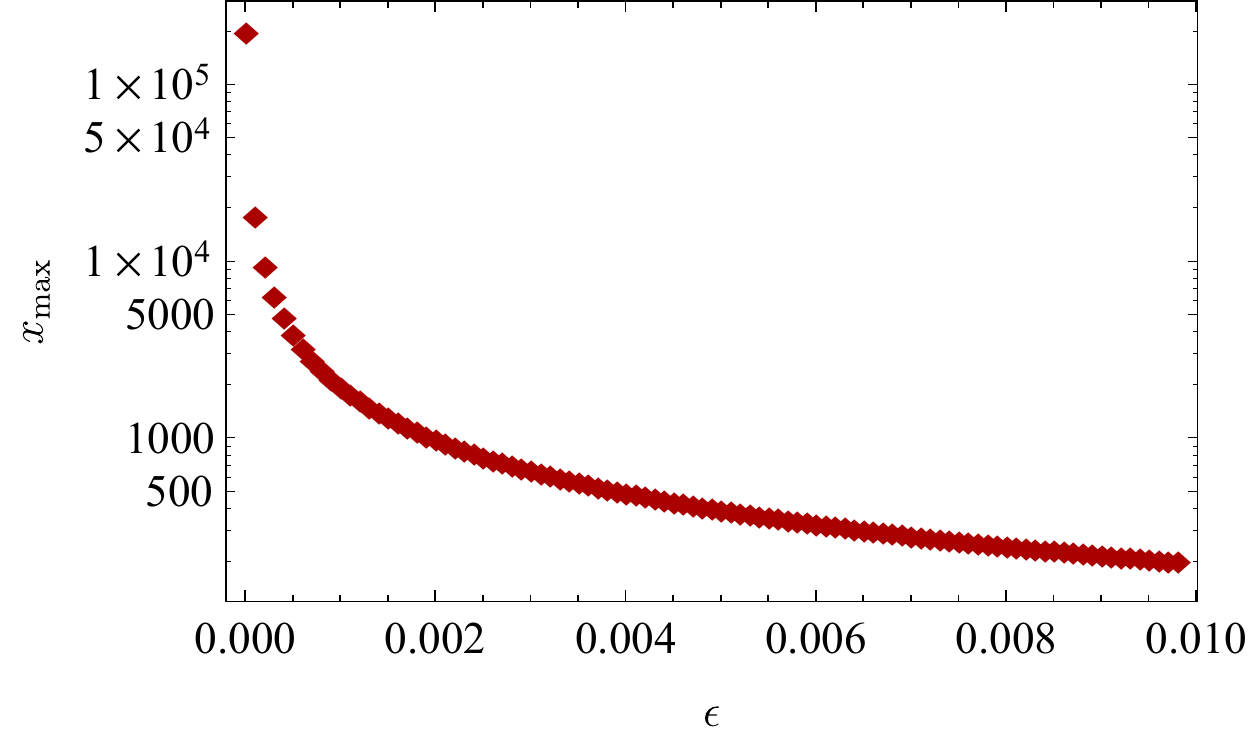}
\caption{\fontsize{9}{12}
Width of the driven LHO resonance - maximum amplitude as a function of $\epsilon=(\omega/\omega_0)^2$ for $f_0=1$.
\label{fig:LHO_width}}
\end{minipage}%
\hspace*{\fill}
\end{figure}

Best start for the regular resonance is linear harmonic oscillator (LHO) with damping and a single harmonic driving force
\beq \label{eq:LHO}
\Ddot{x}+b\Dot{x}+\omega_0^2 x=f_0 \cos{\omega t},
\eeq
with $\omega_0$ being the natural frequency of the undamped system, $b$ is a damping coefficient and $a$ is the amplitude of the driving force. Let us first start with the $b=0$ case. The solution is given by
\beq
x(T)=\begin{cases} \frac{1}{\epsilon} \big(\cos T-\cos \left(T \sqrt{\epsilon +1}\right) \big) & \,,\, \big( \frac{\omega_0}{\omega} \big)^2= 1+\epsilon \\ \frac{1}{2} T \sin T & \,,\, \omega = \omega_0 \,.\end{cases}
\eeq
We rescale time as $T=\omega t$, set $f_0=1$ and show only particular solutions for both undamped and damped scenario. For $\omega=\omega_0$ solution of the system becomes unbounded (Fig. \ref{fig:LHO}). The resonance has some width in the sense that when the driving frequency is close to the natural, there is a corresponding increase in the amplitude of the bounded solution in the form beats - see Figure \ref{fig:LHO_width}. Friction tames the unbounded solution even for $\omega=\omega_0$ as the behaviour is of the form (underdamped case corresponds to $b<2$)
\beq
x(T)=\frac{1}{b}\Big(\sin T-2 e^{-\frac{1}{2} (b T)} \frac{ \sin \left(\frac{1}{2} \sqrt{4-b^2} T\right)}{\sqrt{4-b^2}}\Big).
\eeq

Parametric resonances have a much richer dynamics compared to the regular ones. We will concentrate on the Hill equation
\beq \label{eq:Hill}
\Ddot{x}+f(t)x =0 \,,
\eeq
with the function $f(t)=f(t+T)$ periodic in $T$. For $f(t)=\omega^2_0$ we recover LHO. Particularly, we will be interested in the Mathieu equation, a special case of the Hill equation\footnote{There is a lot of literature on parametric resonances, we used here \cite{arnold_ODE_book} (mathematical standpoint), \cite{LLbookmechanics} (classical mechanics standpoint) and \cite{Santoro:Floquet} (quantum mechanics standpoint).} when
\beq \label{eq:Mathieu_Hill}
f(t)=\omega^2_0+2\epsilon \cos{\omega t} \,,
\eeq
with $\epsilon \in \mathbb{R}$. Hill equations are subject of the Floquet theory and the most famous application of such systems in physics are Bloch waves in solid state physics.

\subsubsection{Basics of Floquet theory} \label{sec:Floquet}

Let us write the equation \eqref{eq:Hill} as the linear system
\beq \label{eq:LinearTP_DE}
\Dot{\bm{x}}=J(t)\bm{x},
\eeq
where in general $\bm{x}$ is a vector of length $N$, while $J(t)=J(t+T)$ is a periodic diagonalizable system operator. For the Hill equation
\beq
\begin{bmatrix}
\Dot{x}_1 \\
\Dot{x}_2
\end{bmatrix}=
\begin{bmatrix}
 0 & 1 \\
-f(t) & 0
\end{bmatrix}
\begin{bmatrix}
x_1 \\
x_2
\end{bmatrix},
\eeq
and $x_1 \equiv x$, $x_2 \equiv \Dot{x}$. As the system is linear we can act with some operator (propagator) on the solution at one time instant (e.g. initial time $\bm{x}_0 \equiv \bm{x}(t_0)$) to obtain the solution at another
\beq \label{eq_Fl_propagator}
\bm{x}(t)=L(t,t_0)\bm{x}_0.
\eeq
One can easily show that the propagator has several intuitive properties
%
\beq
\label{eq:Fl_propagator_prop_1}
 L(t_0,t_0) &=& I\,\\
\label{eq:Fl_propagator_prop_2}
L(t_2,t_1)L(t_1,t_0) &=& L(t_2,t_1)\,\\
\Dot{L}(t,t_0) &=& J(t)L(t,t_0)\,,\label{eq:Fl_propagator_prop_3}
\eeq
where $I$ is an identity operator. Specifically, for time-periodic systems the propagator has to also satisfy
\beq \label{eq:Fl_propagator_prop_32}
   L(t+mT,t_0+mT)=L(t,t_0) \,, m \, \in \mathbb{N}.
\eeq
The last property is the crucial ingredient of the Floquet theorem~\footnote{Generalisation of the Bloch theorem, proven several decades before the work of Bloch. See e.g. \cite{Santoro:Floquet} for the proof.} - under the conditions stated above, the general solution to \eqref{eq:LinearTP_DE} is of the form
\beq \label{eq:parametric_gen_solution}
\bm{x}_i(t)=e^{\Lambda_i (t-t_0)}\bm{u}_i(t) \,
\eeq
where $\bm{u}_i(t+T)=\bm{u}_i(t)$ and $\Lambda_i \equiv (\log{\mu_i})/T$ and $\bm{u}_i(t_0),\,\mu_i$ are eigenvector and eigenvalue of $L_0 \equiv L(t_0+T,t_0)$. Thus, the nature of the propator's eigenvalues decide the question of the system stability. In order to evaluate the $L_0$, note that if we define the fundamental (Wronsky) matrix $W$, whose columns are $N$ linearly independent solutions of the  \eqref{eq:LinearTP_DE} we have
\beq \label{eq:wronsky}
\Dot{W}=J(t)W(t)  \,,\,W(t)=L(t,t_0)W(t_0) \,,
\eeq
as the propagator [system operator] act on the each column as in \eqref{eq_Fl_propagator} [\eqref{eq:LinearTP_DE}]. If we take such initial conditions that $W(t_0)=I$, then
\beq \label{eq:FL_wronsky_calc}
W(T)=L_0.
\eeq
Note also the Liouville's theorem \cite{arnold_ODE_book}
\beq \label{eq:Liouville}
\det(W(t))=\det(W(t_0))\exp{\Big( \int_{t_0}^t \rm{tr}(J(t'))dt' \Big)},
\eeq
where $\det(W(t))$ (Wronskian) is the oriented volume of the parallelepiped spanned by linearly independent solutions.

For the two-dimensional problem (as is the Hill equation), the propagator is a $2 \times 2$ matrix so
\beq \label{eq:FL_eigen}
\mu^2-\rm{tr}(L_0)\mu+\det(L_0)=0,
\eeq
and (Vieta's formulas)
\beq \label{eq:FL_Vieta}
\mu_1+\mu_2=\rm{tr}(L_0) \,,\, \mu_1 \mu_2 = \det(L_0) \,.
\eeq
We will further assume that we are working with the system that preserve the phase space volume (as the Hill equation does), so $\det(W(t))=1$ [eq. \eqref{eq:Liouville}] and hence $\det(L_0)=1$ [eq. \eqref{eq:wronsky}]. In this case \eqref{eq:FL_eigen} has two real solutions when $\rm{tr}(L_0)>2$ and two complex-conjugate solutions if $\rm{tr}(L_0)<2$. From \eqref{eq:FL_Vieta} we find that in the real case, one of the eigenvalues will be larger then $1$ in absolute value so the system is unstable in the sense of Lyapunov\footnote{Equilibrium point $\bm{x}=\bm{x}_e$ is stable in Lyapunov's sense if the phase-space trajectories obtained as a small deviation from it at the initial time stay ``contained'' for all time. Formally:  $\forall \varepsilon>0,\,   \exists \delta(\varepsilon)>0$ such that $\forall \bm{x}_0, |\bm{x}_0-\bm{x}_e|<\delta$ the solution of the system \eqref{eq:LinearTP_DE} with the initial conditions $\bm{x}(t_0)=\bm{x}_0$ satisfies $|\bm{x}(t)-\bm{x}_e|<\varepsilon \, \forall t>t_0$. Proofs regarding the link between the propagator eigenvalues for time-periodic systems and the Lyapunov stability can be found in \cite{arnold_ODE_book}.}. On the other hand, for the complex-conjugate case we have $\mu \mu^{\dagger}=|\mu|=1$ and $\Lambda=i \rm{Im}({\Lambda})$.

Usually, analytical evaluation of the stability is impossible and one has to resort to numerical or approximate methods.

\newpage

\subsection{Stability of the Mathieu equation} \label{sec:Math_stability}

\begin{figure*}[htb]
\begin{tabular}{cc}
\includegraphics[width=8.7cm]{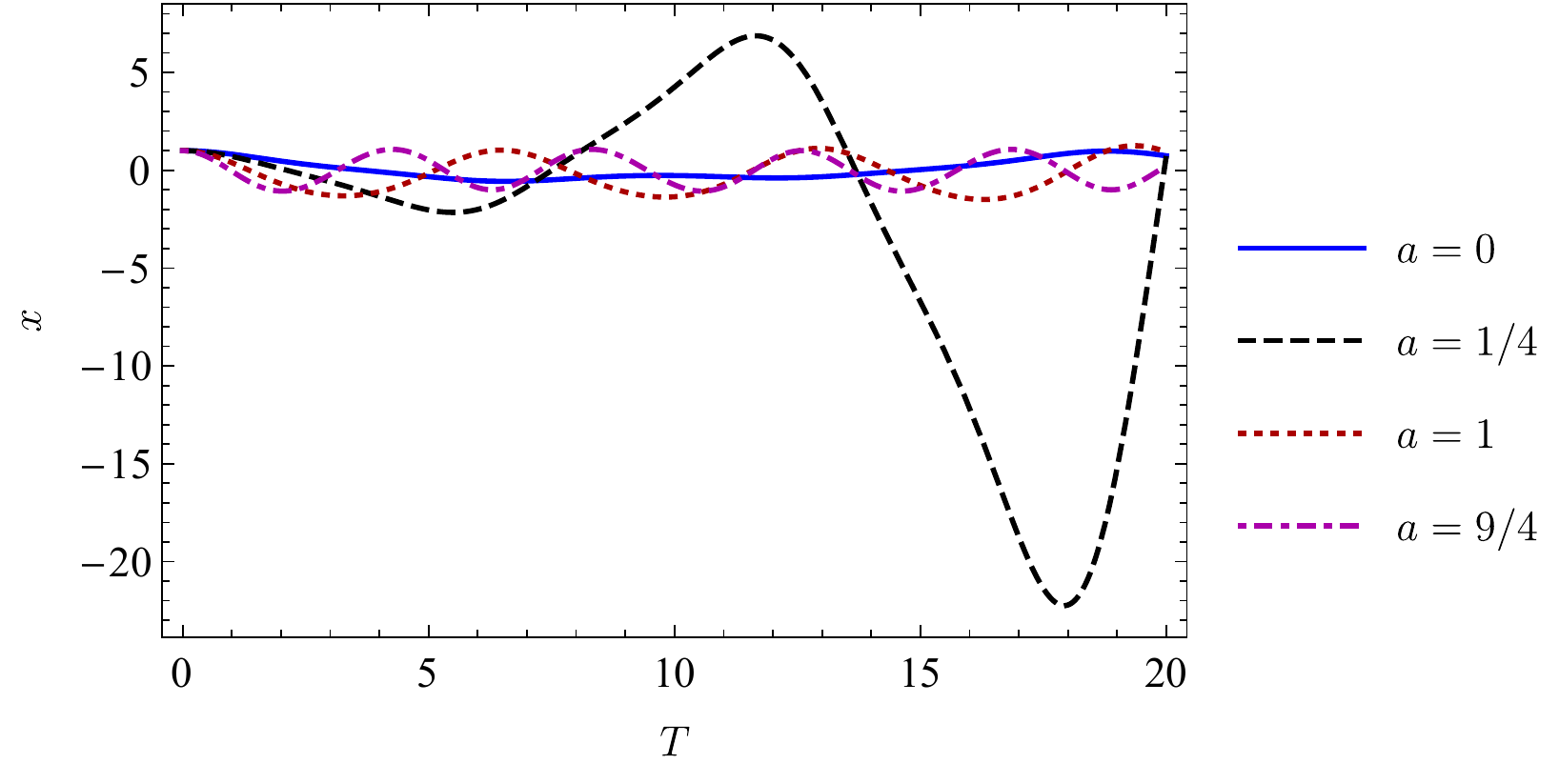}
&\includegraphics[width=8.7cm]{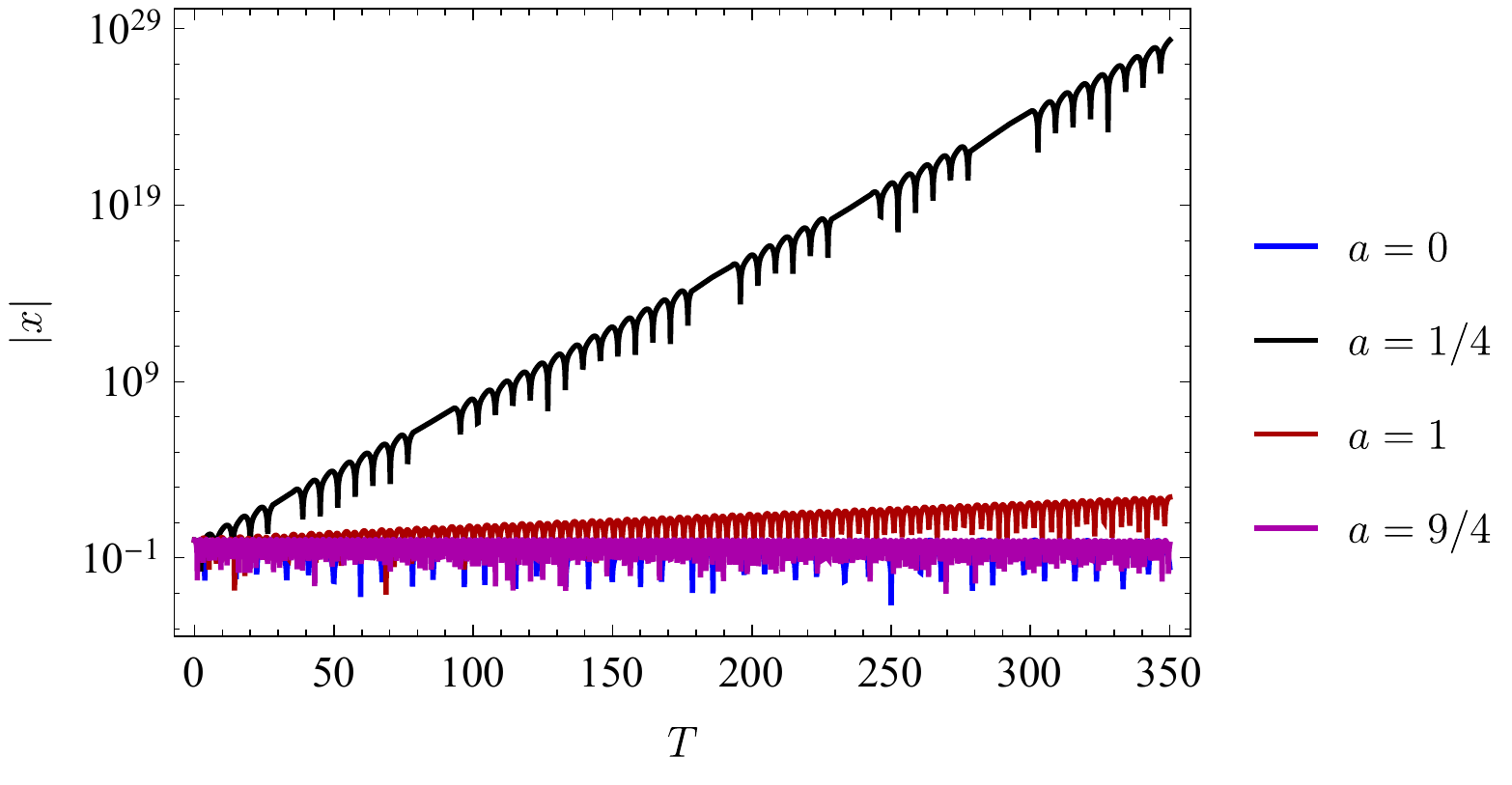}
\end{tabular}
\caption{Solutions to the Mathieu equation for $\epsilon=0.2$.
\label{fig:Mathieu_solution}}
\end{figure*}

We will analyse the stability of Mathieu equation [\eqref{eq:Hill}, \eqref{eq:Mathieu_Hill}] in the limit $\epsilon \to 0$. In this limit, the equation reduces to the LHO. Solution for the initial condition $\{ x(0)=1\,,\,\Dot{x}(0)=0 \}$ is
\beq
x^{(1)}=\cos{\omega_0 t} \, , \, \Dot{x}^{(1)}=-\omega_0 \sin{\omega_0 t} \,.
\eeq
Conversely, for $\{ x(0)=0\,,\,\Dot{x}(0)=1 \}$
\beq
x^{(2)}=(1/\omega_0)\sin{\omega_0 t} \, , \, \Dot{x}^{(2)}=\cos{\omega_0 t} \,.
\eeq
From \eqref{eq:FL_wronsky_calc}
\beq
L_0=\begin{bmatrix}
\cos{\omega_0 T} & (1/\omega_0)\sin{\omega_0 T} \\
-\omega_0 \sin{\omega_0 T} & \cos{\omega_0 T}
\end{bmatrix}
\eeq
and $T=\frac{2\pi}{\omega}$. Thus
\beq
|\Tr(L_0)|=\Big|2\cos{\Big(2\pi \frac{\omega_0}{\omega}\Big)}\Big| < 2 \,\, \text{if} \,  \frac{\omega_0}{\omega} \neq \frac{n}{2} \,, n \in \mathbb{N}.
\eeq
However, if the condition
\beq \label{eq:hom_k_instable}
\frac{\omega_0}{\omega} = \frac{n}{2} \,, n \in \mathbb{N}
\eeq
is met, $|\Tr(L_0)|=2$ and the question of stability depends on the term multiplied by $\epsilon$. Precisely at these points in the parameter space the instability develops in the full Mathieu equation as we will show.

\begin{figure}
\hspace*{\fill}%
\begin{minipage}[c]{0.52\textwidth}
\centering
\vspace{0pt}
\includegraphics[width=1\columnwidth]{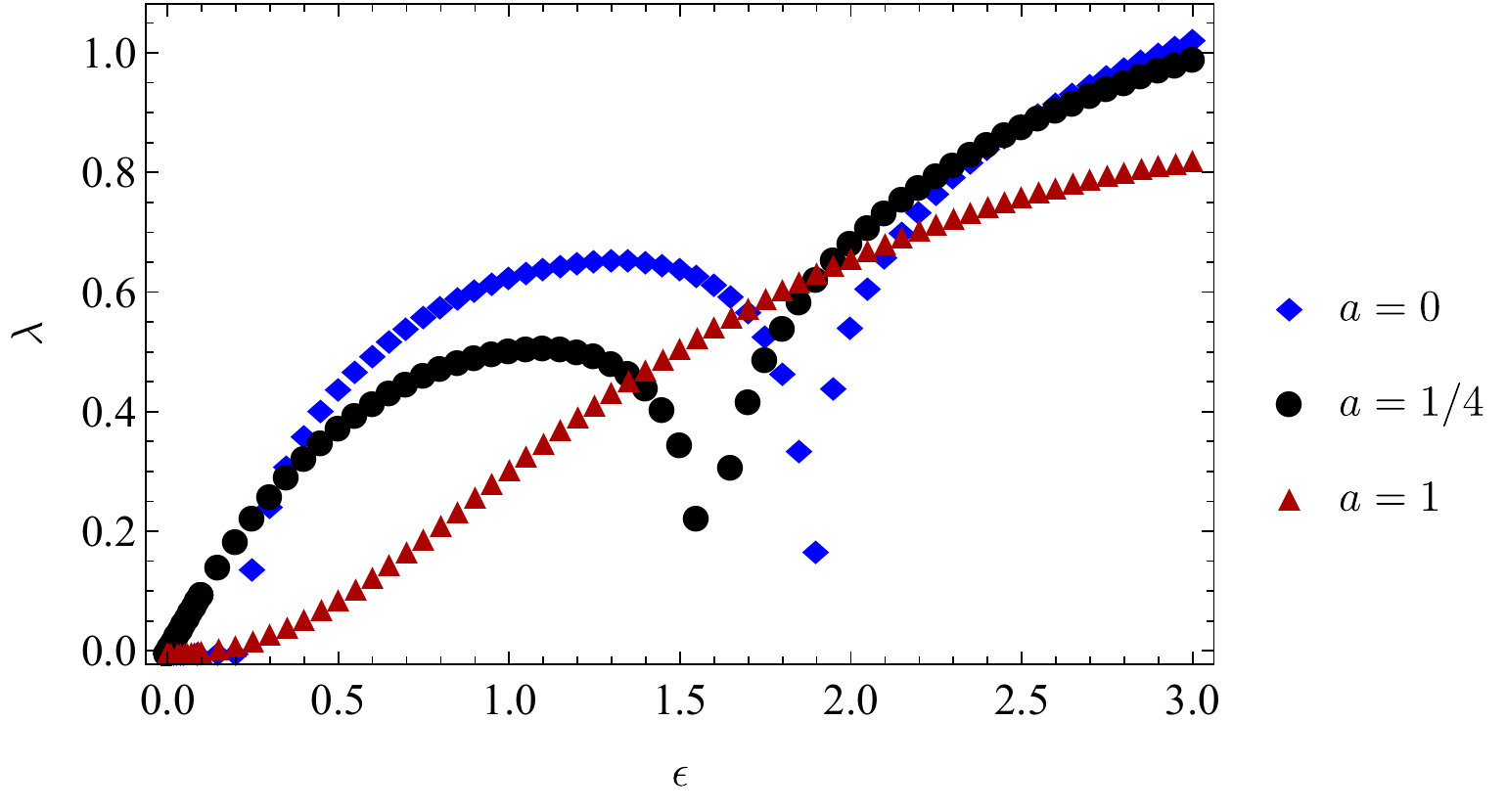}
\caption{ \fontsize{9}{12}
Instability rates $\lambda$ for the first few
\newline
values of  $n$ in \eqref{eq:hom_k_instable} for Mathieu equation.
\label{fig:Mathieu_rate}}
\end{minipage}%
\hfill
\begin{minipage}[c]{0.44\textwidth}
\vspace{0pt}
\centering
\includegraphics[width=1\columnwidth]{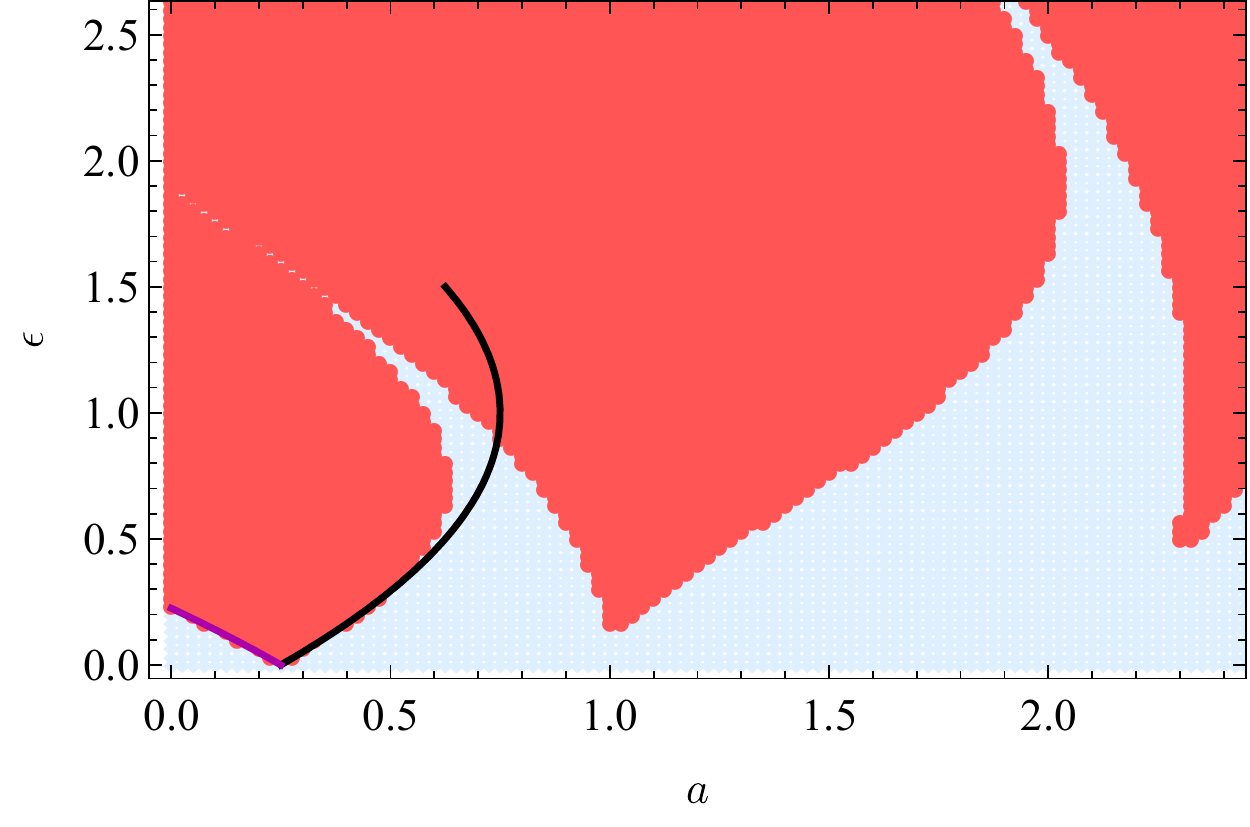}
\caption{\fontsize{9}{12}
Stability diagram for Mathieu equation. Blue dots correspond to stable solutions, while the red dots correspond to the unstable solutions. Black and magenta curves, given by \eqref{eq:Mathieu_zones_border}, correspond to analytical estimates of the borders between (in)stability zones and give good description for small $\epsilon$.
\label{fig:Mathieu_stability}}
\end{minipage}%
\hspace*{\fill}
\end{figure}

We will rescale the time variable as $T=\omega t$ and define $a=\omega_0^2/\omega^2$ so that
\beq
\partial^2_T x+(a+2 \epsilon \cos{T})x=0\,.
\eeq
Applying \eqref{eq:parametric_gen_solution} to Mathieu equation that respects time inversion symmetry we obtain general form of the solution
\beq
x(T)=e^{(\lambda+i\varrho)T} u_1(T)+e^{-(\lambda+i\varrho)T} u_2(T) \,,\,  u_i(T)=u_i(T+2\pi),
\eeq
where $\lambda=\rm{Re}(\Lambda)$ and $\varrho=\rm{Im}(\Lambda)$. Several (numerical) solutions of the Mathieu equation are shown on Fig. \ref{fig:Mathieu_solution}. We have verified that the instability develops for $a=1/4$ even when $\epsilon$ is arbitrarily small, while for intermediate $\epsilon \leqsim 0.3$, $a=2^2/2$ instability also develops with the smaller rate than the $a=1/4$ solution. In the strong-$\epsilon$ regime situation is more complicated as instability rates for the first few $n$ in \eqref{eq:hom_k_instable} interchangeably dominate, as one can see on Fig. \ref{fig:Mathieu_rate}. The stability of Mathieu's equation can be represented on the parametric $\epsilon-a$ (Ince-Strutt) stability diagram and we construct one numerically on Fig. \ref{fig:Mathieu_stability}. From the diagram we can conclude that the non-monotonous nature of the instability rate is probably consequence of the partial overlap of the stability and instability zones of the solutions corresponding to different $n$ in \eqref{eq:hom_k_instable}.  It should be noted that these results do not depend on the sign of $\epsilon$, as Ince-Strutt stability diagram is symmetric under reflections about $a$ axis (see the diagram in \cite{benderbook} and the discussion in Section \ref{sec:Mathieu_zone_border}).

To conclude, in principle instability develops for $a$ given by \eqref{eq:hom_k_instable} for  any $\epsilon \neq 0$. However, for small and even intermediate $\epsilon$ the $a=1/4$ instability is the most pronounced. Contrary to the LHO case, presence of friction will not shut down unbounded behaviour for the whole of the parameter space but the instability zones will be shifted from the $a$ axis. This is intuitive phenomenon as there is a competition between $\propto (-bt)$ exponent from the friction and $\propto \lambda t$ from the instability. Only when the total exponent is positive, instability will develop. Presence of the dissipation further disfavours higher-$n$ instabilities~\cite{arnold_ODE_book}.

\newpage

\subsection{Analytic estimate for the instability timescale of the Mathieu equation} \label{app:RG_time_scales_1st}

Here we derive the instability timescales, using a perturbative expansion in a small coupling parameter $\epsilon$, to be defined in the concrete physical context (see Section \ref{sec:Mathieu_parametric}, Section \ref{sec:mink_axion_hom} and Appendix \ref{sec:mink_axion_hom}). We expect that for specific values of $\omega$, given by Eq.~\eqref{eq:hom_k_instable} regular perturbation theory breaks down. However, one can start from the regular perturbation theory, see how instabilities buildup and use multi-scale \cite{benderbook} or dynamical renormalization group (DRG) \cite{Chen:1995ena} methods to regularize the problem. We will here use the later approach.

We here focus on the dominant instability $\sqrt{a}=1/2$ and denoth the subscript of the solution to the Mathieu equation with $\sqrt{a}$. At zeroth order in $\epsilon$, the solution is given by $x_{1/2}^{(0)}=Ae^{i(1/2)T}+\text{c.c}$. The differential equation for the first order correction is
\begin{equation} \label{eq:MAthieu_1st_DE}
\partial^2_T x_{1/2}^{(1)}+\Big(\frac{1}{2} \Big)^2 x_{1/2}^{(1)}= - \Big(
 A^{\ast}  e^{i(1/2)T} +  A  e^{i(3/2)T} \Big) +\text{c.c.}
\end{equation}
and it's solution is given by
\beq\label{eq:app_Mathieu_1st_bare}
x_{1/2}= \Big( A-\frac{1}{2}A^{\ast} \epsilon \Big)  e^{i(1/2)T} +\epsilon \Big(i
 A^{\ast}  e^{i(1/2)T}(T-T_0) +   \frac{1}{2}   A  e^{i(3/2)T} \Big) +\text{c.c.}\,,
\eeq
where $T_0$ is some arbitrary time where we imposed initial conditions. Higher-order terms will build up a secular terms of the form $(T-T_0)^m$, where $m$ is the order of the expansion, in the limit $m \to \infty$ giving exponential growth. However, this behaviour invalidates our perturbative expansion.

The DRG approach is based on the insight that the invalidation of the regular perturbation theory is a consequence of the big gap between $T_0$ and $T$ \cite{Chen:1995ena, Galley:2016zee}. In order to remedy this problem, we declare the parameters of the solution in Eq. \eqref{eq:app_Mathieu_1st_bare} as ``bare'' and rewrite them as the renormalized ones:
\be \label{eq:renormalized_amplitude}
A(T_0)=Z(T_0, \tau)A(\tau) \,, \,Z(T_0, \tau)=1+\sum^{\infty}_{n=1}a_n\epsilon^n.
\ee
Next, we expand $T-T_0=T-\tau+\tau-T_0$ and choose $a_1$ (``counter-term'') in such a way to cancel secular terms $\propto (\tau-T_0)$. The renormalized solution has the form
\begin{align}\label{eq:app_Mathieu_1st_renor}
x_{1/2}=\Big(A(\tau)-\frac{1}{2}A^{\ast}(\tau) \epsilon \Big)   e^{i(1/2)T}   + \epsilon \Big(i
 A^{\ast}(\tau)  e^{i(1/2)T}(T-\tau)  +\frac{1}{2}   A (\tau) e^{i(3/2)T} \Big) +\text{c.c.}\,, \nonumber
\end{align}
with $a_1 A(\tau)=-iA^{\ast}(\tau)(\tau-T_0)$. Arbitrariness of $\tau$ leads to the RG equation
\begin{equation} \label{eq:RGeq}
\frac{\partial A(t_0)}{\partial \tau}=0.
\end{equation}
Working consistently at the $\epsilon^1$ order and decomposing $A(\tau)=X(\tau)+iY(\tau)$, we find
\begin{equation}
\frac{\partial^2 X }{\partial \tau^2}-\epsilon^2 X(\tau)=0,
\end{equation}
i.e.
\begin{equation}
X(\tau)=e^{\pm \epsilon \tau},
\end{equation}
with $\partial_{\tau} X=\eps Y$. Finally, we choose $\tau=T$ as the ``observational'' time and conclude that the instability rate, to first order in $\epsilon$ is $\lambda=\epsilon$, for  Mathieu equation. 

\subsubsection{Second order results}

\begin{figure}
\centering
\includegraphics[width=0.65\columnwidth]{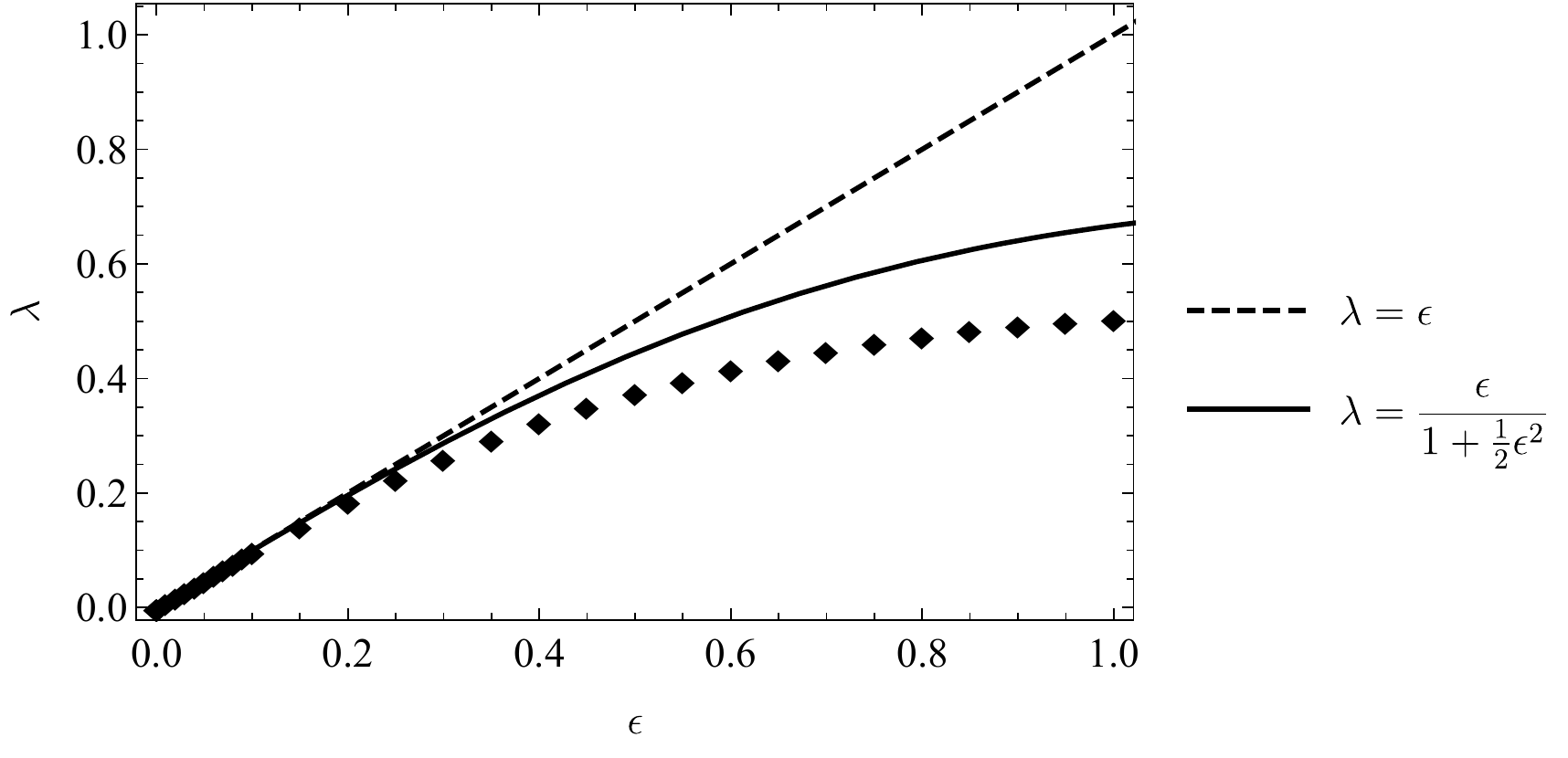}
\caption{Analytical estimates to the first and second order for the instability rates for Mathieu equation with $a=1/4$.}
\label{fig:Mink_axion_rates_Math}
\end{figure}

Here we obtain solution of the Mathieu equation to the second order, using DRG. Differential equation for the second order contribution is of the form
\beq
\partial^2_T x_{1/2}^{(2)}+\Big(\frac{1}{2} \Big)^2 x_{1/2}^{(2)} &=& - (2 \cos{T})x_{1/2}^{(1)}\,.
\eeq
DRG procedure is analogous to the first order case. We will consider only leading order harmonics with $T/2$  as they will give the dominant contribution to the instability. Note that the term of the form $(T-\tau)(\tau-T_0)$  will self-consistently cancel, sign of the renormalizability of the differential equation \cite{Galley:2016zee}. Second-order coefficient in Eq. \eqref{eq:renormalized_amplitude} is
\beq
a_2 A(\tau)=-iA(\tau)(\tau-t_0)-\frac{A(\tau)}{2}(\tau-t_0)^2\,.
\eeq
From the RG equation \eqref{eq:RGeq} we obtain $\lambda^{(R,2)}_{\rm a}=\epsilon \sqrt{1-\eps^2}$. As we should trust this solution to the order of $\mathcal O (\eps^3)$, we perform Pad\'e resummation of the results \cite{benderbook}. As the perturbative result $ \lambda^{(R,2)}_{\rm a}$ is an even function, we used the first non-trivial approximant $(2,1)$ and the final rate estimate is
\beq \label{eq:Mathieu_2nd_final}
\lambda_{\rm a}=\frac{ \epsilon}{1+\frac{1}{2}\epsilon^2},
\eeq
This results gives a very good description of the numerical data for small and intermediate values of $\epsilon$ as shown on Fig. \ref{fig:Mink_axion_rates_Math}.

\subsubsection{Instability zone borders} \label{sec:Mathieu_zone_border}

Approach of this section was originally used to estimate the curves that separate the stable and unstable zones of the $\epsilon-a$ diagram. One would start similarly as for the rate estimate but also treat parameter $a$ as a function of the expansion parameter $a=a(\epsilon)$. To second order in $\epsilon$ critical curves are given by \cite{benderbook, Chen:1995ena}
\beq \label{eq:Mathieu_zones_border}
a_{\star}=\frac{1}{4} \pm \epsilon  - \frac{1}{2} \epsilon^2 \,,
\eeq
as shown on Fig. \ref{fig:Mathieu_stability}. Both \eqref{eq:Mathieu_2nd_final} and \eqref{eq:Mathieu_zones_border} suggest that the stability zones and the rates are even function of $\epsilon$ (note that $e^{\pm \lambda t}$ solution contributions come in pairs due to Floquet theorem).

\subsubsection{Presence of the driving term}

Presence of the driving term as in \eqref{eq:inhomogen_Mathie} doesn't influence the question of stability. We can see that at the perturbative level, adding a particular solution associated with the driving term on the RHS of \eqref{eq:MAthieu_1st_DE} will not quench the instability in any way. We have also checked numerically that the stability diagram doesn't change.

\newpage

\section{Scalar-photon coupling} \label{app:sc-photon}

In the main text we have considered coupling between axions and the Maxwell sector (framework is described in Section \ref{sec:axMaxframe}). As there are several sub-eV candidates for Beyond SM scalar particles \cite{Baryakhtar:2018doz}, we will analyse what changes with regards to the discussion of Part \ref{ch:ax-photon}  if we consider scalar-Maxwell coupling. Instead of \eqref{eq:MFaction}, coupling is realized through
\beq
\label{eq:MFactionS}
{\cal L} \subset - \frac{1}{4} F^{\mu\nu} F_{\mu\nu} -\Big( \frac{k_{\rm s}}{2} \Phi \Big)^\vartheta F^{\mu\nu} F_{\mu\nu} \,, \, \vartheta \in \mathbb{N}.
\eeq
One could also consider higher-order terms for axion coupling but, because of parity, the next lowest one would be $\sim (k_{\rm s} \Phi)^3$. We will consider $\vartheta=1,2$ values for the scalar case, as the quadratic term has been considered in the literature \cite{Olive:2007aj, Stadnik:2015kpa}. Equations of motion \eqref{eq:MFEoMScalar}\,-\,\eqref{eq:MFEoMTensor} (Klein-Gordon and Maxwell equations and energy-momentum tensor for the Einstein equation) are now modified to
%
\beq
\label{eq:MFEoMScalarSC}
\left(\nabla^{\mu}\nabla_{\mu} - \mu^{2}_{\rm s} \right) \Phi &=&\frac{\vartheta\,k_{\rm s}^\vartheta\Phi^{\vartheta-1}}{4}F^{\mu\nu} F_{\mu\nu}\,,\\
\label{eq:MFEoMVectorSC}
\nabla^{\nu} \left(1+k_{\rm s}^\vartheta\Phi^\vartheta\right)F_{\mu\nu} &=& 0\,,\\
T_{\mu\nu} &=&\left(1+k_{\rm s}^\vartheta\Phi^\vartheta\right) F_{\mu}{}^{\rho} F_{\nu\rho} - \frac{1}{4} g_{\mu\nu}\left(1+k_{\rm s}^\vartheta\Phi^\vartheta\right) F^{\rho\sigma} F_{\rho\sigma}  \nonumber \,\\
&& +\nabla_{\mu}\Phi \nabla_{\nu} \Phi  - \frac{1}{2} g_{\mu\nu} \left( \nabla^{\rho} \Phi \nabla_{\rho}\Phi + \mu^{2}_{\rm s}\Phi\Phi \right)
\label{eq:MFEoMTensorSC}
\eeq
%


\subsection{Background scalar field in flat space-time}\label{sec:mink_scalar}

\subsubsection{Homogeneous configuration}
For a non-relativistic scalar field, Maxwell's equations in the Coulomb gauge and a Minkowski background reduce to
\begin{equation}\label{eq:max_sc_coupl_Mink_hom}
\partial_{\mu}F^{\mu \nu}=-g_\vartheta(t)F^{0 \nu}\,,
\end{equation}
where
\begin{equation} \label{eq:g_vartheta_sc}
g^{(\vartheta)}(t)=\frac{\vartheta k_{\rm s}(k_{\rm s}\Phi)^{\vartheta-1}\partial_t{\Phi}}{1+(k_{\text{s}}\Phi)^{\vartheta}}\,,
\end{equation}
and the RHS of Eq. \eqref{eq:max_sc_coupl_Mink_hom} is a well-defined\footnote{This is the case in general, when the RHS of Eq. \eqref{eq:max_sc_coupl_Mink_hom} is proportional to $(\partial_{\mu}\Phi)F^{\mu \nu}$.} current $j^{\nu}$. Note that for $\vartheta=1$, if $k_{\rm s}\Phi(t)=-1$ the current diverges.

The general comments on the Coulomb gauge\footnote{Calculation in the Lorenz gauge can be found in Section III B in \cite{Boskovic:2018lkj}.} from the Section \ref{sec:mink_axion_hom} also apply to this case. Equation \eqref{eq:max_sc_coupl_Mink_hom} reduces to
\begin{equation}
\partial^2_t\bm{A}+g^{(\vartheta)}(t)\partial_t\bm{A}-\bm{\nabla}^2 \bm{A}=0.
\end{equation}
Note that here $\bm{j}_{t}=g^{(\vartheta)}(t)\partial_t \bm{A}$ and $\bm{j}_{l}=-g^{(\vartheta)}(t)\nabla A_0$. Fourier transforming this equation we obtain
\begin{equation} \label{eq:max_scalar_FT_Mink}
\partial^2_t\bm{A}_{\bm{p}}+p^2\bm{A}_{\bm{p}}+\int\frac{d^3\bm{p}'}{(2\pi)^3}g^{(\vartheta)}_{\bm{p}-\bm{p}'}(t)\partial_t\bm{A}_{\bm{p}'}=0,
\end{equation}
with $g^{(\vartheta)}_{\bm{p}}(t)$ being the Fourier transform of $g^{(\vartheta)}(t)$.

If we consider a homogeneous scalar field $\Phi=\Phi_0 \cos{(\mu_{\rm s}  t)}$ and decouple the polarization vectors, with $\bm{A}_{\bm{p}}=\sum_{\lambda}y_{\bm{p}}\bm{\xi}^{(\lambda)}_{\bm{p}}+\text{c.c.}\,$, the previous equation reduces to
\begin{equation} \label{eq:max_scalar_FT_Mink_hom}
\partial^2_t y_{\bm{p}}+p^2y_{\bm{p}}+g^{(\vartheta)}_{\bm{p}}(t)\partial_t y_{\bm{p}}=0.
\end{equation}
The form of this equation is similar to the Ince equation (for the general case see \cite{Hartono:2004}) and we can use a change of variables of the form
\begin{equation}
y_{\bm{p}}=\exp{\Big(- \frac{1}{2}\int^{t}_{0} g^{(\vartheta)}_{\bm{p}}(t')dt' \Big)}f_{\bm{p}}
\end{equation}
to obtain an equation\footnote{Note that: (i) $g^{(\vartheta)}_{\bm{p}}(0)=0$; (ii) the conversion factor between $f_{\bm{p}}$ and $y_{\bm{p}}$ is harmonic and can not change the conclusions regarding stability.} of the Hill type
\begin{equation} \label{eq:HillSC}
\partial^2_t f_{\bm{p}}+(p^2+W_{\vartheta}(t)) f_{\bm{p}}=0,
\end{equation}
where we defined
\begin{equation} \label{eq:Wp_scalar}
W_{\vartheta}(t)=-\frac{1}{2}\partial_t g^{(\vartheta)}_{\bm{p}}-\frac{1}{4} (g^{(\vartheta)}_{\bm{p}})^2.
\end{equation}
We see that $W_{\vartheta}(t)$, i.e. the harmonic term that drives the instability, scales to leading order as $k^{\vartheta}_{\rm s}$.
To the lowest order in $k_{\rm s}$ this equation reduces to the Mathieu equation for both $\vartheta=1$:
\begin{equation}
W_1(t)=\frac{1}{2}\mu_{\rm s}^2 k_{\rm s}\Phi_0 \cos{(\mu_{\rm s}  t)}+\mathcal{O}(k^2_{\rm s} )
\end{equation}
and the $\vartheta=2$ case:
\begin{equation} \label{eq:p2_low_k_Mathieu}
W_2(t)=\mu_{\rm s}^2 (k_{\rm s}\Phi_0)^2 \cos{(2\mu_{\rm s}  t)}+\mathcal{O}(k^4_{\rm s} ).
\end{equation}
From \eqref{} (see Box on page \pageref{} for the derivation) one can then conclude that for $\vartheta=1$
\begin{equation} \label{eq:rate_scalar_p_1}
\lambda_{\vartheta=1}=\frac{1}{4} k_{\rm s}\Phi_0 \mu_{\rm s}+\mathcal{O}(k^2_{\rm s} ) \,,
\end{equation}
and $\vartheta=2$
\begin{equation} \label{eq:rate_scalar_p_2}
\lambda_{\vartheta=2}= \frac{1}{4} (k_{\rm s}\Phi_0)^2\mu_{\rm s}+\mathcal{O}(k^4_{\rm s} ) \,.
\end{equation}
In Box on page \pageref{subsec:scalar_rate} we provide analytical calculation of the instability rate to the higher order in $k_{\rm s}$ and the comparison with the numerical data is analysed in the Box on page \pageref{subsec:scalar_flat_numerics}.

\begin{figure}
\hspace*{\fill}%
\begin{minipage}[c]{0.47\textwidth}
\centering
\vspace{0pt}
\includegraphics[width=1\columnwidth]{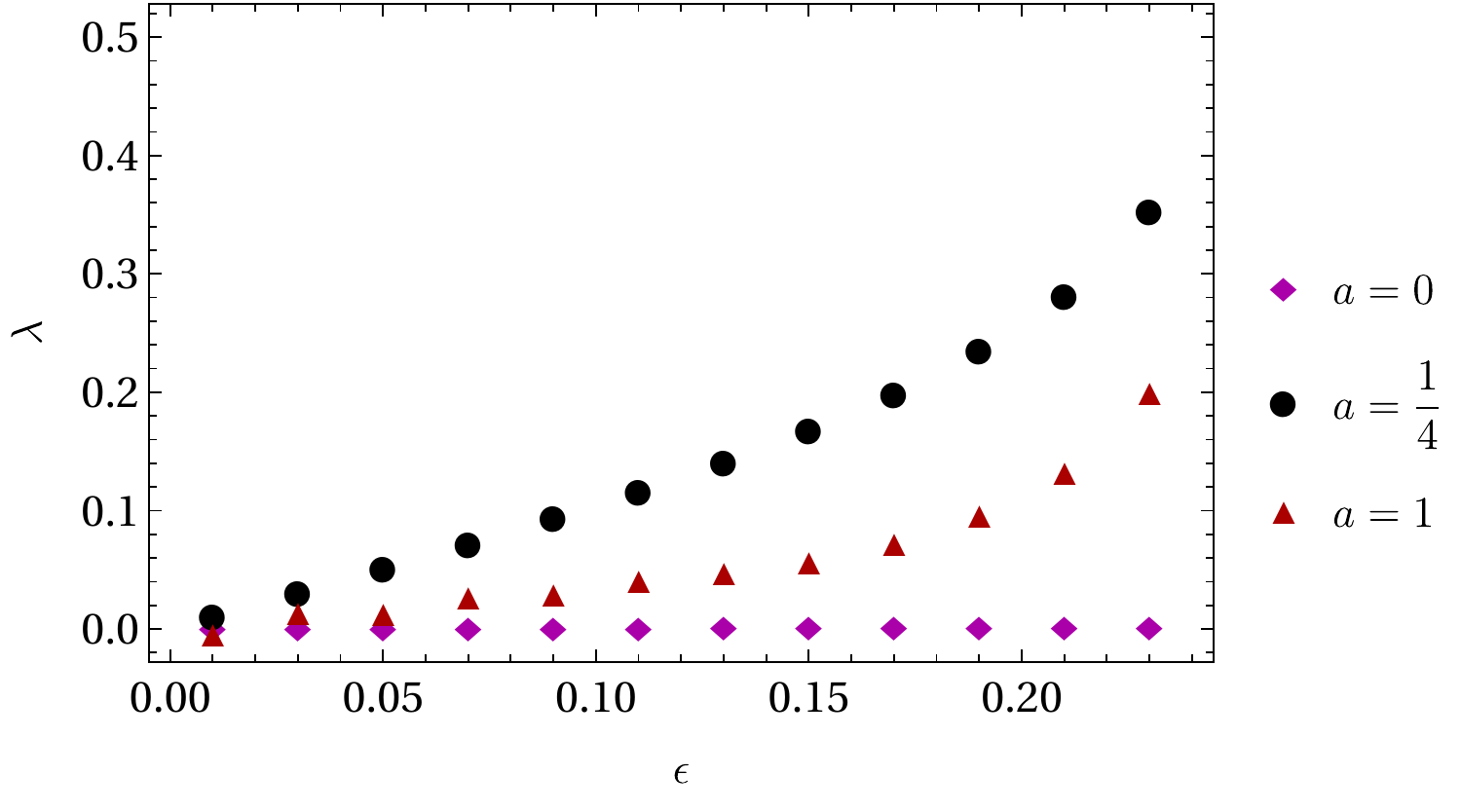}
\caption{ \fontsize{9}{12}
Instability rates $\lambda$ for the first few\newline values of  $n$ in \eqref{eq:hom_k_instable} for \eqref{eq:HillSC} and $\vartheta=1$.
\label{fig:p1rate}}
\end{minipage}%
\hfill
\begin{minipage}[c]{0.47\textwidth}
\vspace{0pt}
\centering
\includegraphics[width=1\columnwidth]{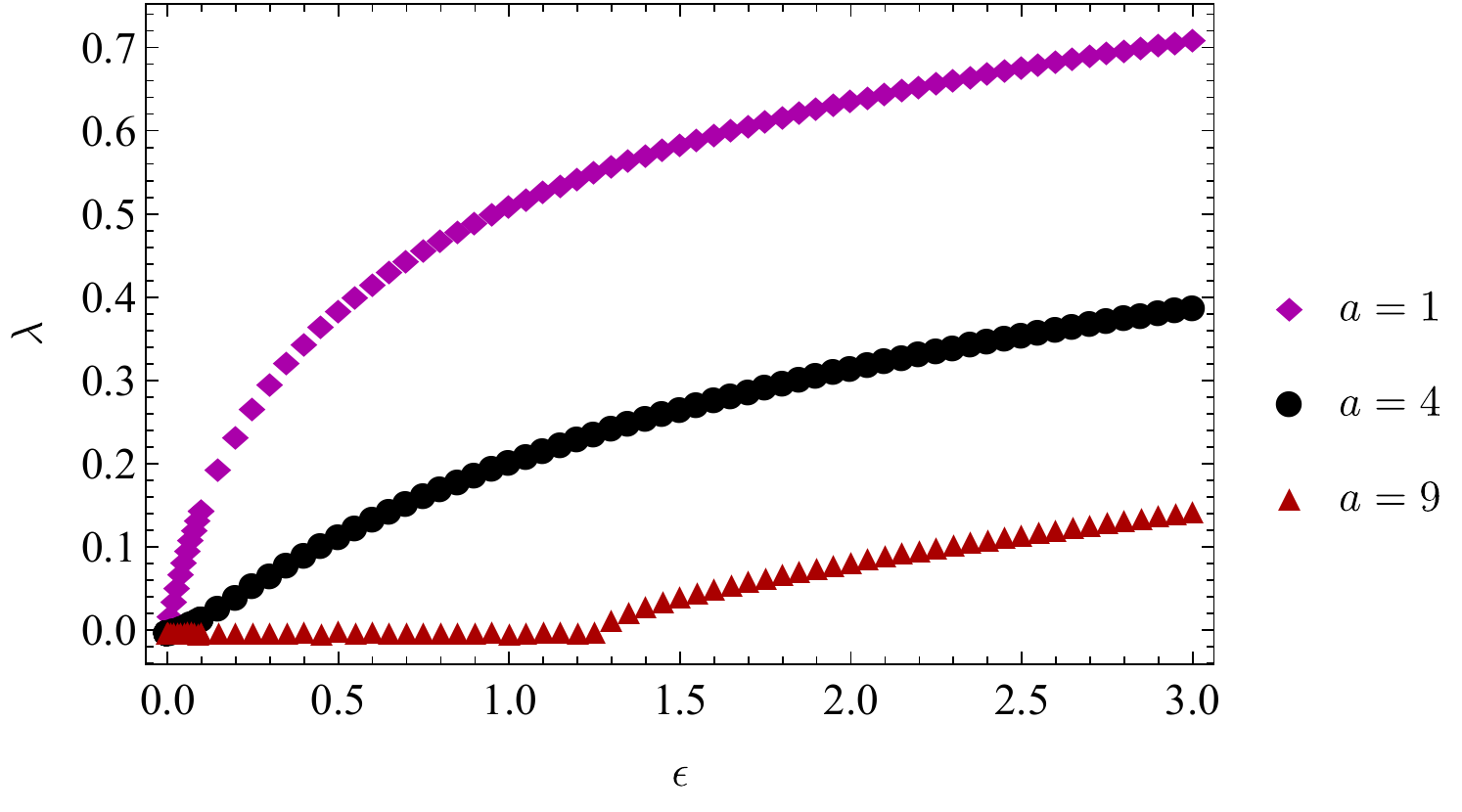}
\caption{\fontsize{9}{12}
Instability rates $\lambda$ for the first few\newline values of  $n$ in $a=n^2$ for \eqref{eq:HillSC} and $\vartheta=2$.
\label{fig:p2rate}}
\end{minipage}%
\hspace*{\fill}
\end{figure}


\begin{framed}
\noindent
{\small {\it Numerical results for the instability of the Maxwell sector coupled to the scalars}\\ ~\label{subsec:scalar_flat_numerics}

Parametric resonances where analysed in Appendix \ref{AppParRes}. In particular in Appendix \ref{sec:Math_stability} numerical estimates of the instability rates where done together with the stability diagram. Here we will perform analogous calculation. Above analysis shows that low-coupling behaviour is similar to the Mathieu equation. On Figs. \ref{fig:p1rate}, \ref{fig:p2rate} we showed rates for several values of $a=(p/\mu_{\rm s})^2$ that correspond to the instabilities. We define $\epsilon=1/2(k_{\rm s}\Phi_0/2)^\vartheta$. Note first that for $\vartheta=2$ instabilities will happen when $a=n^2$. This can be understood from the low-coupling limit as $\vartheta=2$ equation behaves as Mathieu equation with the rescaled time \eqref{eq:p2_low_k_Mathieu}. Secondly, range of $\epsilon$ for the $\vartheta=1$ is small with compared to the regular Mathieu equation and the $\vartheta=2$. As mentioned beneath \eqref{eq:g_vartheta_sc}, for strong coupling constants the equation becomes divergent. Numerical results indicate that around $\epsilon \sim 0.3$ rates develop plateau but it is not clear whether this is a numerical artefact.

On Figs. \ref{fig:p1rate}, \ref{fig:p2rate}  we show the stability diagrams. Interestingly, $\vartheta=2$ doesn't develop instabilities for $a=0$ and in comparison to the Mathieu equation stability diagram (Fig. \ref{fig:Mathieu_stability}) stability zones are pronounced.

}

\end{framed}


\begin{figure}
\hspace*{\fill}%
\begin{minipage}[c]{0.47\textwidth}
\centering
\vspace{0pt}
\includegraphics[width=1\columnwidth]{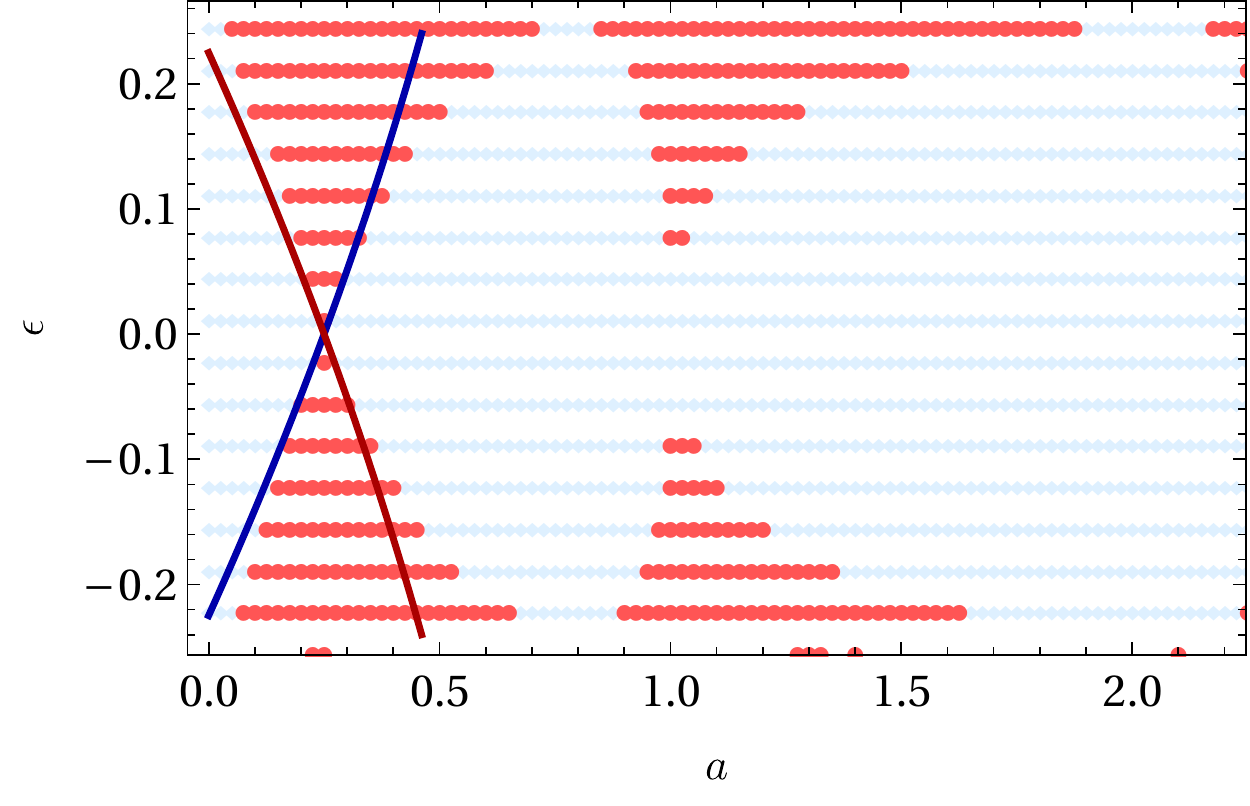}
\caption{\fontsize{9}{12}
Stability diagram for \eqref{eq:HillSC} and $\vartheta=1$. Blue dots correspond to stable solutions, while the red dots correspond to the unstable solutions. Black and magenta curves, given by \eqref{eq:Mathieu_zones_border}, correspond to analytical estimates of the borders between (in)stability zones and give good description for small $\epsilon$ when the equation is approximated by the Mathieu equation.
\label{fig:p1diagram}}
\end{minipage}%
\hfill
\begin{minipage}[c]{0.47\textwidth}
\vspace{0pt}
\centering
\includegraphics[width=1\columnwidth]{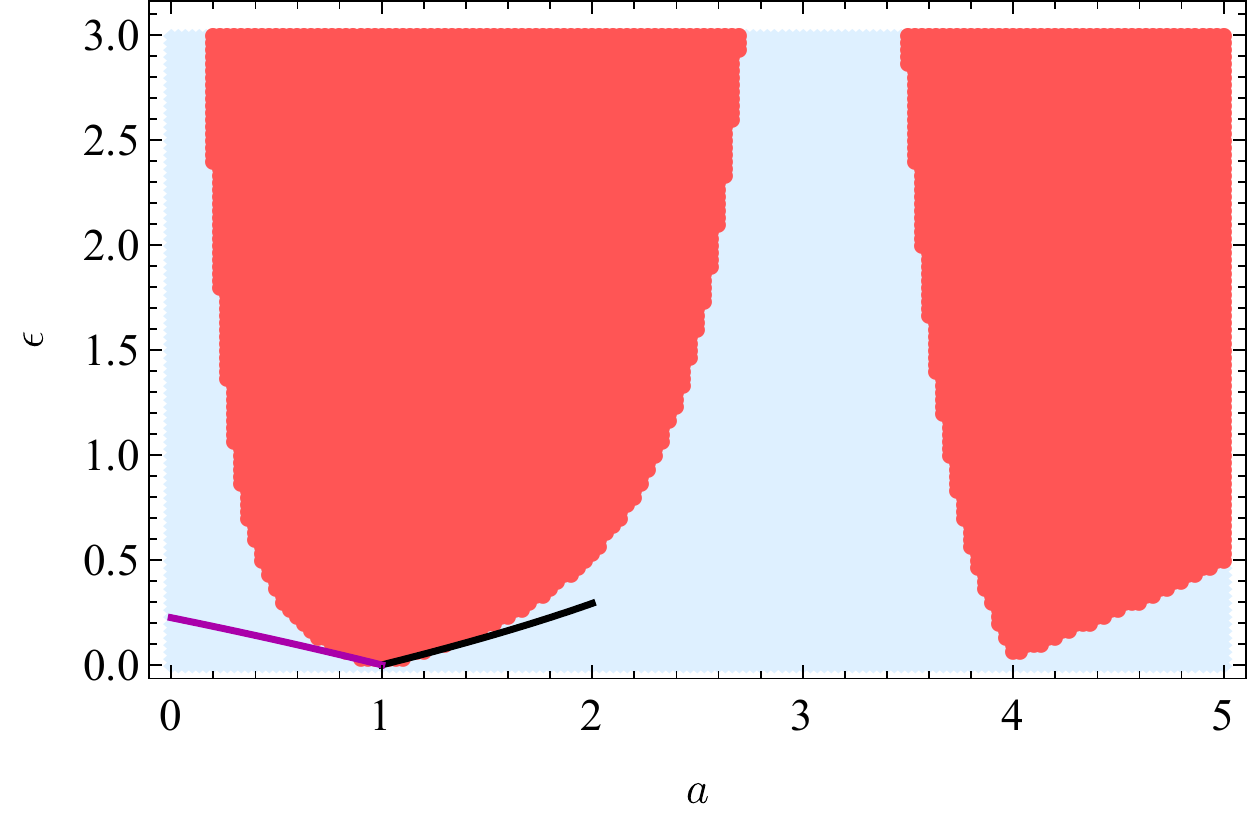}
\caption{\fontsize{9}{12}
Stability diagram for \eqref{eq:HillSC} and $\vartheta=2$. Blue dots correspond to stable solutions, while the red dots correspond to the unstable solutions. Black and magenta curves, given by \eqref{eq:Mathieu_zones_border}, correspond to analytical estimates of the borders between (in)stability zones and give good description for small $\epsilon$ when the equation is approximated by the Mathieu equation. Note that in \eqref{eq:Mathieu_zones_border} $a \to a/4$ as the time needs to be rescaled in order to obtain the Mathieu equation.
\label{fig:p2diagram}}
\end{minipage}%
\hspace*{\fill}
\end{figure}

\newpage

\begin{figure}
\hspace*{\fill}%
\begin{minipage}[c]{0.48\textwidth}
\centering
\vspace{0pt}
\includegraphics[width=1\columnwidth]{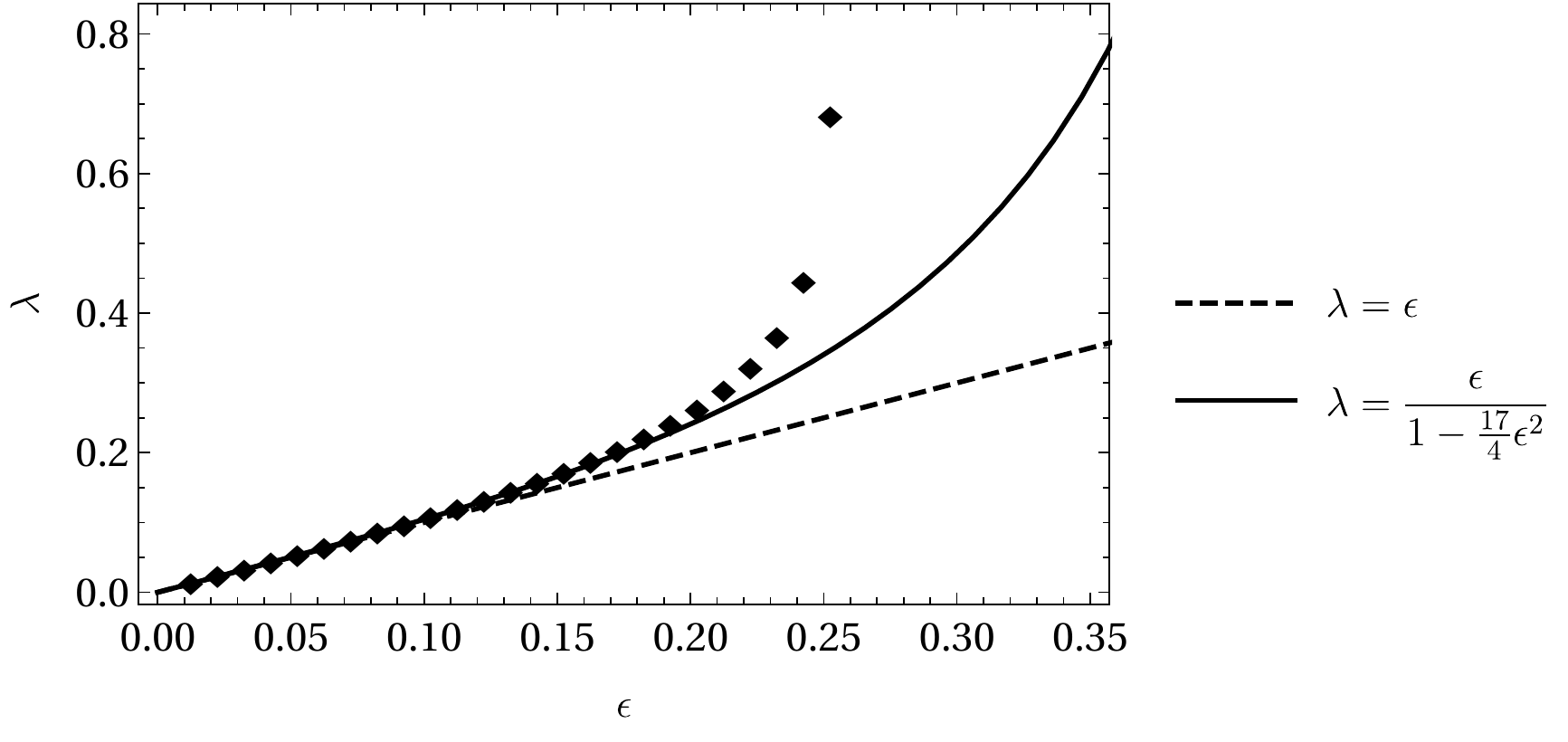}
\caption{\fontsize{9}{12}
Analytical estimates to the first and second order for the instability rates for \eqref{eq:HillSC} and $\vartheta=2$ with $a=1/4$.
\label{fig:p1rate_analytical}}
\end{minipage}%
\hfill
\begin{minipage}[c]{0.48\textwidth}
\vspace{0pt}
\centering
\includegraphics[width=1\columnwidth]{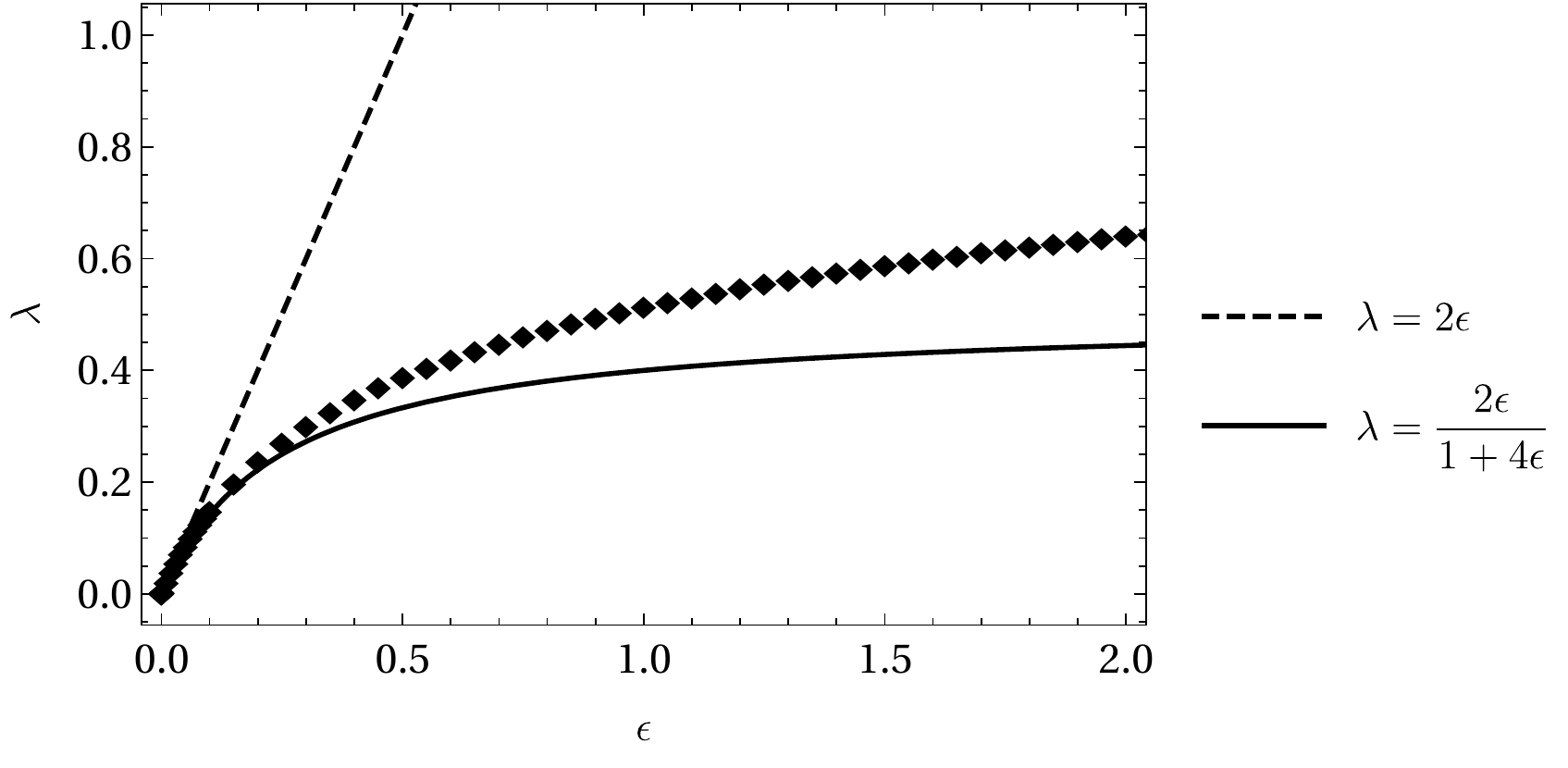}
\caption{\fontsize{9}{12}
Analytical estimates to the first and second order for the instability rates for \eqref{eq:HillSC} and $\vartheta=2$ with $a=1$ (in original notation from Box on page \pageref{subsec:scalar_flat_numerics}).
\label{fig:p2rate_analytical}}
\end{minipage}%
\hspace*{\fill}
\end{figure}

\begin{framed}
\noindent
{\small {\it Analytic estimate of the instability timescale for scalar coupling}\\ ~\label{subsec:scalar_rate}

Here we follow Appendix \ref{app:RG_time_scales_1st} to perform analytical estimates of the instability rates. To lowest order in $k_{\rm s}$ Eq. \eqref{eq:HillSC} reduces to Mathieu equation so that the analysis in Appendix \ref{app:RG_time_scales_1st} is completely applicable. For $\vartheta=2$ we rescale time as $T=2\mu_{\rm s}t$ so that $a/4 \to a=p^2/(2\mu_{\rm s})^2$. We will here provide higher order contribution to the rate estimate.

As will become clear later, we will first consider $\vartheta=2$ case. Equation for the second order correction is
\beq
\partial^2_T \mathcal{X}_{1/2}^{(2)}+\Big(\frac{1}{2} \Big)^2 \mathcal{X}_{1/2}^{(2)} &=& - (2 \cos{T})\mathcal{X}_{1/2}^{(1)} \\ \nonumber
&& -W_2(T;k^{4}_{\rm s})\mathcal{X}_{1/2}^{(0)}+\mathcal{O}(k^6_{\rm s} )\,,
\eeq
where $W_2(T;k^{4}_{\rm s})$ is the Taylor expansion coefficient of the function in Eq. \eqref{eq:Wp_scalar} at the order of $k^{4}_{\rm s}$. We use $\mathcal{X}_{1/2}$ label for the higher order $\vartheta=2$ corrections with $\mathcal{X}^{(1)}_{1/2} \equiv x^{(1)}_{1/2}$ and $\mathcal{X}^{(0)}_{1/2} \equiv x^{(0)}_{1/2}$. For $\vartheta=1$ we use $J_{1/2}$ mutatis mutandis. Performing DRG as in the Appendix \ref{app:RG_time_scales_1st} we obtain $\lambda^{(R,2)}_{\vartheta=2}=2\epsilon \sqrt{(1-3\epsilon) (1-5\epsilon)}$, with $\epsilon=(1/4) k_{\rm s}\Phi_0$. After $(1,1)$ Pad\'e resummation we have
\beq
\lambda_{\vartheta=2}=\frac{2\mu_{\rm s} \epsilon}{ (4\epsilon+1)} \,.
\eeq

For $\vartheta=1$ case, second order correction is governed by the equation
\beq
\partial^2_T J_{1/2}^{(2)}+\Big(\frac{1}{2} \Big)^2 J_{1/2}^{(2)} &=& - (2 \cos{T})J_{1/2}^{(1)} \\ \nonumber
&& -W_1(T;k^{2}_{\rm s})J_{1/2}^{(0)}+\mathcal{O}(k^3_{\rm s} )\,.
\eeq
For the rate estimate we obtain the same results as for the axion case \eqref{}. This result is clearly not a good description as the numerical results indicate (Fig. \ref{fig:p1rate}) that the function $\lambda_{\vartheta=1}(\eps)$ is divergent. Therefore, we go to the third-order contribution
\beq
\partial^2_T J_{1/2}^{(3)}+\Big(\frac{1}{2} \Big)^2 J_{1/2}^{(3)} &=& - (2 \cos{T})J_{1/2}^{(2)} \\ \nonumber
&& -W_1(T;k^{2}_{\rm s})J_{1/2}^{(1)}-W_1(T;k^{3}_{\rm s})J_{1/2}^{(0)}\\ \nonumber
&& +\mathcal{O}(k^4_{\rm s} )\,.
\eeq
Renormalized rate is $\lambda_{\vartheta=1}^{(R,3)}=\eps \sqrt{1+(17/2) \eps^2+(16/3) \eps^3+(2225/144)\eps^4}$, with $\epsilon=(1/2) \left(k_{\rm s}\Phi_0/2\right)^2$. After Pad\'e resummation at the order $(1,2)$ we obtain
\beq
\lambda_{\vartheta=1}=\frac{\mu_{\rm s} \eps}{1-\frac{17}{4}\eps^2}\,.
\eeq

Comparison between the analytic estimates and the numerical results are shown on Figs. \ref{fig:p1rate_analytical}, \ref{fig:p2rate_analytical}.
}

\end{framed}


\subsubsection{Inhomogeneous configuration}

%
\begin{figure*}[htb]
\begin{tabular}{cc}
\includegraphics[width=8.5cm]{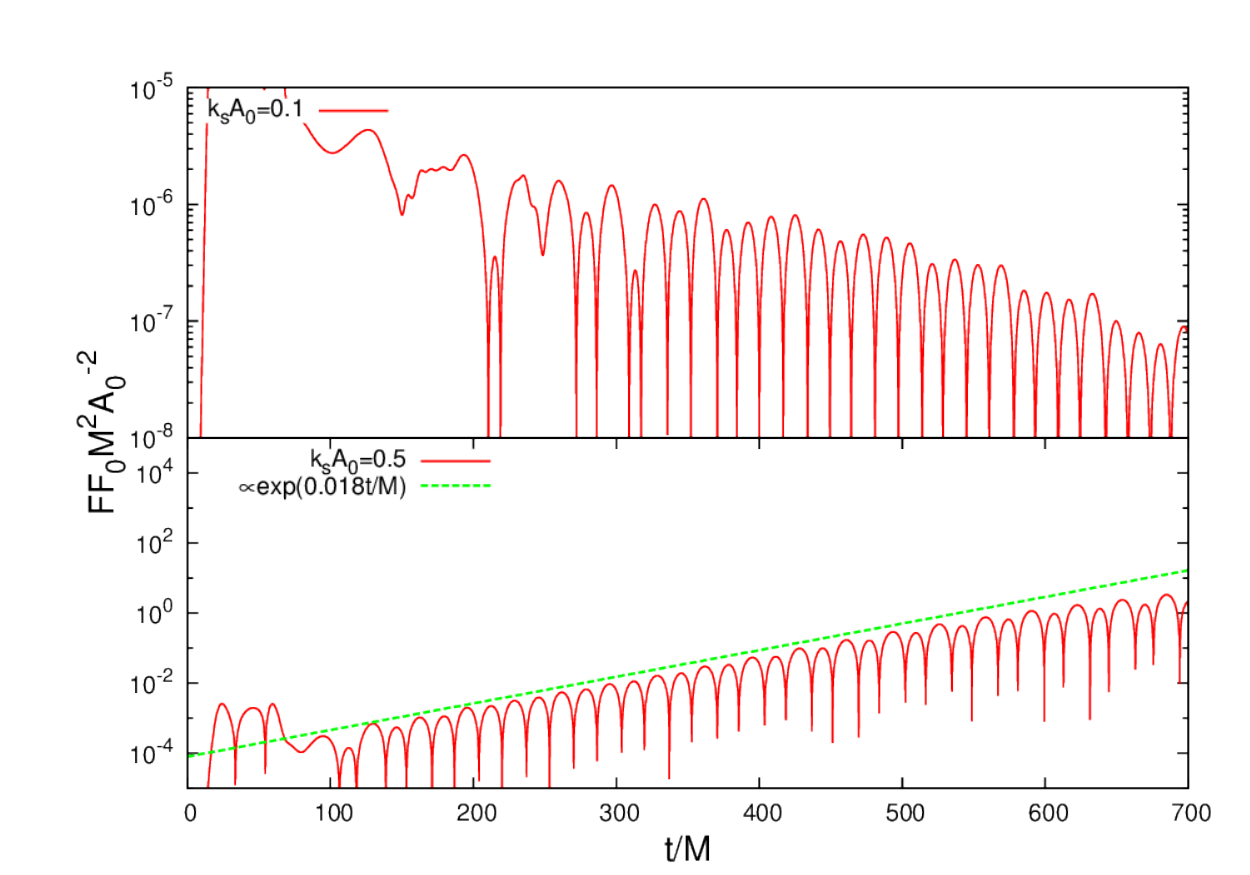}
&\includegraphics[width=8.5cm]{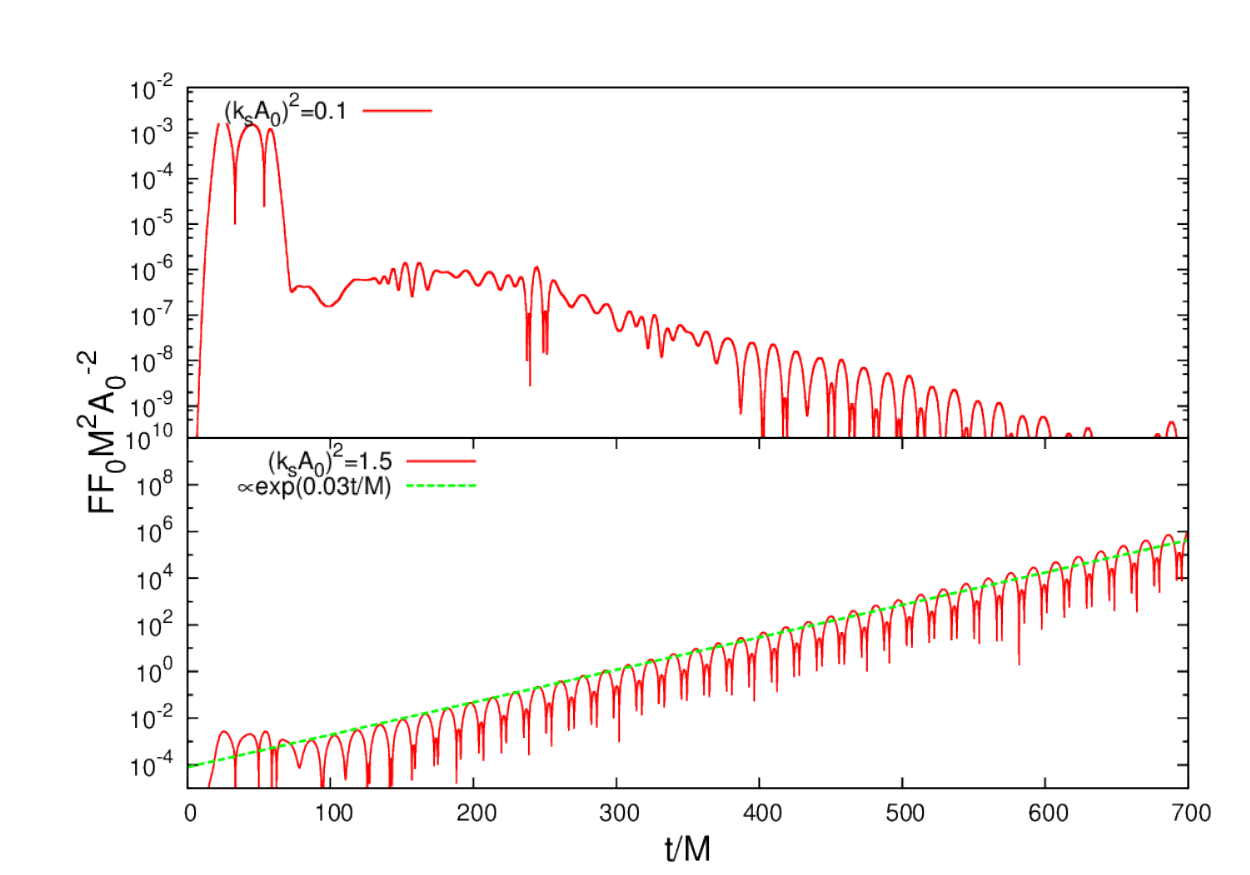}
\end{tabular}
\caption{
Time evolution of the Maxwell scalar $FF_{0}$ (see Box on page~\pageref{subsec:numset}  and on page~\pageref{subsec:numsetSc} for definition of notation) measured
at $r=20M$ for the extended initial-data profiles [Eq.~\eqref{ID_extended}], for $\vartheta=1,2$ (left and right panel respectively). The spacetime is flat and the scalar is not evolved. The initial data corresponds to a
Gaussian EM field with width $w$, gaussian-centered radius $r_0$ and amplitude of $(w,r_{0},E_{0})=(5M,40M,0.001)$. In both panels the mass coupling is $\mu_{\rm s} M=0.2$. Figure credits: \cite{Boskovic:2018lkj}.
\label{graph_fixed_scalar_sample_log_p}}
\end{figure*}
\begin{figure*}[htb]
\begin{tabular}{cc}
\includegraphics[width=8.5cm]{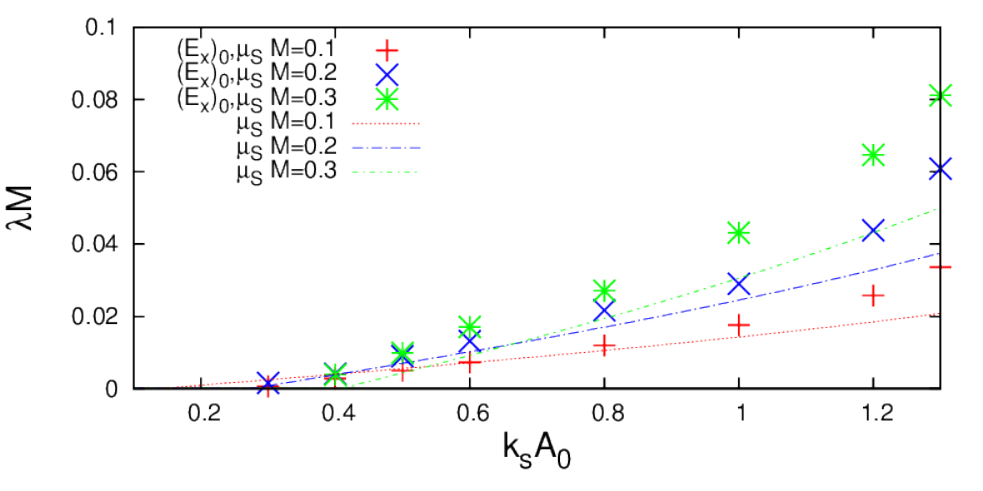}
&\includegraphics[width=8.5cm]{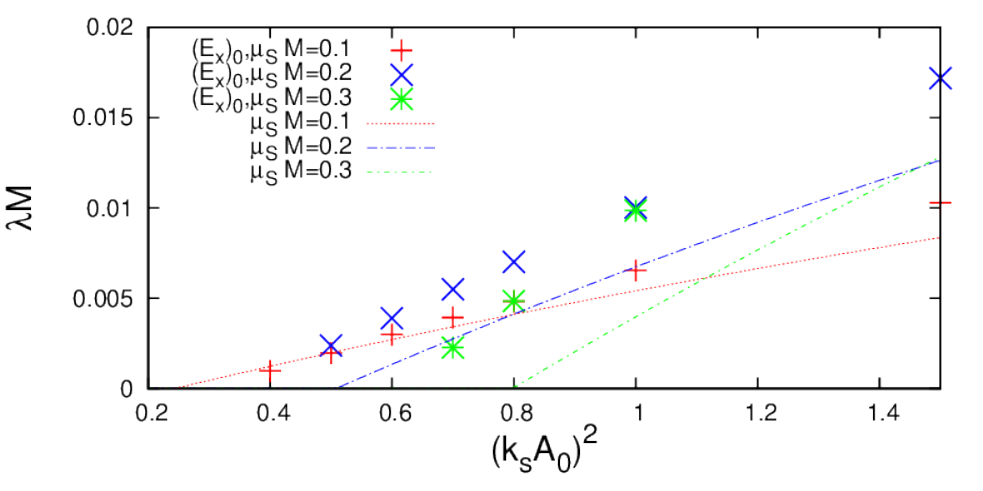}
\end{tabular}
\caption{Growth rates $M\lambda$ as a function of the coupling parameters for $\vartheta=1$ (left) and $\vartheta=2$ (right) for a Minkowski background. Crosses stand for numerically extracted rates, dashed lines are our analytical estimates, which to first order in the coupling are described by Eqs.~\eqref{ana_rate1}-\eqref{ana_rate2} (a full description of the perturbative framework can be found in Box on page \pageref{subsec:scalar_rate}). Our results are consistent with the existence of a critical coupling below which no instability is triggered, and well described by our analytical estimates in the small coupling regime. Figure credits: \cite{Boskovic:2018lkj}.
\label{graph_lambda_p}}
\end{figure*}

The estimates in Section \ref{sec:mink_axion_inhom_an} were worked out for the axionic coupling in the context of inhomogeneous backgrounds, but the underlying physics and mechanism remains the same for scalar-type couplings. Accordingly, we expect that the rate of the dominant instability is given by Eq. \eqref{eq:inst_rate3}. Using Eq.~\eqref{eq:rate_scalar_p_1}, we find,
\begin{equation}
\lambda_\ast[\langle \Phi \rangle]_{\vartheta=1} \approx \frac{1}{4}k_{\rm s}\langle \Phi \rangle\mu_{\rm s}\,,\label{ana_rate1}
\end{equation}
for $\vartheta=1$. Similarly, using Eq.~\eqref{eq:rate_scalar_p_2} for $\vartheta=2$,
\begin{equation}
\lambda_\ast[\langle \Phi \rangle]_{\vartheta=2}  \approx \frac{1}{4}\left(k_{\rm s}\langle \Phi \rangle\right)^2\mu_{\rm s}\,.\label{ana_rate2}
\end{equation}

With regards to the numerical results, in \cite{Boskovic:2018lkj} the EM field was evolved in the background of ``frozen'' scalar cloud for $\vartheta=1, 2$, described by \eqref{eq:211psi}, and on the Miknowski background. Estimates for $\langle \Phi \rangle$ and $d$ are given by \eqref{eq:sc_cloud_av_field} and \eqref{eq:sc_cloud_size}, respectively. Initial data is discussed in Box on page \pageref{subsec:numsetSc}, while the technical details of the numerical setup can be found in \cite{Boskovic:2018lkj}. Scalar mass was varied $\mu_{\rm s} M=0.1,0.2,0.3$, where $M$ is the BH mass that supports the solution~\eqref{eq:211psi} (in this flat analysis this is just the parameter of the scalar profile). Results are summarized in Figs.~\ref{graph_fixed_scalar_sample_log_p}~--~\ref{graph_lambda_p}, and are consistent with the results we obtained for axion couplings ~\cite{Ikeda:2019fvj} (see also Section \ref{sec:mink_axion_inhom}).

As in the axion coupling case, we find the existence of a critical coupling $k_{\rm s} A_0$ below which no instability occurs. This is apparent in Fig.~\ref{graph_fixed_scalar_sample_log_p} (upper panels) for both $\vartheta=1$ and $\vartheta=2$. At large enough couplings, all initial conditions lead eventually to an instability (and exponential growth of the EM field), examples are shown in the bottom panels of Fig.~\ref{graph_fixed_scalar_sample_log_p}. Rate estimates \eqref{ana_rate1} and \eqref{ana_rate2} are shown together with numerical data in Figs.~\ref{graph_lambda_p}.  Notice how such a simple estimate agrees very well with the full numerical evolution
in the small coupling regime where the perturbative approximation is valid for $\vartheta=1$. For $\vartheta=2$ adequacy of analytical estimates is very rough but still captures the qualitative picture.
The growth rate depends very weakly on the initial data and on the coordinate at which the EM field is extracted.

\begin{framed}
\noindent
{\small {\it Numerical setup (scalar coupling): Initial data}\\ ~\label{subsec:numsetSc}

Here we describe scalar initial data as a complement to Box on page ~\pageref{subsec:numsetSc}. Solution of the constraint equation leads to the profile \cite{Boskovic:2018lkj}
\begin{eqnarray}\label{Eq.Initial data of Er}
E^{r}&=&E^{\theta}=\mathcal{A}_{i}=0\,,\label{Eq.Initial data of Er_v2}\\
E^{\varphi}&=&\frac{F(r,\theta)}{1+k_{\rm s}^{\vartheta}\Phi^{\vartheta}}\,,\label{Eq.Initial data of Etheta}
\end{eqnarray}
where $F(r,\theta)$ is an arbitrary function of $r$ and $\theta$.
In \cite{Boskovic:2018lkj} the function
\begin{eqnarray}\label{Eq.F profile}
F(r,\theta)&=&E_{0}e^{-\left(\frac{r-r_{0}}{w}\right)^{2}}\Theta(\theta)\,,
\end{eqnarray}
was used, where $E_{0}$, $r_{0}$ and $w$ are constants, which characterize the strength, the radius, and the width of the Gaussian profile of the electric field, respectively. $\Theta(\theta)$ profiles are the same as in the Box on page ~\pageref{subsec:numsetSc}. These profiles were used in both Minkowski and Kerr scenarios.

Analysis tools used for scalar coupling numerical results (in both Minkowski and Kerr case) are the same as the ones described in Box on page ~\pageref{subsec:numset}.

}

\end{framed}

\subsubsection{Interaction with plasma} \label{app:scalar_plasma}

As in the previous Sections we modify discussion from Section \ref{sec:Mink_plasma} where appropriate and for scalar couplings we find  critical plasma frequencies
\begin{eqnarray}
\omega^{\rm crit}_{\rm plasma}&=&\mu_{\rm s}\sqrt{\frac{1}{4} + k_{\rm s}\Phi_0}\,,\quad \vartheta=1\,,\\
&=&\mu_{\rm s}\sqrt{1 +\frac{1}{2} (k_{\rm s}\Phi_0)^2}\,,\quad \vartheta=2\,.
\end{eqnarray}
%

\subsection{Scalar clouds} \label{sec:Kerr_scalar}

\begin{figure*}[htb]
\begin{tabular}{cc}
\includegraphics[width=8.5cm]{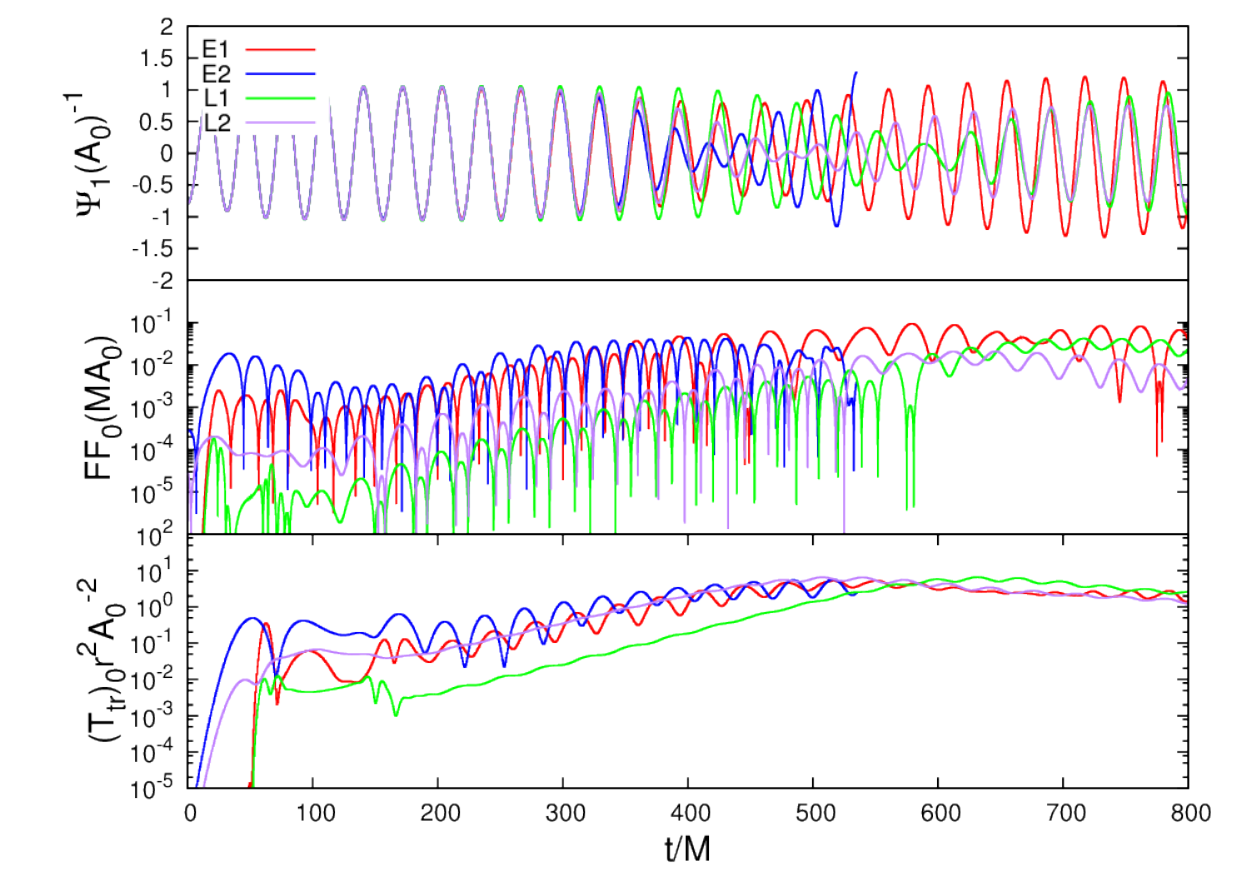}&
\includegraphics[width=8.5cm]{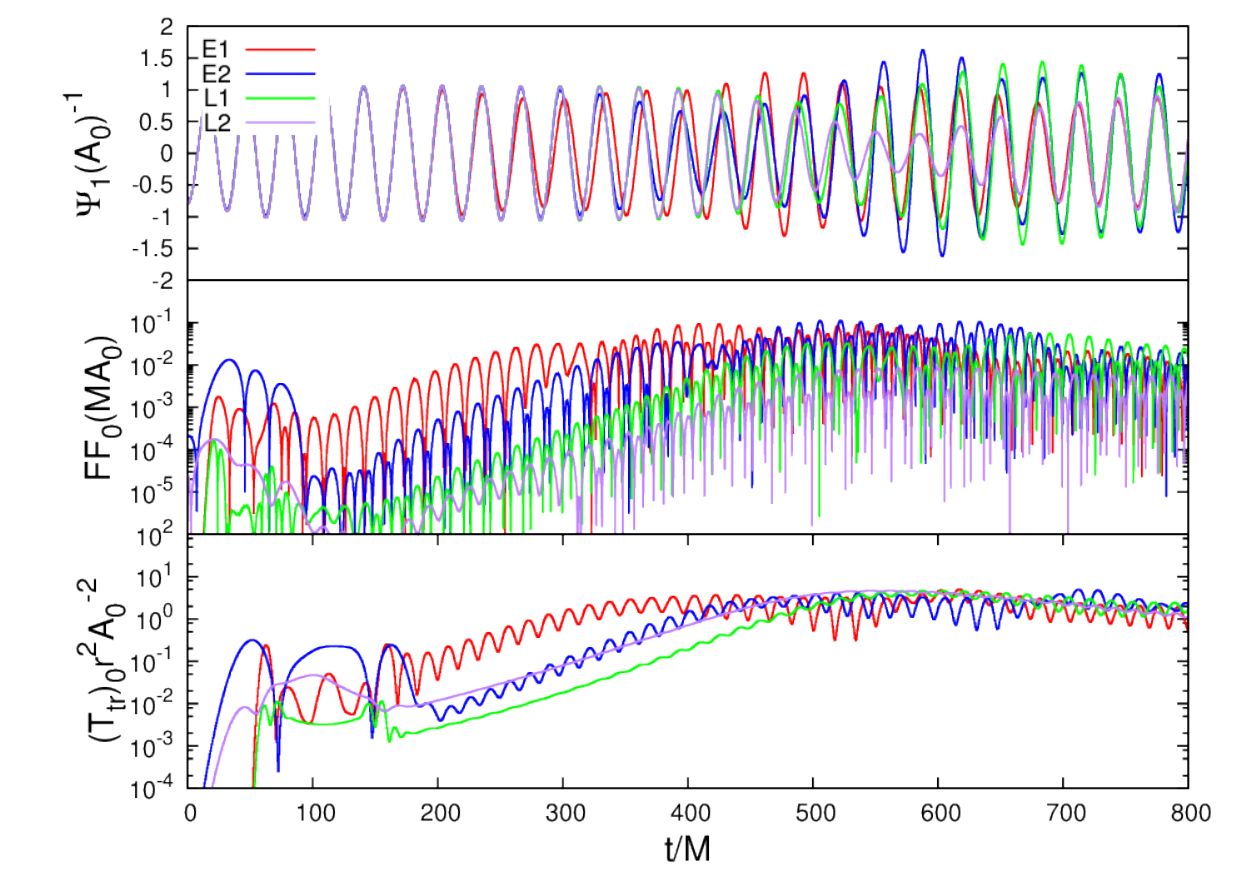}
\end{tabular}
\caption{
Time evolution of the dipolar component of the scalar field, $\Phi_{1}$ (top panel, extracted at $r=20M$, see Box on page~\pageref{subsec:numset}  and on page~\pageref{subsec:numsetSc} for definition of notation; note that the scalar field $\Phi$ is labeled as $\Psi$ on the Figure), of the monopolar component of the Maxwell invariant $FF_{0}$ (middle panel, extracted at $r=20M$), and $(T_{tr})_{0}$ (bottom panel, extracted at $r=100M$) for $\vartheta=1$ (left panels) and $\vartheta=2$ (right panels). The mass of the scalar is $\mu_{\rm s} M=0.2$, the coupling $k_{\rm s}A_{0}=0.5,k_{\rm s}^2A_{0}^2=1.0$ ($\vartheta=1,2$ respectively), and the spin parameter is $a=0.5M$. The initial data is either of the extended (E) or localized (L) type as defined in Eqs.~\eqref{Eq.Initial data of Er}, \eqref{Eq.Initial data of Etheta}, \eqref{Eq.F profile}, \eqref{ID_extended} and \eqref{ID_localized}, and described by a Gaussian centered at $r_0=40M$ and an amplitude of $E_0=10^{-3}$. For $E_1,\,L_1$ the gaussian width is $5M$, for $E_2,\,L_2$ it is $20M$. Figure credits: \cite{Boskovic:2018lkj}.
\label{graph_a05_mu02_ks05_p1_phi_FF_Sr}
}
\end{figure*}

Similarly to the Minkowski case, for the scalar clouds discussion largely parallels one in Section \ref{sec:ax_ph_clouds}. In \cite{Boskovic:2018lkj}
evolution equations were solved for $M\mu_{\rm s}=0.2, a=0.5M$ (we also studied higher spins, the results are qualitatively the same),
the results are summarized in Fig.~\ref{graph_a05_mu02_ks05_p1_phi_FF_Sr} for scalar couplings with $\vartheta=1$ and $\vartheta=2$. As expected from the previous flat-space analysis, for small enough couplings any small EM disturbance dissipates away, and the profile of the scalar cloud is basically undisturbed. On the other hand, when the coupling is larger than a threshold, the EM field grows exponentially. As shown in Fig.~\ref{graph_a05_mu02_ks05_p1_phi_FF_Sr}, for large couplings an instability is indeed triggered. Because the instability acts to produce $p\sim \mu_{\rm s}/2$ vector fluctuations (for $\vartheta=1$), at the nonlinear level these backreact on the scalar field, producing transient clumps of scalar field on these scales. This translates into an increase of the scalar, when observed sufficiently close to the BH, as seen in the upper panels of Fig.~\ref{graph_a05_mu02_ks05_p1_phi_FF_Sr}. On long timescales, the instability extracts energy from the scalar cloud and eventually lowers the effective coupling to sub-threshold values, leading to a now stable cloud. On even longer timescales, superradiance will grow the scalar to super-threshold values and the cycle begins again, as argued in the axion scenario in Ref.~\cite{Ikeda:2019fvj} (see the discussion in Section \ref{sec:blast_leak}).  

We would like to highlight a potential issue with the scalar couplings in general, and that clearly shows up when $\vartheta=1$.
When the effective coupling $k_{\rm s}\Phi$ is of order unity, the kinetic term [left hand side of Eq.~\eqref{eq:MFEoMVectorSC}] for the vector field can vanish and the system becomes strongly coupled. The evolution in such case is ill-defined. In particular, we find for example that we cannot evolve $E2$ (see definition in caption of Fig.~\ref{graph_a05_mu02_ks05_p1_phi_FF_Sr}) in Fig.~\ref{graph_a05_mu02_ks05_p1_phi_FF_Sr} past $t=520M$, for this reason.
It is possible that the dynamics of the gravity sector (neglected in \cite{Boskovic:2018lkj}) cure such anomalies,
for example by producing BHs close to the threshold. Another possibility is that coupling to fermions will ensure that Schwinger-type creation works to prevent the EM field to ever approach such large values. The calculation of the time evolution near the strong coupling was beyond the purpose of \cite{Boskovic:2018lkj} and was left for future work.

\begin{figure}
\centering
\includegraphics[width=8.5cm]{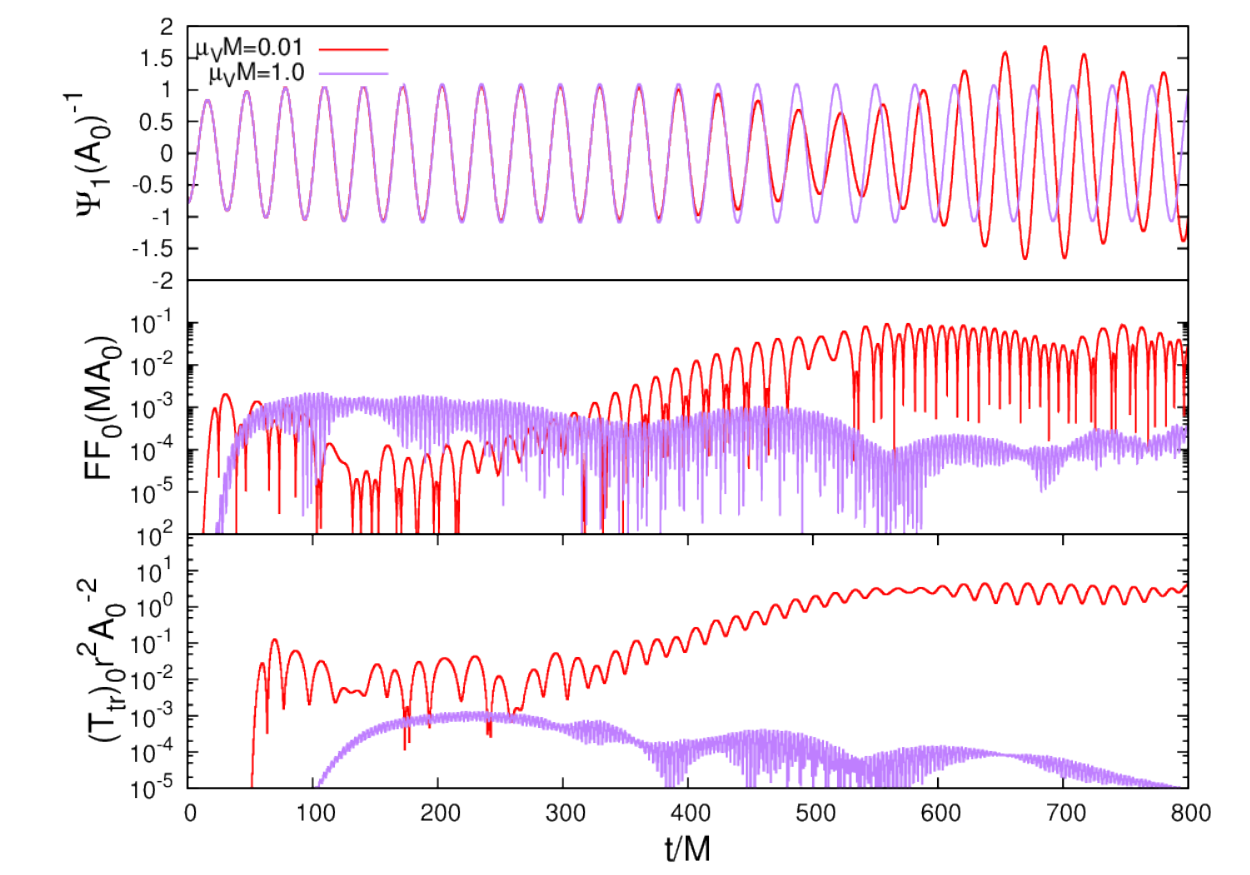}
\caption{
Time evolution of the massive scalar -- massive vector field system around a Kerr BH with $\vartheta=2$ and coupling $k_{\rm s}^{2}A_{0}^{2}=1.0$, in which the initial data is an extended profile with $(r_{0},w,E_{0})=(40M, 5M,0.001)$. Here the scalar mass is $\mu_{\rm s}M=0.2$ and the scalar cloud is evolving around a Kerr BH with $a=0.5M$. The notation for the $y-$axis is explained in Box on page \pageref{subsec:numsetSc}. Figure credits: \cite{Boskovic:2018lkj}.\label{graph_Proca_a05_mu02_ks05_p2_phi_FF_Sr_E1}
}
\end{figure}

Finally, the presence of plasma was modeled as a scalar-Proca system (see Section \ref{sec:plasma_axion_Kerr}). The time domain study is summarized in Fig.~\ref{graph_Proca_a05_mu02_ks05_p2_phi_FF_Sr_E1}, and is consonant with the flat space conclusion.

\newpage


\clearpage


\addcontentsline{toc}{section}{References}

\bibliographystyle{utphys}
\bibliography{references.bib}

\end{document}